\documentclass[12pt]{ucsddissertation}

\usepackage[NoDate]{currvita}
\usepackage{scrextend}
\usepackage{pslatex}
\usepackage{graphicx}
\usepackage{hyperref}
\hypersetup{hidelinks}

\usepackage{microtype}                 
\PassOptionsToPackage{warn}{textcomp}  
\usepackage{textcomp}                  
\usepackage{mathptmx}                  
\usepackage{times}                     
\usepackage{cite}                      
\usepackage{graphics}
\usepackage{comment}
\usepackage{enumitem}
\usepackage{xspace}
\usepackage{color, soul}
\usepackage{amsmath}
\usepackage[title]{appendix}
\usepackage{todonotes}

\usepackage[numbers]{natbib}  

\usepackage[many]{tcolorbox}
\usepackage{xcolor}
\definecolor{main}{HTML}{5989cf}
\definecolor{sub}{HTML}{cde4ff}
\newtcolorbox{boxH}{
    colback = sub,
    colframe = main,
    boxrule = 0pt,
    leftrule = 6pt 
}

\usepackage{amsfonts}
\usepackage{fixltx2e}
\usepackage[flushleft]{threeparttable}
\usepackage{bm}
\usepackage{mathtools}
\usepackage{algorithm}
\usepackage{algorithmic}
\usepackage{makecell}
\usepackage{tabularx,booktabs,ragged2e}
\newcommand\algorithmicprocedure{\textbf{function}}
\newcommand{\algorithmicendprocedure}{\algorithmicend\ \algorithmicprocedure}

\makeatletter
\newcommand\PROCEDURE[3][default]{%
  \ALC@it
  \algorithmicprocedure\ \textsc{#2}(#3)%
  \ALC@com{#1}%
  \begin{ALC@prc}%
}
\newcommand\ENDPROCEDURE{%
  \end{ALC@prc}%
  \ifthenelse{\boolean{ALC@noend}}{}{%
    \ALC@it\algorithmicendprocedure
  }%
}
\newenvironment{ALC@prc}{\begin{ALC@g}}{\end{ALC@g}}
\makeatother

\def\insitu{\textit{in situ}\xspace}
\def\ie{\textit{i.e.,}\xspace}
\def\etal{\textit{et al.}\xspace}
\def\etc{\textit{etc.}\xspace}
\def\eg{\textit{e.g.,}\xspace}
\def\cf{\textit{c.f.}\xspace}

\def\aka{\textit{a.k.a.}\xspace}
\def\vs{\textit{vs.}\xspace}
\def\incl{\textit{incl.}}
\def\first{\textit{first}\xspace}

\def\second{\textit{second}\xspace}

\def\after{\textit{after}\xspace}
\def\before{\textit{before}\xspace}

\DeclareMathOperator*{\argmin}{argmin}

\makeatletter
\gdef\@ptsize{2}
\let\@currsize\normalsize
\makeatother
\usepackage{setspace}
\doublespace
\usepackage[font=small, width=0.9\textwidth]{caption}
\usepackage[T1]{fontenc}
\usepackage{mathptmx}

\usepackage{microtype}  

\usepackage{footnote}
\makesavenoteenv{tabular}
\makesavenoteenv{table}

\usepackage{rotating}  
\usepackage{array}  
\usepackage{booktabs}  

\title{Seamless and Efficient Interactions within a Mixed-Dimensional Information Space}

\author{Chen Chen}
\degree{Computer Science}{Doctor of Philosophy}

\chair{Nadir Weibel}
\committee{William G. Griswold}
\committee{James D. Hollan}
\committee{Cuong Nguyen}
\committee{Haijun Xia}
\committee{Yang Zhang}

\degreeyear{2025}

\begin{document}

\frontmatter
\maketitle
\makecopyright
\makesignature

\begin{dedication}
\setsinglespacing
\raggedright 
\parindent0pt\parskip\baselineskip
    
    To my parents, for taking the risk and supporting me study, research and work abroad --- you are an inspiration to all.
    
    To my friends, colleagues, and mentors, thank you for being a source of unwavering support and encouragement, both during times of triumph and adversity.

\end{dedication}

%
%
\begin{epigraph}
\vskip0pt plus.5fil
\setsinglespacing
\vfil
\noindent ``There is unexpected traffic ahead, but you are still on the right path.''
\vskip\baselineskip
\hskip0pt plus1fil\textit{Google Maps}\hskip0pt plus4fil\null
\vfil

\end{epigraph}

\tableofcontents
\listoffigures
\listoftables

\begin{acknowledgements}

I would like to thank my research advisor, Prof. Nadir Weibel, for his invaluable guidance and support.
I am grateful to my thesis committee members: Prof. James D. Hollan, Prof. William G. Griswold, and Prof. Haijun Xia from the University of California San Diego; Prof. Yang Zhang from the University of California Los Angeles; and Dr. Cuong Nguyen from Adobe Research.
I also want thank my amazing fellow colleagues at the \textbf{H}uman-centered e\textbf{X}tended \textbf{I}ntelligence (HXI) lab: Prof. Matin Yarmand, Manas S. Bedmutha, Weichen Liu, Aaron A. Broukhim and Nishanth Chidambaram; and former colleagues:  Ru Wang, Dr. Janet G. Johnson, Dr. Danilo Gasques, Dr. Steven R. Rick, Dr. Menghe Zhang, Dr. Vishwajith (Vish) Ramesh, Dr. Robert Kaufman and Rohan Y. Bhide for the amazing suggestions and support.

During my internship at Microsoft Research - Redmond, I was fortunate to work with amazing researchers, whom I would like to thank to: Dr. Nicolai Marquardt, Dr. Ken Hinckley, Dr. Andy Wilson, Dr. Bala T. Kumaravel, Dr. Hugo Romat, Dr. Dave Brown and Dr. Michel Pahud and Dr. Payod Panda. 
While not directly involved in my project, I am deeply grateful to those who offered valuable guidance and advice for my research career through various conversations: Dr. Kori Inkpen and Dr. Nathalie H. Riche, Dr. Jina Suh, Dr. Steven M. Drucker and Dr. Gonzalo Ramos. Thank you for the invaluable advice and support throughout my incredible MSR internship and career journey.
Besides this, I am grateful for the encouragement and daily conversations with many of my fellow research interns at Microsoft Research and broader Microsoft, who have since become great friends: Xiajie Zhang, Humphery Curtis, Sneha Gathani, Jialu Gao, Ning Zheng, Kyle Yenkai Huang, Mohit Jain, Seonghee Lee, Dr. Tonmoy Dey, Dr. Mohammad Samin Yasar and Dr. Ananya Bhattacharjee.

During my internship at Adobe Research - San Francisco, I had opportunities to work with amazing research mentors, whom I am grateful to: Dr. Cuong Nguyen, Dr. Jane Hoffswell, Dr. Jeniffer Healey, Dr. Trung Bui, Dr. Thibault Groueix, Dr. Voladimir (Vova) G. Kim, Dr. Alexa F. Siu, and Dr. Dingzeyu Li.
I am also grateful for the suggestions and discussion opportunity offered by many wonderful colleagues at Adobe: Dr. Mira Doncheva, Dr. Anh Truong, Dr. Joy O. Kim, Mary Ann (MJ) Jawili, Dr. Stephen DiVerdi, Dr. Zongze Wu and Dr. Zeyu Jin.
The opportunity of interning at Adobe Research leads to two critical works - PaperToPlace~\cite{Chen2023PaperToPlace, Nguyen2024PaperToPlacePatent} and MemoVis~\cite{Chen2024MemoVis} - that contribute to this dissertation.
Thank you for the amazing research and life advice, alongside the wonderful and exciting opportunity!

Besides members of my thesis committee, I thank to my amazing collaborators and mentors at the University of California San Diego: Prof. Xinyu Zhang, Dr. Emilia Farcas, Prof. Michael V. Sherer, Prof. James D. Murphy, Prof. Alison A. Moore, Prof. Michael A. Hogarth, Prof. Ndapa Nakashole, Prof. Edward J. Wang, Prof. Steven P. Dow, Prof. Scott R. Klemmer, Prof. Dinesh Bharadia, Prof. Haojian Jin and Prof. Yiying Zhang, alongside current and former directors of The Design Lab: Prof. Mai T. Nguyen and Prof. Donald A. Norman; fellow Ph.D. student collaborators: Dr. Janet G. Johnson, Dr. Matin Yarmand, Prof. Ke Sun and Dr. Danilo Gasques; undergraduate and master research assistants: Varun Singh, Ella T. Lifset, Kemeberley Charles, Alice Lee, Yichen Han, and Zhuoqun Xu. 
Thank you for all the time, efforts and dedications! 
I am grateful to be part of The Design Lab community - the birthplace of theory of \textbf{U}ser-\textbf{C}entered \textbf{S}ystem \textbf{D}esign (with the same acronym as UCSD), which later involved into what we now know as \textbf{U}ser-\textbf{C}entered \textbf{D}esign (UCD)~\cite{Norman1986}.
While not included in this dissertation, I have gained valuable lessons and hands-on experience through many exciting research opportunities in the field of mobile computing and sensing~\cite{Chen2021ExGSense, Chen2020Captag, Sun2020, Boovaraghavan2023, Agarwal2019virtual}, design of conversational voice user interfaces for older adults~\cite{Chen2023VOLISetup, Han2022, Chen2023VOLI, Charles2021, Chen2021assets, Chen2021, Lifset2020} and postpartum mothers~\cite{Souza2025}, as well as interactive contouring systems in the field of Radiation Oncology~\cite{Yarmand2021, Yarmand2022astro, Yarmand2024iContourStudy, Yarmand2024iContourDesign}.
Beyond my collaborators, I would also like to thank my friends and fellow graduate students at The Design Lab: Dr. Varun K. Vishwanath, Colin Barry, Dr. Jessica de Souza, Lu Sun, Jude A. Rayan, Yujia Liu, Tony Li, Yinan Xuan, Zhiqing Wang, Qisen Yang, Sirui Tao, Jeongeon Park and Poorva S. Bedmutha; and at the Department of Computer Science and Engineering: Dr. Zhi Wang, Dr. Yi Xu, Geelon So, Yixuan Huang, Victor Diniz and Anthony Yang for the constant support and encouragement.

Externally, I appreciate my wonderful faculty mentors, fellow collaborators and friends from Carnegie Mellon University: Prof. Yuvraj Agarwal, Prof. Christopher Harrison, Prof. Anind K. Dey, Prof. Patrick Carrington, Prof. Mayank Goel, Dr. Sudershan Boovaraghavan, Dr. Neeha D. Arun, Kalyani Tembhe, Franklin M. Li, Kunal Bhuwalka, Saksham Chitkara, Dohyun Kim, Anurag Maravi, Shreyas Nagare, Prasoon Patidar and Haozhe Zhou; from the University of California Los Angeles: Dr. Siyou Pei and Alexander Chen; from the University of Wisconsin Madison: Ru Wang. 
Thank you for the continuous support! It was an amazing experience with you on a wide variety of exciting project.

I want to thank my parents for their unwavering financial, emotional, and academic support of my studies, work, and research abroad - over 6,000 miles away - in the United Kingdom and the United States over the past decade. 
It has been an unforgettable journey to live and pursue academic as well as professional growth in five cities across Europe and North America, to work with brilliant mentors and peers, and to immerse myself in diverse cultures and research communities. 
None of this would have been possible without their constant encouragement and belief in me, even from afar.
I am also grateful to my undergraduate research advisor, Prof. Mark Sumner, as well as the incredible mentors Dr. Richard Davies and Dr. Seksak Pholboon and Dr. Eldar Naghiyev at the Power Electronics and Machines Center (PEMC), University of Nottingham.

Finally, I appreciate financial support from various funding, including First-Year Ph.D. Fellowship from Department of Computer Science and Engineering at the University of California San Diego, the Google Faculty Research Award, Adobe Gift Research Grant, the \textbf{N}ational \textbf{S}cience \textbf{F}oundation (NSF) Grant under CNS-1901048, CNS-192767 and CNS-1952942, the \textbf{N}ational \textbf{I}nstitute of \textbf{H}ealth (NIH)/\textbf{N}ational \textbf{I}nstitute on \textbf{A}ging (NIA) under grant R56AG067393, the support from \textbf{A}gency for \textbf{H}ealthcare \textbf{R}esearch and \textbf{Q}uality (AHRQ), as well as various funded teaching assistant and instruction opportunity from the Department of Computer Science and Engineering, the Program of Engineering Professional Master's Degree, and the Department of Cognitive Science at the University of California San Diego.

Chapter~\ref{sec::memovis}, in full, is a reprint of the material as it appears in the journal of ACM \textbf{T}ransactions \textbf{o}f \textbf{C}omputer-\textbf{H}uman \textbf{I}nteraction~(TOCHI), Volume 31, Issue 5, Article 67 (October 2024). The dissertation author was the primary investigator and author of this journal paper. Co-authors includes Cuong Nguyen, Thibault Groueix, Vladimir G. Kim, and Nadir Weibel. 
This work was also presented in the 2024 Annual ACM Symposium on \textbf{U}ser \textbf{I}nterface \textbf{S}oftware and Technology (UIST 2024) conference.
This work was collaborated with Adobe Research.
The previously published manuscript can be found in \cite{Chen2024MemoVis}.

Chapter~\ref{sec::papertoplace}, in full, is a reprint of the material as it appears in the Proceedings of the 36th Annual ACM Symposium on \textbf{U}ser \textbf{I}nterface \textbf{S}oftware and Technology (UIST 2023). The dissertation author was the primary investigator and author of this conference paper. Co-authors includes Cuong Nguyen, Jane Hoffswell, Jennifer Healey, Trung Bui and Nadir Weibel. 
This work was collaborated with Adobe Research.
The previously published manuscript can be referred to \cite{Chen2023PaperToPlace}.
The associated patent can be referred to \cite{Nguyen2024PaperToPlacePatent}.

Chapter~\ref{sec::vrcontour}, in full, is a reprint of the material as it appears in the Proceedings of the 2022 IEEE \textbf{I}nternational \textbf{S}ymposium on \textbf{M}ixed and \textbf{A}ugmented \textbf{R}eality (ISMAR 2022). The dissertation author was the primary investigator and author of this conference paper. Co-authors includes Matin Yarmand, Varun Singh, Mechael V. Sherer, James D. Murphy, Yang Zhang and Nadir Weibel. 
This work was collaborated with the School of Medicine at University of California San Diego and the Department of Electrical and Computer Engineering at University of California Los Angeles.
The previously published manuscript can be referred to \cite{Chen2022VRContour}.
Readers are encouraged to refer to \cite{Chen2022VRContourWIP, Chen2022PrecisionDrawing} for the related publications.

\end{acknowledgements}

\begin{vita} 
\noindent
\begin{cv}{}
    \begin{cvlist}{}
    
      \item[2025] Doctor of Philosophy in Computer Science, University of California San Diego, La Jolla, California, United States
    
      \item[2024] Certificate, Student-Centered College Teaching and Course Design, Teaching + Learning Commons, University of California San Diego, La Jolla, California, United States
    
      \item[2024] Research Intern, Microsoft Research, Redmond, Washington, United States
      
      \item[2023] Research Scientist/Engineer Intern, Adobe Research, San Francisco, California, United States
    
      \item[2023] Candidate of Philosophy in Computer Science, University of California San Diego, La Jolla, California, United States
      
      \item[2022] Research Scientist Intern, Adobe Research, San Francisco, California, United States
      
      \item[2017] Master of Science in Electrical and Computer Engineering, Carnegie Mellon University, Pittsburgh, Pennsylvania, United States
      
      \item[2016] Bachelor of Engineering (\emph{Hons.}, First Class) in Electrical and Electronic Engineering, The University of Nottingham, Nottingham, United Kingdom
    
    \end{cvlist}
\end{cv}

\awards

    \noindent Distinguished Paper Award, ACM IMWUT 2024, vol. 7
    \\

    \noindent Best Paper Honorable Mention Award, ACM CHI 2024
    \\

    \noindent Microsoft Startup Fund, 2023
    \\

    \noindent UCSD Graduate and Professional Student Association Travel Grant, 2023
    \\

    \noindent Inclusion, Diversity, Equity, and Accessibility (I.D.E.A.) Scholarship, IEEE ISMAR 2022
    \\

    \noindent Adobe Research Fellowship (Finalist), 2021
    \\

    \noindent Qualcomm Innovation Fellowship (Selected Abstract), 2021
    \\

    \noindent Best Poster Award (Runner-up), ACM SenSys 2020
    \\

    \noindent Qualcomm Innovation Fellowship (Selected Proposal), 2020
    \\

    \noindent Qualcomm Innovation Fellowship (Finalist), 2019
    \\

    \noindent Nottingham Advantage Award, 2016
    \\
    
    \noindent British Petroleum Scholarship for Engineering, 2015
    \\

    \noindent Poster Award (Runner-up) for Scientific Content, Engineering Research Placement Scheme, Faculty of Engineering, University of Nottingham, 2015
    \\

\journalpublications

\noindent \textbf{Chen Chen}, Cuong Nguyen, Thibault Groueix, Vladimir G. Kim, and Nadir Weibel, ``MemoVis: A GenAI-Powered Tool for Creating Companion Reference Images for 3D Design Feedback''. ACM Transactions on Computer-Human Interaction (TOCHI 2024), 31, 5, Article 67 (October 2024), 41 pages. DOI: 10.1145/3694681.
\\

\noindent Ella T. Lifset, Kemeberley Charles, Emilia Farcas, Nadir Weibel, Michael Hogarth, \textbf{Chen Chen}, Janet G. Johnson, Mary Draper, Annie L. Nguyen, Alison A. Moore, ``Ascertaining Whether an Intelligent Voice Assistant Can Meet Older Adults’ Health-Related Needs in the Context of a Geriatrics 5Ms Framework'', Gerontology and Geriatric Medicine, Sage Journals (September 2023), DOI: 10.1177/23337214231201138.
\\

\noindent Sudershan Boovaraghavan, \textbf{Chen Chen}, Anurag Maravi, Mike Czapik, Yang Zhang, Chris Harrison, Yuvraj Agarwal, ``Mites: Design and Deployment of a General Purpose Sensing Infrastructure for Buildings'', In Proceedings of the ACM on Interactive, Mobile, Wearable and Ubiquitous Technologies (IMWUT),  7, 1, Article 2 (March 2023), 32 pages, DOI: 10.1145/3580865.
\\

\conferencepublications

\noindent  Jessica de Souza, \textbf{Chen Chen}, Tanya Punater, Ariana Talai, Bárbara Tideman Sartorio Camargo, Ana Carolina Lavio Rocha, Kelly Pereira Coca, Edward Jay Wang, ``Understanding the Challenges and Design Opportunities of Using Voice Assistants to Support Postpartum Mothers in Brazil'', In Proceedings of the 2025 Conversational User Interface Conference (CUI '25), July 8 - 10, Waterloo, ON, Canada, DOI: 10.1145/3719160.3736620.
\\

\noindent  Matin Yarmand, \textbf{Chen Chen}, Michael V. Sherer, Yash N. Shah, Peter Liu, Borui Wang, Larry Hernandez, James D. Murphy, and Nadir Weibel, ``Enhancing Accuracy, Time Spent, and Ubiquity in Critical Healthcare Delineation via Cross-Device Contouring'', In Proceedings of the 2024 ACM Designing Interactive Systems Conference (DIS 2024), July 1 - 5, 2024, IT University of Copenhagen, Denmark, DOI: 10.1145/3643834.3660718.
\\

\noindent Matin Yarmand, \textbf{Chen Chen}, Kexin Cheng, James D. Murphy, Nadir Weibel, ``I'd be watching him contour till 10 o'clock at night: Understanding Tensions between Teaching Methods and Learning Needs in Healthcare Apprenticeship'', In Proceedings of the ACM 2024 CHI Conference on Human Factors in Computing Systems (CHI 2024), May 11 - 16, 2024, Honolulu, HI, United States. DOI: 10.1145/3613904.3642453.
\\

\noindent \textbf{Chen Chen}, Cuong Nguyen, Jane Hoffswell, Jennifer Healey, Trung Bui, Nadir Weibel, ``PaperToPlace: Transforming Instruction Documents into Spatialized and Context-Aware Mixed Reality Experiences'', In Proceedings of the 36th Annual ACM Symposium on User Interface Software and Technology (UIST 2023),  October 29 – November 1, 2023, San Francisco, CA, United States, DOI: 10.1145/3586183.3606832.
\\
  
\noindent \textbf{Chen Chen}, Matin Yarmand, Varun Singh, Michael V. Sherer, James D. Murphy, Yang Zhang, Nadir Weibel, ``Screen or No Screen? Lessons Learnt from a Real-World Deployment Study of Using Voice Assistants With and Without Touchscreen for Older Adults'', In Proceedings of the 25th International ACM SIGACCESS Conference on Computers and Accessibility (ASSETS 2023),  October 23 – 25, 2023, New York, NY, United States, DOI: 10.1145/3597638.3608378.
\\

\noindent Siyou Pei, Alexander Chen, \textbf{Chen Chen}, Franklin Mingzhe Li, Megan Fozzard, Hao-Yun Chi, Nadir Weibel, Patrick Carrington, Yang Zhang, ``Embodied Exploration: Facilitating Remote Accessibility Assessment for Wheelchair Users with Virtual Reality'', In Proceedings of the 25th International ACM SIGACCESS Conference on Computers and Accessibility (ASSETS 2023),  October 23 – 25, 2023, New York, NY, United States, DOI: 10.1145/3597638.3608410.
\\

\noindent \textbf{Chen Chen}, Matin Yarmand, Varun Singh, Michael V. Sherer, James D. Murphy, Yang Zhang, Nadir Weibel, ``VRContour: Bringing Contour Delineation of Medical Structure into Virtual Reality'', In Proceedings of the 21st IEEE International Symposium on Mixed and Augmented Reality (ISMAR 2022),  October 17 - 21, 2022, Singapore, DOI: 10.1109/ISMAR55827.2022.00020
\\

\noindent \textbf{Chen Chen}, Matin Yarmand, Zhuoqun Xu, Yang Zhang, Nadir Weibel, ``Investigating Input Modality and Task Geometry for 3D Delineation in Virtual Reality'', In Proceedings of the 21st IEEE International Symposium on Mixed and Augmented Reality (ISMAR 2022),  October 17 – 21, 2022, Singapore, DOI: 10.1109/ISMAR55827.2022.00054
\\

\noindent \textbf{Chen Chen}, Janet G. Johnson, Kemeberley Charles, Alice Lee, Ella T. Lifset, Michael Hogarth, Alison A. Moore, Emilia Farcas and Nadir Weibel, ``Understanding Barriers and Design Opportunities to Improve Healthcare and Quality of Life for Older Adults through Voice Assistants'', In Proceedings of the 23rd International ACM SIGACCESS Conference on Computers and Accessibility (ASSETS 2021), October 18 - 22, 2021, Virtual Event, USA, DOI: 10.1145/3441852.3471218
\\
 
\noindent \textbf{Chen Chen}, Ke Sun and Xinyu Zhang, ``ExGSense: Toward Facial Gesture Sensing and Reconstructions with Sparse Near-Eye Biopotential Sensing'',In Proceedings of the 20th ACM/IEEE Conference on Information Processing in Sensor Networks (IPSN 2021), May 18 - 21, 2021, Nashville, TN, USA, DOI: 10.1145/3412382.3458268
\\

\noindent Ke Sun, \textbf{Chen Chen} and Xinyu Zhang, ```Alexa, Stop Spying on Me!': Speech Privacy Protection Against Voice Assistant'', In Proceedings of the 18th ACM International Conference on Embedded Sensor Network System (SenSys 2020), November 16 - 19, 2020, Virtual Event, Japan, DOI: 10.1145/3384419.3430727.
\\

\posterpublications

\noindent \textbf{Chen Chen}, Ella T. Lifset, Yichen Han, Arkajyoti Roy, Michael Hogarth, Alison A. Moore, Emilia Farcas, Nadir Weibel , ``How do Older Adults Set Up Voice Assistants? Lessons Learned from a Deployment Experience for Older Adults to Set Up Standalone Voice Assistants'', In the Companion Proceedings of the 2023 ACM International Conference on Designing Interactive Systems (DIS 2023 Companion), July 10 - 14, 2023, Pittsburgh, PA, United States, DOI: 10.1145/3563703.3596640.
\\

\noindent Matin Yarmand, Borui Wang, \textbf{Chen Chen}, Michael Sherer, Larry Hernandez, James D. Murphy, Nadir Weibel, ``Design and Development of a Training and Immediate Feedback Tool to Support Healthcare Apprenticeship'', In the Extended Abstracts of the 2023 CHI Conference on Human Factors in Computing System (CHI 2023 EA), April 23 – 28, 2023, Hamburgh, Germany, DOI: 10.1145/3544549.3585894.
\\

\noindent Yichen Han, Christopher B. Han, \textbf{Chen Chen}, Peng W. Lee, Michael Hogarth, Alison A. Moore, Nadir Weibel, Emilia Farcas, ``Towards Visualization of Time-Series Ecological Momentary Assessment (EMA) Data on Standalone Voice-First Virtual Assistants'', In the Proceedings of the 24th International ACM SIGACCESS Conference on Computers and Accessibility (ASSETS 2022), October 23 - 26, 2022, Athens, Greece, DOI: 10.1145/3517428.3550398.
\\

\noindent Matin Yarmand, Michael V. Sherer, \textbf{Chen Chen}, Larry Hernandez, Nadir Weibel, James D. Murphy, ``Evaluating Accuracy, Completion Time, and Usability of Everyday Touch Devices for Contouring'', In the 2022 Annual Meeting Scientific Program Committee of the American Society for Radiation Oncology (ASTRO 2022), OOctober 23 - 26, 2022, Huston, Texas, USA, DOI: 10.1016/j.ijrobp.2022.07.515.
\\

\noindent \textbf{Chen Chen}, Matin Yarmand, Varun Singh, Michael V. Sherer, James D. Murphy, Yang Zhang, Nadir Weibel, ``Exploring Needs and Design Opportunities for Virtual Reality-based Contour Delineations of Medical Structures'', In the Proceedings of the 14th ACM SIGCHI Symposium on Engineering Interactive Computing Systems (EICS 2022), June 21 - 24, 2022, Sophia Antipolis, France, DOI: doi.org/10.1145/3531706.3536456.
\\

\noindent Kemeberley Charles, \textbf{Chen Chen}, Janet G. Johnson, Alice Lee, Ella T. Lifset, Michael Hogarth, Nadir Weibel, Emilia Farcas, Alison A. Moore, ``How Might An Intelligent Voice Assistant Address Older Adults' Health-Related Needs?'', vol. 69, pp. S243 - S244, In the Journal of the American Geriatrics Society, Wiley, 2021.
\\

\noindent Ella T. Lifset, Kemeberley Charles, Emilia Farcas, Nadir Weibel, Michael Hogarth, \textbf{Chen Chen}, Janet G. Johnson, Alison A. Moore, ``Can an Intelligent Virtual Assistant (IVA) Meet Older Adult Health-Related Needs in the Context of a Geriatric 5Ms Framework?'', vol. 70, pp. S245 - S246, In the Journal of the American Geriatrics Society, Wiley, 2020.
\\

\noindent \textbf{Chen Chen}, Kahlil Mirni, Kemeberley Charles, Ella T. Lifset, Michael Hogarth, Alison A. Moore, Nadir Weibel, Emilia Farcas, ``Toward a Unified Metadata Schema for Ecological Momentary Assessment with Voice-First Virtual Assistants'', In the Proceedings of the 2021 ACM Conversational User Interface Conference (CUI 2021), July 27 - 29, 2020, Virtual Event, Bilbao, AA, Spain, DOI: 10.1145/3469595.3469626.
\\

\noindent Kahlil Mirni, \textbf{Chen Chen}, Ndapa Nakashole, Nadir Weibel, Emilia Farcas, ``Medical Question Understanding and Answering for Older Adults'', The 3rd Southern California NLP Symposium (SoCal NLP Symposium 2021), March 2021, La Jolla, California (Online).
\\

\noindent Matin Yarmand, \textbf{Chen Chen}, Danilo Gasques and Nadir Weibel, ``Facilitating Remote Design Thinking Workshops in Healthcare: the Case of Contouring in Radiation Oncology'' , In the Extended Abstracts of the ACM International Conference on Human Factors in Computing Systems (CHI 2020 EA), Case Study, May 8 - 13, 2020, Virtual Event, Japan, DOI: 10.1145/3411763.3443445.
\\

\noindent \textbf{Chen Chen}, Ke Sun and Xinyu Zhang, ``CapTag: Toward Printable Ubiquitous Internet of Things'', In the Proceedings of the 18th ACM International Conference on Embedded Sensor Network System (SenSys 2020), November 16 - 19, 2020, Virtual Event, Japan, DOI: 10.1145/3384419.3430410.
\\

\patents

\noindent Cuong Nguyen, Trung Bui, Jennifer Healey, Jane Hoffswell, \textbf{Chen Chen}, ``Rendering and Authoring Instructional Data in Augmented Reality with Context Awareness'', US20240320886A1, 2024
\\

\noindent Yuvraj Agarwal, Christopher Harrison, Gierad Laput, Sudershan Boovaraghavan, \textbf{Chen Chen}, Abhijit Hota, Robert Xiao, Yang Zhang, ``Virtual Sensor System'', US20180306609A1, 2018
\\

\teaching

\noindent \textbf{CSE 165} - 3D User Interactions, Department of Computer Science and Engineering, University of California San Diego, Spring 2025
\\

\noindent \textbf{WES 237A} - Introduction to Embedded System Design, Jacobs School of Engineering, University of California San Diego, Winter 2025
\\
    
\noindent \textbf{CSE 118/218} - Ubiquitous Computing, Department of Computer Science and Engineering, University of California San Diego, Fall 2024
\\
    
\noindent \textbf{CSE 165} - 3D User Interactions, Department of Computer Science and Engineering, University of California San Diego, Spring 2024
\\
    
\noindent \textbf{WES 237A} - Introduction to Embedded System Design, Jacobs School of Engineering, University of California San Diego, Winter 2024
\\
    
\noindent \textbf{CSE 118/218} - Ubiquitous Computing, Department of Computer Science and Engineering, University of California San Diego, Fall 2023
\\
    
\noindent \textbf{CSE 118/218} - Ubiquitous Computing, Department of Computer Science and Engineering, University of California San Diego, Fall 2022
\\
    
\noindent \textbf{WES 237A} - Introduction to Embedded System Design, Jacobs School of Engineering, University of California San Diego, Winter 2021
\\
    
\noindent \textbf{CSE 120} - Advanced Software Engineering, Department of Computer Science and Engineering, University of California San Diego, Fall 2020
\\
    
\noindent \textbf{CSE 291H} - Data Center System, Department of Computer Science and Engineering, University of California San Diego, Fall 2019
\\
    
\noindent \textbf{COGS 118} - AI Algorithm, Department of Cognitive Science, University of California San Diego, Summer 2019
\\
    
\noindent \textbf{08-672} - J2EE Web Application Development, Software and Societal Systems Department (previously known as the Institute for Software Research), School of Computer Science, Carnegie Mellon University, Fall 2017
\\
    
\noindent \textbf{08-672} - J2EE Web Application Development, Software and Societal Systems Department (previously known as the Institute for Software Research), School of Computer Science, Carnegie Mellon University, Fall 2016
\\  

\end{vita}

\begin{dissertationabstract}

Mediated by today's visual displays, information space allows users to discover, access and interact with a wide range of digital and physical information.
The information presented in this space may be digital, physical or a blend of both, and appear across different dimensions - such as texts, images, 3D content and physical objects embedded within real-world environment. 
Navigating within the information space often involves interacting with \emph{mixed-dimensional entities}, visually represented in both 2D and 3D.
At times, interactions also involve transitioning among entities represented in different dimensions.
We introduce the concept of \emph{mixed-dimensional information space}, encompassing entities represented in both 2D and 3D.  
A mixed-dimensional information space is promising to harness the strength of entities visually represented in different dimensions.

Interactions within the mixed-dimensional information space should be \emph{seamless} and \emph{efficient}: users should be able to focus on their primary tasks without being distracted by interactions with or transitions between entities.
While incorporating 3D representations into the mixed-dimensional information space offers intuitive and immersive ways to interact with complex information, it is important to address potential seams and inefficiencies that arise while interacting with both 2D and 3D entities.
For example, in a mixed-dimensional information space rendered on a 2D display, navigating the viewing camera with a mouse to explore and identify relevant perspectives - using 2D content such as texts and images - can be tedious.
Creating design feedback on a 3D model can be challenging when trying to find suitable reference images that align with a specific perspective of the 3D model.
When navigating within a mixed-dimensional information space viewed through \textbf{M}ixed \textbf{R}eality~ (MR), seams can arise from suboptimal placement of virtually rendered 2D content within the physical 3D space.
This can cause inefficiencies in an instructional MR experience, where texts and image-based guidance are often rendered in mid-air to guide novices through procedural tasks, as users must repeatedly switch between 2D instructional content and the physical 3D workspace.
Navigation of information space also sees people as existing inside the information space.
In the context of analyzing medical images in the \textbf{V}irtual \textbf{R}eality (VR) headset, seams can arise from the dimensional incongruence between the visually rendered content and the user's mental focus.

Grounded on real-world applications for both general and specialized users including creating reference images for 3D design feedback, carrying out procedural tasks with instructions as well as exploring and contouring medical images, this dissertation introduces new interactive techniques and systems to realize \emph{seamless} and \emph{efficient} interactions within the mixed-dimensional information space.
While 3D is often part of key representations within such space, the interactive experiences introduced in this dissertation includes those rendered on a traditional 2D display, and those experienced through emergent extended reality headset.
Grounded on the user-centered design, this dissertation introduces three interactive systems: 
\textit{MemoVis} which aims to use emergent generative AI to help users create reference images for 3D design feedback;
\textit{PaperToPlace} which demonstrates how paper-based instruction documents can be transformed and spatialized into a context-aware MR experience;
and \textit{VRContour} which explores how contour delineation workflow - an indispensable task in today's radiotherapy treatment planning workflow in the field of radiation oncology - can be brought into VR.
The approaches and design insights presented in this dissertation can guide future efforts in designing interactive workflows that enable users to engage more efficiently with mixed-dimensional information spaces, where content is visually represented in both 2D and 3D.

\end{dissertationabstract}

\mainmatter

\chapter{Introduction}\label{sec::intro}

Since Charles Babbage's invention of the first computer in the early 19th century~\cite{Hyman1985}, digital interfaces have become an integral part to enable our everyday \emph{information space}, a space where users can discover, access and interact with various digital and physical information~\cite{Benyon2001}.
An information space is often made up of different entities that help us undertake activities in the real world~\cite{Benyon2001Online}.
These entities may be digital - rendered on 2D displays or e\textbf{X}tended \textbf{R}eality~(XR) headsets~\cite{Milgram1994} - physical, or a blend of both, appearing across different dimensions such as text, images, 3D content, and physical objects embedded within the real-world environment.
Navigating within the information space~\cite{Benyon2001} often involves interactions with {\it mixed-dimensional} entities, visually represented in both 2D and 3D.
At times, interactions also involve transitioning among entities represented in different dimensions.
We introduce the concept of \emph{mixed-dimensional information space}, defined as such information space that encompasses entities represented in both 2D and 3D.
Representations in 2D and 3D have their own distinct merits:
2D entities such as text and images are easy to create, understand, and share, whereas 3D entities - whether displayed via Web3D~\cite{Web3D} on a 2D screen or rendered in XR - offer intuitive and realistic experiences by conveying a sense of depth.
A mixed-dimensional information space holds promise for leveraging the strengths of both 2D and 3D representations, and should not be limited to digital visual displays. 
For example, a mixed-dimensional information space can include a physical work environment where novice users perform procedural tasks - requiring interaction with 2D instructional manuals and spatially distributed 3D physical objects.

Interactions within the mixed-dimensional information space comprising both 2D and 3D entities should be \emph{seamless} and \emph{efficient}.
Users should be able to focus on their primary tasks without being distracted by interactions with or transitions between entities.
The concept of \emph{seams} has been applied in the early design of videoconferencing systems to describe visible lines or edges that arise when stitching together elements such as remote meeting participants - boundaries that ultimately diminish the telepresence experience~\cite{Buxton1992, Ishii1991, Ishii1990}.
Similarly, within a mixed-dimensional information space, we use the term \emph{seams} to denote the invisible boundaries between entities visually represented in different dimensions.

While incorporating 3D representations into the mixed-dimensional information space offers intuitive and immersive ways to interact with complex information, it is equally important to address potential \emph{seams} and \emph{inefficiencies} that can arise during interactions involving both 2D and 3D entities.
%
For example, in a mixed-dimensional information space rendered on a 2D display, navigating the viewing camera with a mouse to explore and identify relevant perspectives - using 2D content such as texts and images - can be tedious.
Creating design feedback on a 3D model can be challenging when trying to find suitable reference images that align with a specific perspective of the 3D model.
When navigating within a mixed-dimensional information space viewed through \textbf{M}ixed \textbf{R}eality~ (MR), seams can arise from suboptimal placement of virtually rendered 2D content within the physical 3D environment.
This can cause inefficiencies in an instructional MR experience, where texts and image-based guidance are often rendered in mid-air to guide novices through procedural tasks, as users must repeatedly switch between 2D instructional content and the physical 3D workspace.
Navigation of the information space also sees people as existing inside the information space~\cite{Benyon2001}.
In the context of analyzing medical images in XR, seams can arise from the dimensional incongruence~\cite{Darken2005} between the visually rendered content and the user's mental focus.
Buxton advocated for using appropriate devices to simplify interaction syntax~\cite{Buxton1986}. 
Dimensional incongruence arises when the spatial demands of a task do not align with the interaction techniques used to perform it~\cite{Darken2005}.

Grounded on specific applications for both general and specialized users including creating reference images for 3D design feedback, carrying out procedural tasks with instructions as well as exploring and contouring medical images, this dissertation introduces new interactive techniques and systems to realize \emph{seamless} and \emph{efficient} interactions within the mixed-dimensional information space.
While 3D is often part of key representations within such space, the interactive experiences introduced in this dissertation include those rendered traditional 2D display, and those experienced through emergent XR headset.

In this chapter, we first present the overarching research question and thesis statement in Section~\ref{sec::intro::thesis_statement}.
We then outline key challenges in Section~\ref{sec::intro::challenges}, describe our contributions in Section~\ref{sec::intro::contribution}, and acknowledge collaborators in Section~\ref{sec::intro::ack}.

\section{Thesis Statement}\label{sec::intro::thesis_statement}

The purpose of this dissertation is to address the overarching research question: 

\begin{boxH}

How can we design new interactive techniques and systems to realize {\bf seamless} and {\bf efficient} interaction experience within the {\bf mixed-dimensional information space} - encompassing both 2D and 3D visual representations?

\end{boxH}

To address this question, this dissertation adopts a user-centered design approach~\cite{Norman2013} and focuses on three specific applications: \textbf{creating reference images for 3D design feedback}, \textbf{performing procedural tasks with instructions}, as well as \textbf{exploring and contouring medical images}.
Overall, the thesis statement is as follows:

\begin{boxH}

An {\bf seamless} and {\bf efficient} interaction experience within {\bf mixed-dimensional information space} can be achieved by:
{\bf (1)} \textbf{strategically integrating multimodal intelligence} to address the challenge of perspective mismatching, 
{\bf (2)} \textbf{using cues from environmental and behavioral contexts} to address the challenge of spatial misalignment, 
and {\bf (3)} \textbf{combining 2D and 3D representations} to address the challenge of dimensional incongruence.
    
\end{boxH}

\section{Challenges}\label{sec::intro::challenges}

\begin{figure}[t]
    \centering
    \includegraphics[width=\linewidth]{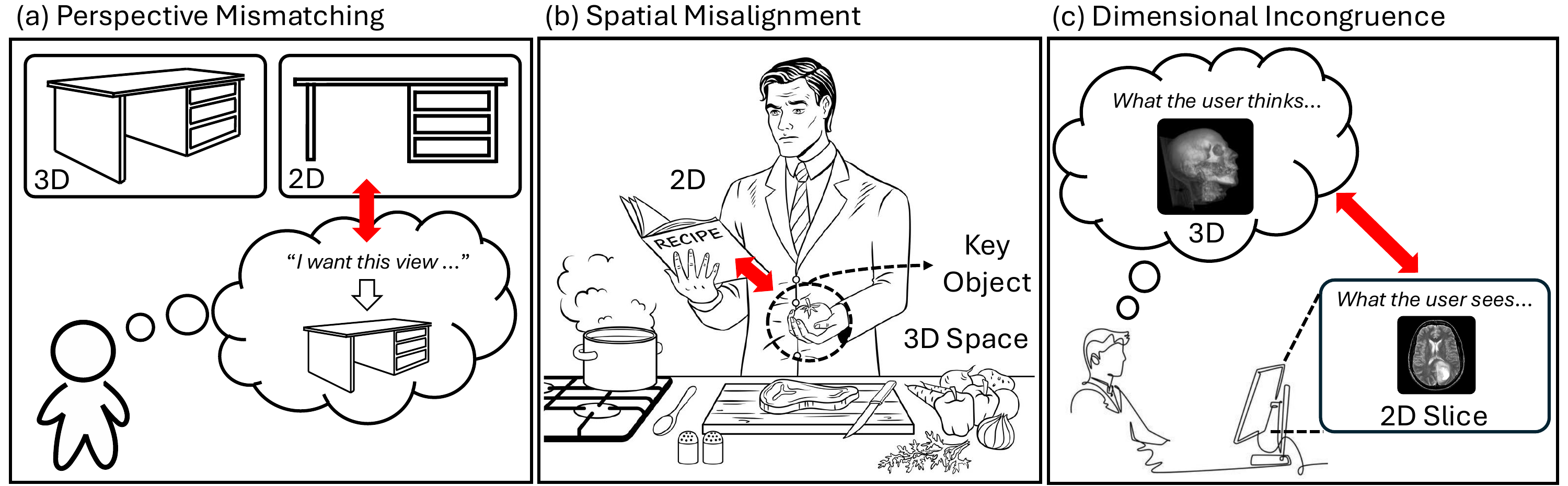}
    \caption[Illustrative figures showing the challenges of interacting within mixed-dimensional information space]{Illustrative figures showing the challenges of interacting within mixed-dimensional information space; (a) perspective mismatching occurs when the view of a 3D model does not match the 2D content; (b) spatial misalignment occurs when the 2D content is not optimally placed within 3D space; (c) dimensional incongruence occurs between the visually rendered content and the user's mental focus. Red arrows highlight the seams between two key entities involved in the user's interaction.}
    \label{fig::intro::challenge}
\end{figure}

Overall, this dissertation addresses three key challenges in designing seamless and efficient interactive experiences within the mixed-dimensional information space.
These three key challenges are illustrated in Figure~\ref{fig::intro::challenge}, where red arrows highlight potential seams between the two primary entities involved in the interaction workflow.

\vspace{+0.1in}
\noindent
$\bullet$~{\bf Perspective mismatching.}
Interactions within the mixed-dimensional information space delivered on today's 2D displays require users to engage with both 3D and 2D entities, which often involve identifying and using relevant perspectives of 3D models - a process that can be inefficient and tedious.
For example, when providing feedback on 3D designs, feedback providers without prior 3D experience often need to create reference images anchored to a specific 3D view, accompanied by textual comments.
However, exploring and identifying a relevant view anchored with textual comments can be tedious.
In the same contexts of creating 3D design feedback, another example is the difficulties of finding online reference images that match the perspective of a selected view of the 3D model.
As illustrated in Figure~\ref{fig::intro::challenge}a, the perspective mismatching between the desired view of the 3D model being commented and the candidate reference image(s) may lead to confusing 3D design feedback.

\begin{figure}
    \centering
    \includegraphics[width=0.6\linewidth]{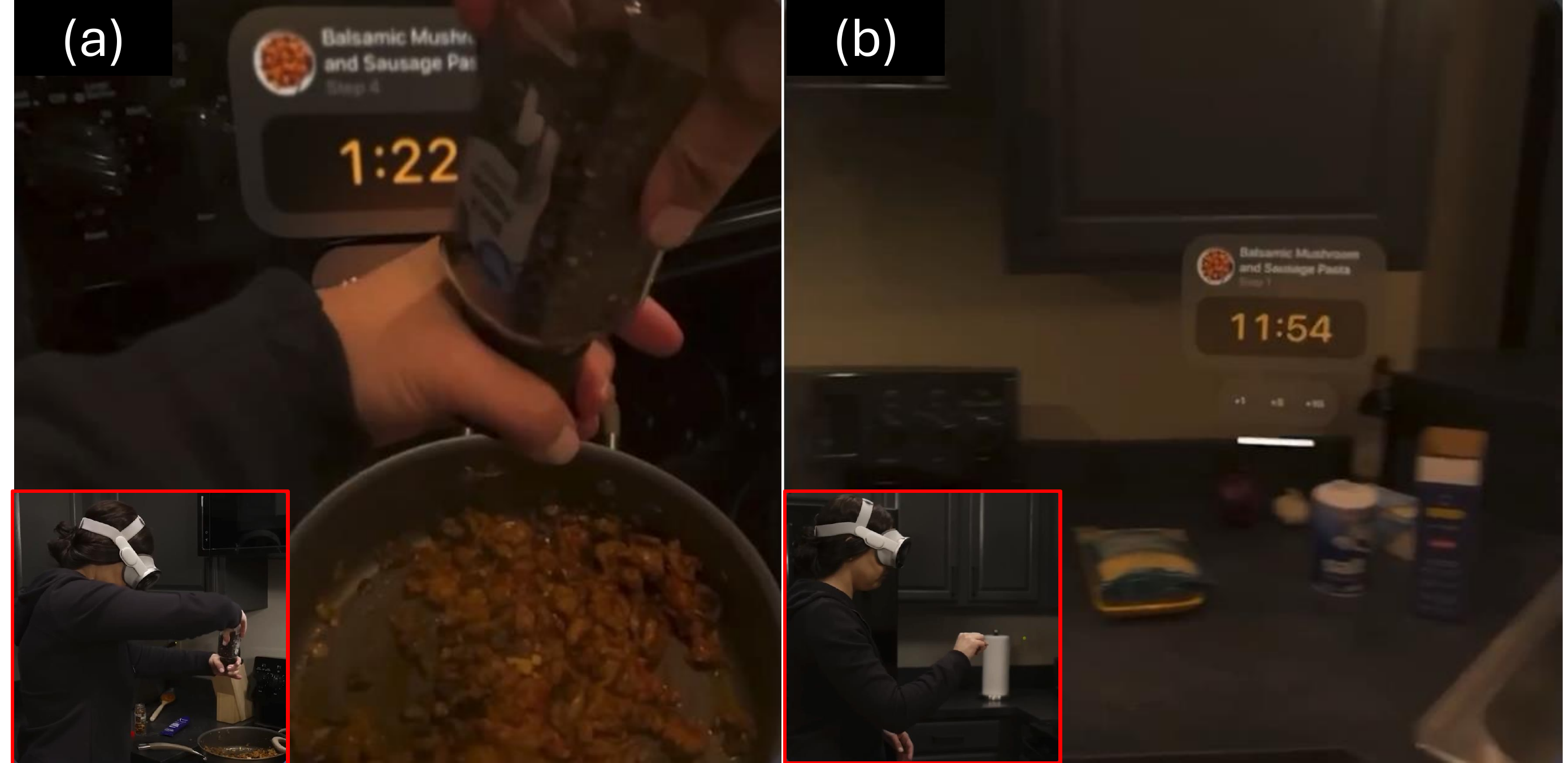}
    \caption[Examples of user experience evaluations of the recent Apple Vision Pro]{Examples of user experience evaluations of the recent Apple Vision Pro, as reported by the Wall Street Journal~\cite{UXAppleVisionPro}, include: (a) virtually rendered widgets obstructing real-world cooking activities, and (b) the need to manually adjust and reposition these widgets in response to changes in real-world activities.}
    \label{fig::intro::mrwsj}
\end{figure}

\vspace{+0.1in}
\noindent
$\bullet$~{\bf Spatial misalignment.}
The spatial placement of 2D content may not always be optimal in relation to the key 3D objects, creating seams between the 2D entities and the 3D space, hindering task efficiency.
In an instructional MR experience, for example, novice users often need to repeatedly engage with instructional guides that consist of 2D content like text and images, and translate them into real-world actions (Figure~\ref{fig::intro::challenge}b).
While today's MR headset allows instruction steps to be rendered in mid-air, overcoming the physical constraints of paper instructions, it remains unclear how these virtually rendered instruction steps can be computationally positioned within 3D space.
Another example regarding the evaluation of user experience of the recently released Apple Vision Pro headset~\cite{VisionPro} highlights the setbacks of occlusion of virtually rendered widgets (Figure~\ref{fig::intro::mrwsj}a) and the need to manually adjust their placement during changes in interactive activities (Figure~\ref{fig::intro::mrwsj}b)~\cite{UXAppleVisionPro}.
The resultant spatial misalignment can create seams between 2D content and 3D space, leading to unnecessary context-switching and potentially reducing task efficiency.

\vspace{+0.1in}
\noindent
$\bullet$~{\bf Dimensional incongruence.}
Dimensional incongruence refers to a mismatch between the spatial demands of a task and the interaction techniques used to carry it out~\cite{Darken2005}.
Seams can arise from dimensional incongruence between visually rendered content and the user's mental focus.
In the context of understanding anatomical structures, users often need to mentally reconstruct medical images in 3D (Figure~\ref{fig::intro::challenge}c).
On the other side, while 3D representation (\eg~rendered in a immersive XR environment) offer intuitive and immersive ways to support consuming of complex information, some tasks may be more efficiently completed using its 2D representation.
For example, while it is intuitive to explore and understand 3D data like medical images in today’s XR headsets, it becomes difficult when users need to annotate individual slices.
In this process, users must focus on 2D slices with detailed anatomical structures of specific cutting plane(s), while still being presented with 3D visualizations.
Dimensional incongruence between the mentally focused information and the rendered content due to dimensional differences may lead to seams and inefficiencies.

\section{Thesis Contributions and Dissertation Structures}\label{sec::intro::contribution}

To address the challenges outlined in Section~\ref{sec::intro::challenges}, this dissertation contributes to the design process of three interactive systems, grounded on three specific applications: creating reference images for 3D design feedback, performing procedural tasks with instructions, as well as exploring and contouring medical images.
We demonstrate how \emph{seamless} and \emph{efficient} interaction experience within the mixed-dimensional information space can look like.
Overall, the three key systems include: 

\vspace{+0.1in}
\noindent$\bullet$~{\bf \ul{MemoVis}: Strategically integrating multimodal intelligence to assist users in creating reference images for 3D design feedback (Chapter~\ref{sec::memovis}).}
Grounded on the workflow of creating reference images in 3D design review, Chapter~\ref{sec::memovis} introduces a novel generative AI driven system designed to enhance interaction experiences with 3D models and their associated 2D feedback rendered on standard 2D display.
MemoVis demonstrates a text-editor interface that assist feedback providers in creating reference images with GenAI driven by the textual comments.
The design of the MemoVis includes a novel real-time viewpoint suggestion feature, based on the emergent vision-language foundation models, to help feedback providers anchor a textual comment with a camera viewpoint.
MemoVis also introduces three types of image modifiers, based on pre-trained 2D generative models, to turn a textual comments into an updated version of the reference images given a specific camera viewpoint.
Ultimately, MemoVis introduces a novel workflow demonstrating how strategically integrating generative AI can help users efficiently create 3D design feedback - minimizing distractions caused by challenges of perspective mismatching, such as between textual comments and relevant views, or between candidate reference images and the intended perspective of the 3D model.

\vspace{+0.1in}
\noindent$\bullet$~{\bf \ul{PaperToPlace}: Transforming instruction documents into spatialized and context-aware MR experiences (Chapter~\ref{sec::papertoplace}).}
Grounded on the scenario of conducting procedural activities guided by an instruction document, Chapter~\ref{sec::papertoplace} demonstrates a novel workflow, PaperToPlace, comprising an \textit{authoring pipeline}, which allows the authors to rapidly transform and spatialize existing paper instructions into an MR experience, as well as a \textit{consumption pipeline}, which computationally places each instruction step at an optimal location that is easy to read and does not occlude key interaction areas.
PaperToPlace demonstrates a computational approach to address the challenge of spatial misalignment by showcasing how virtually rendered instructional steps can be strategically and optimally anchored in physical space using the clues of environmental contexts and users' gestural behaviors.

\vspace{+0.1in}
\noindent$\bullet$~{\bf \ul{VRContour}: Bringing contour delineation workflow into VR (Chapter~\ref{sec::vrcontour}).}
Grounded on the contouring tasks - a critical workflow in \textbf{R}adiation \textbf{T}herapy (RT) treatment planning in the field of radiation oncology where oncologists need to identify and outline malignant tumors and/or healthy organs from a stack of medical images - Chapter ~\ref{sec::vrcontour} contributes to a study that demonstrates how such specialized workflow can be brought into \textbf{V}irtual \textbf{R}eality (VR).
Studies around VRContour demonstrates how interactions with 3D data like medical images can be augmented with today's VR, and how different dimensional representations of medical image information can support various stages of the contouring process.

\section{Acknowledgment}\label{sec::intro::ack}

This thesis introduces three interactive systems that I have led and built during my Ph.D. journey: MemoVis~\cite{Chen2024MemoVis} (Chapter~\ref{sec::memovis}), PaperToPlace~\cite{Chen2023PaperToPlace} (Chapter~\ref{sec::papertoplace}), and VRContour~\cite{Chen2022VRContour, Chen2022VRContourWIP} (Chapter~\ref{sec::vrcontour}).
These works would not have been possible without the support of my advisers, mentors, collaborators and friends from the Department of Computer Science and Engineering, The Design Lab and School of Medicine at University of California San Diego; Department of Electrical and Computer Engineering at University of California Los Angeles; Adobe Research and Nissan Design America.
To recognize and acknowledge their contributions, the pronoun `we' is used throughout this dissertation.

Chapter~\ref{sec::memovis}, in full, is a reprint of the material as it appears in the journal of ACM \textbf{T}ransactions \textbf{o}f \textbf{C}omputer-\textbf{H}uman \textbf{I}nteraction~(TOCHI), Volume 31, Issue 5, Article 67 (October 2024). 
This work was also presented in the 37th Annual ACM Symposium \textbf{U}ser \textbf{I}nterface \textbf{S}oftware and Technology (UIST 2024) conference.
The dissertation author was the primary investigator and author of this journal paper. Co-authors includes Cuong Nguyen, Thibault Groueix, Vladimir G. Kim, and Nadir Weibel. 
This work was collaborated with Adobe Research.
The previously published manuscript can be referred to \cite{Chen2024MemoVis}.

Chapter~\ref{sec::papertoplace}, in full, is a reprint of the material as it appears in the Proceedings of the 36th Annual ACM Symposium on \textbf{U}ser \textbf{I}nterface \textbf{S}oftware and Technology (UIST 2023). The dissertation author was the primary investigator and author of this conference paper. Co-authors includes Cuong Nguyen, Jane Hoffswell, Jennifer Healey, Trung Bui and Nadir Weibel. 
This work was collaborated with Adobe Research.
The previously published manuscript can be referred to \cite{Chen2023PaperToPlace}.
The associated patent can be referred to \cite{Nguyen2024PaperToPlacePatent}.

Chapter~\ref{sec::vrcontour}, in full, is a reprint of the material as it appears in the Proceedings of the 2022 IEEE \textbf{I}nternational \textbf{S}ymposium on \textbf{M}ixed and \textbf{A}ugmented \textbf{R}eality (ISMAR 2022). The dissertation author was the primary investigator and author of this conference paper. Co-authors includes Matin Yarmand, Varun Singh, Mechael V. Sherer, James D. Murphy, Yang Zhang and Nadir Weibel. 
This work was collaborated with the School of Medicine at University of California San Diego and the Department of Electrical and Computer Engineering at University of California Los Angeles.
The previously published manuscript can be referred to \cite{Chen2022VRContour}.
Readers are encouraged to refer to \cite{Chen2022VRContourWIP, Chen2022PrecisionDrawing} for the related publications.

\chapter{Background and Related Work}\label{sec::background}

This dissertation explores the design of interaction experiences within the mixed-dimensional information space, where users engage with information streams visually represented in both 2D and 3D.
This chapter discusses background and related work on the design of visual displays (Section~\ref{sec::background::spatialdisplay}) and the interaction experience within the mixed-dimensional information space~(Section~\ref{sec::background::paperlessinfo}).

\section{Visual Displays for Mixed-Dimensional Information Experience}\label{sec::background::spatialdisplay}
Mixed-dimensional information spaces comprise information entities representations in both 2D and 3D.
While real-world entities such as 2D documents and 3D objects can serve as key elements, a mixed-dimensional information space often integrates with both physical and digital information.
Although rendering 2D content on standard displays is straightforward, interacting with 3D information on 2D screens remains challenging.
The term of ``3D'' often involves creating an illusion of depth and varying distances.
The illusion of depth is often achieved when a flat surface is perceived as containing objects positioned at varying distances from the viewer.
While the systems introduced in this dissertation render 3D content on both standard 2D displays and XR headsets, the contributions are not limited to any specific visual display technology.
This section examines the types of displays that support the delivery of multifarious mixed-dimensional information experience.

\vspace{+0.1in}\noindent
{\bf Interactions with 3D on 2D displays.}
Standard 2D displays have long served as the most basic platform for supporting interactions with 3D content.
Commercially available 3D software like Adobe Substance 3D Collection~\cite{AdobeSubstance} and Blender~\cite{Blender} allows 3D designers to view, interact and edit 3D content.
In domain-specific fields such as healthcare, standalone software like Varian Eclipse~\cite{eclipse} and 3D Slicer~\cite{slicer3D} empowers healthcare professionals to view, interact with, and annotate cross-sectional medical scans as well as 3D-rendered structures.
Emerging Web3D frameworks enable content creators to embed lightweight 3D models that can be rendered directly within web browsers.
For example, DIEL~\cite{Cerbo2010} presents a novel collaborative e-learning platform that enables students to interact with avatars of community members and the 3D environment directly in a web browser.
This setup allows users to perceive proximity relationships with other users and resources, enhancing learning effectiveness.

\begin{figure}[t]
    \centering
    \includegraphics[width=0.8\linewidth]{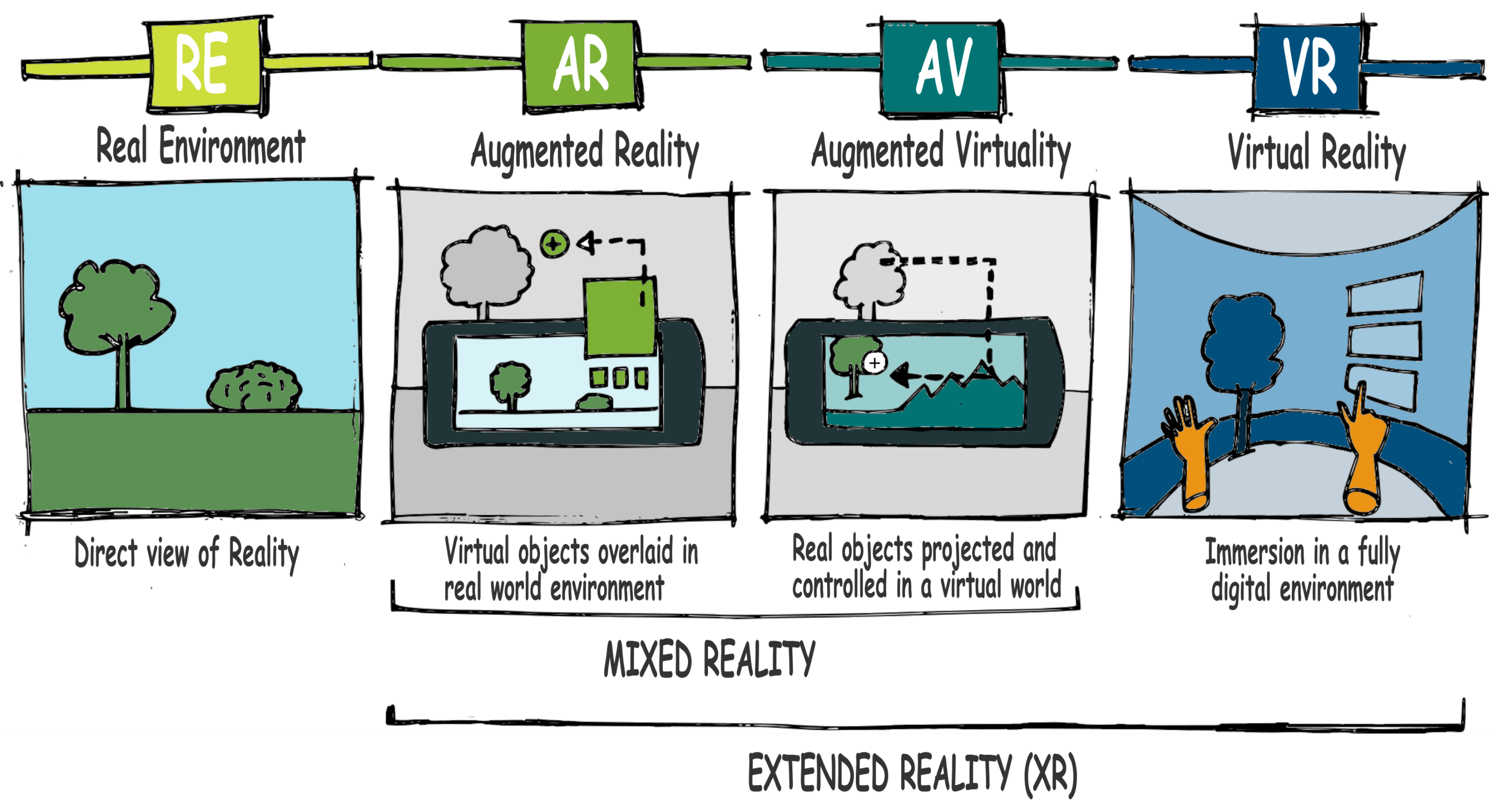}
    \caption[Demonstrative figure of Reality-Virtually Continuum - a spectrum describing the design of MR display that ranges from a completely real environment to a complete virtual environment.]{Demonstrative figure of \textbf{R}eality-\textbf{V}irtually (RV) continuum - a spectrum describing the design of \textbf{M}ixed \textbf{R}eality~(MR) display that ranges from a completely real environment to a complete virtual environment~\cite{Milgram1994}. Demonstrative figure is captured from~ \cite{RVContinuumDemoFigure}.}
    \label{fig::rvcontinuum}
\end{figure}

\vspace{+0.1in}\noindent
{\bf Designing visual displays to support XR experiences with the RV continuum.}
Mixed-dimensional information experience can be realized through today's XR displays.
Since the introducing of ``\textbf{R}eality-\textbf{V}irtuality (RV) Continuum'' - a taxonomy for designing \textbf{M}ixed \textbf{R}eality~(MR) visual displays~\cite{Milgram1994} - the notion of ``XR'' has been used as an umbrella term referring to the \textbf{A}ugmented \textbf{R}eality (AR), \textbf{A}ugmented \textbf{V}irtuality (AV) and \textbf{V}irtual \textbf{R}eality (VR). 
Figure~\ref{fig::rvcontinuum} shows the spectrum of RV continuum, describing three key design streams of XR display, including:
\textbf{AR} implying the way to see extra content being added to the real-world physical environment; \textbf{AV} indicating the interactions when real-world elements are incorporated into virtual world; and \textbf{VR} representing the interactions inside virtual environment~\cite{Milgram1994, Milgram1995}.
The design of XR experience does not restrict to specific display technologies, which can be enabled by a wide range of display technologies~\cite{Azuma2001, Carmigniani2011}.
Despite variations in display technologies, XR displays support the rendering of digital content in both 2D and 3D, either within open spatial environments or constrained spaces.
Noteworthy, a MR display allows users to perceive both digital and physical information.  
Azuma~\etal~\cite{Azuma2001} categorized visual displays supporting XR experiences into  \textbf{H}ead-\textbf{M}ounted \textbf{D}isplays~(HMD), handheld displays and projective or spatial displays.

\vspace{+0.1in}\noindent$\bullet$~{\bf XR experience with headset.}
A HMD is a type of visual display worn on the head that provides visual perception of the physical environment combined with digital information.
HMDs can be designed as either video see-through or optical see-through displays.
A video see-through HMD, such as Quest 3~\cite{Quest3} and Vision Pro~\cite{VisionPro} uses two near-eye displays to render physical environment and digital information, typically by capturing the real world using inside-out cameras.
While targeting a similar goal, optical see-through HMDs, such as HoloLens~\cite{HoloLens} and Magic Leap 2~\cite{MagicLeap2} renders digital information on transparent displays, allowing users to view the physical world \emph{directly} with overlaid digital content~\cite{Normand2012}.
While optical see-through HMD allows a more natural perception of real-world in the contexts of AR experience, video see-through HMD often provides higher-quality rendered graphics, a wider field of view, and greater control over the visual output.

\vspace{+0.1in}\noindent$\bullet$~{\bf XR experience without headset.}
XR experiences can be created \textit{without} headset.
For example, handheld AR utilizes small computing devices such as smartphones and tablets, typically equipped with rear-facing cameras, to enable see-through augmentations~\cite{Rekimoto1997}.
Emergent multi-view 3D displays, such as the Looking Glass~\cite{LookingGlass}, create the illusion of depth by presenting slightly different images to the viewer's left and right eyes.
Recent work has also explored headset-free holographic displays for AR applications, emphasizing their \emph{see-through} capabilities.
For example, a 3D hologram creates the illusion of a floating 3D image by spinning a fan with embedded 3D lights~\cite{Ortega2020}.

\section{Interaction Experiences within the Mixed-Dimensional Information Space}\label{sec::background::paperlessinfo}

A mixed-dimensional information space can be physical, digital, or a blend of both, comprising information streams visually represented as 2D and 3D.
Digital entities with 2D and 3D representations can be rendered on traditional 2D displays, within physical environments through MR displays, or in fully virtual environments using VR.
2D content, including text and images, is easy to create and conveniently portable, but it can be less intuitive for representing and interacting with high-dimensional data.
3D representations provide intuitive ways to visualize high-dimensional data, but interacting with 3D information streams can introduce additional complexities.
Unlike 2D content, which present information through a single layout, interacting with 3D content typically requires an \emph{active} and \emph{exploratory} approach, requiring users navigate a virtual viewing camera to examine key perspectives from different angles.
Kaplan~\etal~\cite{Kaplan1999} posited differences in how viewers perceive 2D and 3D entities during exploration and understanding: the interpretation of 2D content involves rapid assessment of light and dark patterns to extract elements, textures, alongside associated semantic information, whereas the perception of 3D requires inferring depth.

Prior research research has studied how to efficiently convey information using both 2D and 3D entities.
For example, while focusing on canonical tasks such as selection and positioning~\cite{Foley1984}, Darken and Durost~\cite{Darken2005} emphasized the importance of dimensional congruence, advocating for alignment between the dimensional characteristics of a task and the interaction techniques used to perform it.
%
Kumaravel~\etal \cite{Thoravi2022Interactive} demonstrated three contexts, in which interactive mixed-dimensional media can be designed for cross-dimensional collaboration experience in MR environment: MR telepresence for physical task instruction (\eg~Loki~\cite{Kumaravel2019Loki}), video-based instruction for VR tasks (\eg~TutoriVR~\cite{Kumaravel2019TutoriVR}) and live interactions between VR and non-VR users (\eg~TransceiVR~\cite{Kumaravel2020TransceiVR}).

This dissertation focuses on the challenge of designing a seamless and efficient interaction experience within a mixed-dimensional information space.
We extend existing research on designing effective interaction workflows for engaging with entities represented in both 2D and 3D.

\chapter{MemoVis: A GenAI-Powered Tool for Creating Companion Reference Images for 3D Design Feedback}\label{sec::memovis}

\begin{figure}[h!]
    \includegraphics[width=\textwidth]{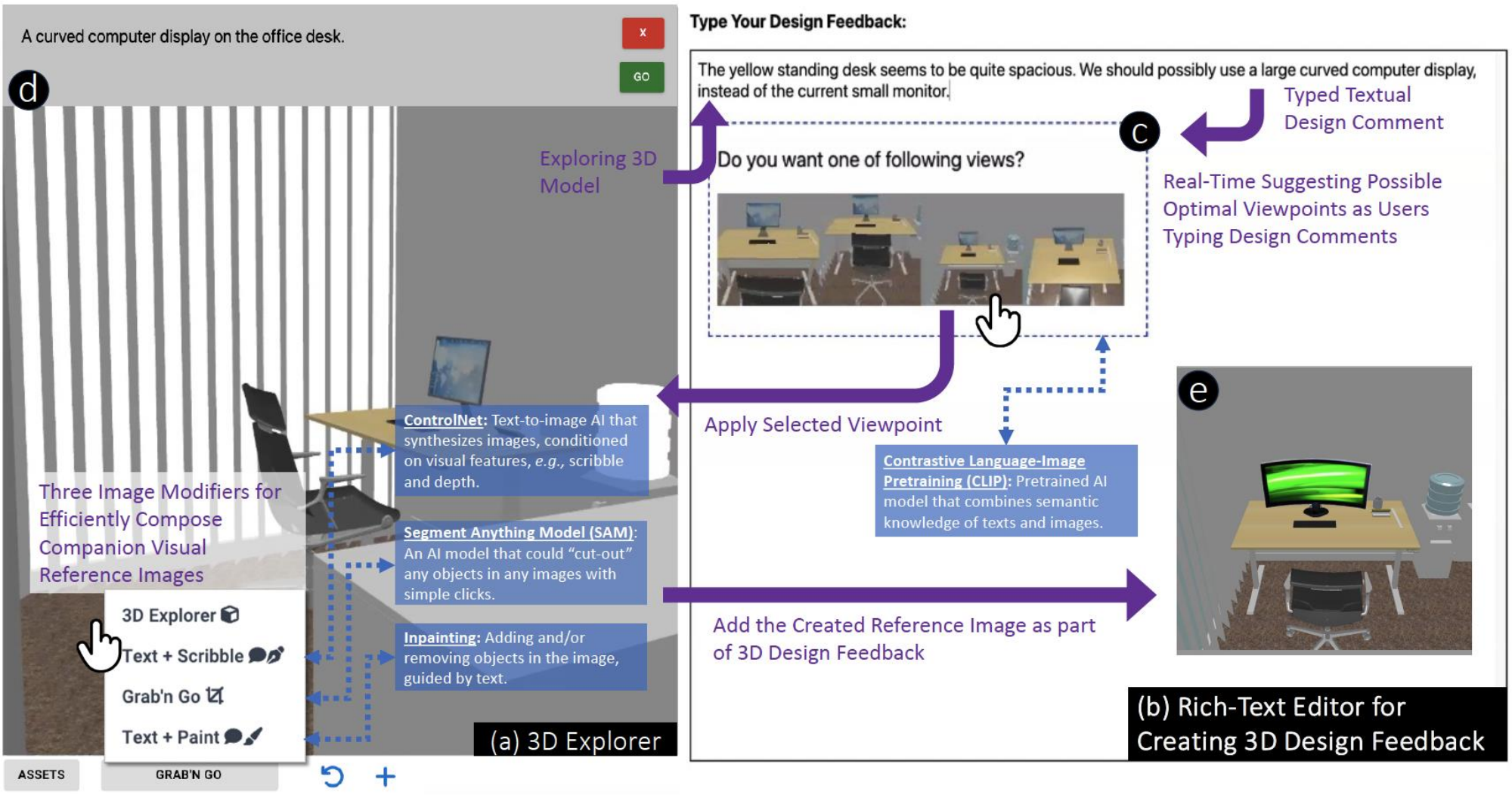}
    \caption[MemoVis is a browser-based text editor interface that assists feedback providers to create companion reference images for 3D design.]{MemoVis is a browser-based text editor interface that assists feedback providers to create companion reference images for 3D design. Feedback providers can (a) explore the 3D model in a 3D content viewer and (b) type the textual feedback comments in a rich-text editor interface; (c) a real-time viewpoint suggestion anchors the textual design comments with a camera viewpoint; (d) a textual prompt could be used to guide the AI generated images. Three types of images modifiers enable feedback providers to efficiently compose companion visual reference images; (e) the visualized reference image reflects the gist of the textual design comments and can be used as part of memo for professional designers to instantiate the feedback.}
    \label{fig::memovis::teaser}
\end{figure}

Providing asynchronous feedback is a critical step in the 3D design workflow.
A common approach to providing feedback is to pair textual comments with companion reference images, which helps illustrate the gist of text. 
Ideally, feedback providers should possess 3D and image editing skills to create reference images that can effectively describe what they have in mind.
However, they often lack such skills, so they have to resort to sketches or online images which might not match well with the current 3D design.
To address this, we introduce \emph{MemoVis}, a text editor interface that assists feedback providers in creating reference images with generative AI driven by the feedback comments. 
First, a novel real-time viewpoint suggestion feature, based on a vision-language foundation model, helps feedback providers anchor a comment with a camera viewpoint. 
Second, given a camera viewpoint, we introduce three types of image modifiers, based on pre-trained 2D generative models, to turn a text comment into an updated version of the 3D scene from that viewpoint. 
We conducted a within-subjects study with $14$ feedback providers, demonstrating the effectiveness of MemoVis. 
The quality and explicitness of the companion images were evaluated by another eight participants with prior 3D design experience.

\section{Introduction}\label{sec::memovis::intro}
Providing asynchronous feedback is a critical step in the 3D design workflow. 
Exchanging feedback allows all stakeholders such as collaborators and clients to review the design collaboratively, highlight issues, and propose improvements~\cite{Gibbons2016, CC3DModel2023, AsyncFeedback2017}.
However, creating effective and actionable feedback is often challenging for 3D design. 
Feedback providers typically convey suggestions about the 3D design via texts.
This practice is similar to doing design review in many 2D domains such as videos~\cite{pavel2016vidcrit,nguyen2017collavr} and documents~\cite{warner2023slidespecs}. 
However, browsing 3D models requires users to navigate information using a viewing camera with six \textbf{D}egrees \textbf{o}f \textbf{F}reedom~($6$DoF). 
Typically, it is more tedious for users to convey where and how changes should be applied on a 3D canvas compared to writing feedback on 2D media.
Additionally, describing certain types of design changes, such as material and texture, can be challenging without extensive 3D knowledge. 
These issues make it especially challenging for individuals with different skill levels to collaborate effectively, like when a designer and a client need to exchange feedback.

To make feedback more instructive, attaching reference images to the textual feedback comments is a common approach for feedback providers to illustrate the texts and externalize the critiques~\cite{Gibbons2016, Barnawal2017}.
The process of creating reference images could also inspire feedback providers to find alternative design problems and generate more insights~\cite{Kang2018, Holinaty2021}.
Ideally, feedback providers should possess basic 3D and image editing skills to create reference images that can effectively describe their thoughts. 
But 3D design is time-consuming, and some users might not have the proficiency to express their thoughts using a 3D software~\cite{CC3DModel2023}. 
As a result, they often resort to sketches or images found on the internet.
For example, when designing the appearance of a 3D bedroom, it might be quicker for a client to search for bedroom images on websites such as Pinterest than to use 3D software like Blender to adjust model's geometry, materials, and textures. 
However, searching for reference images on the internet can also be challenging. 
Finding images that precisely match the viewpoint and 3D structure of the current 3D model is time consuming and often nearly impossible.
The disparity can lead to misunderstandings as designers try to understand the feedback. 
Moreover, online search engines often yield images in similar styles due to various biases in indexing algorithms, potentially causing bias in the feedback and influencing it in unintended ways~\cite{Otterbacher2018}.

Recent \textbf{Gen}erative \textbf{A}rtificial \textbf{I}ntelligence (GenAI) and \textbf{V}ision-\textbf{L}anguage \textbf{F}oundation \textbf{M}odels (VLFMs) offer unique opportunities to address these challenges. 
Text-to-image generation workflow enabled by generative models have been deployed in commercial tools (\eg~Firefly~\cite{AdobeFirefly} and Photoshop~\cite{Photoshop, photoshopAI}) and are being actively studied in HCI with applications in both 3D ~\cite{Liu2023} and 2D ~\cite{Cai2023, Son2023GenQuery} design ideation. 
However, it remains unclear how these text-to-image GenAI models may support the reference image creations in the 3D design \textit{review} workflow.
Using editing tools like Photoshop~\cite{Photoshop, photoshopAI} can create high-fidelity reference images. 
However, this process is time-consuming and requires feedback providers to possess professional image editing skills.
While it is also feasible for feedback providers to generate reference images using existing text-to-image GenAI tools (\eg~\cite{AdobeFirefly, Liu2023, BlenderCopilot}), the synthesized images are usually not contextualized on the current design, making it hard for the designers to interpret the feedback.
Our formative studies indicates that additional controls for 3D scene navigation and alignment of the generated output with the 3D scene structure are crucial for reviewers to create effective reference images.

To explore how text-to-image GenAI can be integrated into the 3D design review workflow, we introduce \emph{MemoVis}, a novel browser-based text editor interface for feedback providers to easily create reference images for textual comments. 
Figure~\ref{fig::memovis::teaser} presents an overview of the workflow. 
MemoVis enables novice users to write textual feedback, quickly identify relevant camera views of the 3D scene, and use text prompts to construct reference images. 
Importantly, these images are aligned with the chosen 3D view, enabling feedback providers to more effectively illustrate their intentions in their written comments.

MemoVis realizes this by introducing a real-time viewpoint suggestion feature to help feedback providers anchor a textual comment with possible associated camera viewpoints.
MemoVis also enables users to generate images using text prompts, based on the depth map of the chosen camera viewpoint. 
To provide users with more controls over the generation process, MemoVis~ offers three distinct modifier tools that complement text prompt input. 
The \textit{text + scribble modifier} allows users to focus the generation on a specific object in the 3D scene. 
The \textit{grab’n go modifier} assists in composing objects from the generated images into the current 3D view. 
Lastly, the \textit{text + paint modifier} utilizes inpainting~\cite{Rombach2022SD} to aid users in making minor adjustments and fine-tuning the generated output.

To understand the design considerations of MemoVis, we conducted two formative studies by interviewing two professional designers and analyzing real-world online 3D design feedback. 
With three key considerations identified from the formative studies, we then prototyped MemoVis by leveraging recent pre-trained GenAI models~\cite{Zhang2023ControlNet, Zhao2023uni, Kirillov2023SAM, Wang2023InstructEdit}. 
Through a within-subjects study with $14$~participants, we demonstrated the feasibility and effectiveness of using MemoVis to support an easy workflow for visualizing 3D design feedback.
A second survey study with eight participants with prior 3D design experience demonstrated the straightforwardness and explicitness of using reference images created by MemoVis to convey 3D design feedback.

In summary, our research around MemoVis explores a potential path and solution to integrate text-to-image GenAI into the 3D design review workflow.
Our key contributions include:

\vspace{+0.1in}
\begin{itemize}[noitemsep, topsep=0pt, leftmargin=*]

\item {\bf Formative studies}, exploring \textbf{(i)} the integration of GenAI into the companion image creation process for 3D design feedback and \textbf{(ii)} the characteristics of real-world 3D design feedback.

\item {\bf Design of MemoVis}, a browser-based text editor interface with a viewpoint suggestion feature and three image modifiers to assist feedback providers to create and visualize 3D design feedback.

\item {\bf User studies}, analyzing the user experience with MemoVis, as well as the usefulness and explicitness of the reference images created by MemoVis to convey 3D design feedback.

\end{itemize}

\section{Related Work}\label{sec::memovis::related}

This section discusses the key related work while designing MemoVis.
We first look into prior research that explored the supporting tools for creating design feedback (Section~\ref{sec::memovis::related::feedback}). 
We then introduce multiple GenAI and VLFMs related to MemoVis (Section~\ref{sec::memovis::related::genai}), and discuss how they could be integrated into 2D and 3D design workflow (Section~\ref{sec::memovis::related::integrate}).

\subsection{Creating Effective Design Feedback}\label{sec::memovis::related::feedback}

Designers often receive feedback along the progress of their design from their collaborators, managers, clients, or even online community~\cite{Krause2017}.
Feedback allows designers to gather external perspectives to avoid mistakes, improve the design and examine whether the design meets the objectives~\cite{Gibbons2016}.
Feedback can also help designers foster new insights and creativity~\cite{Nijstad2006}.

In nearly all creative design processes, asynchronous feedback serves as a critical medium to communicate ideas between designers and various stakeholders~\cite{AsyncFeedback2017}.
The modalities of feedback could affect the communication efficiencies~\cite{Linsey2011}, and the feedback interpretability~\cite{Easterday2007}.
While creating text-only feedback is an easy and widely-adopted method, using visual references is often more effective.
Prior works have explored the integration of reference images with textual feedback.
Herring~\etal~\cite{Herring2009} demonstrated the importance of using reference images during client-designer communications, creating a more effective way to enable designers to internalize client needs.
Paragon~\cite{Kang2018} argued that the visual examples could encourage feedback providers to create more specific, actionable, and novel critique.
This finding guides an interface design that allows feedback providers to browse examples for visual poster design using metadata.
Robb~\etal~\cite{Robb2015} showed a visual summarization system that could crowd-source a small set of representative image as feedback, which could then be consumed at a glance by designers.

Similar to 2D visual design, the system for creating feedback is also urgently needed for \emph{3D design}.
In sectors like the manufacturing industry, the capability to provide feedback and comments is becoming an essential feature in today's 3D \textbf{D}esign \textbf{F}or \textbf{M}anufacturability (DFM) tools.
Many existing research and 3D software have designed features for feedback providers to create textual comments and draw annotations in a specific viewpoint, or on the 3D model directly.
The ModelCraft demonstrates how  freehand annotations and edits can help in the ideation phase during early 3D design stage~\cite{Song2009}.
Professional DFM software, \eg~ Autodesk Viewer, allows adding textual comments of a specific viewpoint and markups to specific parts of 3D assets~\cite{AutodeskViewerFeedback}.
Browser-based tools such as TinkerCAD~\cite{TinkerCAD} also enable feedback providers with less professional 3D skills to add textual comments.
As \textbf{V}irtual \textbf{R}eality~(VR) headsets advance, recent research has also delved into feedback creations within the context of VR-based 3D design review workflow~\cite{Wolfartsberger2019}.

While textual feedback is simple to create, and might be useful in many cases, Bernawal~\etal~\cite{Barnawal2017} showed that the graphical feedback could significantly improve performance and reduce mental workload for design engineers compared to textual and no feedback in manufacturing industry settings. 
However, creating reference images is tedious and challenging.
For certain design with less-common perspective, searching reference images that are well matched with the specific viewpoints is time consuming and sometimes nearly impossible. 
While rapid sketching or using image editing tools might work, such process is tedious and needs feedback providers to have professional image editing and 3D skills --- an impractical expectation for many stakeholders such as managers and clients.
MemoVis shows a novel approach that aims to leverage the power of recent GenAI to help feedback providers easily create companion reference images for 3D design feedback.

\subsection{Generative AI (GenAI) and Vision-Language Foundation Models (VLFMs)}\label{sec::memovis::related::genai}
Language and vision serve as two primary channels for information~\cite{Wu2023}.
Recent AI research has advanced the capabilities of visual and text-based foundation models, many of which have been successfully integrated into commercially available products.
Since the introduction of Generative Adversarial Network~\cite{Goodfellow2014} and Deep Dream~\cite{DeepDream2015, Szegedy2014}, many recent GenAI models, such as Stable Diffusion~\cite{Rombach2022SD}, Midjourney~\cite{Midjourney} and DALL$\cdot$E $3$~\cite{dalle3}, can understand textual prompts and generate images, using models pre-trained by billions of text-image pairs.
Beyond text-to-image generations, several VLFMs attempt to bring natural language processing innovations into the field of computer vision.
The \textbf{C}ontrastive \textbf{L}anguage-\textbf{I}mage \textbf{P}re-training (CLIP) model demonstrates a ``zero-shot'' capabilities to use texts to achieve various image classification tasks, without the need for directly optimizing the model for a specific benchmark~\cite{CLIP, Radford2021}.
The CLIP model also possesses capabilities to represent text and image embeddings in the same space, allowing for direct comparisons between the two modalities~\cite{CLIP}.
The \textbf{B}ootstrapping \textbf{L}anguage-\textbf{I}mage \textbf{P}re-training (BLIP)~\cite{Li2022, Li2023} model is another example of pre-trained VLFMs that can perform a wide variety of multi-modal tasks, such as visual question answering and image captioning.
Grounding DINO~\cite{Liu2023} shows how the transformer-based detector DINO can be integrated into grounded pre-training, to detect arbitrary objects with human input. 
Using Grounding DINO~\cite{Liu2023} and the \textbf{S}egment \textbf{A}nything \textbf{M}odel (SAM)~\cite{Kirillov2023SAM} unlocks the opportunities for inferring segmentation mask(s) based on text input.
Similarly, other pre-trained models, \eg~ Tag2Text~\cite{Huang2023} and RAM~\cite{Zhang2023}, show to generate textual tags with input images.

While text-guided image synthesis is promising, generating images just based on texts may fail to satisfy users' needs due to lack of additional ``control''.
ControlNet~\cite{Zhang2023ControlNet} shows the feasibility of adding additional control to text-to-image diffusion models. 
They show how large diffusion models such as Stable Diffusion can be augmented by additional conditional input such as edges and depths.
This opens possibilities for many follow-up works, \eg~Uni-ControlNet~\cite{Zhao2023uni} and LooseControl~\cite{Bhat2023LooseControl}, that demonstrate the integration of multimodal conditions.
InstructEdit~\cite{Wang2023InstructEdit} and early work like EditGAN~\cite{Liang2021EditGANA} show how users can perform fine-grained editing based on text-instruction.

While innovating GenAI and VLFMs models is \emph{out} of our scope, MemoVis contributes a novel interaction workflow to enable feedback providers easily creating reference images for 3D design feedback by leveraging the strengths of today's GenAI models.

\subsection{GenAI-Powered 2D and 3D Design Workflow}\label{sec::memovis::related::integrate}
Previous research has explored novel techniques for discovering the design space and controlling the generation process of image through \textbf{G}enerative \textbf{A}dversarial \textbf{N}etwork~(GAN)-based GenAI~\cite{Goodfellow2014}, resulting in an increased efficiency of various visual design workflow.
%
For example, Zhang~\etal~\cite{Zhang2021} illustrates how a selection of image galleries may be generated using a novel sampling methods, along with an interactive GAN exploration interface.
\mbox{Koyama~\etal~\cite{Koyama2017, Koyama2020}} proposes the Bayesian optimization-based approach that allows designers to search and discover the design space through a set of slider bars.
Additionally, much prior research proposes novel interaction experience that allow users to input additional control.
%
For example, GANzilla~\cite{Evirgen2022} further demonstrates how iterative scatter/gather techniques~\cite{Pirolli1996} allow users to discover editing directions --- the user-defined control to steer generative models to create content with different characteristics. 
Follow-up research, GANravel~\cite{Evirgen2023}, shows the techniques of global editing (by adding weights to example images) and local editing (with the scribbled masks) for disentangling editing directions (\ie achieve user-defined control while ensuring that unintended attributes remain unchanged in the target image).
GANCollage~\cite{Wan2023} shows a StyleGAN-driven~\cite{Karras2019, Karras2020} mood board that allows users to define possible controls through sticky notes.

Recent advances in text-to-image GenAI have significantly benefited previous research on integrating these technologies into 2D visual design workflows.
For example, the feature of ``Generative fill''~\cite{GenFill2023} introduced by Photoshop shows how texts with simple scribbling could easily in-paint and out-paint the target image (\eg~adding and/or removing objects).
Reframer ~\cite{Lawton2023} demonstrates a novel human-AI co-drawing interface, where the creator can use a textual prompt to assist with the sketching workflow.
The study conducted by Ko~\etal~\cite{Ko2023} demonstrates versatile roles of text-to-image GenAI for visual artists from $35$ distinct art domains to support automating art creation process, expanding ideas, and facilitating or arbitrating in communications.
In another study with $20$ designers, Wang~\etal~\cite{Wang2023} demonstrates the merits of AI generated images for early-stage ideation.
Recent works such as GenQuery~\cite{Son2023GenQuery} and DesignAID~\cite{Cai2023} demonstrate how GenAI could be useful for early-stage ideation during 2D graphic design workflow.
Similarly applied to the ideation stage, CreativeConnect~\cite{Choi2024} further shows how GenAI can may support reference images recombination.
Beyond visual design, BlendScape~\cite{Rajaram2024BlendShape} demonstrates how text-to-image GenAI and emergent VLFMs can be used to customize the video conferencing environment by dynamically changing background and speaker thumbnail layout.

As for 3D design, 3DALL-E~\cite{Liu2023} introduces a new plugin for Fusion $360$ that uses text-to-image GenAI for early-stage ideation of 3D design workflow. 
LumiMood~\cite{Oh2024} shows an AI-driven Unity tool that can automatically adjusts lighting and post-processing to create moods for 3D scenes.
Lee~\etal~\cite{Lee2024} focused on 3D GenAI with different input modalities, and revealed that the prompts can be useful for stimulating initial ideation, whereas multimodal input like sketches play a crucial role in embodying design ideas.
Blender Copilot~\cite{BlenderCopilot} is a Blender plugin that enables designers to easily generate textures and materials by text.
Vizcom~\cite{Vizcom} introduces an early-stage ideation tool for automotive designers that leverages ControlNet to convert designers' sketches into reference images.
In terms of 3D design review, ShowMotion~\cite{Burtnyk2006} demonstrates how a possibly optimal views could be searched from a set of pre-recorded shots by selecting a specific 3D element in the scene.
However, the closed-form searching algorithm only considered the $L2$-distances of each shots to the selected target, neglecting the textual feedback comments.

Inspired by these works, MemoVis demonstrates how to integrate text-to-image GenAI models into the 3D design feedback creation workflow.
Unlike early-stage \emph{ideation}, \emph{modeling}, or \emph{image editing} tools, MemoVis is a \emph{review} tool that aims to enable feedback providers who might not have professional 3D and image editing skills to easily create companion images for the textual comments, contextualized on the viewpoints of initial 3D design.
MemoVis's overarching tenet is to help feedback providers focus on text typing --- the primary tasks while creating design feedback.

\section{Formative Studies}\label{sec::memovis::formative}
We conducted two formative studies to understand the design considerations.
Specifically, we first interviewed with professional 3D designers to understand current practices and the challenges of creating companion images for 3D design review (Section~\ref{sec::memovis::formative::needs}). 
We then analyzed design feedback from an online forum to understand unique 3D design characteristics that feedback writers want to convey (Section~\ref{sec::memovis::formative::online}). 
This process resulted in three design considerations (Section~\ref{sec::memovis::formative::dc}).

\subsection{Formative Study 1: Preliminary Needfinding Study}\label{sec::memovis::formative::needs}

The first study aims to understand the current professional review practices and pain-points in creating companion reference images. 
Professional 3D designers were recruited due to their extensive experience as \emph{both} 3D designers and feedback providers.

\vspace{+0.1in}
\noindent{\bf Participants.}
We recruited FP1 and FP2 as the \textbf{F}ormative study \textbf{P}articipants from Nissan Design America.
FP1 is a modeler and digital design lead, with approximately $20$ \textbf{Y}ears \textbf{o}f \textbf{E}xperience (YoE) of 3D design and $12$ YoE in professional 3D design review process.
FP1 also has extensive experience for 3D game design as a freelance.
FP2 is currently a senior designers specialized in 3D texture design. FP2 has approximately $23$~YoE of 3D design and $15$ YoE in 3D design review.
While both participants have strong experience on using professional 3D software for automotive design such as Autodesk VRED, they also considered themselves as experts in most generic 3D software, \incl~Blender, Adobe Substance Collections as well as Autodesk Maya and Alias.
Although FP2 has tried Midjourney~\cite{Midjourney}, both participants have not used GenAI in the professional settings.

\vspace{+0.1in}
\noindent{\bf Methods.}
Participants first described their demographics including past 3D designs and design review experiences. 
We then conducted semi-structured interviews with each participants, focusing on two guiding questions: {\it ``what does the typical 3D design workflow look like''} and {\it ``what are the potential pain points when creating reference images for design feedback''}. 
For the second question, we further probed participants to discuss how GenAI tools could help with this task. 
The interview was open-ended and participants were encouraged to discuss their thoughts based on their professional experience.
All interviews were conducted remotely, which were then analyzed thematically using deductive and inductive coding approach~\cite{Bingham2021} (see Figure~\ref{fig::memovis::codebook-online-dataa-nalysis-formative-study-1} in Appendix~\ref{sec::memovis::app::codebook} for the generated codebook).
This interview on average took $41$~min with each participants.

\vspace{+0.1in}
\noindent{\bf Findings.} 
Overall, we identified three key findings.

\vspace{0.1in}\noindent$\bullet$~{\bf Creating reference images starts with finding supporting 3D camera views.}
Both participants recognized the importance of finding supporting camera views in the 3D scene to aid written feedback. This task is typically done through manual exploration of the screen and taking screenshots. For example, FP1 described:

\begin{quote}
{\it ``Most of the time, we'll simply take screenshots. We can then markup over the screenshots. Or sometimes the managers will actually take screenshots and markup what they want, or give examples of what they want and send back. [...] I typically use Zoom to record my feedback, because in our regular design software, there is not a lot of functionality like this right now.''}~(FP1) 
\end{quote}

FP1 further emphasized the importance of having these supporting 3D camera views when discussing with non-technical collaborators like managers or clients. for example: {\it ``sometimes, we will just hold an online meeting. They'll talk about like from rear view, or from the top. And we will navigate the design and show them to better understand their feedback.''}

\vspace{0.1in}\noindent$\bullet$~{\bf Conveying changes requires additional work on the reference images.}
When it comes to creating an effective reference image for 3D design feedback, participants reported three main approaches: scene editing, using existing example images, or GenAI.

\vspace{+0.1in}\noindent {\bf (1) Scene editing.}
Participants reported spending time to mock up changes directly on the collected screenshots. 
FP2 described the complexities of preparing reference images for material design: {\it ``for making reference images for interior and definitely for color materials, where they apply fabrics to the seats and the door panels and like a leather grain to the instrument panel, stuff like that, it can be a little bit involved. It's not so quick. So I usually do it quickly in Photoshop. But ideally, you can also do it in the visualization software, like the VRED. I do think preparing these reference images takes the longest.''}

\vspace{0.1in}\noindent {\bf (2) Using existing reference images.}
Some design review stakeholders might resort to using some existing photographic examples of what they want to convey. For example, FP1 mentioned that {\it ``when we get to the more advanced review with the stakeholders outside our studio, sometimes, they might simply show some other example images to demonstrate what they want. But we would generally work with them iteratively, just to make sure there is no miscommunication.''}

\vspace{0.1in}\noindent{\bf (3) GenAI.}
Emerging GenAI tools are seen as a promising way to generate reference images. Both participants recognized the potential benefits for visual reference creations and creativity support that GenAI tools could provide in 3D design review workflow. A key benefit that FP1 emphasized is the low barrier of entry for non-technical users: {\it ``using texts to generate image seem to be flexible and simple for those who do not know image editing software.''}~
FP2 emphasized how images generated using GenAI technologies could inspire new ideas: {\it ``GenAI is like Pinterest on steroids. You're already using Pinterest, but with GenAI you can create even more interesting inspiration. I think you tend to see the same images once in a while on Pinterest. Because if people are picking the same type of images, and you're kind of like in this echo chamber kind of thing. I remember the first few months when I started playing with Midjourney, my brain just got kind of warmed and hot. It was getting massaged! Because I'm seeing these crazy visuals that my brain isn't used to, like these weird combinations of things. I think it was a very good way for ideation. It can be also very stimulating for feedback providers to think and create suggestions.''}

\vspace{0.1in}\noindent$\bullet$~{\bf Controls for image generation.}
Despite the potentials of using GenAI to generate reference images, both participants mentioned that it can be frustrating to generate the right image using \emph{only} text prompts: {\it ``it's like a slot machine. I think designers kind of occasionally like happy surprises. You do 20 and maybe one is cool. But after a while, I think it gets a little frustrating and you really want more control over the output. I know there's a lot of this control on that kind of stuff, for example helping you control a little bit more perspective and painting, and stuff like that, however it is still very hard to visualize the many feedback suggestions, for example, with a small change of a particular assets components.''} ~(FP2)

\vspace{0.1in}\subsection{Formative Study 2: Analysis of Real-World 3D Design Feedback}\label{sec::memovis::formative::online}
The first study only allows us to understand how reference images are used in current 3D design review workflow, we still need to examine what information reviewers typically encode in their feedback, and how visual references are used to convey such information.
While our semi-structured interview offers insightful thoughts from professional designers, it is difficult to analyze real-world design feedback in an ecologically valid settings, as most of design feedback in professional settings are not publicized.
Additionally, the feedback providers may also include those beyond professional 3D designer, who might not have prerequisite skills for using 3D and image editing software.

\vspace{0.1in}
\noindent{\bf Data Curation.}
We collected 3D design feedback from Polycount~\cite{polycount}, one of the largest free online community for 3D professionals and hobbyists. 
Polycount allows its members to post 3D artworks and receive design feedback from the community.
Although Polycount posts tend to be centered around 3D game design, the 3D assets that are discussed broadly can include a wide variety of 3D design cases like characters, objects, and scenes design.
These assets are also common in other 3D design domains.
Therefore, the outcomes yielded from our analysis could also be generalized into broader 3D applications.
Next, We selected threads within one month starting from June 2023.
Since our focus is to understand the characteristics of real-world design feedback, we only focus on the section of ``3D Art Showcase \& Critiques''~\cite{polycountCritiques}.
Irrelevant topics such as announcements and discussions related to 3D software were excluded.
Our data curation led to $15$ discussion threads, including $99$ posts from $15$ creators and $36$ feedback providers. 
Among $15$ threads, eight threads focus on the design of characters (\eg~human and gaming avatars), three threads focus on the design of objects, and four threads focus on the scene design.

\vspace{0.1in}
\noindent{\bf Analysis Methods.}
We used inductive and deductive coding approach~\cite{Bingham2021} to label each design feedback posts thematically.
We aim to understand what were the primary focus of feedback and how the feedback was externalized.
Our codebook (see Figure~\ref{fig::memovis::codebook-online-data-analysis} in Appendix~\ref{sec::memovis::app::codebook}) was generated through five iterations.

\vspace{0.1in}
\noindent{\bf Findings.}
Our analysis leads to findings under five themes (Figure~\ref{fig::memovis::codebook-online-data-analysis}). 
We use OFP\texttt{<Thread\_ID>}-\texttt{<Feedback\_Provider\_ID>} to annotate \textbf{O}nline design \textbf{F}eedback \textbf{P}roviders. 
For example, OFP1-2 indicates the \emph{second} feedback provider in the \emph{first} discussion thread.

\begin{figure*}[t]
    \centering
    \includegraphics[width=\textwidth]{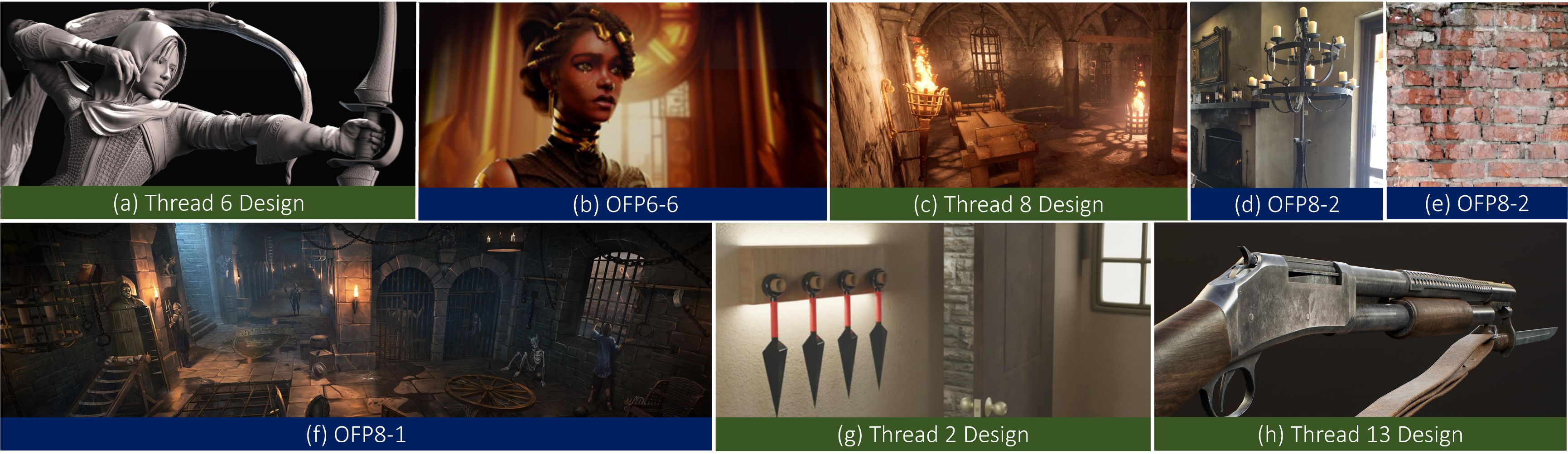}
    \caption[Example reference images from Polycount.]{Example reference images from Polycount~\cite{polycount, polycountCritiques}. (a) Initial design of discussion thread 6 and (b) feedback from OFP6-6; (c) initial design of discussion thread 8 and feedback from OFP8-1 (f) and OFP8-2 (d - e); (g) initial design of discussion thread 2; (h) initial design of discussion thread 13. Green and blue labels indicate the associated reference images are from creators and feedback providers respectively.}
    \label{fig::memovis::polycount-example}
\end{figure*}

\vspace{+0.1in}
\noindent$\bullet$~{\bf Reference images are important to complement textual comments, but creators might need additional ``imaginations'' to transfer the gist of the visual imagery.}
One common approach for creating visual reference is to use internet-searched images. 
However, online images tend to be very different to the original design context, so feedback providers usually  write additional texts to help designers understand the gist of reference images, contextualized on the original design.
For example, while suggesting to change the color tones of the designed character (Figure~\ref{fig::memovis::polycount-example}a), OFP6-6 used a searched figure (Figure~\ref{fig::memovis::polycount-example}b) as a reference image and suggest: {\it ``in terms of tone, maybe look here [refer to Figure~\ref{fig::memovis::polycount-example}b] too, just for its much softer tones across the surface.''}~
Similarly, as for scene design, OFP8-2 used Figure~\ref{fig::memovis::polycount-example}d and Figure~\ref{fig::memovis::polycount-example}e to suggest feedback for a medieval dungeon design (Figure~\ref{fig::memovis::polycount-example}c): {\it ``the rack and some of the barrels and other assets look very new and doesn't have the same amount of wear or appear to have even been used. Need to think of the context and the aging of assets that would have similar levels of wear or damage [...] [for the wall design] I'd probably suggest to use more variation/decals, with some areas of wet or slight moss, or cobwebs.''}~
OFP8-1 also similarly found another internet searched image shown in Figure~\ref{fig::memovis::polycount-example}f as the example to suggest the design of shadow and color tone: {\it ``right now the shadows are way too dark, almost completely black which makes it really hard to tell what is going on, and causes losing a lot of detail in some parts [...] I would look into adding in some cool colored elements into your environment to help balance the hot orange lighting you have. Here is an example with less extreme lighting.''}~
However, the reference images attached by feedback providers were not contextualized on the initial design (\cf~Figure~\ref{fig::memovis::polycount-example}b \vs~Figure~\ref{fig::memovis::polycount-example}a and \cf Figure~\ref{fig::memovis::polycount-example}d - f \vs~Figure~\ref{fig::memovis::polycount-example}c), and could possibly lead to confusions for the creators.

\vspace{+0.1in}
\noindent$\bullet$~{\bf The suggestions conveyed by the design feedback can be either the revision of specific part(s) or the redesign of entire assets.}
While much feedback such as Figure~\ref{fig::memovis::polycount-example}d - f focused on the major revision (or even redesign) of the entire environment, we found much design feedback only emphasize the change of specific part(s) of the design assets, while keeping the rest of design remains unchanged.
For example, for Figure~\ref{fig::memovis::polycount-example}g, OFP2-1 suggested only the change of the knives without commenting on the other part of asset: {\it ``the knife storage seems a bit unpractical. As they are knives, one would probably like to grab them by the handle? Maybe use a belt instead, with the blades going in, or some knife block as can be found in some kitchens.''}~
Similarly, in the design of Figure~\ref{fig::memovis::polycount-example}h, OFP13-3 suggested the additional modeling while satisfying the rest of the design: {\it ``looks cool! With close up shots it would be nice if bolts/screws were modeled.''}~
Although such intended changes are often involved with specific part(s) of the assets, some design feedback explicitly request designers to ``visualize'' how the assets might be integrated into a different environment. 
These feedback help inspire creators to better polish their artwork.
For example, while providing feedback for polishing the tip and edge of the gun in Figure~\ref{fig::memovis::polycount-example}h, OFP13-2 asked the creator to {\it ``just think about how this gun is used, and where wear and damage would be''}.

\vspace{+0.1in}
\noindent$\bullet$~{\bf Although some design feedback provides actionable solutions, others offer potential exploratory directions.}
Despite some design feedback contains specific and actionable steps, many critiques only suggested the problems, without offering suggestions on how to address the problem.
For example, for an ogre character design, OFP1-4 suggested: {\it ``I think it could use some better material separation. The cloth and the metal (maybe even the skin too) seem to have all the same kind of noise throughout.''}~
Some design feedback may indicate the potential exploratory directions that might require the creators to explore by trial and error.
For example, while designing a demon character, OFP3-2 wrote: {\it ``I think that the shape is too round and baroque. More angular form should work better.''}

\subsection{Summary of Design Considerations}\label{sec::memovis::formative::dc}
We highlighted the significance of using reference images in 3D design feedback, applicable to both synchronous and asynchronous design review sessions discussed in Section~\ref{sec::memovis::formative::needs} and Section~\ref{sec::memovis::formative::online} respectively.
Both studies showed that the process of creating good reference images still remains difficult. 
These challenges come from the complexities of 3D design, requiring feedback providers to be well-versed in 3D skills to describe changes efficiently. 
While emerging GenAI are deemed promising to synthesize reference images for 3D design, we have seen from FP2's testimonies that using such tools directly without dedicated interface support can hardly meet the needs of 3D design feedback visualizations.
To integrate text-to-image-based GenAI into 3D design feedback creation workflow, we propose three key \textbf{D}esign \textbf{C}onsiderations (DCs):

\vspace{0.1in}
\noindent {\bf DC1: Design controlled visualization for both local and global changes.}
Our second formative study showed the importance of having a {\it controlled} way to synthesize reference images for both local (\ie~only specific part(s) of the asset need to be changed while keeping the remaining part consistent with current design) and global changes (\ie~major revision or even redesign of the most parts of the artwork).
Demonstrated in Figure~\ref{fig::memovis::polycount-example} and discussed by FP2, although searching for images is commonly used, finding an exact or sufficiently similar design online is difficult. To address this, feedback providers often have to write elaborated and detailed comments explaining the differences and areas of focus.
While vanilla text-to-image GenAI methods offer powerful tools for image generation, the resulting images are usually not contextualized in the initial design. 
Therefore, it is critical to develop a novel interface that could leverage the power of Gen AI to generate in-context imagery at various scales seamlessly integrating this into 3D design.

\vspace{0.1in}
\noindent {\bf DC2: Provide ideation and creativity support for exploratory design feedback.}
We learned that feedback providers often would like to use the reference image as a source of inspiration, as it enables them to think about the alternative ways to  improve to the 3D design.
For example, some design feedback, like OFP13-2, requested the creators to imagine how the artwork would look like after being integrated into a bigger environment. 
Therefore, reference images should also demonstrate how the 3D model might look like when used in difference scenarios, inspiring both creators and feedback providers.

\vspace{0.1in}
\noindent {\bf DC3: Offer ways for feedback providers, including those without 3D and image editing skills to accompany their comments with meaningful in-context visualizations.}
Feedback providers does not have to possess design skills, as shown in Formative Study 1. Thus, feedback providers may also encompass individuals without professional 3D or image editing skills, such as the clients.
Without these skills, one can spend tremendous effort on even the simplest tasks, such as selecting a good view by navigating the scene with an orbit camera, inserting a novel object, or changing a texture or color of an existing object~\cite{CC3DModel2023}.
Therefore, the interface design should provide simple ways for feedback providers to navigate the 3D scene and create reference images without 3D and image editing skills.

\section{MemoVis}\label{sec::memovis::system}
Overall, MemoVis~includes a {\it 3D explorer} (Figure~\ref{fig::memovis::teaser}a), where feedback providers can explore different viewpoints of the 3D model, and a {\it rich-text editor} (Figure~\ref{fig::memovis::teaser}b) where the textual feedback is typed.
Unlike image editing (\eg~Photoshop~\cite{Photoshop, GenFill2023}) and ideation tools (\eg~Vizcom~\cite{Vizcom}), MemoVis~is a {\it review} tool for 3D design.
The key tenet is to enable feedback providers to focus on {\it text typing} --- the primary task for creating 3D design feedback --- while using AI-augmented rich-text editor to create companion images that illustrate the text.
As a {\it review} tool, MemoVis ~ aims to enable feedback providers to create reference images that could efficiently convey the gist of the textual comments, instead of images that are visually aesthetic.
MemoVis's main features include a viewpoint suggestion system and three image modification tools.

\subsection{Real-Time Viewpoint Suggestions}\label{sec::memovis::system::view_suggestion}
In MemoVis, users control an orbit viewing camera with $6$DoF to render the 3D model on screen space similar to mainstream 3D software. 
However, navigating and exploring a 3D scene with a mouse can be tedious for users lacking experience in 3D software.
To enable feedback providers without 3D and image editing skills more efficiently create in-context visualizations for the textual feedback (\textbf{DC3}),
MemoVis automatically recommends viewpoints as the feedback provider continuously typing feedback.
Figure~\ref{fig::memovis::teaser}c shows an example where a closer (and possibly better) viewpoints, shown as thumbnails, of the standing desk are recommended. 
The viewpoint being selected will be anchored with the textual comment and instantly reflected on the 3D explorer.
Feedback providers can ignore the suggestions when the suggested views are less helpful.

\begin{figure*}[t]
    \includegraphics[width=\textwidth]{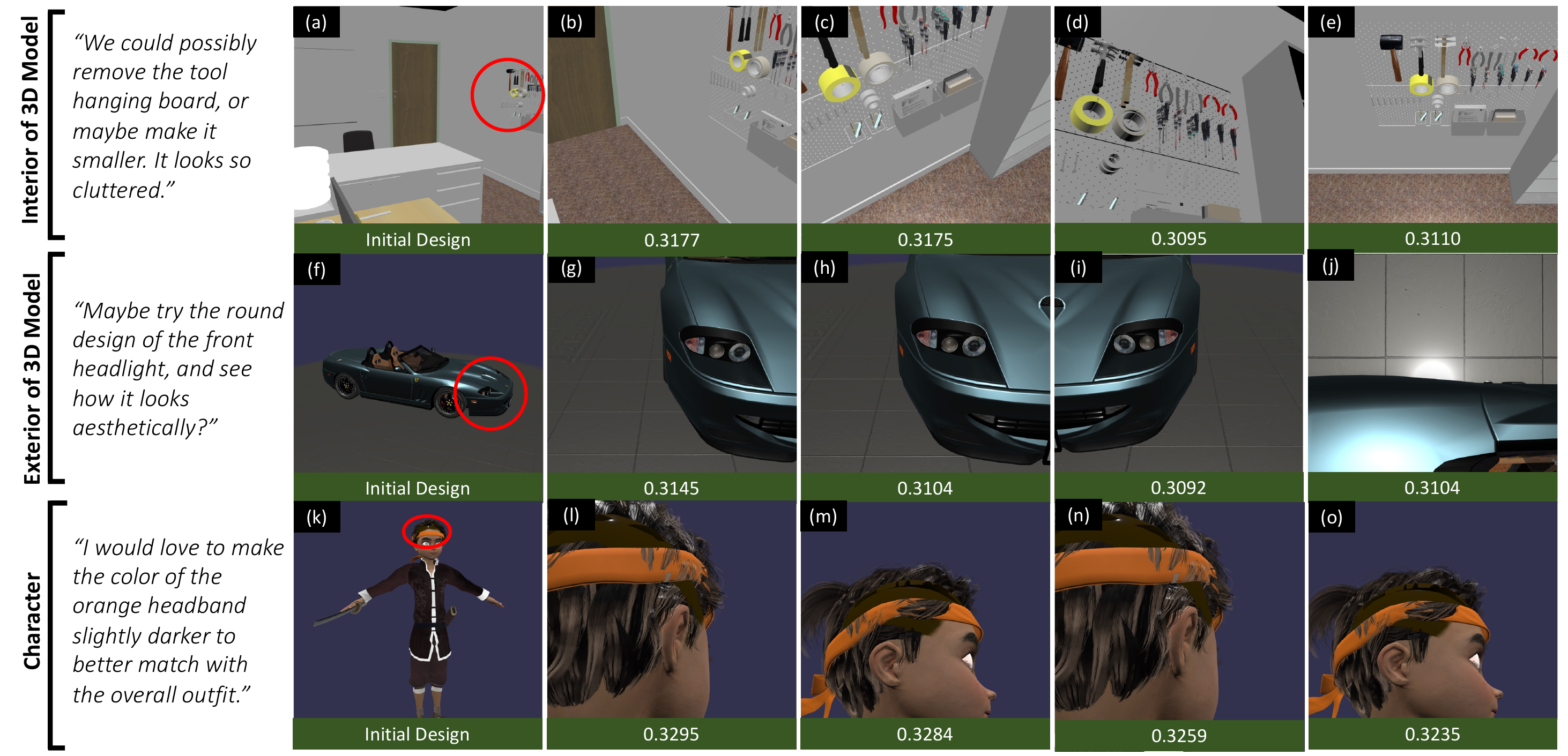}
    \caption[Examples of the suggested viewpoints based on the typed feedback comments.]{Examples of the suggested viewpoints based on the typed feedback comments (leftmost column). We show viewpoints with top-$\bm{4}$ highest CLIP similarity scores for an office 3D model (a - e), a car model (f - j), and a samurai boy model (k - o). a, f, and k show the bird-eye view of the initial 3D model where red circle highlight the focus of textual comments. The cosine similarity scores are shown at the bottom of each suggested viewpoints (b - e, g - j, l - o).}
    \label{fig::memovis::clip-demo}
\end{figure*}

To achieve this, we use the pre-trained CLIP model~\cite{CLIP} to find viewpoints with highest cosine similarities to the textual feedback.
Indeed, CLIP is trained on $\sim 400$~millions text-image pairs with a constrastive loss~\cite{Chopra2005ContrastiveLoss}. 
The system benefits from the human biases in image acquisition~\cite{Hentschel2022, Voigt2023Paparazzi}. For example, despite the vagueness of the text, \eg~{``office desk''}, the pre-trained CLIP model would score the desk from front and top view higher than that from the back and bottom view, as it is more common to take front-facing pictures of desks.
Formally, we parameterize the orbit camera with six parameters: the 3D point that the camera is looking at $(t_x, t_y, t_z)$ as well as its distance $r$ to the camera, and the longitudinal and latitudinal rotation $\alpha$, $\beta$. Each viewpoint can thus be represented by the tuple $\bm{v} = (\alpha, \beta, r, t_x, t_y, t_z)$, with $\alpha \in [0, \pi]$ and $\beta \in [0, 2\pi]$.
Our goal is to search $\hat{\bm{v}} = \mathbf{argmax}_{\bm{v} \in \bm{V}} cos\{f_{text}(\bm{t}) , f_{image}(\bm{I}_{\bm{v}})\}$, where $f_{text}(\cdot)$ and $f_{image}(\cdot)$ represent the CLIP encoding for text ($\bm{t}$) and screen space image ($\bm{I}_{\bm{v}}$) that is associated with a specific viewpoint $\bm{v}$.
During pre-processing, we sample multiple possible viewpoints and encode their corresponding renderings via CLIP to create a database of viewpoints. 
At inference time, we encode the textual comment via CLIP and perform a nearest-neighbor query in the database, which can be done without breaking the interactive flow.

\vspace{+0.1in}
\noindent$\bullet$~{\bf Pre-processing}. 
We first compute the bounding box of the 3D model.
We then discretize the $x$-, $y$-, and $z$-axis into five bins, leading to $5^3 = 125$ sampled target position $(t_x, t_y, t_z)$.
Similarly, we sample $\alpha$ and $\beta$ with $30^{\circ}$ intervals into $12 \times 6 = 72$ possibilities.
We sample $r$ from $\{0.5, 1.0, 1.5\}$ to create \emph{close}, \emph{medium}, and \emph{far} views.
For each 3D model, this pre-processing phase takes around $3$ - $5$ minutes, and results in a $\bm{D} \in \mathbb{R}^{27K \times 500}$ matrix, including $27$~k view points, each encoded via CLIP into a $500$~ dimensional feature vector. 

\vspace{+0.1in}
\noindent$\bullet$~{\bf Real-time inference}. 
As the feedback provider typing, the textual comment ($\bm{t}$) is encoded, and the top-$4$ nearest views under cosine similarity are retrieved. The feature runs every time the user stops typing for $500$~ms and takes under a second to compute.

\vspace{0.1in}
Figure~\ref{fig::memovis::clip-demo} shows examples of suggested viewpoints of the pegboard in an office (Figure~\ref{fig::memovis::clip-demo}a - e, as an interior design example),  the headlight of a car (Figure~\ref{fig::memovis::clip-demo}f - j, as an exterior design example), and the headband of a samurai boy (Figure~\ref{fig::memovis::clip-demo}k - o, as a character design example).
Although MemoVis provides real-time viewpoint suggestions, it still allows feedback providers to manually navigate the view. For instance, they can manually find the view before writing textual comments or adjust the view based on the suggestions.

\subsection{Creating Reference Images with Rapid Image Modifiers}\label{sec::memovis::system::modifiers}
Guided by \textbf{DC2}, providing ideation and creativity support are crucial for creating design feedback, MemoVis therefore uses the recent text-to-image GenAI to create reference images.
However, the generated reference images should match both textual comments and current 3D design.
Critically, MemoVis must be able to generate images with local modification of the scene if the feedback is targeted at a specific part, or images with global edits when the feedback is a global redesign of the scene.
This design goal aims to address \textbf{DC1}, emphasizing on the controlled visualization of the textual feedback for both local and global changes.
MemoVis introduces three image {\it modifiers}, which operate on {\it rapid design layers}.
Feedback providers can use one or multiple {\it modifiers}~ (Figure~\ref{fig::memovis::teaser}a) to easily compose and create reference image, on the associated {\it rapid design layer}, rendered from a selected viewpoint.
We now describe our system design and demonstrate how the {\it modifier} might interact with each {\it rapid design layers}.

\vspace{0.1in}
\noindent{\bf Modifier 1: Text $+$ Scribble Modifier with Scribble Design Layer}

\noindent We leverage ControlNet~\cite{Zhang2023ControlNet, Zhao2023uni} for two scenarios.
In the simpler case, the feedback is just a global texture edit on the scene that does not suggest any geometry modification. In this case,  we use a depth-conditioned ControlNet~\cite{Zhang2023ControlNet} to generate an image. 
The depth guidance ensures that the generated image is anchored in the current design, while the textual prompt generates an image that matches the feedback. 

\begin{figure*}[t]
    \centering
    \includegraphics[width=\textwidth]{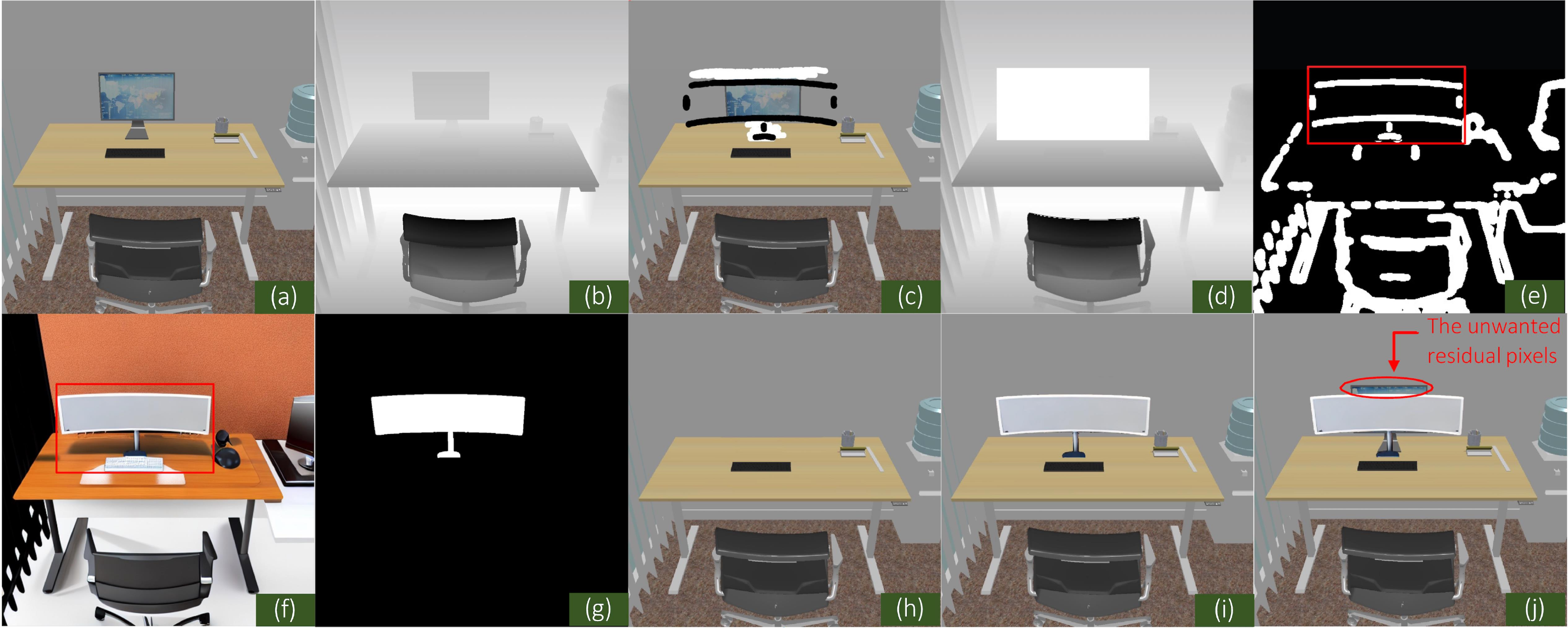}
    \caption[Examples of creating reference image using the text + scribble modifier.]{Examples of creating reference image using the text + scribble modifier. (a) Initial design; (b) associated depth map; (c) manually drawn scribbles (black strokes) with the white strokes indicating the removed geometries; (d) the depth map with the scribbling area being reset; (e) an aggregated scribble from the initial image and the manually drawn scribbles, where the red bounding box shows the scribbling area by feedback providers; (f) synthesized image by ControlNet conditioned by scribble $+$ depth, where the red bounding box shows the area that the feedback providers scribbled; (g) segmented mask generated by SAM; (h) initial design with the primitives describing existing computer display being removed; (i) final composed reference image; (j) final composed reference image without removing objects marked for removal by scribbling.}
    \label{fig::memovis::controlnet-scribble-demo}
\end{figure*}

If the feedback suggests to modify the geometry of part of the scene (\eg~the review from OFP2-1 for the 3D design of Figure~\ref{fig::memovis::polycount-example}g), directly editing the depth image~\cite{Dukor2022} is impractical for feedback providers without graphic knowledge. 
Instead, we leverage a depth- and scribble-conditioned ControlNet in a somewhat more involved strategy\footnote{An inference example using depth- and scribble-conditioned ControlNet could be referred to Figure~\ref{fig::memovis::controlnet-scribble-depth-compare} in Appendix~\ref{sec::memovis::app::compare_controlnet_depth_scribble}.}~\cite{Zhang2023ControlNet, Zhao2023uni}. 
For example, let's say the design feedback is to replace the computer display with a curved one in the scene depicted in Figure~\ref{fig::memovis::controlnet-scribble-demo}a with the depth map shown in Figure~\ref{fig::memovis::controlnet-scribble-demo}b, the feedback provider only has to scribble the rough shape of the new curved computer display to be added, as in Figure~\ref{fig::memovis::controlnet-scribble-demo}c, and provide a text prompt. 
MemoVis starts by creating the input conditions to ControlNet. It extracts the scribble from the initial design (Figure~\ref{fig::memovis::controlnet-scribble-demo}a) using \textbf{H}olistically-nested \textbf{E}dge \textbf{D}etection (HED)~\cite{Xie2015HED} and aggregated it with the manually-drawn scribbles (Figure~\ref{fig::memovis::controlnet-scribble-demo}e). 
The depth map (Figure~\ref{fig::memovis::controlnet-scribble-demo}d), aggregated scribbles and text prompt, are then fed to ControlNet to generate an image~(Figure~\ref{fig::memovis::controlnet-scribble-demo}f).
Empirically, we set the strengths for scribble to $0.7$ and depth condition to $0.3$ emphasizing the higher importance of the newly added object(s).

In addition to changing the geometry, the generated image might modify the texture of the current design which is undesirable. To address this, we leverage automatic segmentation techniques to merge the generated object from the generated image $\bm{I}_{syn}$ \ie~the {``the curved computer display''}, back into the original render $\bm{I}_{init}$.
To achieve this, we compute the bounding box of the user scribbles (red box in Figure~\ref{fig::memovis::controlnet-scribble-demo}e) and find the most salient object within the bounding box using SAM~\cite{Kirillov2023SAM}, leading to a segmentation mask $\bm{I}_{seg}$~(Figure~\ref{fig::memovis::controlnet-scribble-demo}g).
MemoVis~then creates the final reference image shown in Figure~\ref{fig::memovis::controlnet-scribble-demo}j by composition:  $\bm{I}_{syn} \odot \bm{I}_{seg} + \bm{I}_{init} \odot (\mathbf{1} - \bm{I}_{seg}) $, where $\odot$ indicates broadcasting and element-wise multiplication.

This approach allows to visualize the newly added objects, but part of the initial object, \ie~the current display can remain visible, leading to unpleasing visual artifacts, circled in red in Figure~\ref{fig::memovis::controlnet-scribble-demo}j.
To address this, MemoVis detects the mesh primitives to be removed from the image bounding box and depth map, and re-render $\bm{I}^{\prime}_{init}$, which the same image as $\bm{I}_{init}$ but without the object to be removed. Replacing $\bm{I}_{init}$ by $\bm{I}^{\prime}_{init}$ in the composition leads to the final results, displayed to the user, and visualized in Figure~\ref{fig::memovis::controlnet-scribble-demo}i.
Algorithm~\ref{alg::remove_hidden_obj} in the Appendix~\ref{sec::memovis::app::algo} recaps the algorithm.

\vspace{0.1in}
\noindent{\bf Modifier 2 : Grab'n Go Modifier with GenAI Design Layer}

\begin{figure*}[t]
    \centering
    \includegraphics[width=\textwidth]{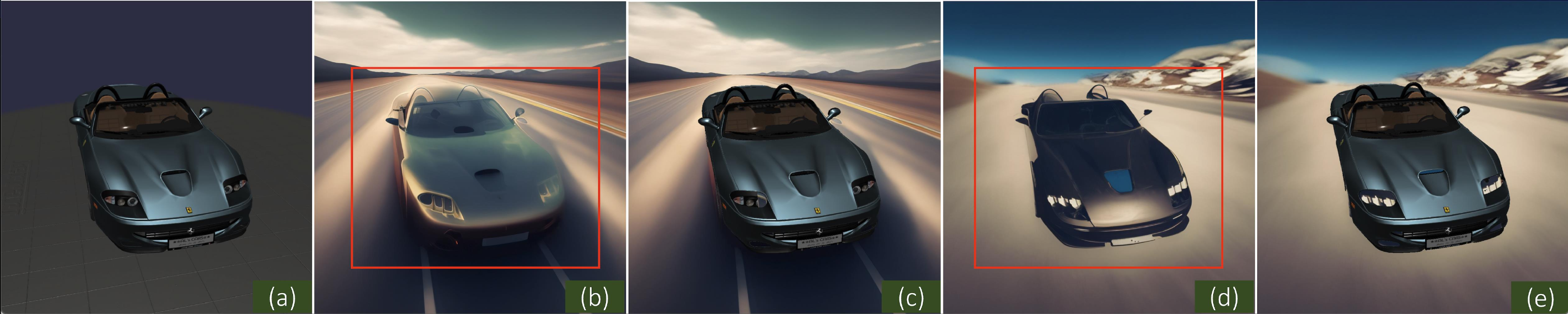}
    \caption[Examples showing how the feedback providers can stage the 3D model into different scene with the grab'n go modifier.]{Examples showing how the feedback providers can stage the 3D model into different scene with the grab'n go modifier. (a) The initial 3D design of a car; (b, d) synthesized image generated by scribble + text modifier with ControlNet conditioned on depth. The prompt {``a Ferrari car driving on the highway''} and {``a Ferrari car driving on a dessert''} were used to synthesize (b) and (d), respectively. The red bounding boxes show the areas drawn by feedback providers; (c) final composed image by bringing initial design into the scene of (b); (e) final composed image by bringing initial design into the scene of (d). }
    \label{fig::memovis::staging}
\end{figure*}

\noindent The {\it grab'n go modifier} is an easy selection tools that allows the user to compose an object from the rendered image into a generated image. 
For instance, considering the car in Figure~\ref{fig::memovis::staging}a, we can generate an image of this car staged in various backgrounds (Figure~\ref{fig::memovis::staging}b and Figure~\ref{fig::memovis::staging}d) using the depth-conditioned ControlNet~\cite{Zhang2023ControlNet}. 
However, the car might have undesirable texture variations. The feedback providers can simply draw red rectangles to select the car object and replace it with the exact car in the current design, thus staging it in the desired environment.

\begin{figure*}[t]
    \centering
    \includegraphics[width=\textwidth]{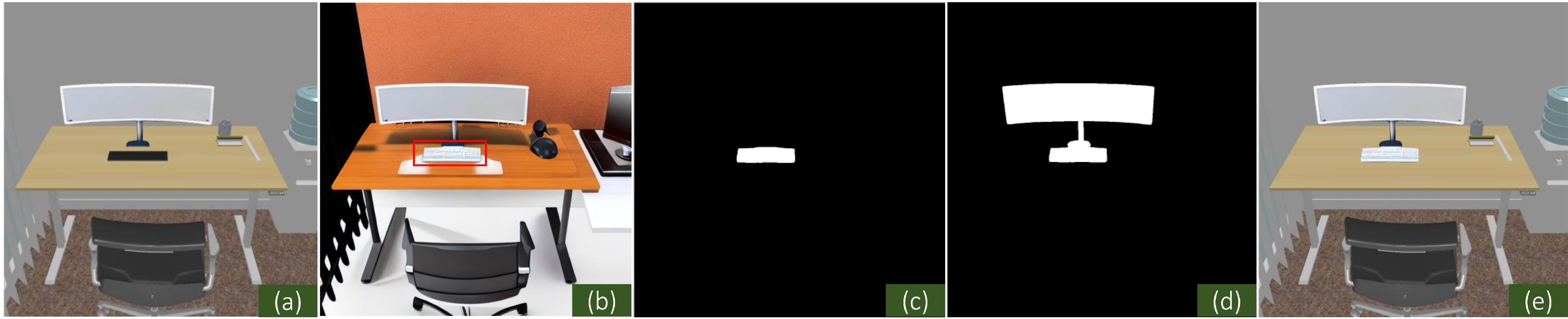}
    \caption[Examples of continuous composing.]{Examples of continuous composing. (a) Reference image of Figure~\ref{fig::memovis::controlnet-scribble-demo}i; (b) feedback provider can draw a bounding box to indicate their intention to add the white keyboard design into the reference image; (c) segmented mask generated by SAM; (d) segmented mask by compute the union of (c) and Figure~\ref{fig::memovis::controlnet-scribble-demo}g; (e) final reference image after including the white keyboard.}
    \label{fig::memovis::controlnet-scribble-continuous-composition}
\end{figure*}

The grab'n go modifier can also be used to do the reverse, \ie~composing objects from the generated images into the current design. For example, if the feedback providers wants to also include the generated keyboard in our previous example with the curved display in Figure~\ref{fig::memovis::controlnet-scribble-continuous-composition}b, they can simply draw a crop on the GenAI design layer. 
To achieve this, similar to the first modifier, we give the bounding box drawn by the user as input to SAM~\cite{Kirillov2023SAM}, and select the highest scored region as the output~(Figure~\ref{fig::memovis::controlnet-scribble-continuous-composition}c). MemoVis then computes a final segmentation mask by \textit{union}-ing Figure~\ref{fig::memovis::controlnet-scribble-continuous-composition}c and Figure~\ref{fig::memovis::controlnet-scribble-demo}g that would be used to create final reference image (Figure~\ref{fig::memovis::controlnet-scribble-continuous-composition}e). 
Formally, the final segmentation mask after $i$ times of applying {grab'n go modifier} could be computed by $\bm{I}_{seg} = \bm{I}_{seg_{i - 1}} \cup \bm{I}_{seg_i}$.
Note how this is significantly simpler than professional image editing software which commonly use the \emph{Lasso} tool~\cite{LassoPhotoshop}. 
As suggested by \textbf{DC3} emphasizing the needs for users without image editing skills, the interactions around grab'n go modifier enables feedback providers without professional image editing skills to efficiently create companion images.

\vspace{0.1in}
\noindent{\bf Modifier 3 : Text + Paint Modifier with Painting Design Layer}

\begin{figure*}[t]
    \centering
    \includegraphics[width=\textwidth]{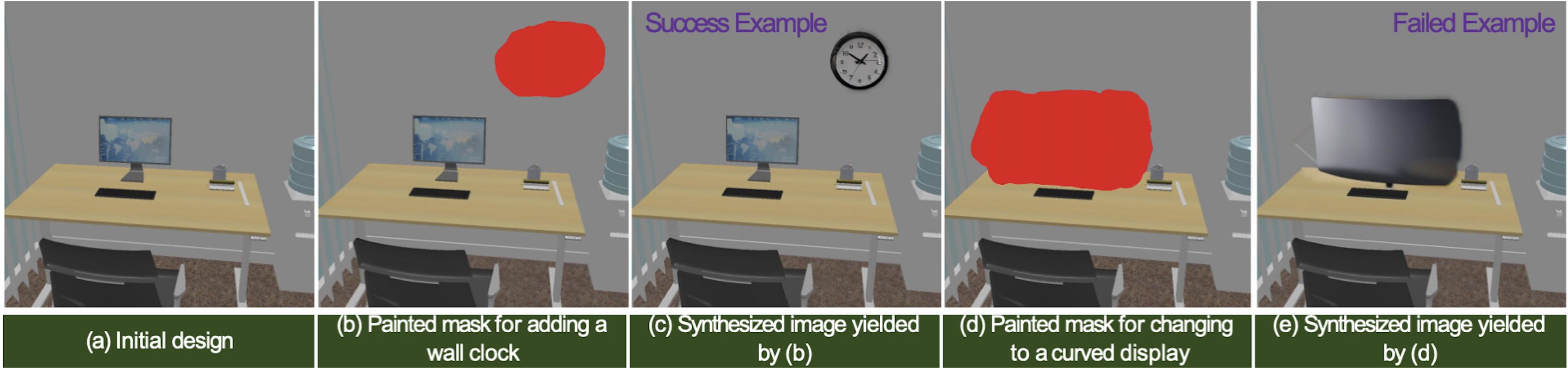}
    \caption[Examples of visualizing feedback that is related to adding new additional object with text + paint modifier.]{Examples of visualizing feedback that is related to adding new additional object with text + paint modifier. (b - c) shows a wall clock is successfully added to the view. (d) aims to add a curved computer display which is more complicated in terms of shapes, geometries and orientations. (e) demonstrates a failed attempt with {text + paint} modifier.}
    \label{fig::memovis::controlnet-inpaint-scribble-compare}
\end{figure*}

\noindent MemoVis integrates Stable Diffusion Inpainting~\cite{Rombach2022SD} as a {\it text + paint modifier}. The user paints a selection mask on the canvas, provides a text prompt and the model generates an image. This can be used to add simple objects; remove objects; or rapidly fix the glitches caused by {text + scribble} and {grab'n go} modifiers.
This tool complements the {text + scribble modifier}.
Figure~\ref{fig::memovis::controlnet-inpaint-scribble-compare} shows an example of adding a wall clock with {text + paint} modifier (Figure~\ref{fig::memovis::controlnet-inpaint-scribble-compare}a - c). 
When trying to add a curved computer display, {text + paint} modifier fails to convey the intention of the design feedback (see Figure~\ref{fig::memovis::controlnet-inpaint-scribble-compare}d). We thus argue for the necessity of user scribbling to add more complicated object, which would be hard to describe in details with text.

\begin{figure*}
    \centering
    \includegraphics[width=\textwidth]{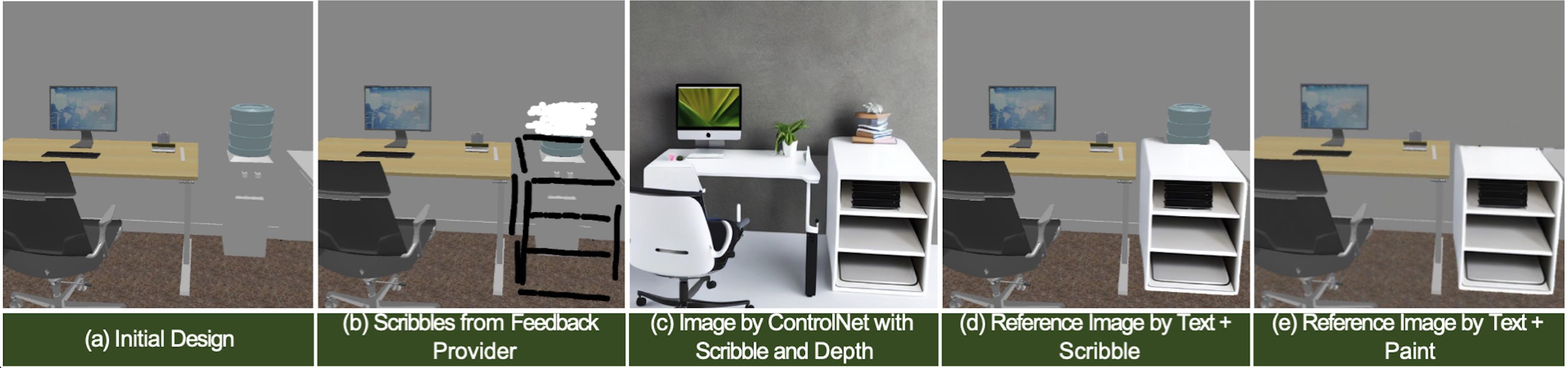}
    \caption{Text + paint modifier can be used to fix the glitches of the images created by text + scribble and grab'n go modifiers.}
    \label{fig::memovis::controlnet-inpaint-fix}
\end{figure*}

MemoVis can also fix artifacts in {text $+$ scribble} and {grab'n go modifiers}.
Those artifact occasionally arises in Algorithm~\ref{alg::remove_hidden_obj}, when the residual area occupies more than $70\%$ of the areas of the corresponding meshes (\ie~$r > r_{th}$).
For example, to change the water dispenser in an office to a storage drawer (Figure~\ref{fig::memovis::controlnet-inpaint-fix}a), the feedback provider might leverage the {text + scribble} modifier to roughly draw the shape of a drawer (Figure~\ref{fig::memovis::controlnet-inpaint-fix}b), leading to the drawer being extracted from synthesized image (Figure~\ref{fig::memovis::controlnet-inpaint-fix}c) and added to the initial design (Figure~\ref{fig::memovis::controlnet-inpaint-fix}d).
While Figure~\ref{fig::memovis::controlnet-inpaint-fix}d may convey the ideas of adding a storage drawer next to the standing desk, the water dispenser is misleading and needs to be removed.
Feedback provider can continuously use the {text + paint modifier} to easily remove the residual areas of the top of water dispenser, leading to a more faithful final reference image (Figure~\ref{fig::memovis::controlnet-inpaint-fix}e).

\subsection{Interactions and Implementations}\label{sec::memovis::system::interaction_implementations}
Figure~\ref{fig::memovis::teaser} shows the design of MemoVis.
With the text $+$ scribble modifier, the feedback provider scribbles while holding the mouse click.
The feedback provider can use the left and right mouse buttons to indicate the geometry of new objects and the unwanted areas, respectively.
With grab'n go modifier, the feedback providers can easily draw a bounding box using either the left or the right mouse button.
The left and right mouse button indicate keeping and removing the key object(s) inside the enclosed box of the synthesized image in the reference image, respectively.
Finally, with the text $+$ paint modifier, MemoVis enables feedback providers to click and drag left and right mouse button to specify the areas for adding (or changing) and removing objects of interests, respectively.

MemoVis was prototyped as a browser-based application to reduce the needs for feedback providers to install large-scale standalone 3D software, as partial engineering efforts to address DC3.
We used Babylon.js v$6.0$~\cite{babylonjs} to implement the 3D explorer for rendering the 3D model.
A flask server was implemented to process the inference workloads, deployed on a GPU-enabled cloud server. 
Further details of the pre-trained models we used can be referred to Appendix~\ref{sec::memovis::app::models}.

\section{Evaluations}\label{sec::memovis::evaluation}
We conducted two user studies to evaluate MemoVis. 
The first study aims to understand the experience of feedback providers while visualizing textual comments using MemoVis, whereas the second study aims to explore how the created images by MemoVis could effectively convey the gist of 3D design feedback while being consumed by designers.
We use PF\# and PD\# to index \textbf{P}articipants for acting as \textbf{F}eedback providers and \textbf{D}esigners, respectively.

\subsection{Study 1: Creating Reference Images}\label{sec::memovis::eval::create_feedback}
Our first study was structured as a {\it within-subjects design}. 
We aim to tackle two \textbf{R}esearch \textbf{Q}uestions (RQs):

\vspace{0.1in}
\begin{itemize}[noitemsep, topsep=0pt, leftmargin=*]
    
    \item \textbf{(RQ1)} How the real-time viewpoint suggestion could help feedback providers navigate the 3D scene while writing feedback?
    
    \item \textbf{(RQ2)} How the three types of image modifiers could help feedback providers visualize the textual comments for 3D design?
    
\end{itemize}
\vspace{0.1in}

\vspace{+0.1in}\noindent{\bf Participants.}
PF1 - PF14 (age, $M = 23.36$, $SD = 2.52$, \incl~seven males and seven females) were recruited as the feedback providers.
While all participants have experience of writing design feedback, most participants had limited experience working with 3D models. We believe that the design feedback providers do not necessary to possess design skills. Among recruited participants, only PF2, PF10, and PF14 were confident of their 3D software skills.
Most of participants did not have experience of creating prompts and using GenAI, although PF1 and PF2 considered themselves as experts of using LLM; PF13 was confident of his proficiency of using text-to-image GenAI. 
Details of participants' demographics and power analysis can be referred to Appendix~\ref{sec::memovis::app::evaluation_participant}.

\vspace{+0.1in}\noindent{\bf Interface Conditions.}
Participants were invited to use two interface \textbf{C}onditions (C1 - C2) to create and visualize textual comments:

\vspace{0.1in}
\begin{itemize}[noitemsep, topsep=0pt, leftmargin=*]

    \item \textbf{(C1)~Baseline.}~Participants were required to create reference image(s) using search and/or hand sketching. We aim to mock up current design review practices based on the findings from Formative Study 1. Participants were not required to use existing GenAI-based image editing tools like Photoshop~\cite{photoshopAI}; while FP2 acknowledged GenAI as a promising approach to generate reference images, its lack of control makes it impractical (Section~\ref{sec::memovis::formative::needs}); unlike FP2 who are professional designers, feedback providers may not possess skills on using image editing software. Participants were instructed to use their preferred search engine for finding well-matched images. PowerPoint was optionally used for sketching and annotations.
    
    \item \textbf{(C2)~MemoVis.}~Participants were invited to use MemoVis to create reference images along with textual comments.
    
\end{itemize}

\vspace{0.1in}
\noindent{\bf Tasks.}
Each participant was instructed to complete three different design critique \textbf{T}asks (T1 - T3) with C1 and/or C2. 
To prevent learning effect, T1 - T3 used \emph{different} 3D models, created by \emph{different} creators. 
For each design critique task, participants were instructed to create at least two design comments, with each comments being accompanied by at least one reference images.
We used T1 to help participants get familiar with C1 and C2, with which participants were instructed to critique a character model of a samurai boy. 
All data collected from T1 was excluded from all analysis. 
For T2 and T3, participants were asked to make the bedroom and the car more comfortable to live in and drive with, respectively.
While T2 and T3 used different 3D models, the skills needed to create design feedback are the same.
Details of the study tasks could be referred to Figure~\ref{fig::memovis::study-models} and Appendix~\ref{sec::memovis::app::study_tasks}.

\vspace{+0.1in}\noindent{\bf Procedures.}
Participants were invited to complete a questionnaire to report demographics, 3D software, 3D design, and GenAI experiences.
They were then introduced to MemoVis, and were given sufficient time to learn and get familiar with both C1 and C2 while completing T1.
For those without prompt writing experience for GenAI, sufficient time was provided to go through examples on Lexica~\cite{lexicaart} and practice using FireFly~\cite{AdobeFirefly}.
Upon feeling comfortable with C1 and C2, participants were invited to complete T2 and T3.
To prevent the sequencing effects, we counterbalanced the order of interface conditions.
Specifically, PF1 - PF7 were required to complete T2 with C2, followed by completing T3 with C1.
Whereas, PF8 - PF14 were required to complete T2 with C1, followed by complete T3 with C2.
Comparing feedback creations for T2 and T3 were out of our scope, we therefore did not counterbalance the order of the tasks.
After each task, participants were invited to rate their agreement of four \textbf{Q}uestions (Q1 - Q4) in a $5$-point Likert scale, with a score $>3$ being considered as a positive rating.

\vspace{0.1in}
\begin{itemize}[noitemsep, topsep=0pt, leftmargin=*]
    
    \item \textbf{(Q1) Navigating the Viewpoint}: {\it ``it was easy to locate the viewpoint and/or target objects while creating feedback.''}
    
    \item \textbf{(Q2) Creating Reference Images}: for the task completed by MemoVis (C2), we used the statement {\it ``it was easy to create reference images with the image modifiers for my textual comments''}. For the task completed by baseline (C1), we used the statement {\it ``it was easy to create reference images with the method(s) I chose''}.

    \item \textbf{(Q3) Explicitness of the Reference Images}: {\it ``the reference images easily and explicitly conveyed the gist of my design comments.''}

    \item \textbf{(Q4) Creativity Support}: {\it ``the reference images helped me discover more potential design problems and new ideas.''}
\end{itemize}
\vspace{0.1in}

A semi-structured interview was also conducted focusing on participants' rationales while evaluating Q1 to Q4.
The study on average took $57.34$ min ($SD = 7.72$~min).  

\vspace{+0.1in}\noindent{\bf Measures and Data Analysis}.
To address RQ1, we measured the {\it navigating time} for each textual comment, defined by the time that feedback providers spent while navigating and exploring the 3D model.
With the Shapiro-Wilk Test~\cite{Shapiro1965}, we verified the normal distribution of the measurements under each condition ($p > .05$).
One-way \textbf{An}alysis \textbf{o}f \textbf{Va}riance (ANOVA)~\cite{Girden1992} ($\alpha = .05$) was therefore used for statistical significance analysis.
The eta square ($\eta^2$) was used to evaluate the effect size, with $.01$, $.06$ and $.14$ being used as the empirical thresholds for {\it small}, {\it medium} and {\it large} effect size~\cite{Cohen1998}.
We used thematic analysis~\cite{Braun2012} as the interpretative qualitative data analysis approach, along with deductive and inductive coding approach~\cite{Bingham2021} to analyze participants' responses during semi-structured interviews, to better understand participants' thoughts and uncover the reasons behind the measurements and survey responses.
We used the initial codes from Q1 - Q4.
We first read the transcripts independently and identified repeating ideas using initial codes derived from Q1 to Q4. 
Next, we inductively come up with new codes and iterate on the codes as sifting through the data.
The final codebook can be found in Figure~\ref{fig::memovis::codebook-study-1} in Appendix~\ref{sec::memovis::app::codebook}.

\begin{figure*}[t]
    \centering
    \includegraphics[width=\textwidth]{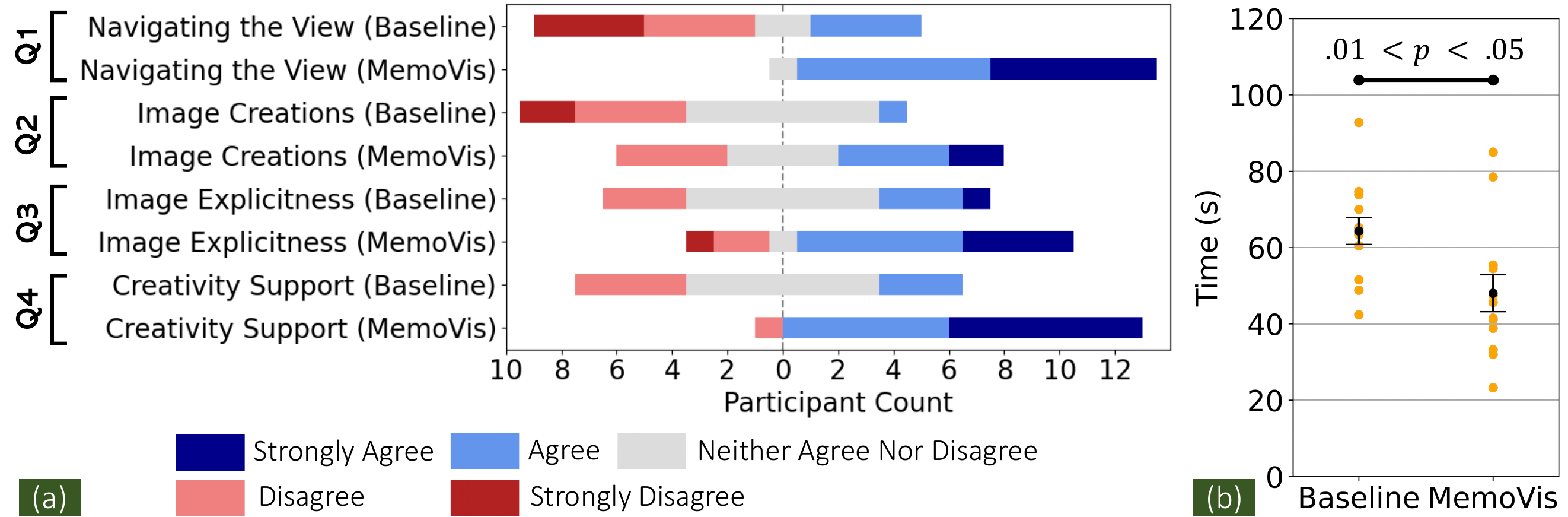}
    \caption[Survey responses and navigating time measurements from Study 1.]{Survey responses and navigating time measurements from Study 1. (a) Participants' responses of Q1 - Q4; (b) the total navigating time of baseline and MemoVis. PF14 was excluded from Figure~\ref{fig::memovis::study-1-results}b, as the participant did not use the viewpoint suggestion feature.}
    \label{fig::memovis::study-1-results}
\end{figure*}

\vspace{+0.1in}\noindent{\bf Results and Discussions}

\noindent Most participants found our viewpoint suggestion features useful~(RQ1) and the image modifiers easy to use to visualize textual comments~(RQ2). Most participants also believed the reference images created with MemoViscould easily and explicitly convey the gist of 3D design feedback. Patterns of usages of each image modifiers could be referred to Figure~\ref{fig::memovis::study-1-modifier-usages} in Appendix~\ref{sec::memovis::app::modifier_usages}.

\vspace{+0.1in}\noindent$\bullet$~{\bf How the real-time view suggestions could help feedback providers navigate inside 3D explorer (RQ1)?}
Overall, $13/14$ participants (except PF14) leveraged the viewpoint suggestion features while visualizing the textual comments.
More participants positively believed that it was easy to locate the viewpoints and target objects with MemoVis, compared to the baseline~($13$~\vs $4$, Figure~\ref{fig::memovis::study-1-results}a). 
Figure~\ref{fig::memovis::study-1-results}b shows a significant reduction ($F_{1, 24} = 7.398$, $p = .018$, $\eta^2 = .236$) of navigating time while using MemoVis ($M = 44.06$s, $SD = 18.94$s) \vs~the baseline ($M = 66.67$s, $SD = 16.94$s).
Notably, the normality of the measurement was verified ($p_{baseline} = 0.54$, $p_{MemoVis} = 0.17$).
Our qualitative analysis suggested two benefits brought by the viewpoint suggestion feature.

\vspace{+0.1in}
\noindent{\textbf{\textit{Providing guides to find the viewpoint contextualized on the design comments.}}}
Most participants appreciated the help and guides brought by the viewpoint suggestion feature.
For example, 
{\it ``the view angle is good that it's trying to give you a context''}~(PF1) 
and {\it ``it helped me a lot to find a nice view where I could create reference image''}~(PF2).
PF8 also suggested the potential benefits for feedback providers to make faster decision: {\it ``it helps me to make faster decisions. Sometimes I don't know which views might be better. And it gives me options that I could choose''}~(PF11).
After trying baseline, PF11 emphasized: {\it ``[without viewpoint suggestion] I have to decide on how to look. I have to make bunch of decisions in the way. It is pretty cognitively demanding''}.
PF7, who did not have prior 3D experience, initially felt {\it``confused about the directions of the viewpoint''}, while attempting to navigate the 3D scene using the 3D explorer.
After using the viewpoints suggested by MemoVis, she felt {\it ``it's a much better view, as the bed could be seen from a nice view angle''}.

\vspace{+0.1in}
\noindent{\textbf{\textit{Locating target object(s) in a scene.}}}
Despite occasional failures and the needs of minor adjustments, most participants appreciated the benefits of being able to quickly locating target objects.
For example, PF8 acknowledged: {\it ``I feel that around 85\% of the time that the system could give me the right view that I expected, although I sometimes might still need to adjust like a zoom''}.
PF4 justified the reason of not {\it strongly agreeing} with Q1: {\it ``although it was helpful, like the system gave me a nice view suggestion. But I had to move it manually. Although that was a good starting point, I still need to make adjustments by myself''}.
During the interview, PF2 believed that {\it ``the view suggestions would be more useful for the bigger scene''}.
With the past experience of designing 3D computer games, he further commented: {\it ``sometimes I work in video games. And video game maps could become really large. And there are multiple things. For example, there's a very specific area that I find to go to, and to edit it. And then if I can type and say, for example, the boxes on the second floor of the map. And it instantly teleport me to there. Then, there is gonna save me a lot of time to find it in the hierarchy''}.
After PF14 critiqued a car design with MemoVis, he commented: {\it ``I think moving around with car was easier just because it's a car instead of the room. But I believe for the bedroom it would be much helpful to have the view suggestion feature that kind of guide''}.

\vspace{0.1in}\noindent$\bullet$~{\bf How MemoVis could better support feedback providers to create and visualize textual comments for 3D design (RQ2)? }
Figure~\ref{fig::memovis::study-1-results}a shows that compared to the baseline, more participants rated positively in terms of reference image creations ($6$~\vs~$1$ for Q2), explicitness of the images ($10$ \vs~$4$ for Q3), and creativity support ($13$ \vs~$3$ for Q4).
Feedback provider participants have overall used $48$~times ($42.11$\%) of text $+$ scribble modifier, $30$~times ($26.32$\%) of grab'n go modifier, and $36$~times ($31.57$\%) of text $+$ paint modifier (see Figure~\ref{fig::memovis::study-1-modifier-usages} in Appendix~\ref{sec::memovis::app::modifier_usages} for the detailed usages of the image modifiers). 
While completing the baseline task, despite the challenges of searching well-matched internet images, participants used three main strategies to make the reference images more explicit. 
Specifically, four participants used annotations to highlight the areas that the textual feedback focus;
One participant (PF3) used two reference images to demonstrate a good and bad design, and expected the designers to understanding gist by contrasting two extreme examples;
And two participants (PF2, PF13) provided multiple reference images and hoped the designers could extract different gist from separate images.

\vspace{0.1in}
\noindent{\textbf{\textit{Image explicitness.}}}  
Participants appreciated the explicitness of the reference images created by MemoVis, while keeping the contexts of the initial design.
For example, {\it ``I like how it \underline{keeps the context} around this picture. It could be much easier for people to understand my thoughts''}~(PF2), {\it ``it allows me to \underline{generate reference image in the same scenario}, and not some random scenario that I've pulled from the internet''}~(PF9), and {\it ``this image references are pretty easy and explicitly. It just \underline{conveys my point}. That's what matters. Now it's up to the person to make the decision to how have to make it better''}~(PF11).
Some participants like PF6 prefer to the MemoVis-created reference images, compared to the searched and hand-sketched images.
PF11 highlighted the easy and convenience workflow: {\it ``this process was pretty easy. The reference image is just like pop up to me! This is something I love. Like when I provide feedback while I was teaching, I didn't explicitly tell the students like, your typography is really hard to read, you should change it to this font. But something like a bigger front, a different style, just like that.''}~
In contrast, after creating feedback with baseline task, PF13 complained: {\it ``you can find tones of image on the Google. They are very realistic. They are very decorative. But it's just not related to my model, it's not in the context [...] when I use an internet image, there are more details, but there is even more confusing part. So many times, I just tweak my textual feedback, to minimize the possible confusions for the designers [...] I think if I were doing by myself, I would just spend some more time and use Photoshop to edit the internet images.''}~
P14 also commented: {\it ``I typical have a specific image in mind. I think to come up with that design, [the MemoVis] is much much better. When I search for something, it is typically very generic. I never search something that is very specific like, a bedroom with blue walls or something. It's easy if you're coming in with a specific design.''}

\begin{figure*}[t]
    \centering
    \includegraphics[width=\textwidth]{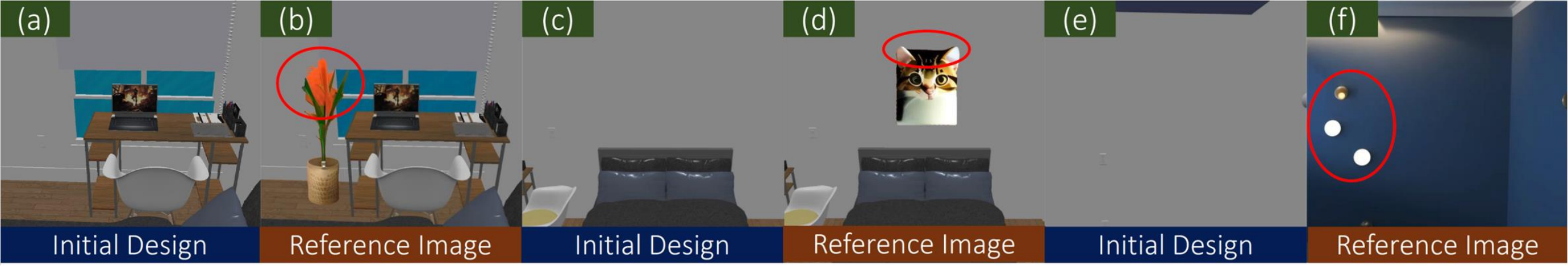}
    \caption[Examples of inspirations and creativity support.]{Examples of inspirations and creativity support. (a, c, e) show the selected viewpoints of the initial design; (b, d, f) show the created visual references by PF4. The reported unexpected components are highlighted by red circle.}
    \label{fig::memovis::study-examples}
\end{figure*}

\vspace{0.1in}
\noindent{\textbf{\textit{Unexpected inspirations and creativity support.}}}
Most participants recognized the benefits of MemoVis for inspiring new ideas, which is similar to FP2's comments (Section~\ref{sec::memovis::formative::needs}).
For example, {\it ``there are just so many possibilities. If you search through Google, your thoughts are limited by your experience. But this tool could give me so much unexpected surprises, which is a good thing and they are many times actually better than what I thought. I think it works quite well to help inspire more new ideas.''}~(PF12)
PF13 particularly enjoyed the mental experience of getting inspired iteratively while refining the textual prompts: {\it ``when I was typing like the description, it was just a text like a description. I don't really have a solid image in my brain. But [MemoVis] helps me shape my idea, and helps me better think iteratively [...] it's like when you're writing papers, instead of starting from the scratch, you have a draft, so it's easier to discuss and revise based on it [...]''}~ 
Figure~\ref{fig::memovis::study-examples} demonstrates three examples with unexpected creativity that were appreciated by PF4.
For example, upon seeing Figure~\ref{fig::memovis::study-examples}b, PF4 thought out loud that: {\it `` I wasn't expecting the plant to have like the little orange leaves, but I think it was a good idea for the final design.''}

\vspace{+0.1in}
\noindent{\textbf{\textit{Simple and easy image creation workflow.}}}
Most participants appreciated the simple workflow for creating reference images with MemoVis, and felt {\it ``very easy to use''}~ (PF14).
This was also evident by the responses for Q2 reported in Figure~\ref{fig::memovis::study-1-results}.
Participants highlighted the merits of easy workflow facilitated by the conveniences of all image modifiers.
For example, {\it ``the system is very easy to use, although it might require a trial or two to become familiar with the modifiers. Once I understand, it is really handy and convenient to instantly create reference images that I want, which also match the design and my comments. All image modifiers are really useful, essential and indispensable!''} (PF1)~
PF9 particularly appreciated the option of {\it ``select and extract''} supported by grab'n go modifier: {\it ``I think the workflow is pretty good. [...] The image generation is not the only part. But I'm getting the option to select and extract [the new objects that I am trying to suggest], instead of just using some random image on Google [PF9 refers to the baseline interface condition].''}
Participants also valued the simplicity and effectiveness of composing text-to-image prompts. 
For example, even without prior GenAI experience, PF14 commented {\it ``it was really easy to write prompt. I love the flexibility to write prompts. It also helped me think.''}

\begin{figure}
    \centering
    \includegraphics[width=0.9\textwidth]{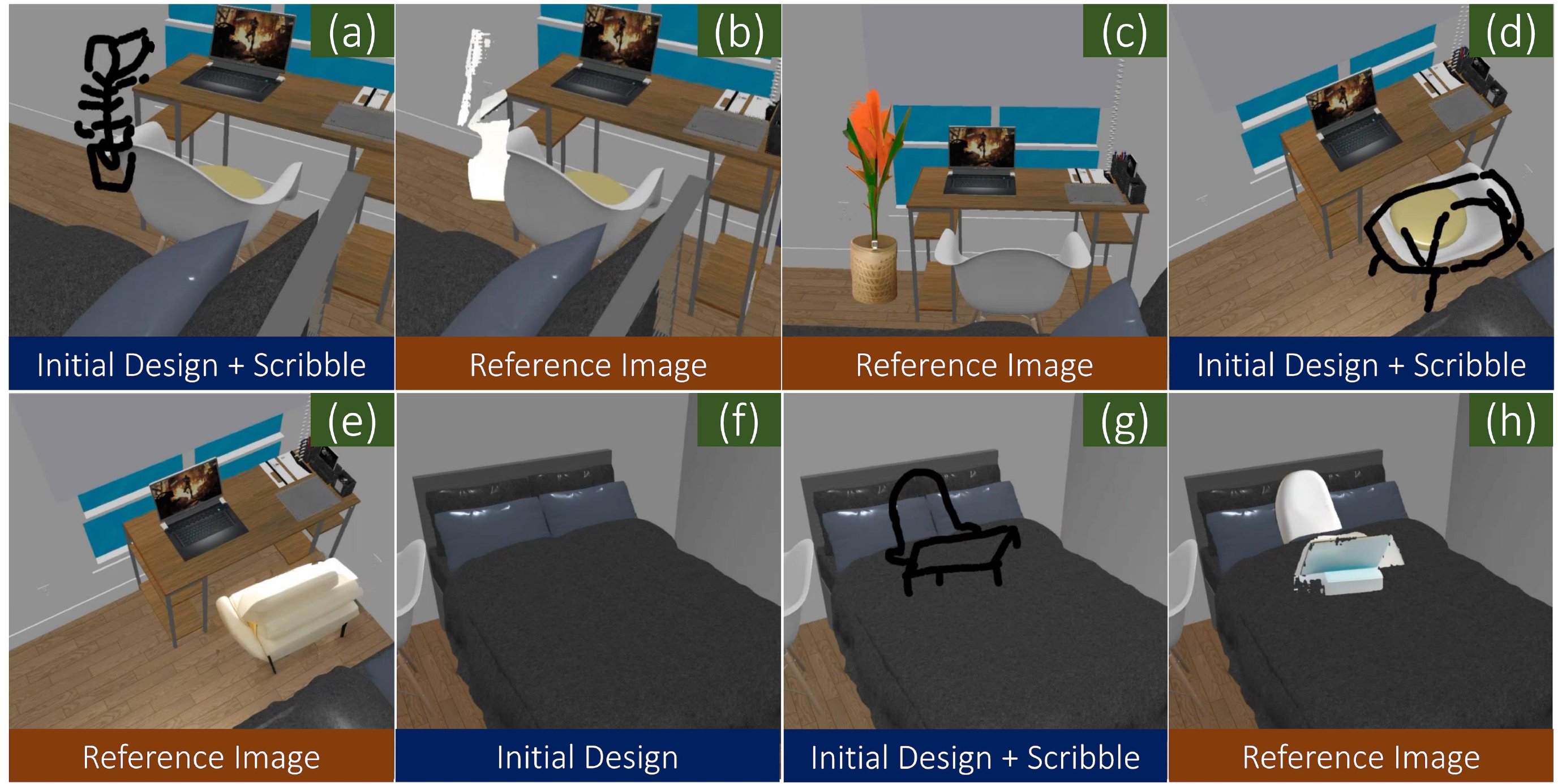}
    \caption[Examples of unsatisfactory reference images.]{Examples of unsatisfactory reference images. (a) initial sketch from PF4 to describe an indoor plant. (b) MemoVis failed to generate the desirable image. (c) with a different camera view, PF4 successfully created the expected reference image. (d) initial sketch from PF2 to describe a new chair in the office. (e) MemoVis was able to generate a new chair but it didn't match PF2's expectation. (f - h) initial design, scribble, and final created reference image by PF7.}
    \label{fig::memovis::study-examples-failed}
\end{figure}

\vspace{0.1in}\noindent\textbf{\textit{Scenarios when feedback providers fail to create explicit and satisfying companion reference images. }}
We observed multiple scenarios where participants failed to create satisfying visual references that can explicitly convey the textual feedback. 
Our analysis unveiled three key reasons.

\vspace{0.1in}
\noindent{\textbf{{(1)~Unsatisfactory generation.}}}
Text-to-image generation is still an emerging technology and therefore is not perfect~\cite{Liu2022}. MemoVis could fail completely to generate a reasonable shape (Figure~\ref{fig::memovis::study-examples-failed}b) or was able to generate a reasonable image but failed to match what the users wanted (Figure~\ref{fig::memovis::study-examples-failed}d - e).
These setbacks may result in less explicit reference images, potentially causing misunderstandings for the designers and contributing to the negative ratings shown in Figure~\ref{fig::memovis::study-1-results}a.
For example, PF4 thought aloud while examining Figure~\ref{fig::memovis::study-examples-failed}b: {\it ``it doesn't look like a plant. I don't think this is explicit enough for the designer to understand''}.
In these cases, we observed that participants tend to try for several attempts or with a different view angle until they can generate the desirable image (Figure~\ref{fig::memovis::study-examples-failed}c).
While PF4 positively rated the image explicitness, she {\it disagreed} that it was easy to create reference images.
PF2 made a similar comment with respect to Figure~\ref{fig::memovis::study-examples-failed}e: {\it ``the chair does not look like the one I wanted.''}

\vspace{0.1in}
\noindent{\textbf{(2)~ Difficulty of writing prompts}.}
Few participants emphasized the importance of prompts and the challenges in writing them. 
For example, {\it ``sometimes, I need several times of revision on the prompt to make the generated image better. [...] So while my feedback takeaways could be visualized, it still needs several attempts.''}~(PF12)~
A small number of participants also described the needs for extracting the knowledge of the existing design: {\it ``if I make a prompt, it should have the knowledge of this existing design. For example, if I say like, keep the same blanket, then it should be same.''}~(PF5)~
PF11 suggested that MemoVis should automatically create prompts based on the textual feedback: {\it ``I thought when I write the feedback, the prompt will be automatically generated so that it can bring me the reference image. But it's more like I write the feedback and then I also create the prompt [...] It was initially hard for me to distinguish the prompt and the feedback itself, that I need to tell the AI versus also tell the designer.''}~
PF7, a non-native English speaker, found it challenging to phrase the prompt of ``lap desk'', leading to the complete failure of the final reference image created by MemoVis (Figure~\ref{fig::memovis::study-examples-failed}f - h).
This led him to rate the implicitness of the reference images (Q3) as \emph{strongly disagree}.

\vspace{+0.1in}
\noindent{\textbf{(3)~Lack of alternative design exploration}.}
While most participants believed that the images created by MemoVis are more explicit and could provide inspirational support compared to the baseline, some participants commented on the necessity of being able to see multiple inference results similar to mainstream search engines.
For example: {\it ``compared to the Google image, I feel there's something very inspiring about seeing like 20 images all at once from like different creators''}~(PF6),
and {\it ``when you search for an image on Google, it has the big long list of different images. That helps me to see different ideas. And because it's Google, it's pulling from a bunch of different websites. So I think that also helps me to think like, oh, this is what someone else thought. So yeah, I think if I'm thinking about it that way.''}~(PF4)~

\subsection{Study 2: Assessing Reference Images}\label{sec::memovis::eval::eval_feedback}
Our second study aims to tackle the key RQ: {\it how the reference images created by MemoVis could convey the gist of the 3D design feedback, compared to the images created by the baseline condition (\ie~internet searched images and/or hand sketches)?}

\vspace{+0.1in}\noindent{\bf Participants.}
We recruited PD1 - PD8 (age, $M = 23.75$, $SD = 2.55$, \incl~four males and four females) as the designer participants, with prior 3D design experience, from an institutional 3D design \& e\textbf{X}tended \textbf{R}eality~(XR) student society.
No designer participants were involved in the formative studies and the first study.
Details of the demographic backgrounds of designer participants can be referred to Appendix~\ref{sec::memovis::app::evaluation_participant}.

\vspace{+0.1in}\noindent{\bf Procedures.}
The second study was structured as a {\it survey study}, where participants were invited to complete an online questionnaire.
We first collected all reference images collected from first study, including $44$ and $39$~design feedback, created by C1 and C2, respectively.
Each design feedback contains a textual comment, and one (or multiple) reference image(s).
We also captured the viewpoints of the initially designed 3D models for each reference image.
Next, we randomly shuffled all design feedback, which were then divided into eight groups.
All groups, except for the last one containing six design feedback, consist of $11$ design feedback each.
On average, each group comprises $46.21\%$ ($SD = 9.54\%$) design feedback created by MemoVis, and $74.05\%$ ($SD = 9.21\%$) design feedback contributed by different feedback provider participants from the Study 1.
We then assigned the eight surveys to PD1 to PD8 for evaluation.
Participants were invited to rate each reference image in a $5$-point Likert scale, regarding {\it how well the reference image can convey the gist of the textual comment}.
Participants were encouraged to put their justifications as the textual comments for each rating.
Each questionnaire takes approximately \mbox{$10$ - $15$~min} to complete.

\vspace{+0.1in}\noindent{\bf Analysis.}
We analyzed the Likert-scale score for each reference image rated by one designer participants.
Qualitatively, we also collected the textual comments that participants provided to justify their ratings.
These include $33$~and $31$~textual comments for design feedback created by baseline and MemoVis, respectively.
Inductive coding approach~\cite{Bingham2021} were used to analyze participants' qualitative responses.

\vspace{+0.1in}\noindent{\bf Results and Discussions}

\begin{figure*}[t]
    \centering
    \includegraphics[width=\textwidth]{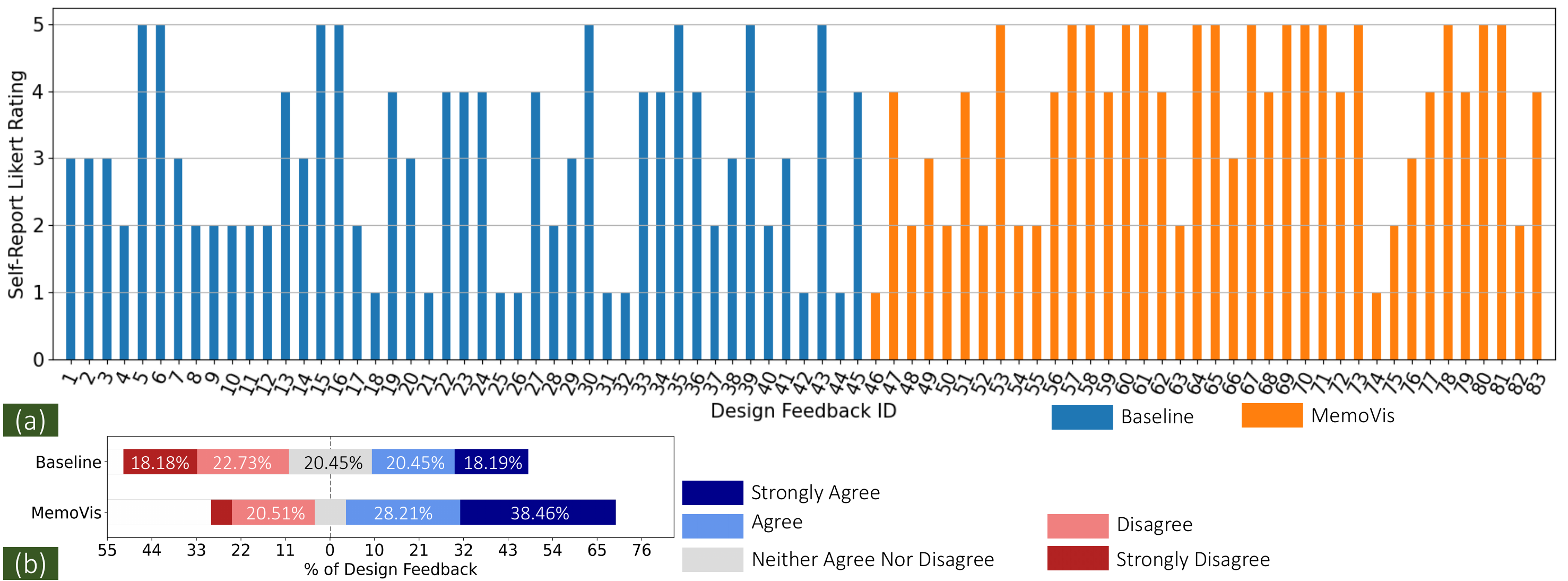}
    \caption[Survey responses of Study 2.]{Survey responses of Study 2. (a) Participants' response of each 3D design feedback, in a scale of \bm{$1$} to \bm{$5$} where \bm{$1$} indicates \emph{strong disagree} and \bm{$5$} indicates \emph{strongly agree}; (b) cumulative analysis of the \% of 3D design feedback for each Likert scale by two interface conditions. Note that Figure~\ref{fig::study-2-results}b was generated based on the survey responses shown in Figure~\ref{fig::study-2-results}a.}
    \label{fig::study-2-results}
\end{figure*}

\noindent Figure~\ref{fig::study-2-results}a shows the Likert rating for each feedback.
We found that around $66.67\%$ of the reference images created by MemoVis were rated positively, versus only around $38.64\%$ of images created by baseline approach were rated positively (Figure~\ref{fig::study-2-results}b). 

\begin{figure}[t]
    \centering
    \includegraphics[width=\textwidth]{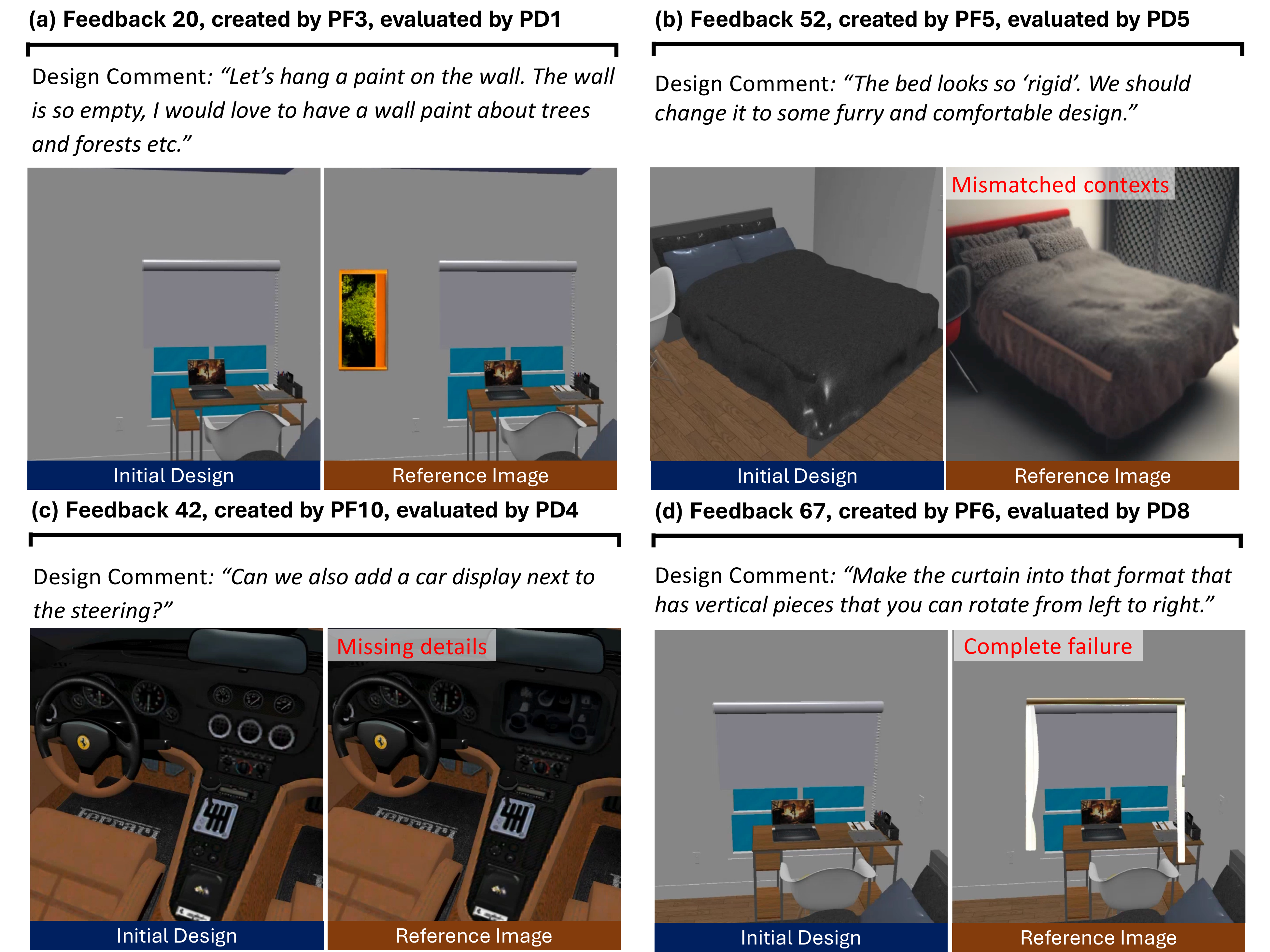}
    \caption[Examples of 3D design feedback created by MemoVis.]{Examples of 3D design feedback created by MemoVis. The participants \emph{strongly agreed}, \emph{neither agree nor disagree}, \emph{disagreed} and \emph{strongly disagreed} that the reference image from the 3D design feedback (a), (b), (c) and (d) can convey the gist of the textual comment, respectively.}
    \label{fig::memovis::study-2-baseline-example}
\end{figure}

When evaluating reference images and feedback produced using MemoVis, participants liked the explicitness of the generated reference images, \eg~ {\it ``clear and focused''}~(PD6), {\it ``the reference image perfectly captures all the elements mentioned in the design comments''}~(PD4) and {\it ``the image clearly shows what the text is trying to say''}~(PD1). 
Figure~\ref{fig::memovis::study-2-baseline-example}a shows an example of how PD1 believed that {\it ``it is easily to identify the new picture and the desired location''} in the initial bedroom 3D model.

\begin{figure}
    \centering
    \includegraphics[width=\textwidth]{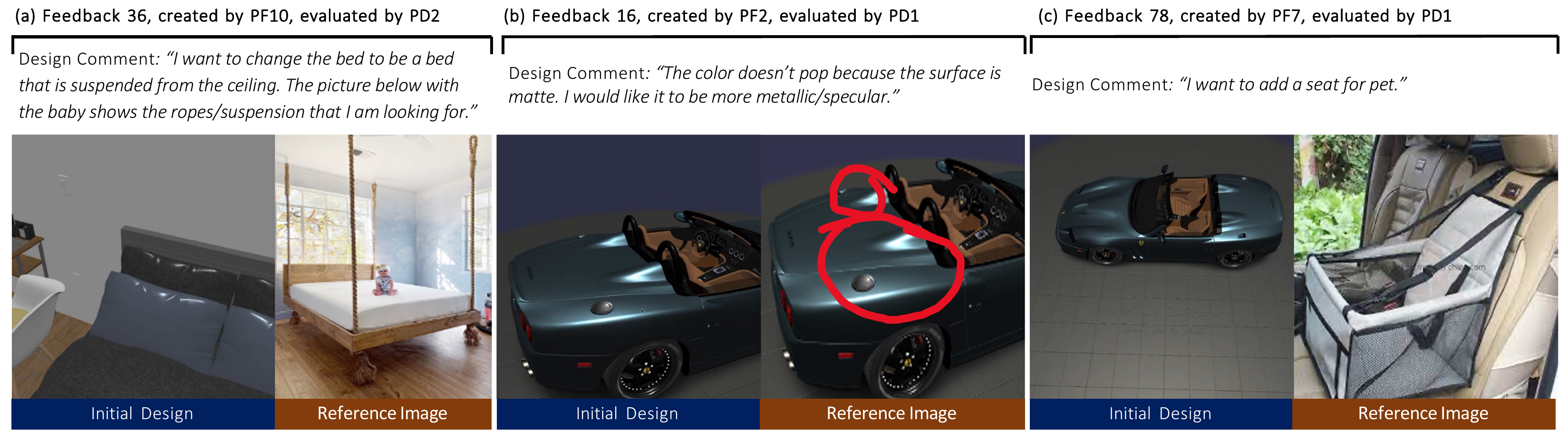}
    \caption{Examples of 3D design feedback created by the baseline interfaces. The red traces indicate the annotations drawn by the feedback provider participant PF2.}
    \label{fig::memovis::study-2-baseline-example-compare}
\end{figure}

For the images produced in the baseline condition, comments from the designer participants point to a common drawback of mismatched contexts: the contextual difference between the reference images and the textual comments can cause confusion. 
For example, Figure~\ref{fig::memovis::study-2-baseline-example-compare}a shows an example of the reference image created by PF10, where PD2 judged: {\it ``the structure between the bed and the floor can easily be mistaken for legs upon a cursory glance''}. 
Figure~\ref{fig::memovis::study-2-baseline-example-compare}c shows an example when both the viewpoint and the requested change in the reference image are dramatically different to the initial design. The designer participant did not feel fully confident in grasping the feedback: {\it ``the idea of a pet seat is clear, but it seems so different from the original image that it would be confusing.''}~(PD1)~
We also found that some feedback provided attempted to reduce ambiguity by annotating on the initial design to highlight the changes (Figure~\ref{fig::memovis::study-2-baseline-example-compare}b). This approach, however, is not as effective when the requested change is not explicit. In this case, the feedback provider was asking for a new material design on the car body. PD1 commented: {\it ``I'm not sure what the circles are trying to show.''}

Although most of participants favored the reference images created by MemoVis, participants also highlighted few setbacks of the MemoVis-generated images. 
We identified three key reasons from the designers' perspective.

\vspace{+0.1in}
\noindent{{\bf (1) ~Mismatched contexts}}.
Similar to the baseline condition, occasionally, designer participants pointed out context mismatches, although these did not affect their understanding of the design feedback.
Figure~\ref{fig::memovis::study-2-baseline-example}b shows an example of how the overall bedroom design was changed though the focus of the textual feedback is the bed, which might be caused by the failure of the grab'n go modifier.
Despite this, PD5 noted: {\it ``even the background is a little different, the reference image still preserves the angle of the view, and the new environment setting.''}

\vspace{+0.1in}
\noindent{{\bf (2) ~Missing details}}.
While some reference images can reflect the suggested edits, designer participants believed that the missing details might cause misunderstanding.
For example, regarding Figure~\ref{fig::memovis::study-2-baseline-example}c, PD4 \emph{disagreed} that the reference image can convey the textual comment because {\it ``while the display has been added as per the design comments, some panels have been removed. I don't know if these removals align with the design comments.''}~
Such confusion might result in incorrect translations of the design feedback into the final 3D model.
While mismatched contexts were identified as common problems, designer participants did not highlight missing details for the reference images created by baselines.

\vspace{0.1in}
\noindent{{\bf (3) ~Complete failure}}.
In very few cases, designer participants highlighted that the reference image can be entirely unsuccessful. For example, PD8 commented on the design feedback of Figure~\ref{fig::memovis::study-2-baseline-example}d: {\it ``I don't understand what is that white thing, doesn't look like a curtain.''}~
Such reference images might cause designers to misunderstand the gist of the design feedback, potentially necessitating further communication with the feedback providers.

\section{Discussions}\label{sec::memovis::discussion}
Having demonstrated the MemoVis~ system as an effective GenAI-powered tool for creating companion reference images for 3D design feedback, this section discusses the practical implications~(Section~\ref{sec::memovis::discuss::app}) and future improvements~(Section~\ref{sec::memovis::discuss::genai}) informed by our explorations.

\subsection{Practical Implications}\label{sec::memovis::discuss::app}

Overall, our studies showed that MemoVis can assist feedback providers in efficiently creating companion reference images for 3D design feedback.
The design of MemoVis validates the feasibility of using GenAI and VLFMs to support a more simple and efficient workflow of asynchronous 3D design review. 
This section discusses key practical implications drawn from our findings.

\vspace{+0.1in}\noindent{\bf Real-time viewpoint suggestions}.
To assist in creating reference images, MemoVis needs to first allow feedback providers to efficiently locate 3D camera viewpoints pertinent to the written textual comments.
Our study showed that the real-time viewpoint suggestions can support this task by analyzing the written comment and suggesting semantically-relevant views in the 3D scene. 
The effectiveness of this feature has also been demonstrated by $11$ feedback providers {\it without} proficient 3D software skills.
Despite having 3D skills, few participants like PF2 found this feature valuable when it comes to seeking relevant views on a potentially large-scale 3D model.
While earlier studies suggested closed-form solutions for viewpoint selection based on area~\cite{Plemenos1996}, silhouette~\cite{Feldman2005, Vieira2009}, and depth~\cite{Blanz1999} attributes, along with their combinations~\cite{Secord2011}, the relationships between the chosen view and textual semantics are often absent.
MemoVis introduces a novel paradigm, enabling feedback providers to effortlessly identify relevant views for contextualizing the reference images they intend to create.
Further, looking at the qualitative feedback from Section~\ref{sec::memovis::eval::create_feedback}, a number of participants, exemplified by PF2, believed that viewpoint suggestion would be even more useful in larger-scale 3D models such as those used in video game design.
While prior research, such as IsoCam~\cite{Marton2014}, has explored the viability of employing a touch-based controller for navigating the camera in large projection setups, it is impractical to apply such complex hardware setups to 3D design review workflow. 
Instead, MemoVis~ leverages VLFMs to infer what types of views that the feedback providers might be interested in exploring, enabling feedback providers to spend less time on maneuvering the viewing camera and more time on the main {\it feedback writing} task.

\vspace{+0.1in}
\noindent{\bf Image modifiers.}
We highlighted the advantages of MemoVis's rapid image creation workflow with three types of image modifiers, for facilitating creating companion reference images while writing 3D design feedback.
It is worthwhile to emphasize that MemoVis is a review and feedback creation tool, instead of an image editing tool like the recent GenAI-powered Photoshop~\cite{Photoshop, photoshopAI}, or an ideation tool like Vizcom~\cite{Vizcom}.
Hence, advancing techniques to realize aesthetic and high-quality images are \emph{beyond} our scope. Rather, our key focus is to prioritize the clarity of the reference images and their alignment with textual comments.
Our results have shown the explicitness of the reference images created and the capabilities to maintain the contexts of the anchored viewpoint, confirmed by both feedback providers (Section~\ref{sec::memovis::eval::create_feedback}) and designers (Section~\ref{sec::memovis::eval::eval_feedback}).
Additionally, MemoVis's approach can also  help with creative thinking. The images generated with the modifiers can sometime offer new ideas and inspirations for feedback writers.
We also demonstrated that a larger number of participants found the reference image creation workflow using MemoVis to be easier compared to today's methods involving image searching and/or sketching (Figure~\ref{fig::memovis::study-1-results}a).
While the design of MemoVis is intended to encourage feedback providers to focus on the feedback typing task, we showed that it is still viable to request feedback providers to engage in simple rough sketching and painting directly on top of the 3D explorer.
Our discovery indicates that incorporating simple non-textual input modalities like sketches and painting will not significantly raise the interaction cost~\cite{Raluca2013}. Instead, it gives feedback providers added control to ensure consistency in the reference image within the context of the initial design, as highlighted in our formative studies.
In theory, this finding could be linked to Kohler's recommendations~\cite{Kohler2022} for generic user experience design: granting users autonomy through customization may enhance the sense of ownership, potentially improving the overall interaction experience. Nevertheless, dedicating time and resources to customization development could also elevate interaction costs, possibly diminishing the user experience~\cite{Raluca2013, Heidi2008, Norman2013, Bennett2023}. 
MemoVis presents an exemplary design that balances low interaction costs while affording feedback providers additional controls in the creation of reference images.

\vspace{+0.1in}
\noindent{\bf Integration with the State-Of-The-Art (SOTA) models.}
We demonstrated the feasibility of using the MemoVis system, powered by year-2023's pre-trained models, to assist feedback providers in efficiently creating reference image for 3D design feedback.
With continuous advancement of SOTA performance of today's VLFMs~\cite{Morris2023AI, Chen2023AI}, we believe our contribution toward designing a novel interaction workflow and experience for efficient 3D design review will continue to hold.
We consider the underlying GenAI and VLFMs as the engineering primitives to drive the novel interaction experience, where the feedback providers could efficiently create companion reference images for the feedback comments, while focusing on text typing.
The enhancements of the inference quality of recent text-to-image SOTA models could help generalizing the practical applicability of MemoVis through potentially more photorealistic synthesized images~\cite{Balaji2023}, simpler and more intuitive prompts ~\cite{Hao2022}, as well as a reduced inference latency ~\cite{Yang2023}.
For example, SOTA pipelines, such as Promptist~\cite{Hao2022} for optimizing text-to-image GenAI prompts, could potentially reduce the failures when feedback providers write low-quality prompts.
Other 3D-related GenAI SOTA pipelines like InseRF~\cite{InseRF} that enables text-driven 3D object insertions might also be integrated to enhance features for feedback providers \textit{with} prior 3D experience, enabling deeper exploration.

\vspace{+0.1in}
\noindent{\bf Integration with real-world workplace applications.} 
The interaction designs of MemoVis can be integrated into larger workplace applications, as a form of lightweight plugins.
Feedback providers could focus on their primary task, {\it feedback typing}, instead of editing images or 3D models (Section~\ref{sec::memovis::system}).
These MemoVis-translated plugins enable feedback providers to inspect 3D models, create and visualize textual feedback comments, and send them to designers just like the memo notes. 
Such interaction experience would be simple and fluid~\cite{Elmqvist2011} that facilitates smooth discussion and design collaborations between feedback providers and designers, by encouraging the feedback providers focusing on thinking and typing textual comments.
For example, MemoVis can be integrated as an add-on for Gmail.
This integration allows feedback providers to effortlessly create accompanying reference images while typing textual comments within an email.
MemoVis can also be implemented as a plugin for iMessage.
Feedback providers could use this plugin to conveniently examine 3D models and create reference images for textual comments while  engaging in text conversations with their designers, making the process as straightforward as creating a memoji~\cite{memoji}.
Although MemoVis was contextualized in an {\it asynchronous} 3D design review workflow, this plugin is invaluable in expediting the creation of reference images during {\it synchronous} conversations in \textbf{I}nstant \textbf{M}essaging~(IM) applications, effectively minimizing the duration of silence~\cite{Li2023, Kim2017}.
Beyond supporting 3D design feedback, MemoVis~ can still be generalized to broader multi-modal GenAI-based applications that requires efficient visualizations of texts. For example, MemoVis ~ can be integrated with existing collaborative writing tools like Notion. MemoVis's capabilities of efficiently visualizing textual content shows promise in enhancing synchronous discussions without disrupting the flow of conversation.

\subsection{Improving MemoVis}\label{sec::memovis::discuss::genai}

After exploring the practical implications of MemoVis, this section explores potential key directions for future improvements, drawing insights from our findings.

\vspace{+0.1in}
\noindent{\bf Boosting AI inferences with human feedback.}
In Study 1, we observed that the feedback providers might need to adjust the viewpoint based on MemoVis's suggestion and/or make multiple attempts to create the reference images that could convey the gist of the design feedback. 
One future direction is to understand how we could leverage the behavioral actions from feedback providers to boost future AI inferences.
Similar ideas have been successfully used in many large language model applications through designing prompts with few-shot learning~\cite{fewshot} and integrating the strategies of reinforcement learning from human feedback~\cite{Ziegler2020}.
For example, instead of searching possible viewpoints solely based on pre-trained CLIP model~\cite{CLIP}, MemoVis might incorporate the {\it past view preferences} from the feedback providers to find the most likely viewpoint, with which the feedback providers want to anchor the textual comments.
Realizing this may require the design of an effective cost function with respect to CLIP inferences and the preferences from feedback providers that the MemoVis could optimize.

\vspace{+0.1in}
\noindent{\bf Eliminating manual prompt writing toward simple and fluid reference images creation experience.}
MemoVis ~ necessitates feedback providers to create additional prompts for creating reference images, which can be redundant.
Sometimes, prompts may require feedback providers to include additional context beyond the immediate objects of focus. 
Although most participants were satisfied with the experience of using MemoVis~to create companion reference images for textual feedback, we found participants occasionally found it difficult (PF5) and tedious (PF12) to write additional prompt, or confused of the difference between feedback comments and prompts (PF11).  
While prior research, \eg~\cite{Hao2023, Manas2024}, has demonstrated novel prompt optimization algorithms for vanilla text-to-image GenAI, we opted not to incorporate this feature into the current implementation due to lack of validations on creating photorealistic images using conditioned text-to-image GenAI (Section~\ref{sec::memovis::related::genai}).
Additionally, granting flexibility to feedback providers to write prompts also enables them to iteratively enhance the prompts upon the less optimal reference images.
Although Liu~\etal~\cite{Liu2022} have discussed the key guidelines of crafting text-to-image prompt, future work might explore how to optimize the text-to-image prompts for conditioned text-to-image GenAI, and how to integrate broader contexts from the written feedback, without laborious trial-and-error process for creating prompts, as highlighted by PF10.
While advancing techniques for prompt creations and explorations is beyond our scope, future work may explore the feasibility of integrating interactive prompt engineering techniques~\cite{Feng2024, Wang2024} to assist feedback providers in streamlining the feedback creation process while allowing the exploration of the nuances of prompt creation.
Ultimately, we envision MemoVis being able to automatically generate the prompts for text-to-image GenAI without the awareness of feedback providers, leading to a simple and fluid interactions experience~\cite{Elmqvist2011}, where the candidate reference images could be responsively created and updated as the feedback providers typing the textual comments.

\vspace{+0.1in}
\noindent{\bf Integrating GenAI with searching.}
While our research on MemoVis shows how \mbox{text-to-image} GenAI can be a potential path to enhance the 3D design review workflow, the Study 1 indicated that GenAI alone might be insufficient.
Despite most participants are satisfied with the experience of using MemoVis to create reference image for textual comments, few participants (\eg~PF6) emphasized the opportunity to {\it integrate} rather than to {\it replace} internet search and hand annotation approaches with MemoVis.
While MemoVis is helpful when feedback providers do not have a specific design in mind, few participants (\eg~PF4, PF11) emphasized the usefulness of using online images when a specific design suggestion in the mind.
With these observations, a compelling research path naturally emerges: how could we woven today's image searching approach into MemoVis's pipeline. 
Although this direction is similar to recent works like GenQuery~\cite{Son2023GenQuery}, which showed how to integrate GenAI and search to help instantiate designers' early-stage abstract idea, and DesignAID~\cite{Cai2023}, which emphasized the importance of {``augmenting humans rather than replacing them''} while attempting to use GenAI to help exploring visual design spaces, reviewing and providing feedback for 3D design are fundamentally different from 2D graphic ideation workflow due to the complexities of 3D models and the convoluted thinking process while feedback providers are attempting to create reference images (Section~\ref{sec::memovis::eval::create_feedback}).
Future direction might consider how to integrate existing search engine into MemoVis, opening new paradigms of efficiently co-visualizing textual feedback alongside human interactions, internet searches and GenAI.
For example, despite the imperfections and the potential risk of causing misunderstandings, our results indicate that a few feedback providers still prefer to use online images tailored to their specific design needs.
While tools like Photoshop~\cite{Photoshop, photoshopAI} enable feedback providers to edit online images, such workflows are often not streamlined, time-consuming and inefficient for those lacking image editing skills.
Although MemoVis empowers feedback providers to use image modifiers to create reference images primarily from text, future work may further explore how these image modifiers can be extended and integrate the contexts of the searched image(s).

\section{Limitations}\label{sec::memovis::limitation}
Having demonstrated the promising of MemoVis, we also acknowledge multiple key limitations with respect to system design and evaluations.

\vspace{+0.1in}
\noindent\textbf{Inference latency.}
MemoVis took around $30$ seconds before generating a synthesized image.
While MemoVis is asynchronous, through which the feedback providers could continuous explore the 3D model, we found most participants still prefer to shorten the wait time for a more streamlined feedback creation workflow.
Although reducing latency is beyond our scope, some participants (\eg~PF6) found it as a critical setbacks in terms of software engineering design perspective, compared to the internet searching.
We speculate future advancement of SOTA GenAI and GPU parallel computing research would help overcome this limitation.

\vspace{+0.1in}\noindent\textbf{Participants.}
Our first formative study was conducted with only two participants as it required individuals with extensive design experience.
The evaluation studies were conducted with only $14$ feedback providers and eight designer participants, due to the lengthy duration in Study 1 and limited resources for Study 2 which requires participants to have prior 3D design experience.
Despite the validity of our results that were mainly qualitative based, future work might further explore the usability of MemoVis with more participants coming from different backgrounds, to minimize the bias from the recruited participants.
For instance, in our study tasks, the feedback provider participants recruited can only represent the clients engaged with the workflow of design feedback creations. Future research may recruit feedback provider participants who can represent other types of feedback providers like collaborators and managers.

\vspace{+0.1in}\noindent\textbf{Evaluation conditions and tasks.}
First, our studies were only based on a bedroom and a car model (Appendix~\ref{sec::memovis::app::study_tasks}).
Future researchers might explore the generalizability of our system to a wider ranges of 3D models.
As discussed in Section~\ref{sec::memovis::discuss::app} and speculated by PF2 that the viewpoint suggestions could be {\it ``much helpful''} for a larger scene, future work might also investigate how MemoVis~ could help feedback providers to navigate and create reference images for a larger environment 3D models.
Second, the baseline condition in the Study 1 (Section~\ref{sec::memovis::eval::create_feedback}) required participants to create reference images using searching and/or hand sketching.
Although, through our formative studies, it is possible to simulate the prevailing practices of most participants in creating reference images for 3D design feedback, future research might explore comparing MemoVis with a broader range of GenAI-powered baseline, such as creating reference images using vanilla text-to-image GenAI tools~\cite{AdobeFirefly} and professional image editing software~\cite{Photoshop}.

\vspace{+0.1in}\noindent\textbf{Evaluations in an ecologically valid 3D design review workflow.}
Despite the potentials of integrating MemoVis as part of larger collaborative design workflow and workplace applications (Section~\ref{sec::memovis::discuss::app}), our current evaluation was based on a monolithic browser-based application in a controlled laboratory setting. 
Future research might deploy MemoVis, and investigate the user experience in a realistic real-world 3D design review workflow.
One future direction might investigate the feasibility and effectiveness of MemoVis, after being integrated and deployed into today's mainstream workplace and IM applications such as Gmail and iMessage~(Section~\ref{sec::memovis::discuss::app}).

\section{Conclusion}\label{sec::memovis::conclusion}
We designed and evaluated MemoVis, a browser-based text editor interface that assists feedback providers in easily creating companion reference images for textual 3D design comments. MemoVis~ integrates several AI tools to enable a novel 3D review workflow where users could quickly locate relevant design context in 3D and synthesize images to illustrate their ideas.
A within-subject study with $14$~feedback providers demonstrates the effectiveness of MemoVis.
The quality and explicitness of the companion images were evaluated by another eight participants with prior 3D design experience.

\section{Acknowledgment}
We thank the insightful feedback from the anonymous reviewers.
We appreciate the discussions with fellow researchers from Adobe Research, including Mira Dontcheva, Anh Truong, Joy O. Kim, and Zongze Wu, as well as Rima Cao from the Department of Cognitive Science at the University of California San Diego.
We thank Zeyu Jin for providing the text-to-voice GenAI pipeline to synthesize narrations in the companion videos.

Chapter~\ref{sec::memovis}, in full, is a reprint of the material as it appears in the journal of ACM \textbf{T}ransactions \textbf{o}f \textbf{C}omputer-\textbf{H}uman \textbf{I}nteraction~(TOCHI), Volume 31, Issue 5, Article 67 (October 2024). The dissertation author was the primary investigator and author of this journal paper. Co-authors includes Cuong Nguyen, Thibault Groueix, Vladimir G. Kim, and Nadir Weibel. 
This work was also presented in the 2024 Annual ACM Symposium on \textbf{U}ser \textbf{I}nterface \textbf{S}oftware and Technology (UIST 2024) conference.
This work was collaborated with Adobe Research.
The previously published manuscript can be found in \cite{Chen2024MemoVis}.

\chapter{PaperToPlace: Transforming Instruction Documents into Spatialized and Context-Aware Mixed Reality Experiences}\label{sec::papertoplace}

\begin{figure}[h!]
    \includegraphics[width=\textwidth]{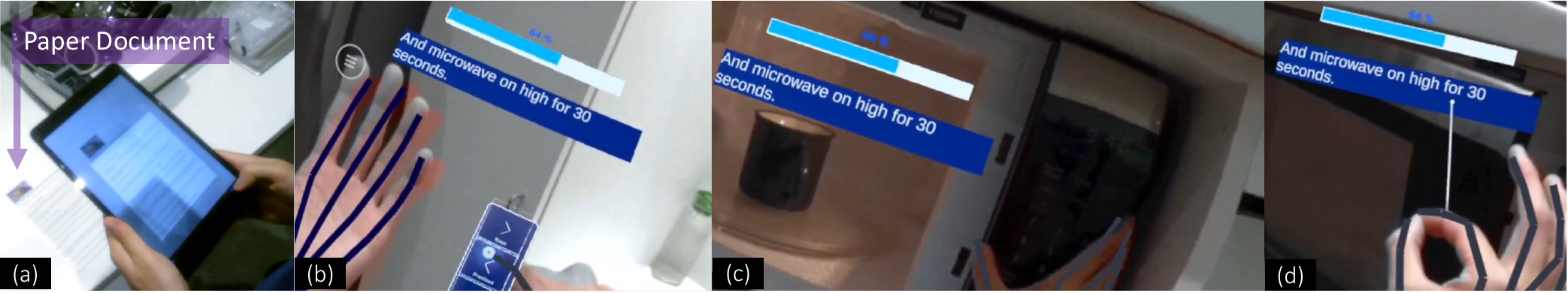}
    \caption[Overview of PaperToPlace]{Overview of PaperToPlace: (a)~The author creates an MR experience by taking a snapshot of a paper document, with an optional ML-supported pipeline for associating key objects with each instruction step; (b)~The consumer can browse the spatialized instruction steps using a hand menu; (c)~The step is placed at an optimal location to minimize context switching and prevent occlusion of important interaction areas (\eg~not occluding the touchpad while setting the time on a microwave); (d)~The consumer can ``pinch-and-drag'' the step to refine the system placement. Steps (b - d) show the first-person MR view.}
    \label{fig::papertoplace::teaser}
\end{figure}

While paper instructions are a mainstream medium for sharing knowledge, consuming such instructions and translating them into activities can be inefficient due to the lack of connectivity with the physical environment.
We propose PaperToPlace, a novel workflow comprising an {\it authoring pipeline}, which allows the authors to rapidly transform and spatialize existing paper instructions into an MR experience, and a {\it consumption pipeline}, which computationally places each instruction step at an optimal location that is easy to read and does not occlude key interaction areas.
Our evaluation of the authoring pipeline with $12$ participants demonstrates the usability of our workflow and the effectiveness of using a machine learning based approach to help extract the spatial locations associated with each step.
A second within-subjects study with another $12$ participants demonstrates the merits of our consumption pipeline to reduce context-switching effort by delivering individual segmented instruction steps and offering hands-free affordances.

\section{Introduction}\label{sec::papertoplace::introduction}
Paper-based instructions are common for knowledge sharing.
Such instructions are often related to tasks that require users to interact with multiple objects spatially distributed in an environment.
For example, when following a recipe, a user may need to interact with multiple kitchen appliances like the cooktop, fridge, and microwave. %
When following a safety manual, a compliance manager may need to interact with various machines on the factory floor.

However, performing a task while consuming instructions can be tedious as the text is typically \textit{disassociated} from the user's physical environment. 
Thus, a user has to balance reading the instructions, figuring out what they mean in the environment, and performing the task, which can be cognitively demanding~\cite{Eiriksdottir2011}.
For example, when frying a piece of steak while following a cookbook, one needs to frequently switch between the cookbook and the pan to check the searing technique, temperature,~\etc 
This switch can be costly if the user places the cookbook somewhere peripheral so that it does not obstruct the task area. 
The user might end up spending more time trying to navigate the text and environment, than performing the task.
This problem is made worse if the user forgets some important information like temperature or duration, and has to repeatedly come back to the instruction to double check. 

Consumer \textbf{A}ugmented \textbf{R}eality~(AR) and \textbf{M}ixed \textbf{R}eality~(MR)\footnote{While AR and MR are frequently interchangeable in literature, our contribution lies under the MR paradigm that focus on interactivity and context awareness~\cite{Skarbez2021}.} offer a unique opportunity to address this document-activity disassociation by overlaying digital elements on top of the environment.
These approaches are becoming more accessible, and studies have demonstrated their potentials for training workers to conduct tasks that are spatial in nature~\cite{Guntur2020}.
While prior works investigated the affordances of virtual guidance for conducting spatial tasks~\cite{Johnson2021}, and how to integrate such guidance in MR~\cite{Chidambaram2021}, they have not explored how document contents and their associated consumption experience could be constructed in MR. 
To that end, Microsoft Dynamic~$365$ Guides~\cite{Dynamic365} is an industry solution to help enterprise users manually create instructions and anchor them in an MR experience.

Across these MR instruction systems, the placement of the virtual instructions is often decided by the authors, and cannot be dynamically adapted to real-world contexts. 
This assumption could lead to undesirable experiences in both the consumption phase and the authoring phase. 
For the consumption phase, a static MR instruction could be mistakenly placed at a location too far from the user's task, at a distance that is difficult to read, or at a position that occludes key objects that the user is interacting with. 
For authoring, the author has to spend time associating and placing an instruction with its corresponding physical object. This process is time consuming and has to be repeated for every new set of instructions, even though the physical layout of objects might not change much over time.
While some prior works, \eg~FLARE~\cite{Gal2014Flare}, showed the usefulness of creating a persistent AR layout, real-world instructional activities are frequently changing (\eg~users might move from one place to another depending on the procedural step in the instructions), causing such a static layout to be infeasible.

We propose PaperToPlace, a novel end-to-end workflow that transforms paper instructions into a context-aware instructional MR experience by {\it segmenting} monolithic documents; {\it associating} instruction steps with real-world anchoring objects; and optimally {\it placing} the virtual instruction steps so that they are easy to read and revisit while completing the tasks.
To realize this goal, PaperToPlace consists of an {\it authoring} and a {\it consumption} pipeline.
With the \mbox{authoring} pipeline, the author can simply take a snapshot of the \mbox{paper} document by leveraging a mobile camera~(Figure~\ref{fig::papertoplace::teaser}a).
Our system then segments the text in the document into individual instruction steps. 
The author can manually edit these segments, and \textit{associate} each step with the spatial location where the relevant activities will occur.
To help with this association task, we designed a \textbf{M}achine \textbf{L}earning~(ML) approach, where a fine-tuned language model was used to suggest the relevant spatial locations to the author. 
Our consumption pipeline is designed to render these instruction steps and place them in the associated spatial locations.
To ensure users could easily consume the spatialized instruction, the placement of each step in MR is optimized using a probabilistic optimization approach based on pre-created environmental models and the tracked gaze and hands.
Figure~\ref{fig::papertoplace::teaser}c shows an example of a spatialized instruction step, where the step ``\textit{and microwave on high for 30 seconds}'' is tagged with ``{\it microwave}'' and is optimally rendered in front of the user while not occluding their view as they set the heating~time.

We prototyped PaperToPlace on Meta Quest Pro~\cite{QuestPro}, and conducted two within-subjects studies to evaluate the authoring pipeline with $12$ participants, and the consumption pipeline with another $12$ participants.
We demonstrated the usability of our authoring pipeline, and the effectiveness of using an ML-based approach to help the authors extract the spatial location associated with each step.
We then illustrated the effectiveness of the consumption pipeline for reducing context-switching effort, delivering the segmented instruction steps, and offering hands-free affordances.

With the assumption that the spatial profiles (\ie~the environmental geometry and associated semantic labels, see Section~\ref{sec::papertoplace::system_overview::design_insights}) are available, we contribute the design and evaluations of:

\vspace{4px}

\noindent$\bullet$ {\bf An authoring pipeline} that allows users to transform paper instructions into a spatialized MR experience;

\noindent$\bullet$ {\bf A consumption pipeline} that can computationally place the virtual instruction steps in the optimal place without either occluding the user's view or leading to large degrees of context switching.

\section{Related Work}\label{sec::papertoplace::related}
This paper is motivated by prior work on incorporating instruction experiences into MR and designing context-aware MR experiences.

\subsection{Integrating Instruction Experiences into MR}\label{sec::papertoplace::related::instruction_doc}
MR has been widely used for augmenting document consumption experiences~\cite{Qian2022DuallyNoted, Chen2022VRContourWIP, Chen2022VRContour}.
Augmenting instructional documents, however, is still challenging due to the need to connect and integrate with real-world scenes and activities~\cite{Ganier2004, Schriver1997}.
Many prior works have explored the use of MR to augment a procedural instruction experience --- an important asynchronous collaboration task.
For example, ProcessAR~\cite{Chidambaram2021} proposed \insitu procedural AR instructions that could be rapidly created by experts, and used to teach novices through spatial and temporal demonstrations (\eg information about how to move a tool in the temporal domain and orient it in the spatial domain).
However, the placement of textual instructions was not explored.
CAPturAR~\cite{Wang2020} introduced a MR tool that helps users rapidly author context-aware applications, by referring to recorded activities.
Commercial tools such as Microsoft Dynamic~$365$ Guides~\cite{Dynamic365} enables experts to author a MR instruction experience by enacting the guidance, placing the instruction in the designated space, and recording the tool operations.

Although these works explored the design of MR-based instruction experiences, existing paper instructions are usually left behind, resulting in unnecessary time and effort to redesign a usable instruction workflow.
Additionally, existing MR instruction experiences are often not able to dynamically adapt to the changing environmental context of real-world activities (\eg~\cite{Dynamic365, Chidambaram2021}), causing user frustration when virtual graphics occlude interaction tasks.
PaperToPlace is novel in that it supports reusing existing paper documents that are designed by professional writers in a reader-centered way~\cite{Weller1986}, and can transform such documents into a spatialized and context-aware MR experience that is adapted to both the user's needs and the environmental characteristics.

\subsection{Designing Context-Aware MR Experiences}\label{sec::papertoplace::related::context_ar}
Context-aware MR systems aim to show {\it ``the `right' information, at the `right' time, in the `right' place, in the `right' way''}~\cite{Fischer2012}, which requires understanding both human and environmental contexts.

Prior research explored various computational approaches to realize this goal.
For example, Lindlbauer~\etal~\cite{Lindlbauer2019} used the real-time cognitive load, estimated by pupil dilation, to decide when and where the application should be shown, as well as how much information should be delivered (\ie~level of detail) in an MR system.
Lang~\etal~\cite{Lang2019} used simulated annealing to place virtual agents by considering key anchoring surfaces in the environment identified by pre-trained mask R-CNN.
Liang~\etal~\cite{Liang2021} used a similar approach to build a scene-aware virtual pet which could behave naturally in the real-world (\eg~respond to a food bowl).
Yu~\etal~\cite{Yu2021} proposed an interactive and context-aware furniture recommendation MR system by considering the real-world scene (\ie~spatial context) and the learnt furniture compatibility in a latent space (\ie~category context).
ScalAR enabled designers to author a semantically adaptive AR experience~\cite{Qian2022ScalAR}. 
Liu~\etal~\cite{Liu2021} attempt to generate suggestions for the arrangement of work surfaces in HoloLens, by capturing users' habitual behaviors of interacting with objects on the work surface.
Similar to our work, SemanticAdapt~\cite{Cheng2021} used an optimization approach to automatically adapt MR layouts between different environments by considering the virtual-physical semantic connections.
However, the target applications were only related to information consumption (\eg~consuming news feeds) and did not consider those related to real-world activities. 

Inspired by this prior work, PaperToPlace demonstrates a novel consumption pipeline that leverages a similar computational approach to analyze the tracked gaze and hand position, as well as the anchoring surfaces of the key objects in a target environment.

\section{Preliminary Needs-Finding Study}\label{sec::papertoplace::prelim}
To understand the pain-points for consuming existing paper instructions (\ie~{\it monolithic}, \textit{non}-segmented, \textit{non}-spatialized, and \textit{non}-context-aware), we conducted a needs-finding study, rooted in participant observations~\cite{Bogdewic1992, Spradley2016} and semi-structured interviews.

\subsection{Participants, Tasks, and Procedure}\label{sec::papertoplace::prelim::procedures}
We recruited four participants (age, $M = 24.75$, $SD = 2.87$, two~females, two~males).
Participants were required to complete the designated task using paper instructions.
Specifically, PP1 and PP4 were required to make coffee with a coffee machine by following the user manual~\cite{CoffeeManual}~(Figure~\ref{fig::papertoplace::prelim_study_tasks}a).
PP2 and PP3 were required to make a chocolate microwave cup cake from an existing recipe~\cite{ChocolateMugCake}~(Figure~\ref{fig::papertoplace::prelim_study_tasks}b).
These tasks were chosen since they are common activities in an office kitchen; require participants to read the textual instructions; and could be completed in a reasonable time for an unpaid study.
We also used the existing instructions~\cite{CoffeeManual, ChocolateMugCake} created by professional writers to minimize the impact of non-professional writing styles.
All participants reported little (PP3, PP4) to no (PP1, PP2) prior knowledge for the designated task.
Next, we conducted a semi-structured interview, focusing on {\it ``what are the pain-points while performing the designated tasks using the given paper instruction, and why?''}
Finally, we brainstormed potential designs of a MR experience for consuming instruction documents. 
We explained the concept of MR for PP3 and PP4 who were not familiar with it.
Participants were encouraged to sketch their imagined design on an iPad Canvas.
The study took on average $30$~min ($SD = 3.25$~min).

\subsection{Findings}\label{sec::papertoplace::prelim::findings}
Overall, we identified three pain-points through the study.

\begin{figure}[t]
    \centering
    \includegraphics[width=0.9\textwidth]{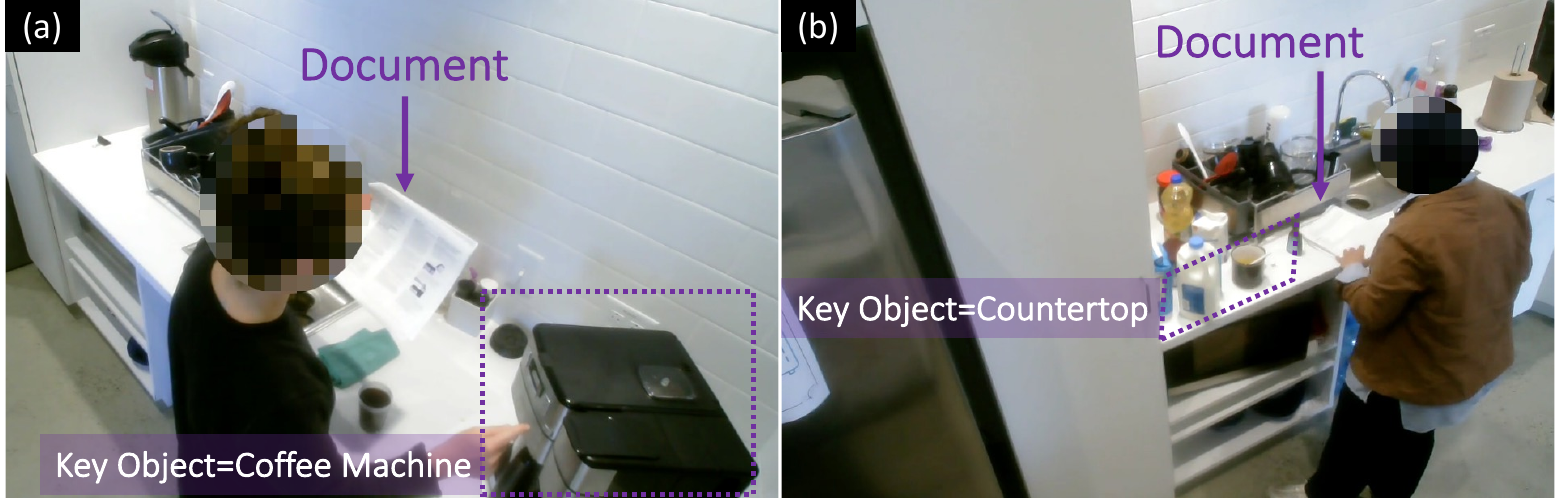}
    \caption[Preliminary needs-finding tasks.]{Preliminary needs-finding tasks. (a) Making a cup of coffee using a coffee machine (PP1). (b) Making a chocolate cake in a mug with a microwave (PP3).}
    \label{fig::papertoplace::prelim_study_tasks}
\end{figure}

\begin{figure}[t]
    \centering
    \includegraphics[width=\textwidth]{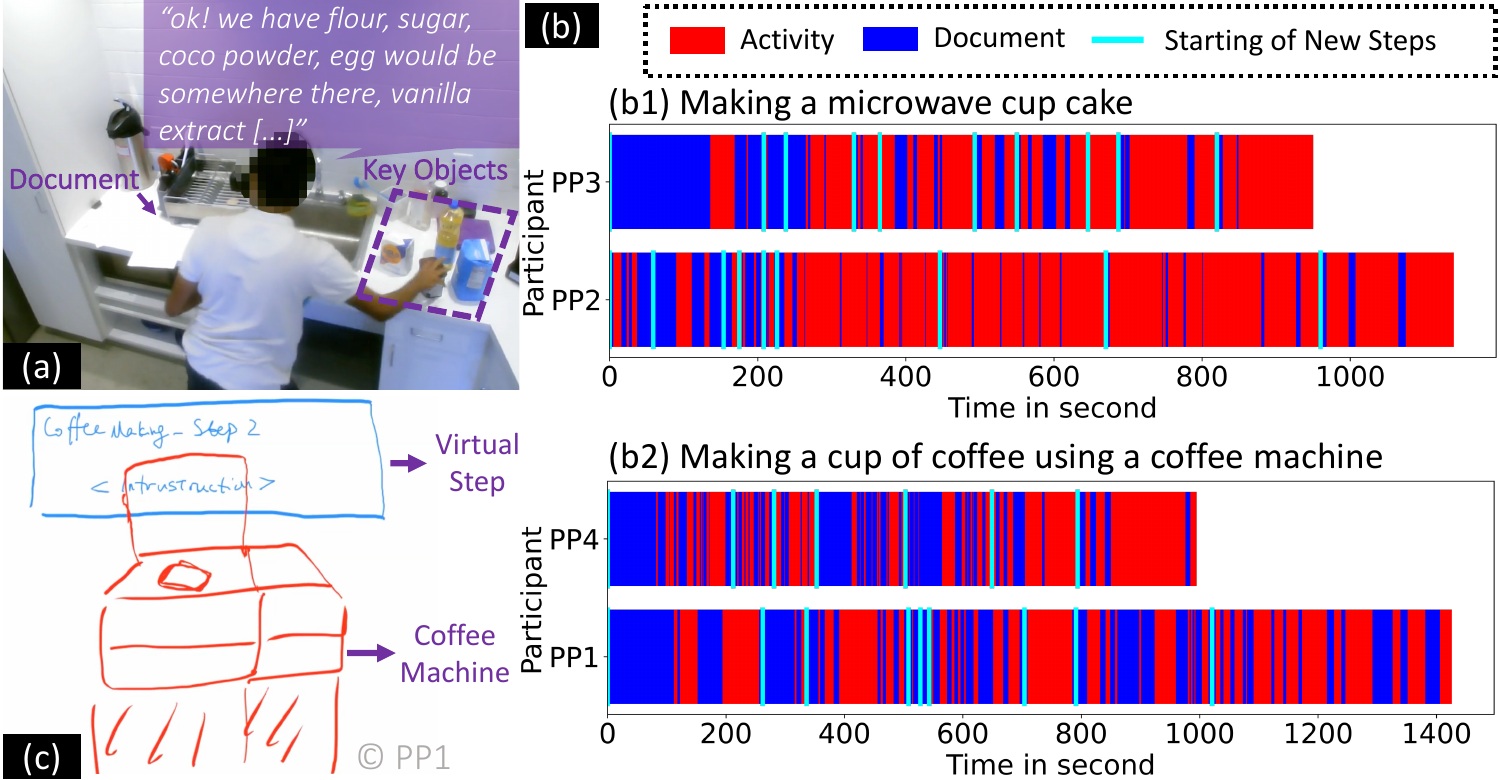}
    \caption[Preliminary needs-finding study results.]{Preliminary needs-finding study results. (a) Example context switching while PP2 was attempting to map instructions with real-world objects. (b) The annotated timestamps showing participants' current focus as either the document or real-world activities. (c) PP1's design for an instructional MR experience while using a coffee machine.}
    \label{fig::papertoplace::prelim_study_results}
\end{figure}

\vspace{0.1in}
\noindent{\bf The overwhelming amount of information or lack of necessary details in the instructions can impact the usability.}
During the semi-structured interviews, two participants~(PP1, PP3) pointed out that the sometimes overwhelming amount of information and irrelevant content could be distracting. 
For example: {\it ``there was a lot of information on the document. And it wasn't easy for me to know where and what information I should be looking at''}~(PP1) and {\it ``the first setback was too much information''}~(PP3).
To handle the potentially overwhelming amount of information, PP2 first skimmed the document in search of relevant content: 
{\it ``I am first attracted to see where the bullet points are. [...] And then if I just skim through the first two or three points, I understand that this is not relevant. So I'm just skipping those sections completely.''} 
While designing the MR experience, PP1 incorporated such insights into her design (Figure~\ref{fig::papertoplace::prelim_study_results}c) and commented:
{\it ``I would rather the MR just gives me one small step every time. For example, I am making coffee and reaching the step two, and in the virtual instruction, it will say like coffee making step two, and then here will be just a instruction with just a few sentences.''}

On the other hand, PP2 and PP4 believed the lack of details for certain steps could impact their ability to perform the sub-task.
For example: {\it ``I've never cracked an egg so I don't know how to do it. [...] If it is some things that I've never actually done [and the instruction document does not tell me how], I might actually be confused''}~(PP2).
In essence, different users likely require different levels of information based on their prior experience with the intended task.

\vspace{0.1in}\noindent{\bf There are missing connections between the instruction step and real-world activities.}
All participants identified a need for establishing spatial connections between instructions and real-world objects or activities.
For example, during the design phase of the study, PP2 emphasized that {\it ``having [a virtual] arrow [in MR] that can help me and connect me to the object is helpful.''}

On the other hand, most participants suggested that overwhelming spatial guidance might be unnecessary and could lead to visual disturbance, similar to their concerns about the potentially overwhelming amount of information already in some instructional documents.
Reflecting further on PP2's suggestion to include virtual arrows, he highlighted some potential downsides of this approach, noting that
{\it ``[having useless spatial indicators] is going to be overloaded. I know these basic things, and I don't need pointers to see like a spoon or a mug.''}
PP1's design sketches implied an alternative approach to establish connections between the instruction step and real-world objects by placing the virtual step close to the coffee machine, without occluding the user's view (Figure~\ref{fig::papertoplace::prelim_study_results}c).

\begin{figure}[t]
    \centering
    \includegraphics[width=\textwidth]{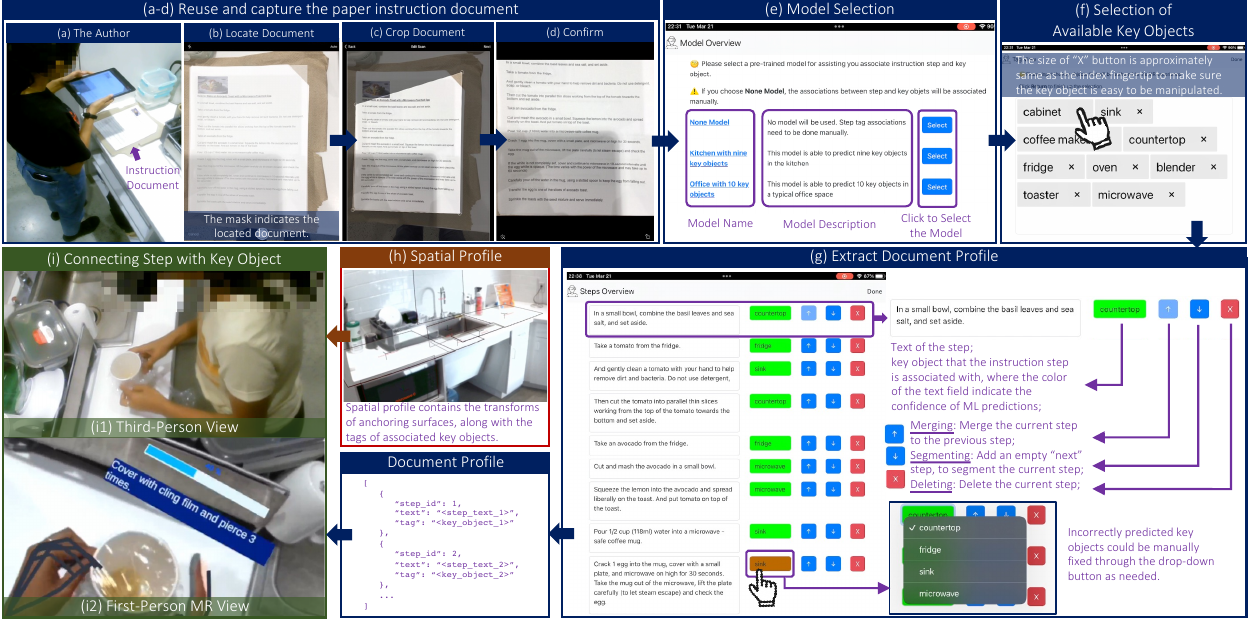}
    \caption[PaperToPlace system overview.]{PaperToPlace system overview. We assume a spatial profile \mbox{(h, red block)} was pre-created. The author uses the authoring pipeline \mbox{(a - g, blue blocks)} to extract the document profile for the MR experience. With consuming pipeline \mbox{(i, green block)}, the instruction steps are displayed based on the environmental (loaded via the spatial profile) and user's contexts.}
    \label{fig::papertoplace::system_overview}
\end{figure}

\vspace{0.1in}
\noindent{\bf Frequent context-switching between the instruction document and real-world activities should be minimized.}
By analyzing the video recording, we found that users perform frequent context-switching between documents and real-world activities.
For example, while PP2 was conducting the task, he naturally commented: {\it ``ok! we have flour, sugar, coco powder, egg would be somewhere there, vanilla extract [...]''}, thereby demonstrating frequent context-switching between instructions and real-world objects  as he checked off items from the recipe list~(Figure~\ref{fig::papertoplace::prelim_study_results}a).
Participants generally used two types of strategies to minimize context-switching: {\it holding} and {\it switching to} the document while performing the task.
Specifically, we found that PP1 tended to hold the documents while performing tasks, although occasionally this approached was inconvenient for steps requiring two hands (Figure~\ref{fig::papertoplace::prelim_study_tasks}a).
In contrast, other participants tended to place the document on the countertop while reading the content, and move the document to a new place when switching to other steps (Figure~\ref{fig::papertoplace::prelim_study_tasks}b).
To understand how participants used the documents to perform individual steps, we manually labeled the timestamp of the recorded video while participants were switching between documents and real-world activities, or otherwise. 
Through this process, we observed that the participants highly relied on the documents to perform tasks. 
Specifically, while participants were attempting to complete a particular step, the document was still frequently referenced even though the participants had already read it at the beginning of each step (Figure~\ref{fig::papertoplace::prelim_study_results}b).

\subsection{Design Considerations}\label{sec::prelim::considerations}
We identified three fundamental \textbf{D}esign \textbf{C}onsiderations (DCs) by analyzing the data from our preliminary needs-finding process.

\vspace{0.1in}\noindent
{\bf DC1: Only delivering the segmented instruction could enhancing information consumption experience.}
We show that the participants expect to consume relevant information corresponding to their current activities, yet existing paper documents usually deliver all information to users at the same time.
An improved MR instruction consumption experience could create novel and flexible ways to segment the document, such that only relevant information is delivered to users for each associated step.

\vspace{0.1in}\noindent
{\bf DC2: Optimally placing instruction texts next to the areas of interactive activities might be helpful for the intermittent and repetitive information consumption experience.}
While participants generally read the entirety of an instruction step before performing the corresponding actions, the instruction step may need to be consumed {\it repeatably} (Figure~\ref{fig::papertoplace::prelim_study_results}b).
Therefore, the placement of the virtual instruction step in MR should consider the spatial location where the relevant task would occur.

\vspace{0.1in}\noindent
{\bf DC3: The right level of spatial guidance could help users associate instructions with spatialized key objects.}
While few existing works~(\eg~\cite{Johnson2018, Johnson2021}) suggested the usefulness of spatial guidance for MR-based instructional experiences, our participants emphasized the importance of {\it moderate} spatial guidance, with limited visual disruptions.
Thus, usable spatial guidance, without causing overwhelming visual disturbance, should be provided at the beginning of each instruction step.

\section{PaperToPlace System Overview}\label{sec::papertoplace::system_overview}
Based on Section~\ref{sec::papertoplace::prelim::findings}, we designed PaperToPlace (Figure \ref{fig::papertoplace::system_overview}), comprising of two pipelines:
\textbf{(1)}~{\bf an authoring pipeline} for an {\bf author} to rapidly and easily create a spatialized MR instruction experience from existing paper-based instructions (Figure~\ref{fig::papertoplace::system_overview}a - g,~Section~\ref{sec::papertoplace::authoring}), 
and \textbf{(2)}~{\bf a consumption pipeline} for enabling a {\bf consumer} to explore context-aware, spatialized instruction steps in MR (Figure~\ref{fig::papertoplace::system_overview}i,~Section~\ref{sec::papertoplace::consuming}).
While we use cooking tasks in an office kitchen as a running example for the design and evaluations, our approach could be transferred to other types of instruction documents.

\subsection{Assumptions and System Walkthrough}\label{sec::papertoplace::system_overview::design_insights}
We consider an {\bf environment} to be a typical workspace (\eg~the kitchen) for supporting a procedural {\bf task}~(\eg~baking a cake). 
Each environment contains multiple physical {\bf key objects}, which are defined as the important, stationary objects that are usually attached to the environment permanently (\eg~the fridge and microwave).
We did not consider non-stationary objects (\eg~a~mug) due to the lack of support for real-time arbitrary object tracking with Quest Pro~\cite{QuestPro}.
Each key object contains one or multiple {\bf anchoring surfaces}, which are virtual surface(s) that describe the approximated geometry of the objects.
We use these surfaces to determine the placement of instruction in MR.

\begin{figure}[b]
    \centering
    \includegraphics[width=0.8\textwidth]{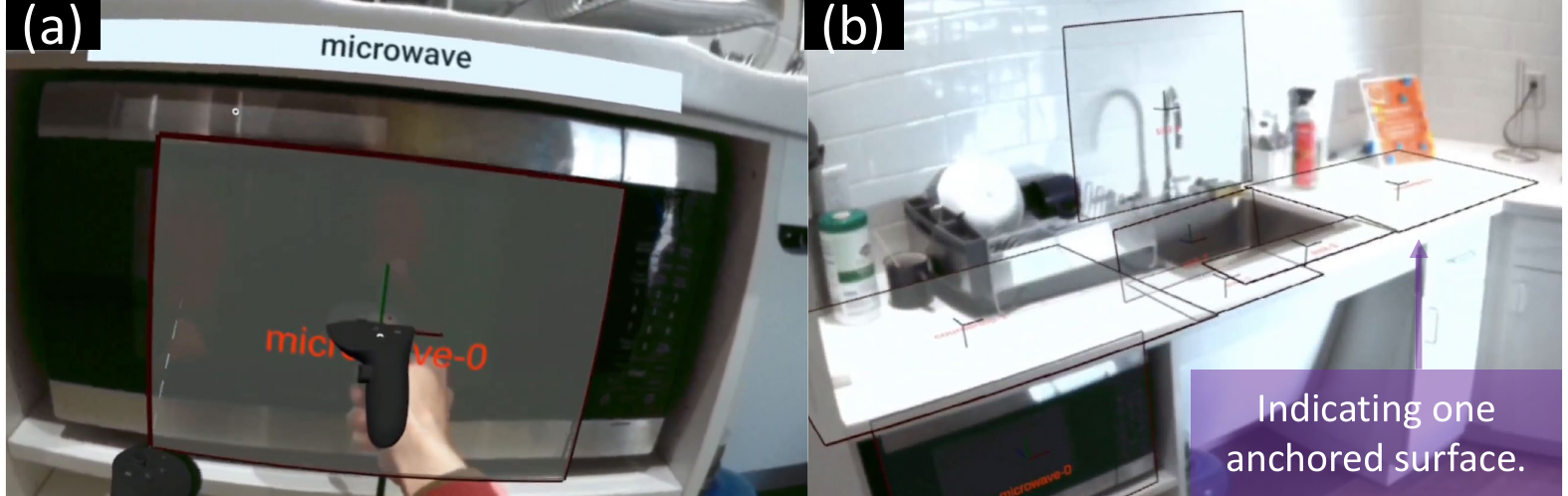}
    \caption{(a)~Creation of an anchoring surface, visualized as a semi-transparent mask, using touch controllers. (b)~Examples of anchoring surfaces in our experimental kitchen. Both scenes were captured as first-person MR views.}
    \label{fig::papertoplace::anchoring_surfaces}
\end{figure}

We also define a \textbf{spatial profile} as a collection of the labels of these key objects and their anchoring surface(s). 
We assume that a spatial profile could be created offline, either through automatic geometry processing (\eg~\cite{AppleRoomPlan}) or manual means (Figure~\ref{fig::papertoplace::anchoring_surfaces}).
Example environment where spatial profiles could be created in advance include office kitchen, a rental house, and a factory.
To realize this assumption, we implemented a MR interface in the Quest Pro~\cite{QuestPro} that allow each anchoring surface (represented by a 2D plane) and the associated key object could be easily declared (Figure~\ref{fig::papertoplace::anchoring_surfaces}a). 
The spatial anchor APIs~\cite{OculusSpatialAnchors} were used to ensure the declared anchoring surfaces are persistent in the environment~(Figure~\ref{fig::papertoplace::anchoring_surfaces}b).

With these assumptions, the goals of the two pipelines of PaperToPlace are described as below:

\vspace{0.1in}\noindent$\bullet$~{\bf Authoring Pipeline.}
Given an existing paper-based instruction document, the author would first segment the document content into smaller steps, each of which will only be associated with one key object.
For each step, the author needs to identify the {\bf metadata}, including: \textbf{(i)} the text of each instruction step, and \textbf{(ii)} the key object that the step should be anchored on.
Together, such metadata makes up our \textbf{document profile} --- the essential elements to recreate a spatialized MR experience from existing paper documents.

\vspace{0.1in}\noindent$\bullet$~{\bf Consumption Pipeline.}
While consuming the document in MR, the instruction steps float in mid-air and will be optimally attached to one of the anchoring surfaces of the associated key object, based on {\bf user context} - real-time user interactions data in MR such as the tracked eye gaze and hand joints.
For example, the step {\it ``boil a cup of water in the microwave for 5 min''} should be attached to one of the anchoring surfaces of the {\it``microwave''}~key object and not impact the user's interactions.

\subsection{Application Scenarios}\label{sec::papertoplace::system_overview::users}
We consider two user roles, where the {\bf author} use the authoring pipeline to rapidly create an instruction MR experience and the {\bf consumer} use the consumption pipeline to consume the authored MR experience while completing tasks. 
We target on colocated and asynchronous collaborations~\cite{Johansen1988}, where the instructions are authored and consumed in same environment.
Specific application scenarios are described below.

\vspace{0.1in}
\noindent{\bf Author and Consumer are Different Users.}
For a specific procedural task, PaperToPlace could be used to facilitate asynchronous collaborations between experts and novices.
For example, company administrators could use PaperToPlace for training new employees to use the provided facilities (\eg~coffee machines and fridges in the shared office kitchen). 
Chidambaram~\etal~\cite{Chidambaram2021} also demonstrated the usefulness of using a similar instruction MR experience to teach novices assembly mechanics.

\vspace{0.1in}\noindent{\bf Author and Consumer are the Same User.}
While we differentiate two user roles, it is possible that the author and consumer are the {\it same} user.
As the cognitive processes for consuming instructions usually occur in working memory, which is constrained by both time and processing capacity, it is often necessary for one to revisit the procedures repeatably for the {\it same} task~\cite{Ganier2004, Anderson2013}. 
For example, while cooking the same meal, it is common for the user to refer back to the cookbook each time upon starting a new step. 
Figure~\ref{fig::papertoplace::prelim_study_results}b confirmed such patterns, where participants repeatedly refer to the instructions while completing a specific step.
In this scenario, the user could author a personalized and spatialized MR experience, which s/he could use repeatably when preparing the same meal in the future; for example, a user could customize a recipe to take into account his/her preferences for spice level.

\section{Authoring Pipeline}\label{sec::papertoplace::authoring}

\noindent The authoring pipeline extracts the document profile from an existing paper document to create an MR experience {\it rapidly} and~{\it easily}.

\subsection{Document Capture and Parsing}\label{sec::papertoplace::authoring::snapshot}
\noindent One question for document reuse is {\it how to enable users to rapidly capture and extract the document profile from an existing instruction document}?
Inspired by mobile applications that allow users to capture and analyze scanned documents (\eg~Adobe Scan~\cite{AdobeScan} and Tab~\cite{Tab}), we similarly enable authors to simply take a snapshot of the instruction document to generate a document~profile (Figure~\ref{fig::papertoplace::system_overview}a - d).

The author can then adjust the scanned region to crop out unnecessary components (\eg~titles, \etc) as needed (Figure~\ref{fig::papertoplace::system_overview}c). 
PaperToPlace then leverages OCR services by Google Vision API~\cite{GoogleVisionAPIOCR} to parse the scanned image into machine readable text, due to its ability to extract paragraph structure in the parsed text using full text annotations~\cite{GoogleVisionAPIFullTextAnnotation}.
By default, we segment each paragraph as one step in the instructions. However, the author can re-segment the steps and fix errors in a dedicated mobile~interface (Figure~\ref{fig::papertoplace::system_overview}g).

\subsection{Selecting the Model and Key Objects}\label{sec::papertoplace::authoring::selecting_key_object}
To extract the document profile, we designed a manual and ML-assisted approach to help authors rapidly and easily associate key objects with each step.
Our ML-assisted approach leverages a pre-trained language model for a specific environment to predict the key object that is associated with each step.
After transforming the existing paper document into machine readable text, the authors need to select the model for the target environment (Figure~\ref{fig::papertoplace::system_overview}e).
For environments without a pre-trained model, the manual approach enables authors to manually extract the metadata of each step.

The author then selects the key objects that exist in the target workspace (\ie~the set of {\it available} key objects) (Figure~\ref{fig::papertoplace::system_overview}f). 
First, this step ensures the key objects contained in the extracted document profile aligns with the spatial profile of the intended environment. 
For example, a cooking instruction step such as {\it ``boil a cup of water''} could be executed either in a typical household kitchen on a {\it cooktop}, or in an office kitchen that only provides a {\it microwave}. 
Second, setting the available key objects also provides prior knowledge to help increase the accuracy while predicting key object associations.
For example, if an office kitchen only has a {\it microwave}, the key object for {\it ``boil a cup of water''} should not be predicted as {\it oven}, even though {\it oven} is a possible label for the pre-trained model (Section~\ref{sec::papertoplace::authoring::edit}).

\subsection{Creating Document Profile}\label{sec::papertoplace::authoring::edit}
Creating a document profile requires two types of metadata:

\vspace{4px}\noindent{\bf (1) The text of each procedural step.}
While by default we consider and segment each sentence as one instruction step (Section~\ref{sec::papertoplace::authoring::snapshot}), the author could overwrite the system segmented results by {\it segmenting}, {\it merging} and {\it deleting} specific step(s) (Figure~\ref{fig::papertoplace::system_overview}h). 
When a specific step is modified or a new merged step is generated, the associated key objects will be re-predicted (if the ML supported mode is used).
Although some instruction steps might be associated with multiple or no key objects, such flexibility allows the author to split or merge the target step(s).
In response to \textbf{[DC1]}, additional flexibility is also provided for the author to modify the generated text of each step to ensure that the right information with right level of details could be delivered to the consumers.

\vspace{4px}\noindent{\bf (2) The key object that the step is associated with.}
The authors can use either a manual or ML-assisted approach to determine the key object associated with each step.
Figure~\ref{fig::papertoplace::system_overview}g shows a dedicated interface with segmented instructions, where the authors can use the drop downs to select the associated key objects, with the color scale indicating the confidence of our ML predictions (if applicable).

While manually assigning each step to a key object could work robustly, predicting key objects using a ML-assisted approach that requires a pre-trained model is challenging, due to the needs for a dataset and ground truth labels.
Creating such a dataset that could be generalized to {\it all} procedural instructions is not realistic, and labeling each step with a ground truth key object is also difficult and time consuming.
Instead of preparing such dataset for \textit{all} instruction documents, we chose to focus on domain-specific dataset that is publicly available.
Alternatively, the dataset could be created via vendors or crowdsourcing.

We describe our methods for generating such a pre-trained model below.
Although our running example is based on cooking instructions, the overall approach could be transferred to other type of instructions documents, provided the unlabeled dataset is available. 

\begin{figure}[t]
    \centering
    \includegraphics[width=\textwidth]{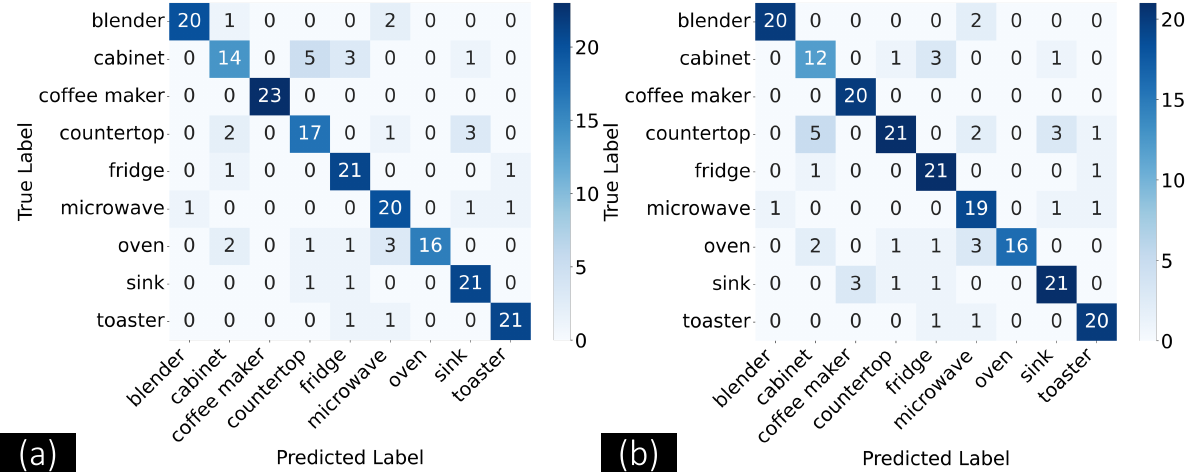}
    \caption{Confusion matrices of the fine-tuned BERT model, with the ground truth generated by rule-based method (a) and manually labeling (b).}
    \label{fig::papertoplace::model_performance}
\end{figure}

\vspace{4px}\noindent{\bf Dataset:}
We used RecipeNLG~\cite{Bien2020RecipeNLG} for training purposes, which contains more than $2.2$M cooking recipes where each recipe includes multiple ordered instruction steps.
The process of aggregating all steps from all recipes yielded an unlabeled dataset with $19.5$M steps, with an average of $11.54$ ($SD = 7.13$) words~per~step.

\vspace{4px}\noindent{\bf Rule-Based Labeling of Training Instruction Steps:} 
Instead of manually labeling each step, we used a rule-based approach to label steps that contain the \textit{exact} words of the predefined key objects.
For example, we label the step {\it ``boiling a cup of water in the \underline{microwave} for 5 min''} as {\it ``microwave''}, yet the step {\it ``boiling a cup of water for 5 min''} will not be labeled, and thus will not be included as part of our training dataset.
We iteratively selected nine key objects that exist in a typical kitchen: {\it blender}, {\it cabinet}, {\it coffee maker}, {\it countertop}, {\it fridge}, {\it microwave}, {\it oven}, {\it sink}, {\it toaster}.
Such selections also ensures a reasonable amount of instruction steps for subsequent model fine-tuning purposes.
We generated a dataset where each of the nine labels is associated with $218$ instruction steps (\ie~the \emph{labeled} dataset contains $218 \times 9 = 1962$ instruction steps).

\vspace{4px}\noindent{\bf Training:}
Due to the limited dataset size, we fine-tuned the model using the output classification layer of a $12$-layer BERT model for uncased vocabulary, which has been used for generating contextual language embeddings~\cite{Devlin2018}.
We used Adam optimizer with the learning rate, $\epsilon$, and batch size set to $2\times10^{-5}$, $10^{-8}$, and $32$, respectively, recommended by Devlin~\etal~\cite{Devlin2018}.
$80\%$, $10\%$ and $10\%$ of the dataset are used for training, validation and testing, with the amount of steps for each key object balanced across the sets.

\vspace{4px}\noindent{\bf Model Performance:}
We demonstrated an overall $83.57\%$ testing accuracy by considering the label generated by our rule-based approach as the ground truth (Figure~\ref{fig::papertoplace::model_performance}a).
Additionally, we manually label the associated key objects on the testing dataset to limit the impact from any errors generated by our rule-based approach, which lead to a $82.13\%$ overall accuracy (Figure~\ref{fig::papertoplace::model_performance}b).

\vspace{4px}\noindent{\bf Model Execution: }
The pre-trained model is used for predicting key objects from the segmented text of each step. 
To enhance the accuracy of the predicted key objects, we use prior knowledge provided while specifying the available key objects (Section~\ref{sec::papertoplace::authoring::selecting_key_object}).
Specifically, the final assigned key object is the ML predicted key object with the highest confidence score that also belongs to the set of available key objects of the target environment.

\section{Consumption Pipeline}\label{sec::papertoplace::consuming}
The consumption pipeline aims to spatialize each steps by anchoring them at the optimal position next to the key object.
For example, consider how the instruction {\it``microwave on high for 30 seconds''} should be attached to a microwave. An ideal location would be at the front surface of the microwave door. A less idea location would be at the front of the input panel because the instruction might get in the way when the user tries to set the timer (see examples in Figure~\ref{fig::papertoplace::teaser}c, Figure~\ref{fig::papertoplace::system_overview}i, and Figure~\ref{fig::papertoplace::consumption::examples} in Appendix~\ref{sec::papertoplace::app::study_results::consuming}).

\subsection{Interaction Design}\label{sec::papertoplace::consuming::interactions}
Our consumption pipeline provides dedicated interaction metaphors based on the preliminary findings (Section~\ref{sec::papertoplace::prelim::findings}).

\vspace{4px}\noindent{\bf Navigating Between Individual Steps.}
Consumers can use hand menus to easily and rapidly switch between steps (Figure~\ref{fig::papertoplace::teaser}b).
We adopted the suggestions from \textbf{DC1} and the conceptual design of Figure~\ref{fig::papertoplace::prelim_study_results}c that advocate the idea of delivering the right level of information only at the right time.
Therefore, PaperToPlace only renders the current instruction step along with a task completion progress bar.
When a new step is triggered, PaperToPlace first anchors the virtual label in front of the consumer, since a initial instruction step consuming is usually required before consumers proceeding on execute the associated steps (Figure~\ref{fig::papertoplace::prelim_study_results}b).

\vspace{4px}\noindent{\bf Animating Spatial Guidance.}
To address \textbf{DC3}, we decided not using the persistent visual guidance (\eg~virtual arrows)~\cite{Johnson2018, Johnson2021} that might cause unnecessary visual disturbance.
Instead, we use a animated flying effect where the virtual step could {``fly''} toward the key object after initial instruction step consuming.
Such design leverage the fact that a motion effect could direct the consumers' attention, and could implicitly and rapidly offer visual guidance of the spatialized key object without causing overwhelming disturbance while consumers are executing the steps~\cite{Harley2014}.

\vspace{4px}\noindent{\bf Placement of Instruction Steps.}
PaperToPlace places and anchors the instruction step on one of the anchoring surfaces of the key objects while not occluding the important region.
This design emphasizes \textbf{DC2} suggesting the connections between instruction and real-world contexts, and could bring convenience while the consumers are attempting to refer back to the instruction step repeatedly while completing the step (Figure~\ref{fig::papertoplace::prelim_study_results}c).
If the consumer dislikes the label position, they can request a new position update on-demand using a mid-air pinch gesture.
We also allow the consumers to use pinch-and-drag gestures to manually move the step to their preferred place (Figure~\ref{fig::papertoplace::teaser}d).
Such feedback action will in turn help on future decisions while placing instruction steps.

\begin{table}[t]
    \small
    \centering
    \caption{Notations of the key parameters and functions.}
    \begin{tabular}{ll} 
     \hline
     \textbf{Notations} & \textbf{Descriptions} \\
     \hline\hline 

     $W$, $H$ & \makecell[l]{Number of discretized cells along the width and \\ height of the anchoring surface.} \\ 
    
     $rot_s$ & \makecell[l]{Rotation, represented by quaternion, of the surface $s$.}  \\
     
     \makecell[l]{$du_s$, $dr_s$, $df_s$} & \makecell[l]{The up, right, and forward direction of the surface $s$.}\\
     
     $p_{eye}$ & \makecell[l]{The midpoint of left and right eye in world coordinate.}\\ 
     
     $df_{eye}$ & \makecell[l]{The forward direction of the gaze, averaged by left \\ and right eye gaze.}\\ 
      
     $a = (r, c, s)$ & \makecell[l]{Representation of an instruction step placement with \\ respect to anchoring surface $s$, with index of $r$ and $c$ \\ along width and height, where $r \in [0, W)$, $c \in [0, H)$.}\\
     
     $p_{a}$ & \makecell[l]{The position of the instruction step in world coordinate.}\\

     $\theta_{x, y}$ & \makecell[l]{The angle between vector $x$ and $y$.}\\
     
     \hline
    \end{tabular}

    \label{table:notations}
\end{table}

\subsection{Problem Formulation}\label{sec::papertoplace::consuming::problem}
We formulated the process of optimally placing the instruction step as an optimization problem, where we used the tracked hands and gaze, as well as the anchoring surfaces defined by the spatial profile to search the optimal placement for each step.
Table~\ref{table:notations} summarizes the notations of key parameters.

\vspace{4px}\noindent{\bf Representations of an Instruction Step Placement.}
We first discretized each anchoring surface into $W \times H$ virtual cells where each cell has a dimension of $3cm \times 3cm$.
We assumed that the center of each virtual step should be aligned with the center of the cell on the anchoring surface.
We used $a = (r, c, s)$ to indicate the placement of a step, where $r$ and $c$ indicate the index of cell along the width and height of the surface $s$.
The world position of the attempted step placement is $p_{a} = p_{topLeft} + 0.03 \cdot dr_s \cdot r - 0.03 \cdot du_s \cdot c$, where $p_{topLeft}$ is the position of the top left vertex of surface $s$. 

\begin{figure}[t]
    \centering
    \includegraphics[width=0.8\textwidth]{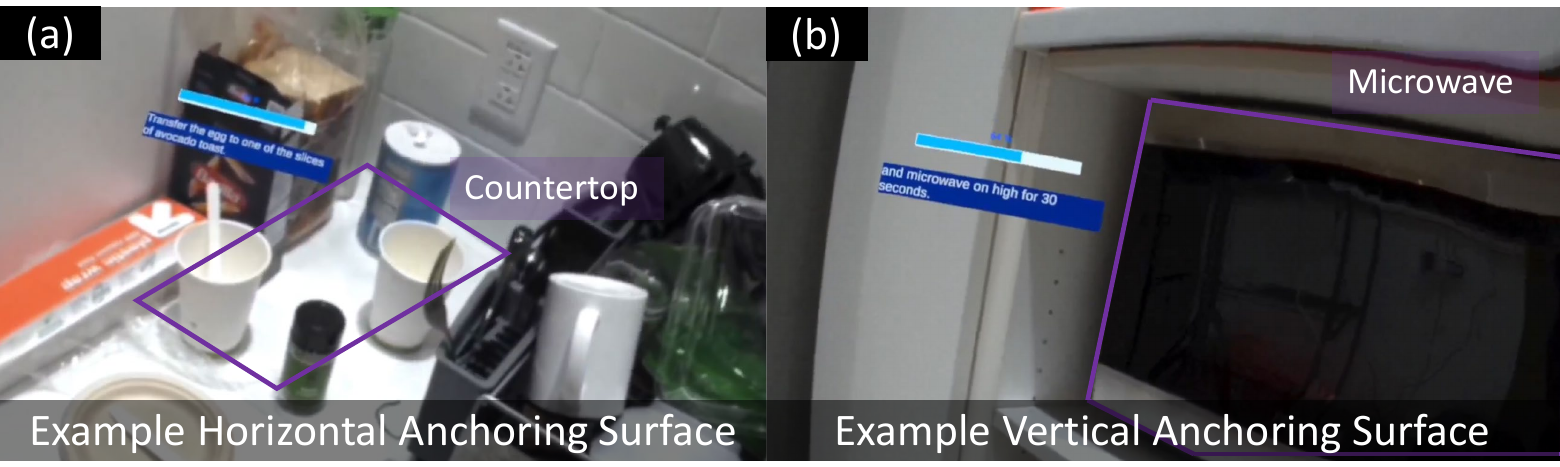}
    \caption{First-person view of rotating instruction step for (a) horizontal anchoring surface (\eg countertop) and (b) vertical anchoring surface (\eg microwave).}
    \label{fig::papertoplace::label_rotation}
\end{figure}

\vspace{4px}\noindent{\bf Representations of the Step Label Rotation.}
Reading angles is a critical factor for consuming document~\cite{Morris2007}.
It is therefore important to determine the rotation of the step, such that the text is always perpendicular to user's looking direction (\ie~the virtual texts should be delivered facing toward user's eye).
To address this, we rotated the anchoring surfaces by using the potential looking direction~($p_{a} - p_{eye}$).
Algorithm~\ref{alg:lable_rot} shows how we compute the rotation of the step for horizontally (\eg~countertop) and vertically placed anchoring surfaces (\eg~the front surface of fridge).
Figure~\ref{fig::papertoplace::label_rotation} demonstrates two examples where the steps are appropriately rotated.

\begin{algorithm}[t]
   \small 
   \caption{Computing the rotation of instruction step.} \label{alg:lable_rot}
   \begin{algorithmic}[1]
        \PROCEDURE{GetRotation}{$rot_{s}$, $p_a$, $p_{eye}$}
            \STATE $dir \gets p_a - p_{eye}$  \COMMENT{Approximate potential looking direction.}
            \STATE $\alpha_{up} \gets Angle(du_s, dir)$
            \STATE $\alpha_{right} \gets Angle(dr_s, dir)$           
            \IF{$s$ is a horizontal anchoring surface}
               \STATE \textbf{return} $rot_{s} * Quaternion.Euler(90^{\circ} - \alpha_{up}, 0, 90^{\circ} - \alpha_{right})$
            \ELSE[$s$ is a vertical anchoring surface.]
                \STATE \textbf{return} $rot_{s} * Quaternion.Euler(90^{\circ} - \alpha_{up}, \alpha_{right} - 90^{\circ}, 0)$
            \ENDIF
        \ENDPROCEDURE
    \end{algorithmic}
\end{algorithm}

\vspace{4px}\noindent{\bf Importance Map.}
One goal while attempting to place an instruction step onto associated anchoring surface(s) is to find the optimal placement that will not occlude consumer's interactions with the key object.
While Lang~\etal~\cite{Lang2019} presumed that the centroid of a key object is the most important area and should not be occluded by virtual MR agents, such assumption is invalid in our problem as the real-world interactions are highly dynamic.
For example, while interacting with microwave, the critical areas might be the region on top of keypad when inputting cooking time, or the center areas when the user is checking whether the food is cooked.
We used {\bf importance map}~($imap$) to represent the importance of each possible cell on anchoring surface, where a larger $imap$ value implies a higher probability that the corresponding position being interacted, and therefore should not be occluded by the virtual step. 
The values of $imap$ are determined using near real-time data provided by the MR headset.
By leveraging the pre-created spatial profile, we used the tracked gaze and hands to infer the importance of each possible position on anchoring surface(s).
Intuitively, the areas near the hands, which usually imply the regions that are interacted by the consumers, might be more important and should not be occluded by the steps.
Section~\ref{sec::papertoplace::consuming::map_and_surface} describes the computations of $imap$.

\subsection{Importance Map on Anchoring Surfaces}\label{sec::papertoplace::consuming::map_and_surface}

\begin{algorithm}[t]
   \small 
    \caption{Approximating the importance of individual frame.}
    \label{alg:frame_weights}
    \begin{algorithmic}[1]
        \PROCEDURE{GetFrameWeights}{$v_{left}[\;]$, $v_{right}[\;]$, $t[\;]$}
            \STATE $N \gets Length(t)$  \COMMENT{$N$ indicates the frame size.}
            \STATE $w, w_{speed}, w_{time} \gets [0] * N, [0] * N, [0] * N$
        
            \FOR{$i \gets 1$ to $N$}
                \STATE $w_{speed}[i] \gets (|v_{left}[i]| + |v_{right}[i]|)/2$
                \STATE $w_{time}[i] \gets (t[i]- t[1]) / (t[N] - t[1]) $
            \ENDFOR
        
            \STATE $w_{speed} \gets MinMaxNormalize(w_{speed}) $
            \COMMENT{Normalize to $0$ to $1$.}
            
            \STATE $w_{time} \gets MinMaxNormalize(w_{time}) $
            \COMMENT{Normalize to $0$ to $1$.}
            
            \STATE $w \gets Normalize((1 - w_{speed}) w_{time}) $ \COMMENT{$\Sigma_1^{N}(w[i])$ should be $1$.}
        
            \STATE \textbf{return} {$w$}
        
        \ENDPROCEDURE
    \end{algorithmic}
\end{algorithm}

\begin{figure}[t]
    \centering
    \includegraphics[width=\textwidth]{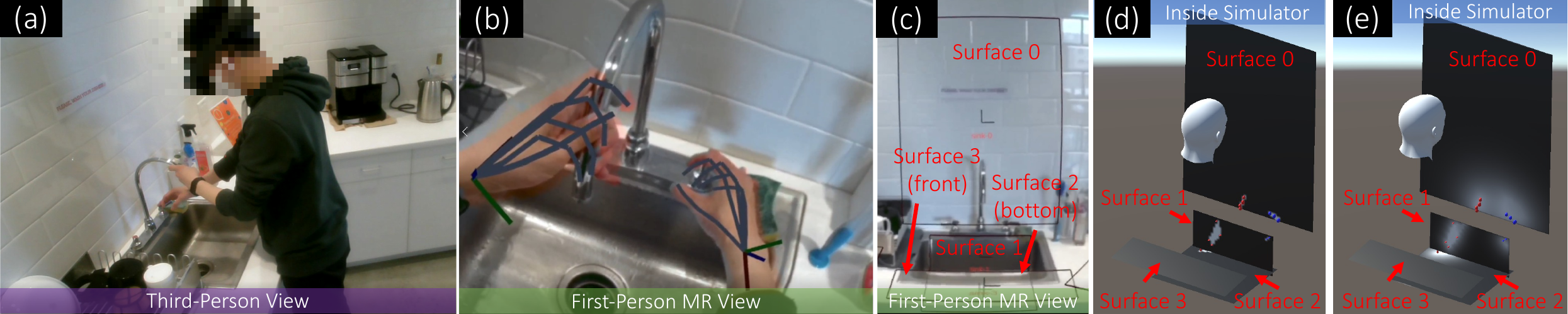}
    \caption[Examples of the importance map.]{Examples of the importance map. (a) Third-person view; (b) First-person view through MR; (c) First-person view of the key object (sink) containing four anchoring surfaces; (c) Importance map from single frame; (d) Overall importance map from a set of frames. The {red} and {blue} points in (d) and (e) indicate the projected key joints of the tracked left and right hand on the anchoring surfaces, respectively.}
    \label{fig::papertoplace::importance_map}
\end{figure}

\noindent{\bf Approximate the Importance of Each Frame.} 
The contextual data from each {\bf frame} refers to the tracked gaze and hand joints, at a specific time instant.
Inferring $imap$ from contextual data on single frame is less reliable, as real-world activities are highly dynamic.
Yet, contextual data collected from a set of frames are practically not equally important.
Therefore, while aggregating contextual data across a set of frames, it is important to consider the relative importance of individual frames.
PaperToPlace uses the tracked eye behaviors to approximate the relative importance of contextual data of each frame.
The \first intuition is based on the instantaneous angular speed of gaze, where a slower angular speed of gaze implies a higher importance.
For example, while attempting to input time when using microwave, the consumer might be \textit{fixing} at the keypad, during which the contextual data could offer meaningful clues for approximating $imap$.
Whereas, the consumer might rapidly \textit{saccade} around the environment while finding ingredients, during which the contextual data might be less meaningful.
While \cite{Nystrom2010, Mould2018} attempted to design closed-form solution for classifying saccade and fixation using the speed of gaze, such eye behaviors usually varies across users and tasks.
Our \second intuition is based on the observation that the contextual data from a more recent frame might be more useful to indicate the interactions in the subsequent task episode.
Algorithm~\ref{alg:frame_weights} shows the computation of the {\bf frame weight}~($w$) that is used to quantify the relative importance of contextual data at each frame.
Experimentally, we set $N = 90$, which is approximately $1$ second of past contextual data.
We approximated $w$ based on instantaneous angular speed of left and right gaze~($v_{left}, v_{right}$), and the timestamps ($t$) of each frames.
Remarkably, a slower eye moves (\ie~smaller $v_{left}$ and $v_{right}$) and a more recent timestamp (\ie~ larger $t$) would lead to a more important frame weight.
The $MinMaxNormalize(\cdot)$ and $Normalize(\cdot)$ represent the {\it min}-{\it max} normalization (to $[0, 1]$), and the normalization process such that the summation of the list is $1$.

\begin{algorithm}[t]
    \small
    \caption{Algorithms for computing importance map.}
    \label{alg:map}
    
    \begin{algorithmic}[1]

        \PROCEDURE{GetMap}{$trackedJoints[]$, $W$, $H$}
            \STATE $hits, mask \gets [], zeros(W, H)$
            \FOR{$p_{joint}$ in $trackedJoints$}
                \STATE $hit \gets RayCast(start:p_{eye}, direction:p_{joint} - p_{eye})$
                \IF{$hit \neq null$}
                    \STATE $w \gets ToWidthIndex(hit.x)$
                    \STATE $h \gets ToHeightIndex(hit.y)$
                    \STATE $mask[w, c] \gets 1$
                    \STATE $hits.add([w, c])$
                \ENDIF
            \ENDFOR
            
            \STATE $p_{hull} \gets ConvexHull(hits)$
            \STATE $mask \gets FloodFill(mask, p_{hull})$
            \STATE \textbf{return} {$mask$}
        \ENDPROCEDURE
        \STATE  
        \PROCEDURE{GetOverallMap}{$frames$, $W$, $H$}
        
            \STATE $map \gets zeros(W, H)$
            \FOR{$i \gets 1$ to $N$}
                \STATE $imap_{left} \gets GetMap(frames[i].left.joints, W, H)$
                \STATE $imap_{right} \gets GetMap(frames[i].right.joints, W, H)$
                \STATE $map \gets map + w[i] * (Soft(imap_{left}) + Soft(imap_{right}))$
            \ENDFOR
            \STATE \textbf{return} {$MinMaxNormalize(map)$}     \COMMENT{Normalize to $0$ to $1$.}
        \ENDPROCEDURE
    \end{algorithmic}
\end{algorithm}

\vspace{4px}\noindent{\bf Approximate the Importance Map from Tracked Hands.}
To compute the overall $imap$ from a set of frames, we first approximated the $imap$ from the contextual data collected at each frame.
This could be realized by $GetMap(\cdot)$ (Algorithm~\ref{alg:map}), which $FloodFill$ the $ConvexHull$ of the projected points of $15$~tracked hand joints~\cite{OculusHandTracking} on the associated anchoring surfaces. 
Figure~\ref{fig::papertoplace::importance_map}d shows an example of the $imap$ from single frame, approximated by tracked hands in Figure~\ref{fig::papertoplace::importance_map}b.

\vspace{4px}\noindent{\bf Generate the Overall Importance Map.} 
While we assumed that the areas that are not occluded by hands might be less important, the importance assigned to each cell on the anchoring surfaces might not be equally same.
For example, while inputting time using keyboard of microwave, the further the step being placed from the areas occluded by hands could lead to lower probability that the important areas being occluded by hands. 
Therefore, $imap$ should be expected to model {\it how important of a specific pixel on the anchoring surface}, instead of {\it whether the particular pixel is important}.
Inspired by Lang~\etal~\cite{Lang2019}, we used $Soft(\cdot)$ to approximate the importance at each possible placement on the anchoring surface(s) generated by left and right hand respectively (Algorithm~\ref{alg:map}). 
Equation~\ref{alg:blur_map} describes this process, where $e_{min}^{r, c}$ indicates the $L2$-distance from placement $(r, c)$ to the closest placement(s) where the computed importance of individual frame from $GetMap(\cdot)$ (Algorithm~\ref{alg:map}) is $1$.

\begin{equation}
    \label{alg:blur_map}
    imap(r, c) = 1 - \frac{e_{min}^{r, c}}{max_{r', c'}e_{min}^{r', c'}}
\end{equation}

The overall $imap$ is finally computed by aggregating the $imap$ generated by left ($imap_{left}$) and right ($imap_{right}$) hands on each frame using previous computed weight ($w$) in Algorithm~\ref{alg:frame_weights}.
This process could be demonstrated in $GetOverallMap(\cdot)$ (Algorithm~\ref{alg:map}).
Notably, if there is no area being occluded by hands, we set $imap = \bm{O}_{W, H}$. 
Figure~\ref{fig::papertoplace::importance_map}e shows an example overall $imap$.

\subsection{Constraints and Costs}\label{sec::papertoplace::consuming::cost}
\noindent To solve the optimization problem for placing the instruction step on the anchoring surface(s), we need to model the constraints such that {\it the placed step will minimally occlude the user's view and will not be too far from the user's focused attention}.

\vspace{4px}\noindent{\bf Total Cost.} 
We designed the overall cost ($C_{total}(a)$, Equation~\ref{eq::total_cost}) as a weighted sum of the visibility cost ($C_V({a})$), the readability cost ($C_R({a})$), hand angle cost ($C_{HA}({a})$), and preference cost ($C_P({a})$).
$\lambda_V$, $\lambda_R$, $\lambda_{HA}$, and $\lambda_P$ are the weights associated with each of designed cost.
Experimentally, we set them to $0.24$, $0.24$, $0.24$ and $0.28$, with slight emphasis on consumers' preference.
We provide rationales of the design of each costs.

\begin{figure}[t]
    \centering
    \includegraphics[width=0.8\textwidth]{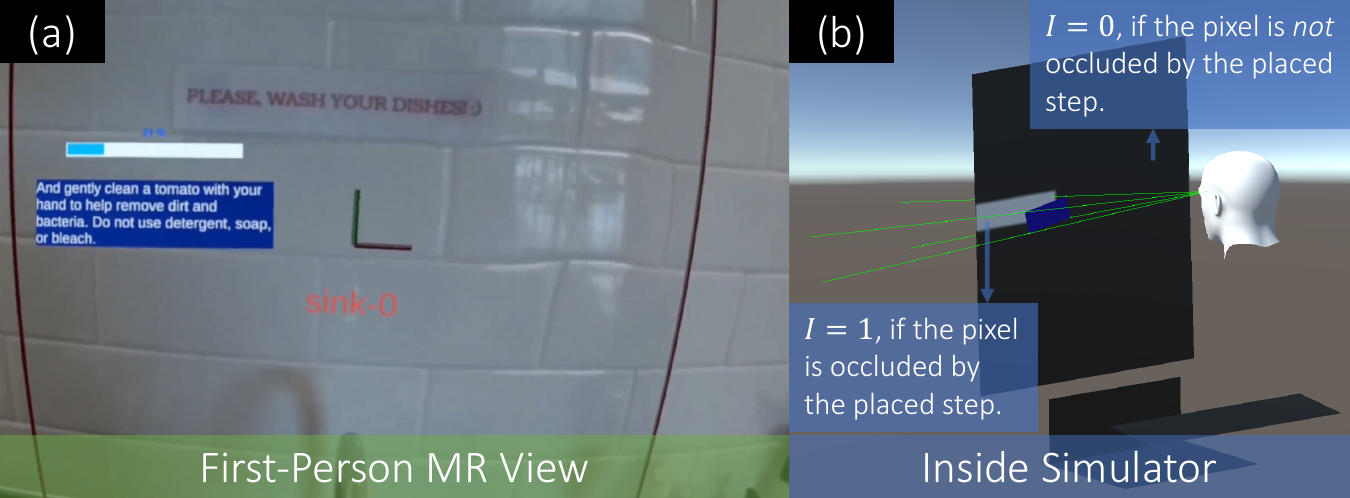}
    \caption[Example of occlusion map.]{Example of occlusion map. (a)~An example placement of the instruction step anchored next to the sink; (b)~Visualization of the generated occlusion map, see Figure~\ref{fig::papertoplace::importance_map}c for the corresponding real-world scene.}
    \label{fig::papertoplace::occlusion_map}
\end{figure}

\begin{equation}
    C_{total}(a) = \lambda_V C_{V}(a) + \lambda_R C_{R}(a) + \lambda_{HA} C_{HA}(a) + \lambda_P C_{P}(a) 
    \label{eq::total_cost}
\end{equation}

\vspace{4px}\noindent{\bf (i) Visibility Cost.}
$C_V$ aims to to measure how much key areas of the anchoring surfaces are occluded by a step placement $a$, and to penalize the situation while the step occluding the important areas.
Equation~\ref{eq::visibility_cost} defines $C_V$, where $I$ indicates {\bf occlusion map} and $imap$ indicates the relative importance of each discretized cells on the anchoring surfaces computed by Algorithm~\ref{alg:map}.

\begin{equation}
    C_{V}(a) = \frac{[\sum_{r, c} I(r, c) \cdot imap(r, c)]^2}{||imap(r, c)||_2 \cdot \sum_{r', c'}I(r', c')}  
    \label{eq::visibility_cost}
\end{equation}

Notably, the $I(r, c)$ is assigned as $1$ when the pixel $(r, c)$ is occluded by the step from the center of both eye (Figure~\ref{fig::papertoplace::occlusion_map}).
Figure~\ref{fig::papertoplace::occlusion_map}b shows an example occlusion map when the instruction step is placed next to the wall behind the sink.
We finally normalized $C_V(a)$ to be independent of dimensions of anchoring surface(s).

\vspace{4px}\noindent{\bf (ii) Readability Cost.}
We penalized the solution when the delivered step is too far from the user's attention.
To model this constraints, we used the eye tracking results and measure the $L2$-distance from the placed step to the weighted average of looking direction of both eye ($df_{eye}$).
Equation~\ref{eq::readability_cost} defines $C_R(a)$, where $d_{max}$ represents the maximum $L2$-distance between two arbitrary solutions on the anchoring surfaces, which is computed by the maximum distance of the convex hull consisting of all vertices of the anchoring surfaces.
Additionally, $C_R(a)$ need to enforce the instruction step is placed within binocular vision (\ie~approximately $\pm 60^{\circ}$)~\cite{Pettigrew1986}, to minimize the needs of moving head in order to read the instructions.
We used a coefficient $k$ to penalize the cost function, where $k$ is set to $1$ if $\theta_{p_a - p_{eye}, df_{eye}} < 60^{\circ}$, otherwise we set  $k = \theta_{p_a - p_{eye}, df_{eye}}$.

\begin{equation}
    C_{R}(a) = \frac{k \cdot ||({p}_{a} - {p}_{eye}) \cdot {df}_{eye} \cdot {df}_{eye} - ({p}_{a} - p_{eye})||_2}{d_{max}} \label{eq::readability_cost}
\end{equation}

\vspace{4px}\noindent{\bf (iii) Hand Angle Cost.}
We modeled the observation that the instruction documents are usually held by and placed in front of the consumers' hands (Figure~\ref{fig::papertoplace::prelim_study_tasks}).
We first computed the angle between forward direction of the hand and the direction vector pointing from hand to the attempted solution ($\theta_{df_{hand}, p_{a} - p_{hand}}$), for left and right hand respectively (noted as $\theta_{leftHand}(i)$, $\theta_{rightHand}(i)$ computed from frame $i$).
We then formulated the overall hand angle cost, by aggregating the angle cost generated by each frame (Equation~\ref{eq::hand_angle_cost}).

\begin{equation}
    C_{HA}(a) = \sum_{i} \frac{w[i] \cdot [\theta_{leftHand}(i) + \theta_{rightHand}(i)]}{360^{\circ}} \label{eq::hand_angle_cost}
\end{equation}

\vspace{4px}\noindent{\bf (iv) Preference Cost.}
While PaperToPlace could determine the optimal step placement based on near-real-time context data, we retained the flexibility for users to specify their preferred placements.
Eqn.~\ref{eq::preference_cost} defines $C_P(a)$, where $p_{pref, S_i}$ refer to the preferred step placement in the world coordinates for step $S_i$.
We used $p_{pref, S_i} = null$ to indicate that the user has not manually fix the step placement.
Similar to $C_R$, we used $d_{max}$ to normalize preference cost.

\begin{equation}
    \label{eq::preference_cost}
     C_{P}(a) = 
    \begin{cases}
       \frac{||p_{a} - p_{pref, S_i}||_2}{d_{max}} & p_{pref, S_i} \neq null\\
       0 & p_{pref, S_i} = null
    \end{cases}
\end{equation}

\begin{figure}[h!]
    \centering
    \includegraphics[width=0.7\textwidth]{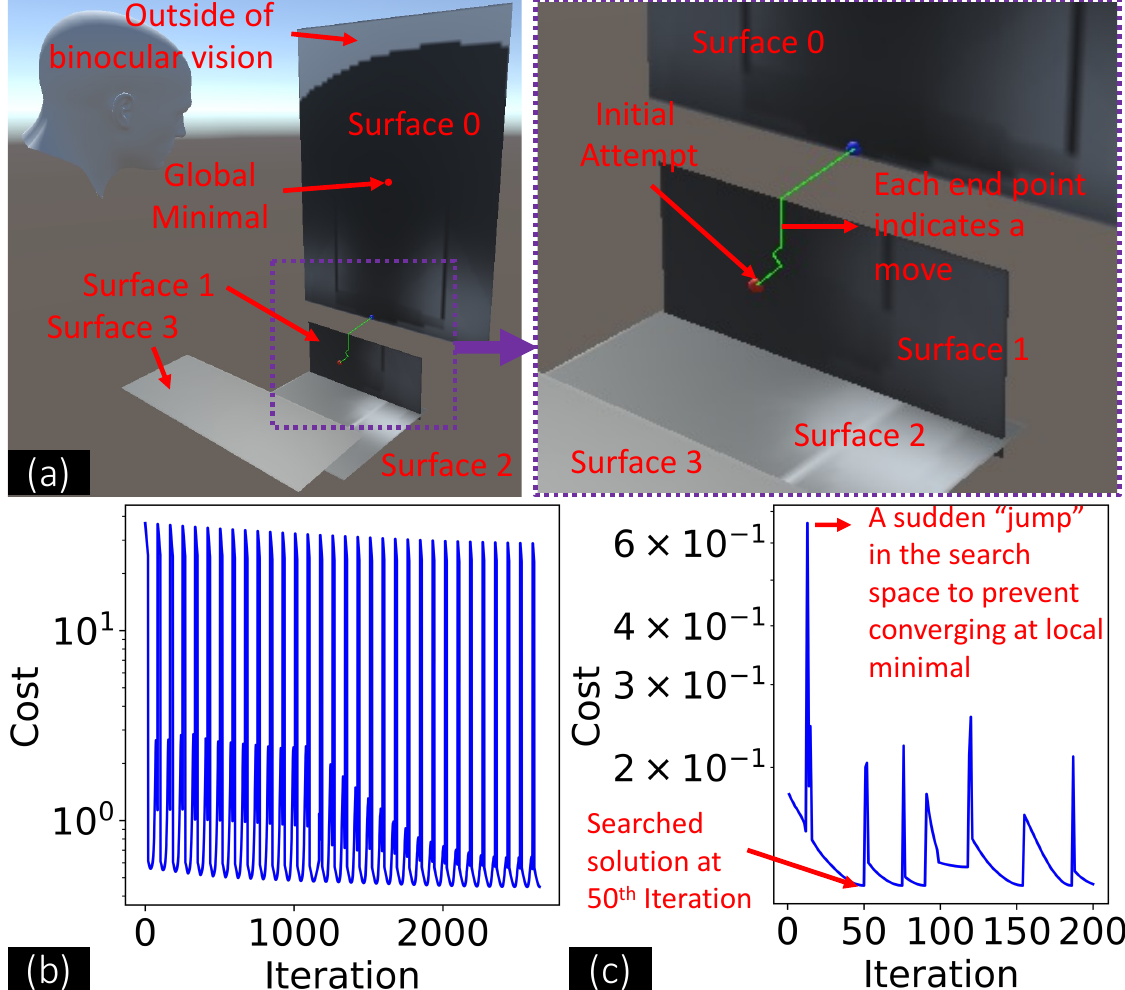}
    \caption[(a) Example placement attempts. (b - d) Example optimized cost over each iteration using greedy algorithm~(b) and simulated annealing approach~(c).]{(a) Example placement attempts (the green trace). The darker color of the anchoring surface indicates a lower $C_{total}$; (b - d) Example optimized cost over each iteration using greedy algorithm~(b) and simulated annealing approach~(c). To increase readability, $log$ scale is applied for $y$-axis. We only showed the traces before current cost reaching global minimal.}
    \label{fig::papertoplace::example_energy_traces}
\end{figure}

\begin{figure}[t]
    \centering
    \includegraphics[width=\textwidth]{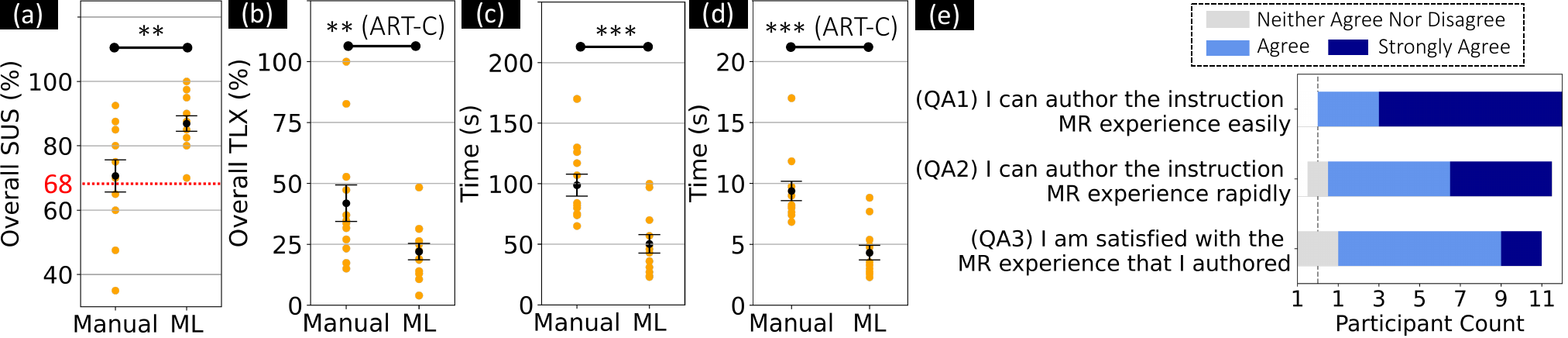} 
    \caption[Results of authoring pipeline evaluations.]{Results of authoring pipeline evaluations. (a - d) The overall SUS, weighted TLX scores, TCT of extracting document profiles, and the average task completion time for deciding each instruction step with segmented instruction step; (d) Survey results of how participants assessed the overall authoring pipeline and authored MR experience.}
    \label{fig::papertoplace::authoring_eval}
\end{figure}

\subsection{Optimizations}
We aimed to minimize $C_{total}$ by searching optimal placement for a specific step $\hat{a}$ (\ie~$\hat{a} = \argmin_{a} C_{total}(a)$).
Finding optimal placement by computing $C_{total}$ for each possibilities is impractical due to unnecessary latency and computational overhead.
We instead used simulated annealing to approximate the global optimal~\cite{Kirkpatrick1983}.

\vspace{4px}\noindent{\bf Make a Move by Choosing the Neighbor Solution.}
We chose the neighbour solution $a^{i+1} = (r^i + {\delta_r}^{i}, c^i + {\delta_c}^i)$ by making a move of current solution $a^{i} = (r^i, c^i)$ at iteration $i$.
Equation~\ref{eq::move} describes our approach to determine the change over width (${\delta_r}^{i}$) and height (${\delta_c}^{i}$) of the anchoring surface at iteration $i$ that yields smallest $C_{total}$, where $({\delta_r}^{i}, {\delta_c}^{i}) \in \{({\delta_r}, {\delta_c}) | {\delta_r}, {\delta_c} \in \{ \pm1, 0\} \land {\delta_r}{\delta_c} \neq 0\}$.

\begin{equation}
    \delta_r, \delta_c = {argmin}_{\delta_r', \delta_r'} {C_{total} (a^{i + 1} = (r^i + \delta_r', c^i + \delta_r'))} 
    \label{eq::move}
\end{equation}

Notably, to ensure the sampled neighboring solution is on the target anchoring surface, Equation~\ref{eq::move_constraints} should be satisfied.

\begin{equation}
    0 \leq r^{i+1} \leq W \quad 0 \leq c^{i+1} \leq H \quad r^{i+1}, c^{i+1} \in \mathbb{N}    
    \label{eq::move_constraints}
\end{equation}

Although such method performs well while making moves \textit{within} single anchoring surface, the attempted solution will not be made \textit{across} the surfaces.
To address this, we specify that the placement on the neighbor anchoring surface, which is closest to the current placement attempt, would be chosen upon Equation~\ref{eq::move_constraints} being violated.  
Figure~\ref{fig::papertoplace::example_energy_traces}a demonstrates an example of how the neighbor solution is chosen while placement moving across the surfaces.
To prevent converging at local minimal, we choose a random move in the global search space if $C_{total} (a^{i + 1}) > C_{total} (a^{i}) $~\cite{Kirkpatrick1983}.

\vspace{4px}\noindent{\bf Metropolis-Hastings Sampling and Simulated Annealing.}
We first selected a random placement on one of anchoring surface(s) randomly and used the Metropolis-Hastings algorithm~\cite{Robert1999} to sample the subsequent attempt.
The probability for accepting the new attempt $a^{i+1}$ is determined by Metropolis criteria.
Equation~\ref{eq::samping} defines the the transition kernel of the Markov chain, where $T(i)$ indicates the temperature that will decay over the iteration.

\begin{equation}
    p(a^{i+1} | a^i; i) = min \{1, exp(\frac{C_{total}(a^i) - C_{total}(a^{i+1})}{T(i)})\}\label{eq::samping}
\end{equation}

Notably, we used the empirical definition of $T(i)$, where $T(i) = \frac{T_1}{i + 1}$~\cite{Kirkpatrick1983, Lang2019}.
Experimentally, we set $T_1 = 100$, and the number of iterations $i_{max} = 200$.

\vspace{4px}\noindent{\bf Comparisons of Optimization Performance.}
Figure~\ref{fig::papertoplace::example_energy_traces} demonstrates an example optimization result while attempting to place a step in front of sink that consist of four anchoring surfaces (Figure~\ref{fig::papertoplace::importance_map}c).
To better demonstrate the merits, we used greedy approach that tries each possibilities~(Figure~\ref{fig::papertoplace::example_energy_traces}b) and simulated annealing approach (Figure~\ref{fig::papertoplace::example_energy_traces}c).
Figure~\ref{fig::papertoplace::example_energy_traces}a visualizes the $C_{total}$ at each pixel on the anchoring surfaces with greedy approach, where darker area indicating a lower $C_{total}$.
We showed that the greedy approach and simulated annealing approach need to make $2650$ and $49$ attempts respectively before finding the global minimum.

\section{Implementations}\label{sec::papertoplace::implement}
We implemented the authoring pipeline on an iPad ($9$th generation) and the consumption pipeline on the Quest Pro~\cite{QuestPro} due to its colored passthrough, as well as eye~\cite{OculusEyeTracking} and hand tracking~\cite{OculusHandTracking, Chen2022PrecisionDrawing} capabilities.
We also implemented an optimization server using Unity 2021.3.9~\cite{unity20210309} on a separate machine, and a Flask server for managing document (Section~\ref{sec::papertoplace::authoring}) and spatial profiles (Section~\ref{sec::papertoplace::consuming}).

\section{User Studies}\label{sec::papertoplace::study}

\noindent Two \textit{within-subject} studies were designed to evaluate PaperToPlace.
$12$ participants (PA1 - PA12) were recruited for evaluating the authoring pipeline  and another $12$ participants (PC1 - PC12) were recruited for evaluating consumption pipeline.
During the evaluation, participants either authored or consumed the MR experiences for three cooking tasks (T1 - T3) that could be easily conducted in a typical office kitchen.  
Each study consists of two sessions where participants need to complete the designated tasks that involves with four key objects: {\it microwave}, {\it fridge}, {\it sink}, and {\it countertop}~(Figure~\ref{fig::papertoplace::study_design}).
We used T1 as the training task, through which participants could get familiar with the designed interfaces.
T2 and T3 were used for formal evaluations, each of which could be completed within $10 \sim 15$~min.
Appendix~\ref{sec::papertoplace::app::tasks} provides the details of instructions.

\subsection{User Study 1: Authoring Pipeline}\label{sec::papertoplace::study::authoring}
\noindent The first study aims to understand how well our authoring pipeline can support authors in easily and rapidly creating an MR instruction experience with manual or ML-assisted mode (Section~\ref{sec::papertoplace::authoring::edit}).
We aim to tackle three Research Questions~(RQs): 

\begin{itemize}[leftmargin=*]
    
    \item \textbf{(RQ1)} How the proposed authoring pipeline usable?
    
    \item How the ML could support \textbf{(RQ2)} a {\it faster} and \textbf{(RQ3)} an {\it easier} authoring experience?
    
\end{itemize}

\vspace{0.1in}
\noindent{\bf Participants and Procedures}.
PA1 - PA12 (age, \mbox{$M = 25.33$}, $SD = 2.81$, \incl~ eight males and four females) were recruited.
Six participants disagreed that they are experts of the designated tasks, with the remaining participants held a neutral opinion.
After training participants to use both of interfaces with T1, participants then completed one task for each interface condition (manual and ML-assisted). 
The tasks and interface conditions were counterbalanced across participants (Figure~\ref{fig::papertoplace::study_design} in Appendix~\ref{sec::papertoplace::app::study_results}). 
Participants were finally instructed to briefly test the MR experience that was authored, to see if they were satisfied with the authoring outcomes inside MR.
After each session, participants were asked to rate how strongly they agree with three prompts, shown in Figure~\ref{fig::papertoplace::authoring_eval}e, in a $5$-point Likert scale. 
Participant were then invited to fill out the NASA TLX~\cite{nasatlx}, followed by System Usability Scale (SUS)~\cite{Brooke1996sus}, as approximations of perceived workload and level of system usability.
A semi-structured interview was conducted to understand participants' responses.
The study on average took $37.87$~min ($SD = 5.45$~min).

\vspace{0.1in}
\noindent{\bf Measures.}
We analyzed the differences in the overall SUS, weighted TLX scores, and Task Completion Time~(TCT) while extracting document profiles using both the manual and ML-supported modes.
Shapiro-Wilk test~\cite{Shapiro1965} was used to check the normality of data in each catalogue.
Repeated Measure Analysis of variance (RM-ANOVA)~\cite{Girden1992} with Tukey's HSD~\cite{Abdi2010} ($\alpha = 0.05$) were used for analyzing statistical significance and \textit{post-hoc} comparisons.
Upon a failure of normality check, the Aligned Rank Transform~(ART)~\cite{Wobbrock2011}, followed by ART contrast test~(ART-C)~\cite{Elkin2021} were used as the nonparametric approach for statistical significance analysis and \textit{post-hoc} test\footnote{Notations for indicating the \textit{post-hoc} test results shown in Figure~\ref{fig::papertoplace::authoring_eval}, \ref{fig::papertoplace::consumption::evaluation_measure}, and \ref{fig::papertoplace::consumption::overall}: $*$ ($p < .05$), $**$ ($p < .01$), $***$ ($p<.001$).}.
The partial eta square ($\eta_p^2$) was used to evaluate the effect size for ART and RM-ANOVA, with $.01$, $.06$ and $.14$ indicating the thresholds for small, medium and large effect size~\cite{Cohen1998}.
We used thematic analysis~\cite{Braun2012}, and deductive and inductive coding~\cite{Elo2008} to analyze qualitative data, to understand participants' experience.

\vspace{0.1in}\noindent
{\bf Results and Discussions}

\noindent
Overall, most participants found the authoring pipeline easy to use (RQ1), and agreed that the ML-assisted mode could help the authoring experience faster (RQ2) and easier (RQ3).

\vspace{0.1in}
\noindent{\bf Is the Proposed Authoring Pipeline Usable (RQ1)?}
Both manual ($M = 70.61$, $SD = 17.23$) and ML-assisted ($M = 86.88$, $SD = 8.33$) modes demonstrate good system usability, where an overall SUS score $\geq$ $68$ is interpreted as ``good''~\cite{Sauro2011, Sauro2016}.
All participants agreed that both pipelines are easy to use (Figure~\ref{fig::papertoplace::authoring_eval}a). 
For example:
{\it ``they are easy to use! [... and] integrated very good!''}~(PA1) and {\it ``it has a really good workflow''}~(PA9).
While $11$ participant believed that they could author such MR experience with both interfaces rapidly (Figure~\ref{fig::papertoplace::authoring_eval}e), PA10 held a neutral opinion, as she expected a fully automated system to extract the key objects associated with each instruction step.
Overall, $10$ participants were satisfied with the MR experience they authored (Figure~\ref{fig::papertoplace::authoring_eval}e).
Examples include:
{\it ``a cool way to transfer knowledge''}~(PA2) and {\it ``useful to see [the instruction] while cooking''}~(PA12).
Few participants suggested the reason(s) for being not fully satisfied with the authored MR experience.
For example,
{\it``I think I made some mistake when I author it. So it guided me to the wrong spot''}~(PA5)
and {\it ``I have to do [the tasks with authored experience] by myself, to see what it is like for me to experience that first before having a novice do it so''}~(PA12).
These implied the lack of ways to enable the authors to revise the authored experience inside MR iteratively.

\begin{figure}[t]
    \centering
    \includegraphics[width=0.9\textwidth]{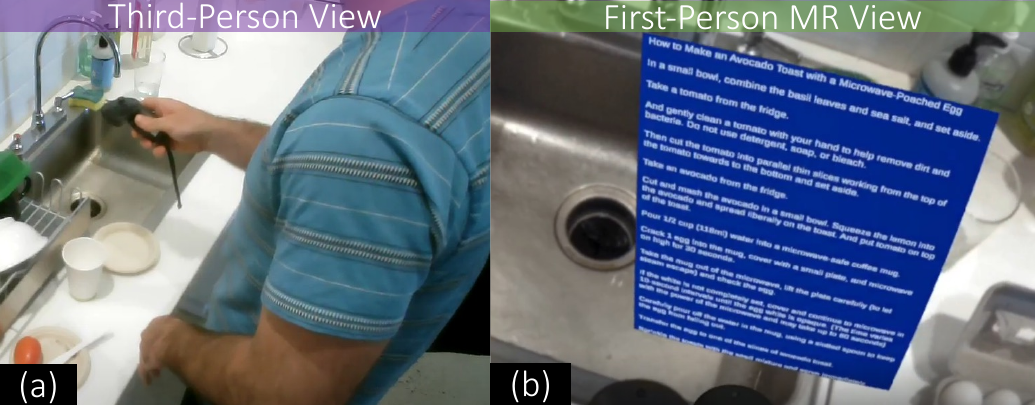}
    \caption{Baseline scene. Third-person (a) and first-person view through MR (b).}
    \label{fig::papertoplace::consumption::baseline}
\end{figure}

\vspace{0.1in}\noindent{\bf Can ML Support a Faster Authoring Experience~(RQ2)?}
The RM-ANOVA showed a reduced overall TCT ($F_{1, 22} = 16.66$, $p < .001$, $\eta_p^2 = .43$, Figure~\ref{fig::papertoplace::authoring_eval}c) and average TCT (ART: $F_{1, 22} = 35.98$, $p < .001$, $\eta_p^2 = .77$, Figure~\ref{fig::papertoplace::authoring_eval}d) for authoring each step while using ML-assisted mode, versus manual approach.
Most participants echo such observations.
For example:
{\it ``it could help me saving time, because [the key objects have] already [been] filled out [...] it speeds up by maybe a half a second [for each instruction]''} (PA3)
and {\it ``it helps me to save a lot of time! And that makes it a lot more convenient!''}~(PA4)

\vspace{0.1in}\noindent{\bf Can ML Support an Easier Authoring Experience (RQ3)?}
The RM-ANOVA demonstrated a higher SUS ($F_{1, 22} = 8.65$, $p = .008$, $\eta_p^2 = .28$, Figure~\ref{fig::papertoplace::authoring_eval}a) and a lower TLX score (ART: $F_{1, 22} = 17.24$, $p = .002$, $\eta_p^2 = .61$, Figure~\ref{fig::papertoplace::authoring_eval}b) of the ML-assisted mode, compared to the manual counterpart.
Most participants appreciated the convenience and helpfulness brought by the predicted key objects.
First, nearly all participants believed that the ML helped on reducing effort for tagging key objects.
For example: 
{\it ``ML brought less effort, I just need to check if the predicted key object is correct or not [...] Even if I still need to check it, I don't have to pay $100$\% of attention. I don't have to do all the thing. I just have to do part of the thing''}~(PA3), 
Particularly, the features of real-time predicting the new key object while modifying a specific step were favored by some participants.
For example:
{\it ``when I saw one of the step to be very long and [are associated with two key objects] [...] capable of predicting key object after being modified is obviously helpful! And also the opposite feature where you could just like combine two tasks, followed by generating predicted key object! [...] it gives more flexibility while segmenting the instruction step''}~(PA10).

Second, some participants highlighted the helpfulness for the mental thought process for ML-assisted mode. 
{\it ``I was able to create a mental map of how I will spatially move across at different instances [by looking at the predicted key objects]''}~(PA4)
and {\it ``it could help me make a decision''}~(PA3).
Particularly, PA5 appraised the feature of predicting key object in real-time while revise the instruction: {\it ``[while adding or editing the steps based on existing paper instruction], the real-time predicted location could help create a more clear instruction step. For example, if I type `heat the water', and the predicted location is oven for somehow, then I might just type `heat the water in the microwave' to make the instruction more clear''}
However, PA10 held an opposite opinion: {\it ``I just read the instruction step and then check if this assigned [key object] was countertop or not, and changed it to countertop rather than going and checking the other options. [...] but I did not read [the predicted key object] first.'' }

Finally, few participants suggested the merits of using color scale to visualize the confidence of predicted key object.
For example: 
{\it ``color could be helpful for conveying uncertainty''}~(PA9)
and {\it``color confidence is important to me. If it's red, I would be more aware of checking whether this is actually correct or not''}~(PA12).

\subsection{User Study 2: Consumption Pipeline}\label{sec::papertoplace::study::consuming}
The second study aims to evaluate the consumption pipeline and attempts to address:
{\it ``how the PaperToPlace could help the consumers to complete the designated activities {\it faster} and {\it easier}?''}

\begin{figure}[t]
    \centering
    \includegraphics[width=0.7\textwidth]{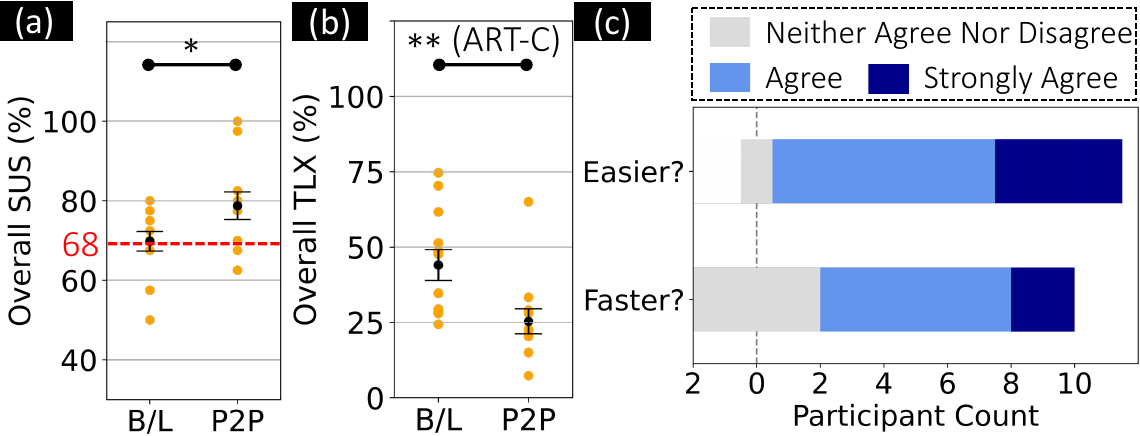}
    \caption[Consumption pipeline evaluation results.]{Consumption pipeline evaluation results. (a - b) Overall SUS and TLX score; (c) Survey results of how participants considered the overall consumption experience of PaperToPlace faster and easier, versus baseline. ``B/L'' and ``P2P'' indicate baseline and PaperToPlace condition.}
    \label{fig::papertoplace::consumption::overall}
\end{figure}

\begin{figure*}[t]
    \centering
    \includegraphics[width=\textwidth]{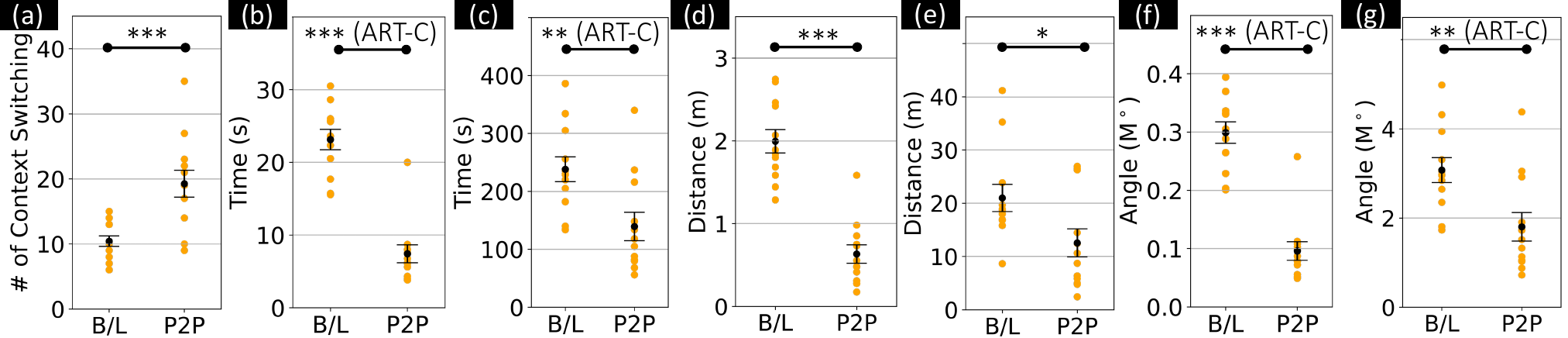}
    \caption[Results of context switching evaluations.]{Results of context switching evaluations. (a) The total number of context switching while using the monolithic document and PaperToPlace. The average time (b), $d_{head}$ (d), and $\theta_{head}$ (f) during each episode; The total time (c), $l_{head}$ (e), and $\theta_{head}$ (g) during all episode while completing the task. ``B/L'' and ``P2P'' indicate the baseline and PaperToPlace~conditions.}
    \label{fig::papertoplace::consumption::evaluation_measure}
\end{figure*}

\vspace{0.1in}\noindent{\bf Participants and Procedures}.
PC1 - PC12 (age, $M = 27.83$, $SD = 6.55$, \incl~ six males and six females) were recruited.
All participants were {\it not} familiar with the designated tasks.
We also built a baseline experience where the consumers could read existing monolithic instruction document inside MR (Figure~\ref{fig::papertoplace::consumption::baseline}).
Instead of asking participants to read a paper document, rendering a virtual monolithic document attached to the touch controller could minimize the impacts of confounding factors caused by uncomfort of the headset. 
We designed two sessions (Figure~\ref{fig::papertoplace::study_design}) with counterbalancing being considered to minimize the impacts of prior learning experience and task familiarity.
Before each session, T1 was used to help participants learn and familiarize with the system.
During the session, participants were instructed to complete \mbox{T2/T3} using the baseline and PaperToPlace.
Participants were invited to fill out NASA TLX~\cite{nasatlx} and SUS~\cite{Brooke1996sus} at the end of each session, followed by a semi-structured interview.

\vspace{0.1in}\noindent{\bf Measures. }
To understand the performance of context switching, we defined each {\bf episode} as {\it the interval between the time when participants stopped a task to seek instructions and when they returned to the task}.
During each episode, we analyzed the \textbf{(i)} time; \textbf{(ii)} the distance of path of head movement ($d_{head}$); \textbf{(iii)} the angular changes of the forward direction of the head ($\theta_{head}$).
The data lies outside of such intervals are out of our scope, as the performance of real-world activities could be affected by participants' prior cooking experience.
Same approaches in Section~\ref{sec::papertoplace::study::authoring} were used to analyze the questionnaire responses and participant's qualitative feedback.
The study on average took $59.40$~min~($SD = 5.80$~min).

\vspace{0.1in}\noindent{\bf Results and Discussions. }
Overall, most participants believed that the PaperToPlace could help the consumers to complete the designated tasks faster and easier (Figure~\ref{fig::papertoplace::consumption::overall}c). 
Quantitatively, we demonstrated a higher overall SUS score ($F_{1, 22} = 4.44$, $p = .046$, $\eta_p^2 = .17$, Figure~\ref{fig::papertoplace::consumption::overall}a) and a lower perceived workloads (ART: $F_{1, 22} = 18.52$, $p = 0.001$, $\eta_p^2 = .63$, Figure~\ref{fig::papertoplace::consumption::overall}b).
Based on participants' feedback, we now discuss how PaperToPlace could help participants complete the designated tasks faster and easier.

\vspace{0.1in}\noindent{\bf Context Awareness Reduces the Effort of Context-Switching.}
PaperToPlace reduces the average time (ART: $F_{1, 22} = 49.18$, $p < .001$ $\eta_p^2 = .82$, Figure~\ref{fig::papertoplace::consumption::evaluation_measure}b), $d_{head}$ ($F_{1, 22} = 57.29$, $p < .001$ $\eta_p^2 = .72$, Figure~\ref{fig::papertoplace::consumption::evaluation_measure}d) and $\theta_{head}$ (ART: $F_{1, 22} = 51.19$, $p < .001$ $\eta_p^2 = .82$, Figure~\ref{fig::papertoplace::consumption::evaluation_measure}f) on each episode.
While PaperToPlace on average leads to frequent document readings ($F_{1, 22} = 15.77$, $p < .001$, $\eta_p^2 = .42$, Figure~\ref{fig::papertoplace::consumption::evaluation_measure}a), the accumulated time (ART: $F_{1, 22} = 9.20$, $p = .020$ $\eta_p^2 = .42$, Figure~\ref{fig::papertoplace::consumption::evaluation_measure}c), $d_{head}$ ($F_{1, 22} = 5.30$, $p = .030$ $\eta_p^2 = .19$, Figure~\ref{fig::papertoplace::consumption::evaluation_measure}e) and $\theta_{head}$ (ART: $F_{1, 22} = 7.48$, $p = .019$ $\eta_p^2 = .40$, Figure~\ref{fig::papertoplace::consumption::evaluation_measure}g) of all episodes are reduced.

First, participants suggested the convenience for referring back to the instructions repeatedly with PaperToPlace.
For example: {\it ``[with baseline], I need to check it back and forth every time while trying to grab food from the fridge [...] [PaperToPlace] gives me feeling like [the instruction] is just on my side. It's like always on my side, like right beside my head''}~(PC9) and 
{\it ``because [the step] would always be right there with just a little bit of information I need, I think it'd be very useful''}~(PC11).
Most participants explicitly highlighted the benefits of finding optimal placements, without causing occlusions to the key interaction areas.
For example: {\it ``it is useful to bring the instruction step to me by just a pinch.''}~(PC10), {\it ``I like the position of instruction step!''}~(PC4), and {\it ``I think it is useful! And especially the function of where you pinch again, it will move to another location, so it can ensure [the step] will never block your sight''}~(PC8).
More example of instruction step placements could be referred to \mbox{Figure~\ref{fig::papertoplace::consumption::examples}e - h}.
However, PC6, who unveiled her ADHD~\cite{adhd}, suggested: 
{\it ``I was distracted with [the PaperToPlace], because [when the step occasionally was not anchored on the optimal position] the information was here and I was there. It reminded me of the moments that I forgot what I was supposed to do, or what I have to do.''}~

Second, most participants acknowledged the helpfulness by establishing connections between instructions and key objects.
For example: {\it ``I like how it took me to the sink because this activity has to be near the sink. That's a very helpful on spatial understanding!''}~(PC4), {\it ``Although we know where the fridge is, having that is really convenient to just not give anything a thought and do things as per the instructions''}~(PC10).

Finally, five participants also mentioned the merits of hands-free of PaperToPlace, compared to the baseline where the consumers need to hold the virtual document.
For example {\it ``[baseline] is more cumbersome, because I need to free one hand and make the hand very clean to make sure that the hand is clean to touch the controller''}~(PC5).

\vspace{0.1in}\noindent{\bf Segmented Instruction Helps Findings the Relevant Information Easier.}
All participants suggested that the segmented instructions is helpful, \eg~{\it ``[instructions] need to be as concise and as short as possible to be read at the same time. [PaperToPlace] did its job!''} (PC5).
Most participants suggested the merits of reducing stress while translating instruction into real world activities.
For example: {\it ``I have more calmness [with PaperToPlace], because [with baseline] I was seeing everything all at once, and that was giving me the feeling I'm in a hurry.''} (PC7),
and {\it ``looking at the entire document at once was so hard that I forgot where I have to keep following from start to finish to find where I was. But [PaperToPlace] gave me one by one instructions, which is super easy!''}~(PC3)

\section{Limitations and Future Works}\label{sec::papertoplace::future}

We identify our limitations from four perspectives.

\vspace{0.1in}
\noindent{\bf (1) Enabling an Iterative Authoring Process.}
We observed that during authoring, participants intended to segment the instruction steps by {\it only} considering whether only one key object is associated with the specific step, without synthesizing other factors (\eg~the density of the information contained by single step while viewed inside MR).
However, this cannot ensure a satisfied MR experience from the perspective of the consumers.
Future work might investigate potential {\it iterative} authoring workflow that allows the authors to refine their authored document profile inside MR while piloting the created instructional MR experience.

\vspace{0.1in}\noindent{\bf (2) Transforming Richer Metadata into MR Experiences.}
We consider that the metadata of each instruction step {\it only} contains the text of the step and the key objects that the virtual step should be anchored on (Section~\ref{sec::papertoplace::system_overview::design_insights}).
This might not be realistic for real-world instruction documents with heterogeneous kinds of metadata, such as the duration information, the caveats that usually requires the consumers' attention, and the notifications from the environmental sensors.
For example: {\it ``I would like to have warning text, like `do not use detergent', maybe show up in a different color or something''}~(PC4).
Future research might investigate the richer metadata that need to be augmented inside the MR experience, and the methods to use existing language models to extract such metadata as well as transform them into spatialized and context-aware MR experience.

\vspace{0.1in}
\noindent{\bf (3) Automatically Switching between Instruction Steps and Triggering the Position Update of the Instruction Label.}
We currently required consumers to explicitly click the virtual button to switch to the next instruction step, and to pinch to update the current position of the instruction step on demand~(Section~\ref{sec::papertoplace::consuming::interactions}).
While participants (\eg~PC10) with some prior MR experience felt it is {\it ``easy and useful''} to use the virtual hand menu and pinch gesture, others (\eg~PC1) suggested the frustrations of occasional failures of pinch gesture detections and the virtual button clicking.
Future work might consider designing a state machine, which could specify how to switch to the subsequent step {\it automatically} based on user's activities that might be inferred from face (\eg~\cite{Chen2021ExGSense}), body (\eg~\cite{OculusBodyTracking}) and environmental (\eg~\cite{Boovaraghavan2023, Agarwal2019virtual, Chen2020Captag}) sensor data. 

\vspace{0.1in}\noindent{\bf (4) Supporting a Broader Range of Applications. }
While many participants believed {\it ``cooking demonstrates [PaperToPlace] very well''}~(PC4), we only evaluated on cooking instructions, due to the poor quality of passthrough capabilities of Quest Pro~\cite{QuestPro}; availability of dataset to fine tune language model for alternative instruction activities; and limited study resources.
Future work might explore other activities with more powerful language model such as GPT and prompt engineering techniques being used for creating document profiles. 
Participants also emphasized the values of adaptive placements (\eg~{\it ``the adaptive placement of instruction is definitely useful for paper cutting! I don't want to cut my hand. And I wanted the instruction to be always besides my hand''}~(PC9)) and reduced context switching (\eg~{\it ``in the gym, where I need some instructions to teach me how to use the equipment, such as how you hold the gears with good postures. [...] with [baseline], it is less efficient and [I have] to stop in between and read the instructions''}~(PC2)) that might be transferred to other activities.
Another direction is to investigate the support for finer grained tasks that might be involved with moving objects, leading to a dynamically changed spatial profile.
For example, {\it``if you are doing PCB soldering, it might be hard to track that tiny component and to pinpoint the exact location on the PCB board. But if [PaperToPlace] can do that, it will be super helpful!''}~(PC8).
This requires high quality passthrough and capabilities to track real-time location of the electronic components which are considered as {\it non}-static key objects.
Instead of using Quest Pro, future researchers might consider a more recent higher-end headset, \eg~Vision Pro~\cite{VisionPro}.
\section{Conclusion}\label{sec::papertoplace::conclusions}
We present and evaluate PaperToPlace, comprising {\it an authoring pipeline}, which allows authors to rapidly transform existing paper instructions into a MR experience, and {\it a consumption pipeline}, which enables consumers to view spatialized instructions using a context-aware approach.
Two within-subject studies with two different cohorts of $12$~participants demonstrate the usability and effectiveness of the proposed authoring and consumption workflows.

\section{Acknowledgment}
We thank the insightful feedback from the anonymous reviewers. 
We appreciate the discussions with fellow researchers from Adobe Research, including Mira Dontcheva, Alexa Sui, Stephen DiVerdi, Ryan A. Rossi, Chang Xiao, Geonsun Lee, Mustafa Doğa Doğan, and Shakiba Davari. We thank Matin Yarmand for the help on the narration of the accompanion video.
%


Chapter~\ref{sec::papertoplace}, in full, is a reprint of the material as it appears in the Proceedings of the 36th Annual ACM Symposium on \textbf{U}ser \textbf{I}nterface \textbf{S}oftware and Technology (UIST 2023). The dissertation author was the primary investigator and author of this conference paper. Co-authors includes Cuong Nguyen, Jane Hoffswell, Jennifer Healey, Trung Bui and Nadir Weibel. 
This work was collaborated with Adobe Research.
The previously published manuscript can be referred to \cite{Chen2023PaperToPlace}.
The associated patent can be referred to \cite{Nguyen2024PaperToPlacePatent}.
\chapter{VRContour: Bringing Contour Delineation of Medical Structure into Virtual Reality}\label{sec::vrcontour}
Contouring is an indispensable step in \textbf{R}adiation \textbf{T}herapy~(RT) treatment planning. However, today's contouring software is constrained to only work with a 2D display, which is less intuitive and requires high task loads. 
\textbf{V}irtual \textbf{R}eality~(VR) has shown great potential in various specialties of healthcare and health sciences education due to the unique advantages of intuitive and natural interactions in immersive spaces.  
VR-based radiation oncology integration has also been advocated as a target healthcare application, allowing providers to directly interact with 3D medical structures.
We present VRContour and investigate how to effectively bring contouring for radiation oncology into VR.
Through an autobiographical iterative design, we defined three design spaces focused on contouring in VR with the support of a tracked tablet and VR stylus, and investigating dimensionality for information consumption and input~(either 2D or \mbox{2D $+$ 3D}). 
Through a within--subject study~(\mbox{N = $8$}), we found that visualizations of 3D medical structures significantly increase precision, and reduce mental load, frustration, as well as overall contouring effort.
Participants also agreed with the benefits of using such metaphors for learning purposes. 

\section{Introduction}\label{sec::vrcontour::intro}
\begin{figure*}[t]
    \centering
    \includegraphics[width=\linewidth]{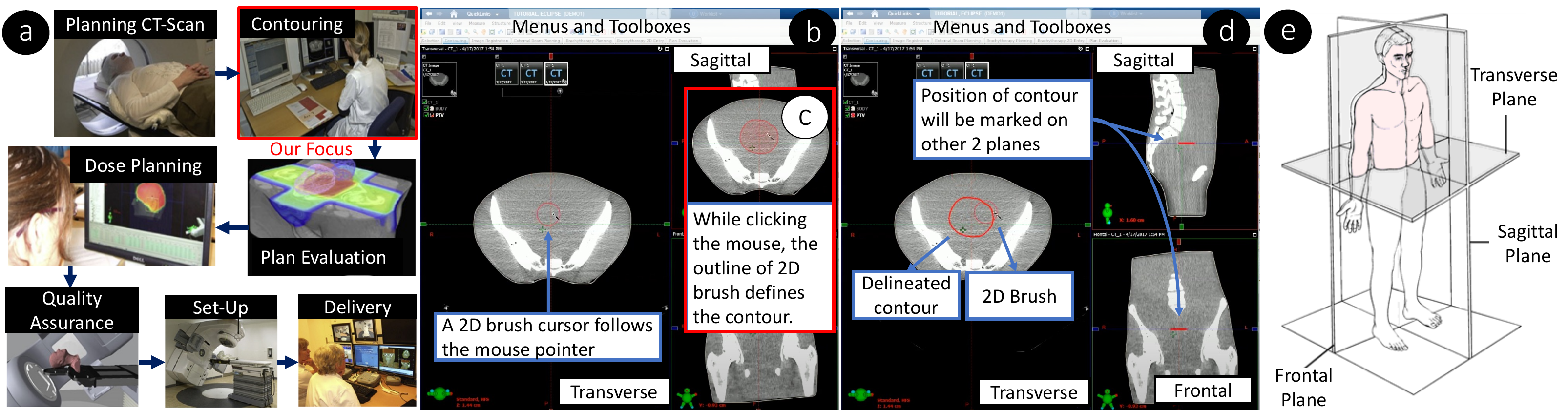}
    \caption[(a) Workflow of today's RT treatment. (b - d) Contouring using Eclipse. (e) Three planes of motions.]{(a) Workflow of today's RT treatment \cite{Boejen2011}. (b - d) Contouring using Eclipse~\cite{eclipse}; For delineating contours, a 2D brush is used (b); while clicking the mouse, an area is colored (c); upon releasing the mouse, the contour is defined by the colored area (d). (e) Three planes of motions.}
    \label{fig::vrcontour::example_interface}
\end{figure*}

\textbf{R}adiation \textbf{T}herapy~(RT) treatment is indispensable in cancer management.
Contouring is a critical step in the RT treatment planning workflow where oncologists need to identify and outline malignant tumors and/or healthy organs from a stack of medical images (Figure~\ref{fig::vrcontour::example_interface}).
Inaccurate contouring could lead to systematic errors throughout the entire treatment course, leading to missing the tumors or overtreating the healthy tissues~\cite{Zhai2021}, and could cause increased risks of toxicity, tumor recurrence and even death~\cite{Wuthrick2015}.
Oncology software tools, \eg~Eclipse~\cite{eclipse}, have been introduced to simplify the contouring workflow: oncologists can identify and segment key structures by analyzing and delineating outlines on a stack of 2D scans using a general 2D desktop display (Figure~\ref{fig::vrcontour::example_interface}b - d). 
This is sometimes not intuitive.
Mentally reconstructing 3D physiological structures could also introduce significant cognitive load while analyzing 2D images~\cite{He2017}.

3D immersive experiences in \textbf{V}irtual \textbf{R}eality~(VR) have unlocked a variety of applications in healthcare~(\eg~\cite{ARTEMIS2021}) and workspaces~(\eg~\cite{Chen2021ExGSense}).
Because of its unique advantages of intuitive and natural interaction, VR has shown great potential in different specialties of healthcare and health science education~\cite{Pelta2022}.
``Radiation oncology integration'' has also been advocated as a target healthcare application for providers in VR~\cite{Williams2018, Jin2017, Boejen2011}.
Such value has been widely recognized since Hubbold \etal~\cite{Hubbold1997} built the first 3D displays for visualizing radiation treatment plans.
Prior research has evaluated the affordances of VR while being applied to visualize RT treatment plans~\cite{Patel2007, Su2005} and enhance understanding of dose distributions~\cite{Phillips2008}. 
However, the delineation of contours - the initial phase of the workflow (Figure~\ref{fig::vrcontour::example_interface}a) - has not been explored.
A few recent works (\eg~\cite{elucis, Chen2022VRContourWIP, DicomVR}) attempted to bring radiation planning procedures into VR. 
However, the lack of ways to enable precise drawing of contours hinders the practical uses.
While Williams \etal~\cite{Williams2018} showed that $83$\% of oncologist participants found it useful to use VR for visualizing and contouring medical images, the kinds of merits that VR could offer are still unclear.
%

In this work, we present {\it VRContour} and investigate how to bring radiation oncology contouring into head-mounted VR by leveraging the merits introduced by 3D \textbf{U}ser \textbf{I}nterfaces~(UI). 
%
%
Since contouring is a challenging and highly specialized task in oncology, we spent more than nine months with domain experts and applied a user-centered design approach to consider their practical needs (\eg~precision and lead time).
%
%
%
Through an iterative and autobiographical design process with healthcare domain experts at different experience levels (\incl~a MD student, a resident and an attending physician), we decided to use a tracked tablet and Logitech VR stylus~\cite{LogitechVRInk} for its high drawing precision, and defined three design spaces by considering the dimensionality (\ie~either 2D or \mbox{2D $+$ 3D}) of {\it information consumption}~(\ie~2D: only visualizing cutting-planes on 2D tablet; \mbox{2D $+$ 3D}: visualizing both cutting-planes on 2D tablet and 3D medical structure) and {\it information input}~ (\ie~2D: only support contouring on 2D tablet with VR stylus; \mbox{2D $+$ 3D}: support contouring on both 2D tablet and direct drawing in 3D).
We populated essential components and metaphors~(\eg~menu designs in immersive environment, 3D object manipulations, and inter-slice interpolations) based on prior works and through our iterative design.

We built a proof-of-concept prototype to evaluate three design spaces on the HTC Vive Eye Pro~\cite{HTCVivePro}.
A separate prototype used to emulate today's contouring on a desktop-based 2D display was also implemented on the same headset as a baseline reference.
To minimize the impact of prior experience with 2D image interpretation, we intentionally involved only junior MD students and other health science students~(\eg~pre-med) who have been trained in anatomy, but without extensive exposure to 2D medical images. 
%
%
Through a within-subject study with eight participants, we showed how 3D visualizations could increase contouring precision, although the capabilities of direct contouring in 3D did not reduce lead time.
%
Our user experience evaluations suggested that VRContour leads to reductions in mental load, frustration, and overall effort when introducing 3D visualizations, and all participants agreed that contouring in 3D is promising for learning purposes.
%
We believe our work will benefit and direct future efforts on designing VR-based oncology software.

\section{Related Work}\label{sec::vrcontour::related}
\subsection{Contouring in Radiotherapy (RT) Treatment Planning}\label{sec::vrcontour::related::contouring}
RT treatment aims to deliver high doses of radiation to a tumor while limiting the dose that is received by the surrounding healthy tissue, which can be categorized into seven steps (\mbox{Figure~\ref{fig::vrcontour::example_interface}a})~\cite{Boejen2011}.
Providers need to interpret CT scans~(MRI and PET if needed) to mentally reconstruct a ``virtual anatomy'' and precisely locate and segment tumors as well as surrounding healthy tissues on each slice.
Oncology software, \eg~Eclipse~\cite{eclipse}, is widely used in today's clinics to streamline this process, allowing oncologists to easily align the scans and draw outlines with mouse and/or stylus. 
%
Figures~\ref{fig::vrcontour::example_interface}b - d show an example of the Eclipse~\cite{eclipse} interface, where oncologists can locate and contour the targets by rapidly switching slices across the preferred plane of motion (\mbox{Figures~\ref{fig::vrcontour::example_interface}e}) using scroll wheels.
A 2D brush is commonly used as the delineation tool for drawing and refining contours (Figure~\ref{fig::vrcontour::example_interface}b - d)~\cite{Werner2012}.
Advanced features, \eg~\mbox{inter-slice interpolations~\cite{Albu2008, Zukic2016nd}} are also used to reduce the number of slices to be manually contoured, leading to less lead time.

Despite these tools, contouring is still a lengthy process, as the oncologists need to examine all slices, each requiring manually distinguishing the tumoral volume from surrounding healthy tissues~\cite{Dowsett1992}. 
%
Existing works attempted to realize the paradigm of ``auto-contouring'', enabled by \textbf{A}rtificial \textbf{I}ntelligence~(AI). 
For example, Nikolov~\etal~\cite{Nikolov2021} demonstrated a novel 3D U-Net architecture that could achieve expert-level performance while delineating $21$~distinct head-and-neck targets.
While AI-assisted contouring can effectively empower clinicians beyond their current practices~\cite{Zhai2021}, their performance is considered less favorable than manual approaches. 
Additionally, the limited availability of training datasets could impact the scalability and inclusive design of AI-based contouring.
\mbox{Ramkumar~\etal~\cite{Ramkumar2013}} investigated the possibilities of integrating novel interaction modalities into the today's contouring workflow. 
They found that tablet-PC, with multi-touch and gesture-based input are the optimal ones, and could be integrated into the existing workflow for enhancing contouring efficacy.

%
Contouring on today's 2D interface is also error-prone, especially for clinicians with less experience or in lower tier medical centers~\cite{Abrams2012, Duhmke2001, Eisbruch2010}.
%
This is because it requires experience for oncologists to mentally reconstruct the medical structure for target localization by looking at a stack of planar scans, the only support provided by existing tools.
\mbox{Dowsett~\etal~\cite{Dowsett1992}} showed that more than $30$\%~of errors are introduced while contouring spinal canal and lungs using a mouse and lightpen on 2D displays.
%
%
Due to the difficulties of mastering contouring skills, existing research has proposed accessible learning supports using web-based \cite{eContour,Sherer2019} and mobile applications~\cite{Yarmand2022astro}.
While enhancing contouring training might be helpful, we believe that the new immersive 3D visualization technology can potentially help oncologists simplify and streamline the contouring procedure, and therefore might lower the barriers and efforts of contouring procedures in the long term.
Our research around VRContour aims to investigate the affordances of an immersive 3D UI, and explore how participants are able to understand and interact with the tumoral target directly in 3D. 

\subsection{Precise Drawing in VR}\label{sec::vrcontour::related::vr_drawing}
%
Enabling oncologists to engage in precise VR delineations is challenging.
We thus examine the existing research of evaluating human's drawing ability and designing drawing tools in VR.
While most existing works focus on drawing for 3D modeling, many design components are transferable to our context.

Early studies have investigated the issues of drawing inaccuracies in 3D~\cite{Schmidt2009}.
Fitzmaurice~\etal~\cite{Fitzmaurice1999} indicated how orientations of pen and paper could affect the 2D drawing performance.
\mbox{Arora~\etal~\cite{Arora2017}} showed how the lack of drawing surface support is the major cause of inaccuracies, which also depends on the orientations of drawing surface.
\mbox{Cannav\`{o}~\etal~\cite{Cannavo2020}} and \mbox{Chen~\etal~\cite{Chen2022PrecisionDrawing}} demonstrated the merits of using stylus for 2D and 3D drawing respectively in VR while being compared to traditional VR controller.
Similar study conducted by Pham~\etal~\cite{Pham2019} also indicate the better performance of using VR pen for pointing tasks.
\mbox{Barrera Machuca~\etal~\cite{Barrera2018Multiplanes}} demonstrated how spatial ability could affect the performance of 3D drawing and indicated such ability could affect the drawing shapes, yet not the line precision.

%
With the aforementioned VR based drawing evaluations, several existing works proposed drawing tools by integrating 2D tablets for general designers.
For example, SymbiosisSketch~\cite{Arora2018} proposed a 3D drawing application by combining the interaction metaphors of mid-air drawing and surface drawing, as as to create detailed 3D designs with head-mounted AR.
They showcased how designers could create a 3D canvas with mid-air stroke, and subsequently refine and add details with 2D tablet.
%
%
VRSketchIn~\cite{Drey2020VRSketchIn} demonstrated how tablet and VR could be integrated for better 3D drawing.
%
Other AR authoring tools such as PintAR~\cite{Gasques2019demo}, Pronto~\cite{Leiva2020} and Rapido~\cite{Leiva2021} also leveraged similar ideas for supporting drawing in 3D. 

While focusing on designing contouring applications for oncologists, our work borrowed several design ideas from these works.
%
However, due to the uniqueness of contouring requirements and the needs of practicing oncologists, directly transferring the techniques from applications facing general users to contouring applications does not necessarily lead to a usable system by radiation oncologists.
We therefore co-designed with domain experts iteratively to better understand our system requirements~(Section~\ref{sec::vrcontour::prelim}).

\begin{figure*}[t]
    \centering
    \includegraphics[width=0.9\linewidth]{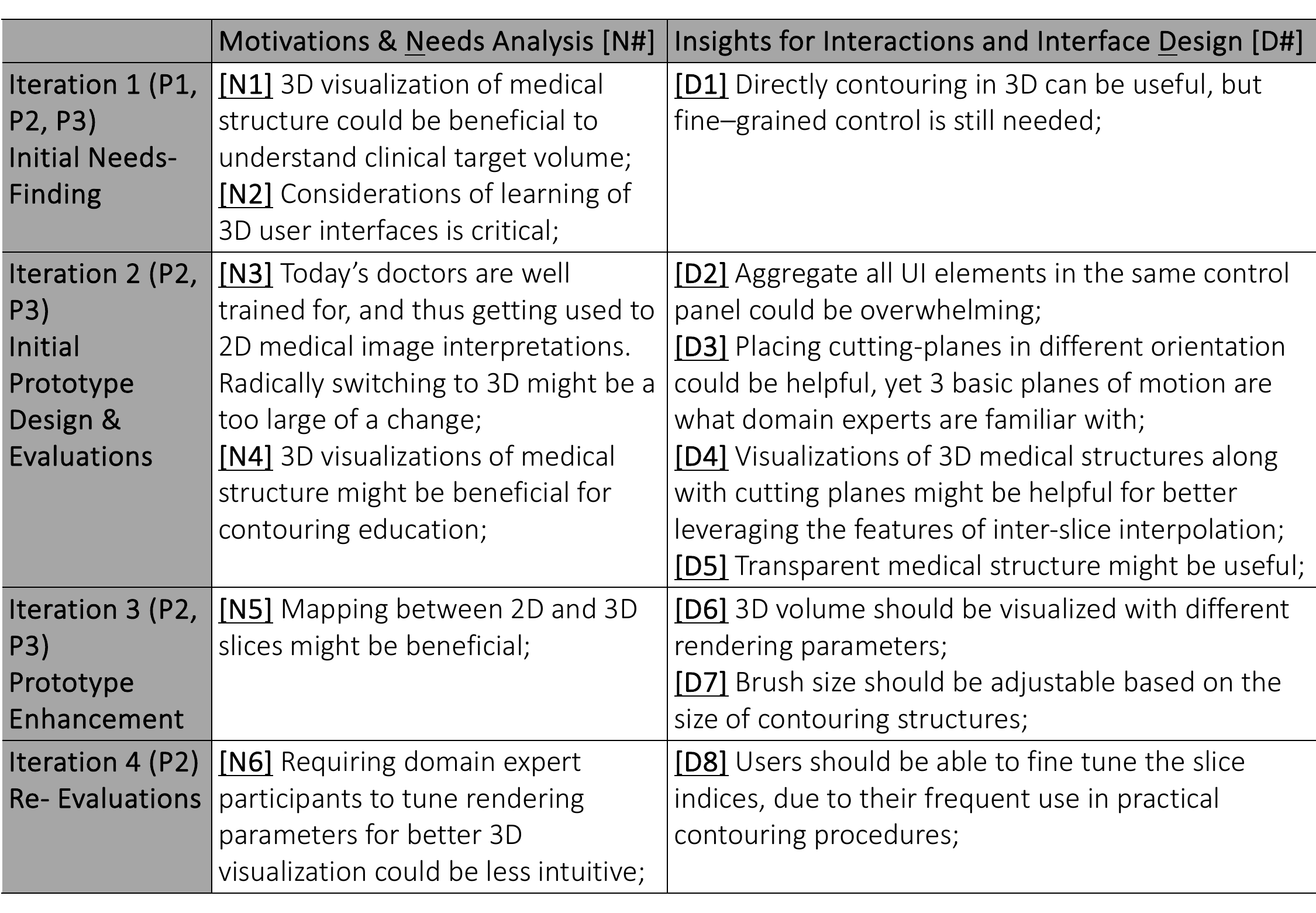}
    \caption{Insights and takeaways from preliminary interviews and iterative design.}
    \label{fig::vrcontour::iterative_design_takeaway} 
\end{figure*}

\subsection{eXtended Reality (XR) in Healthcare and Health Science Educations}\label{sec::vrcontour::related::vr_health}
E\textbf{X}tended \textbf{R}eality~(XR) has been advocated as the next frontier for future healthcare and health science education~\cite{Balasubramanian2021}.
Existing works have explored the affordances of immersive 3D visualization of medical structures and their real-world applications.
Zhang~\etal~\cite{Zhang2021} built a server-aided 3D DICOM viewer using mobile \textbf{A}ugmented \textbf{R}eality~(AR), with real-time rendering, interactions and full transfer functions editing.
%
%
However the lack of contouring support limits its use for RT treatment planning.
Anatomy Studio~\cite{Zorzal2019} proposed an AR tool for virtual dissection that combines tablets with stylus to assist anatomists by easing manual tracing and exploring cryosection images.
While supporting manual segmentation, the limited resolution and the non-transparent isosurface rendering prevents anatomists from exploring finer grained structures.

As for health sciences education, Nicholson~\etal~\cite{Nicholson2006} demonstrated significant merits of using VR for anatomy education, with an average of $18$\%~of score improvement.
Ma~\etal~\cite{Ma2016} designed an AR applications by overlaying the CT dataset on top of real-world participants, and demonstrated $86.1$\% approval in terms of educational value.
%
Boejen~\etal~\cite{Boejen2011, Su2005} summarized how VR could benefit today's RT training by enabling students to simulate and train clinical situations without interfering with practical clinical workflow and without the risks of making errors. 
Zhang~\etal~\cite{Zhang2023VR4Accupuncture, Zhang2024AcuVR} explored how to bring the acupuncture training workflow into VR.

In terms of RT treatment planning, Patel~\etal~\cite{Patel2007} showed how the stereoscopic visualization could foster the understanding of \mbox{spatial} relationships between anatomy and calculated dose distributions.
Williams~\etal~\cite{Williams2018} suggested that 3D visualization could be advantageous when the definition of the clinical target volume is based on a 3D spread of disease along anatomical features with intricate topology.
Such volumes could be difficult to observe and evaluate as they usually do not lie neatly in any of the motion of planes~\cite{Williams2018}. 
DICOM VR~\cite{Williams2018, DicomVR} showed a system that allows oncologists to manually segment the target directly in 3D space.
Early results of using this system show that while the contouring time seems to be reduced compared to 2D display based software, the drawing precision might be sacrificed to some extent~\cite{Williams2018}.
Although $83$\% of oncologists found DICOM VR useful~\cite{Williams2018}, the underlying reasons are still obscure. 
In addition, the lack of physical surface support and use of VR controllers naturally hinders precise delineations while refining outlines on a 2D virtual plane~\cite{Arora2017}.
We aim to understand how VR can benefit contouring, and how to design such a system by fully exploiting the affordances that this medium offers.

\begin{figure*}[t]
    \centering
    \includegraphics[width=\linewidth]{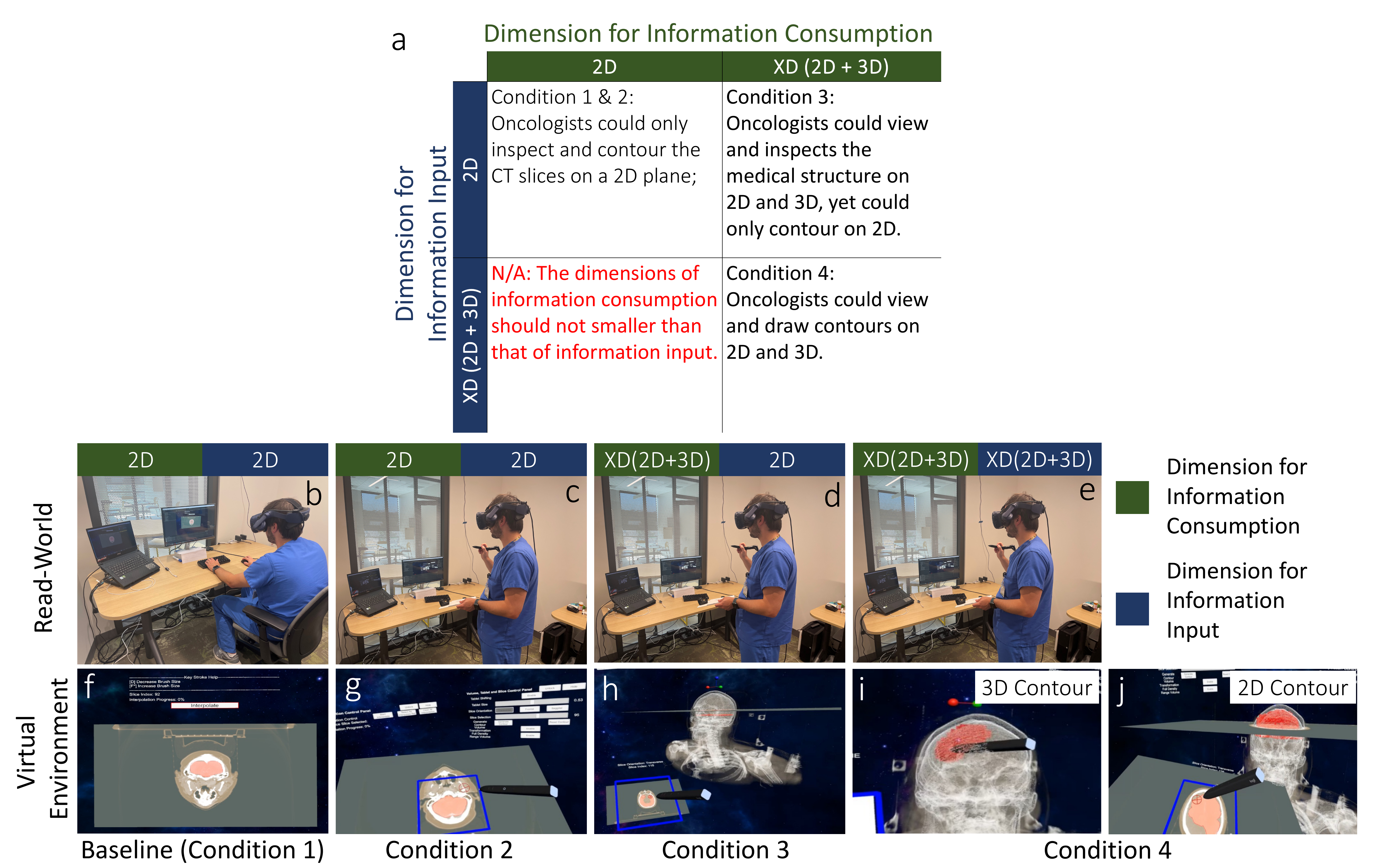}
    \caption[Design of VRContour.]{Design of VRContour. (a) Explorations of the design space by considering the dimension of information input and consumption. (b - j) Final prototype of four conditions generated by our preliminary autobiographical and iterative design process.}
    \label{fig::vrcontour::design-space-explorations}
\end{figure*}

\section{Preliminary Exploratory and Iterative Design}\label{sec::vrcontour::prelim}
%
%
We used an autobiographical approach and user-centered design, to explore and then implement as well as refine our prototype {\it iteratively} with domain experts~\cite{Neustaedter2012DIS, Chen2022VRContourWIP}. 
This helps us to analyze and better understand the complex real-world workflows with a user-centered design tenet~\cite{Norman1986}.

\subsection{Participants}\label{sec::vrcontour::prelim::participants}
Our participants include a third year medical student (P1, F), a senior resident fellow (P2, M), and an attending oncologist (P3, M).
P2 and P3 have extensive experience teaching contouring, and all participants are affiliated to the \href{http://health.ucsd.edu}{UC San Diego Health System}.
No participant had prior VR experience.
%
As today's contouring on 2D display is highly dependent on familiarity and understanding of 2D radiology images, the results from P2 and P3 are highly affected by their past experience regarding interpretations of medical scans and 2D contouring workflow.
To minimize such impacts we therefore included P1 intentionally, who is familiar with human anatomy, but not with 2D medical image interpretations.

\subsection{Procedure}\label{sec::vrcontour::prelim::procedure}
Our preliminary design consisted of four iterations. 
Each interview (or co-design workshop) lasted for $30$ - $90$~minutes (\mbox{$M$ = $53$~min.}, \mbox{$SD$ = $30$~min.}).
All iterations were video and audio recorded, and thematic analysis was used for data interpretation.

\vspace{+0.1in}
\noindent{\bf [I1] Iteration 1.} 
We first conducted three remote semi-structured interviews with all participants~\cite{Yarmand2021}.
We explained basic VR concepts by showing example VR based medical image visualization applications, to help them brainstorm conceptually. 
We focused on two guiding questions: \mbox{\textbf{(i)}} {\it ``What are the benefits that VR might bring to today's contouring?''} and \mbox{\textbf{(ii)}} {\it ``What are the features that might be useful to augment during contouring procedure?''}

\vspace{+0.1in}
\noindent{\bf [I2] Iteration 2.}
We then built a simple prototype based on \mbox{[I1]}, where a tracked tablet and Logitech VR stylus was used as the input tool (Figure~\ref{fig::vrcontour::tablet_sizing}). 
Participants could draw contours on a cutting-plane rendered on a tracked tablet using a 2D brush metaphor~(Figure~\ref{fig::vrcontour::drawing}a - d), which is consistent with today's 2D contouring~(\mbox{Figure~\ref{fig::vrcontour::example_interface}b - d}). 
We also designed a 3D brush metaphor by extending the existing 2D brush, which is essentially a semi-transparent sphere attached to the tip of the VR pen.
Participants could then directly draw a colored volume inside 3D medical structures using a 3D brush~(Figure~\ref{fig::vrcontour::drawing}e-f).
All control menus were attached to the tracked tablet.
Section~\ref{sec::vrcontour::design} outlines the details of our design.
P2 and P3 were invited to a co-design workshop where they could try out the system; P1 was excluded from this second iteration due to the lack of familiarity with practical contouring procedures.

\vspace{+0.1in}
\noindent{\bf [I3] Iteration 3.}
We improved the system design based on lessons learnt from \mbox{[I2]}, and organized remote interviews with P2 and P3. 
The goal of the session was to understand the feasibility and setbacks of the provisional design.

\vspace{+0.1in}
\noindent{\bf [I4] Iteration 4.}
We finally invited P2 to experience the enhanced system.
We also discussed extensively with P2 to ensure domain experts participants could inspect the target structure in both 2D and 3D through our rendering algorithm. 
P3 was not involved during [I4] due to clinical duties.

\subsection{Findings}
Figure~\ref{fig::vrcontour::iterative_design_takeaway} shows key takeaways during each iteration.
\textbf{[N\#]} and \textbf{[D\#]} denote {\it the motivation \& \underline{N}eeds analysis}, reflecting experts' thoughts of how 3D contouring might be beneficial to future oncology diagnosis and educations, as well as suggestions on {\it interaction and interface \underline{D}esign}.
The lessons learned are summarized into three areas:

\vspace{+0.1in}
\noindent \textit{\textbf{Inspecting Medical Images in 3D.}}
All participants recognized the challenges of learning how to analyze 2D medical images, especially for small organs ({\it``things like vasculature can be harder to identify''}~[P1]).
Participants explained the approach to locate and understand a specific target from a set of medical images: 
{\it `` You get a stack [of planar medical images]. And then you can go [back and forth], and change the different [cutting plane] to see where you are. [...] We don't get to see only one image. You can go up and down and then kind of orient yourself in.''}~[P1]~
Key benefits include the ``overall view'' capabilities for helping determining the boundary between tumors and surrounding healthy tissues - {\it``[...] often you're trying to decide where does that cancer stop? [...] Being able to visualize everything in 3D could help you see those borders better.''}~[P2] - and having flexibility for placing and orienting cutting-planes for inspecting target as well as inter-slice interpolations [P2, P3].

\vspace{+0.1in}
\noindent\textit{\textbf{Annotating \& Contouring Medical Images in 3D.}}
All participants, during I2 and I4, agreed that directly drawing in 3D could be helpful for identifying hard tissues (\eg~bones), yet not for the soft ones (\eg~blood vessels).
P3 emphasized the importance of preserving the features of 2D contouring, yet using 3D as an assistive visualization technique:
{\it ``as you're drawing in 2D, you can see a 3D rendering of a structure [...] that actually could be a really interesting concept and might improve contouring accuracy. To really see the structure, like a prostate cancer, being able to see the course of the rectum, see where the prostate sits, see how the seminal vesicles go over [...] and kind of understand that anatomy, and actually see that 3D, that actually might result in more accurate contouring, especially for people that are just learning.''}
P2 further extrapolated the usefulness of mapping 3D annotations back to 2D slices:
{\it [It would be nice to] have like a bright star, and you can put it somewhere. And then when you switch over [to 2D] and start scrolling up and down, [you see] a star on the excellent position where you put it. } [P2]

\vspace{+0.1in}
\noindent\textit{\textbf{Learning about 3D UI and Interactions.}}
During [I1], P2 and P3 took on average $10$~minutes to get used to the prototype.
They believed such learning effort should \textit{not} be problematic for general practicing oncologists (\eg~{\it ``I don't think it would take that long to get used to pushing the buttons''}~[P2]).
Participants also emphasized the importance of leveraging the skills of inspecting and contouring medical slices on 2D displays that oncologists already have in today's practice: {\it``we work with 2D images, but have to think in 3D. It's a little bit of paradox.''}~[P3], {\it``No doctor will ever see it as 3D, we're always scrolling a plane [...] but it's always on 2D.''}~[P2].

\subsection{Discussions and Design Space Explorations}\label{sec::vrcontour::prelim::design_space}
We finally defined a design space, leading to three evaluation conditions alongside a baseline condition for emulating today's 2D contouring~(Figure~\ref{fig::vrcontour::design-space-explorations}a).
The prototypes of the four conditions are demonstrated in Figure~\ref{fig::vrcontour::design-space-explorations}b - j.
We re-use the concept of information flow as the dimensions of our morphological analysis~\cite{Sears2009HCI}.
Specifically, the {\it dimension of information consumption} indicate the ways that the users consume the output information from VR (\eg~inspecting scans), and the {\it dimension of information input} refers to how users outlines contours inside VR (Figure~\ref{fig::vrcontour::design-space-explorations}a).
While each dimension can be either 2D or \mbox{2D $+$ 3D}, the dimension for information consumption should not be lower than the one of the information input.
Based on \mbox{[N3]} and \mbox{[D1]}, we also intentionally excluded any design that would enable contouring {\it purely} on 3D, as mid-air drawing in VR makes it challenging to be precise due to the lack of physical support and perception of depth~\cite{Arora2017}.
The four conditions are summarized below:

\vspace{+0.1in}
\noindent{\bf [C1] Contour on a 2D desktop display (Baseline).}
We built a prototype to emulate today's contouring on 2D desktop display using a keyboard and mouse~(Figure~\ref{fig::vrcontour::design-space-explorations}b and f). 
Since enhancing the comfort of the VR headset is beyond our scope, we re-created this scenario to establish a baseline, such that the confounding factors introduced by comfort of VR headset could be minimized.
While wearing the VR headset, oncologists need to go through each transverse slices to mentally understand the medical structures, followed by outlining key structures on each slices.
During \mbox{[I4]}, we confirmed with P2 that the prototype could sufficiently emulate today's 2D contouring.

\vspace{+0.1in}
\noindent{\bf [C2] Contour on a 2D tablet {\it \textbf{without}} visualizations of 3D medical structure.}
C2 leverages the idea of direct interaction where the drawn stroke is aligned with the physical tip of the VR pen~(Figure~\ref{fig::vrcontour::design-space-explorations}c and g). 
Participants are required to inspect and contour on 2D slices rendered on a tracked tablet, using a VR stylus.
3D visualizations were not provided in this condition.

\vspace{+0.1in}
\noindent{\bf [C3] Contour on a 2D tablet {\it \textbf{with}} visualizations of 3D medical structure.}
C3 offers a 3D rendered medical structure hovering in the mid-air (Figure~\ref{fig::vrcontour::design-space-explorations}d and h). 
This, hypothetically, could reduce the cognitive task load for thinking in 3D while inspecting different cutting-planes in 2D. 
As oncologists are not allowed to draw in 3D, we consider the dimensions for information input as 2D only.

\vspace{+0.1in}
\noindent{\bf [C4] Contour in the 3D medical structure {\it \textbf{and}} on tablet.}
Participants are expected to contour on both 3D and 2D (Figure~\ref{fig::vrcontour::design-space-explorations}i and j). 
Besides the hypothesized benefits in C3, we extrapolated that the direct 3D contour could reduce lead time, due the potential elimination of slice-wise drawing on the 2D interface.
After 3D contouring, oncologists could continue refine the outlines with pen and tablet.

\section{System Design}\label{sec::vrcontour::design}

\begin{figure}[t]
    \centering
    \includegraphics[width=\linewidth]{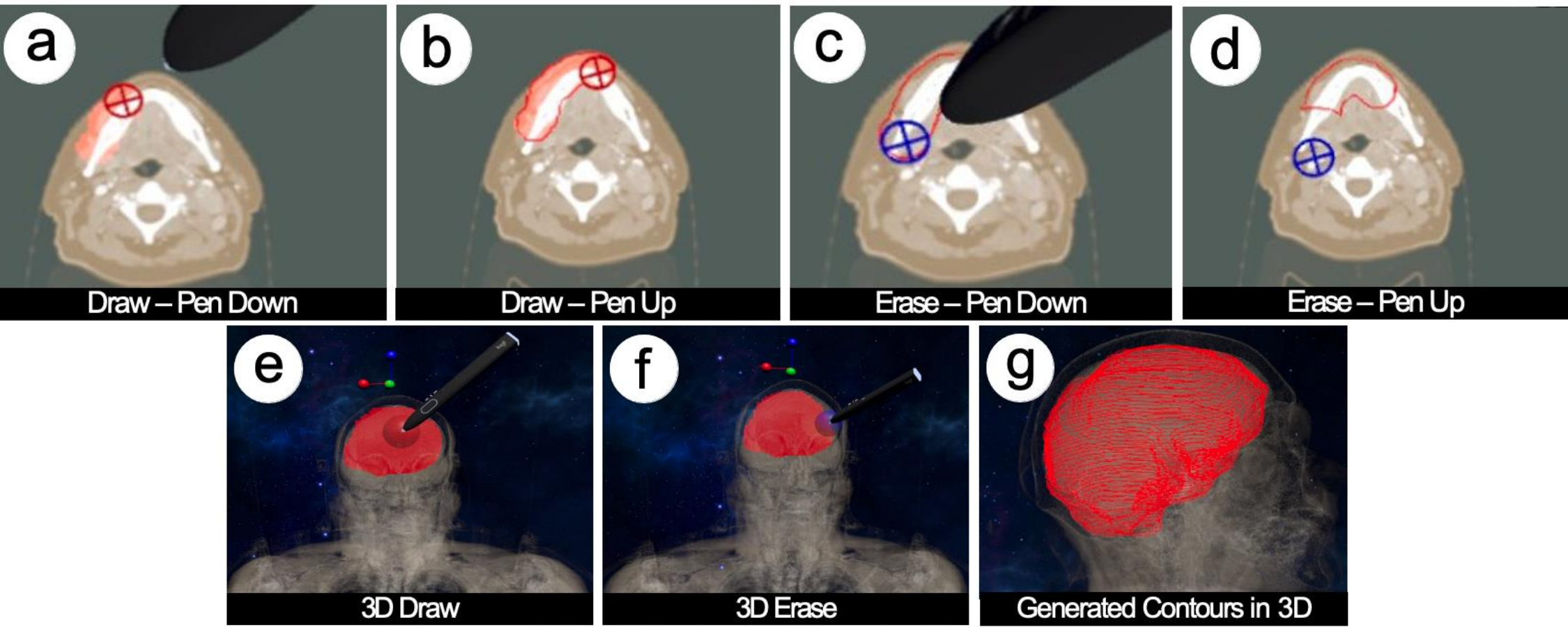}
    \caption[Drawing/erasing contours using brush metaphors.]{Drawing/erasing contours using brush metaphors. (a - b) The contour is automatically generated after the VR pen is lifted from the tablet; (c - d) the new contour is re--generated upon the completion of erasing; (e - f) contours, visualized as a volume, could also be drawn/erased directly inside 3D volume; (g) generated contours in 3D (side view).}
    \label{fig::vrcontour::drawing}
\end{figure}

\subsection{Volumetric Rendering}\label{sec::vrcontour::sys::rendering}
\begin{figure}[h!]
    \centering
    \includegraphics[width=\linewidth]{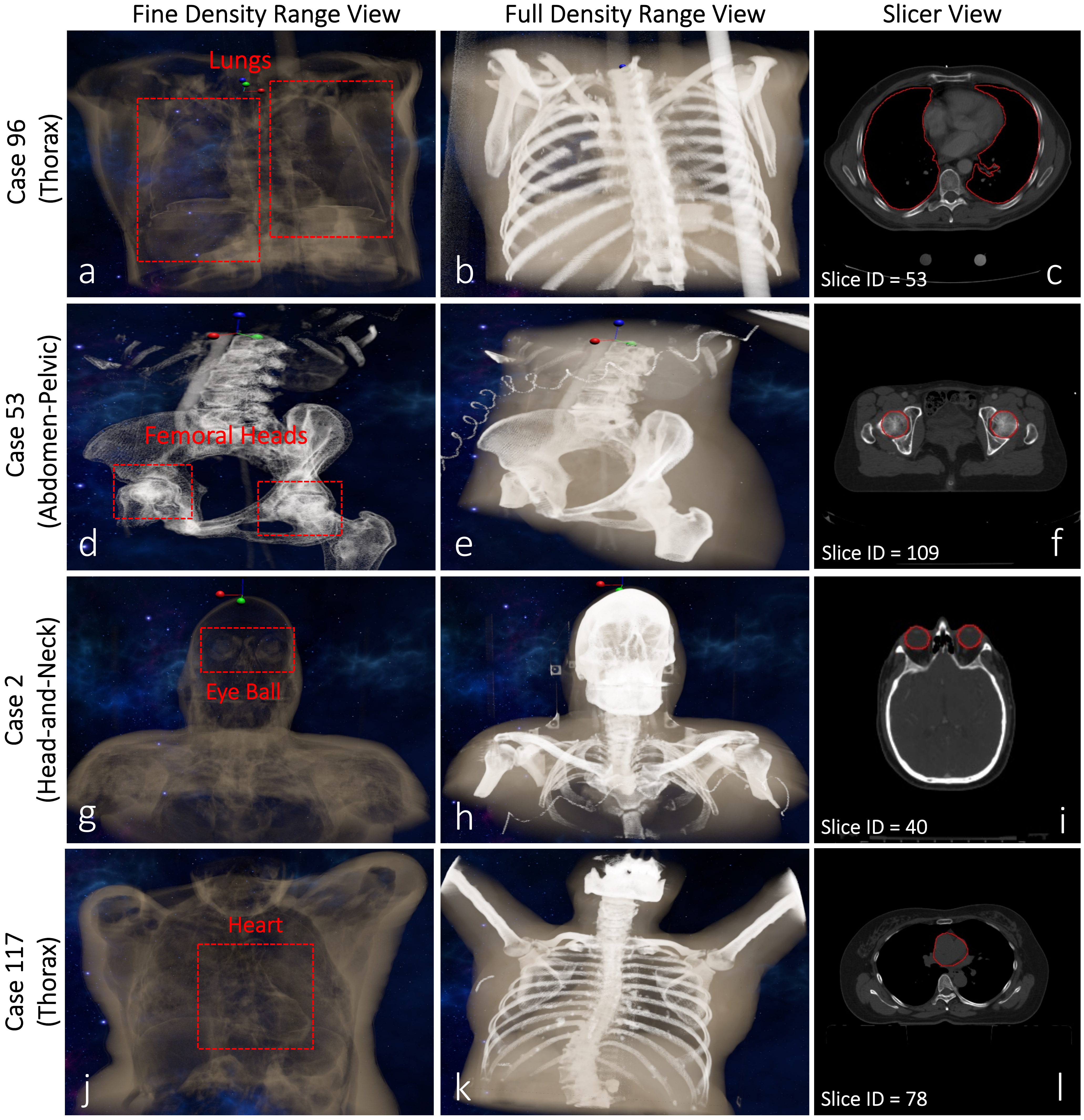}
    \caption[Contour tasks.]{Contour tasks: (a - c) lungs from the thorax structure; (d - f) femoral heads from the abdomen-pelvic structure; (g - i) eye balls from the head--neck structure; (j - l) heart from the thorax structure. Red contours in c, f, i, and l indicate expert contours. Displayed data are extracted from \href{https://econtour.org}{eContour}~\cite{eContour}, cases \href{http://econtour.org/cases/96}{$96$}, \href{http://econtour.org/cases/53}{$53$}, \href{http://econtour.org/cases/2}{$2$}, and \href{http://econtour.org/cases/117}{$117$}.}
    \label{fig::vrcontour::rendering}
\end{figure}

Volumetric rendering refers to the process that creates 2D projections from discretely sampled 3D dataset.
Although investigating optimal rendering techniques is beyond our scope, existing works (\eg~\cite{Zhang2021}) have shown that there is no single algorithm that could be applied to every 3D task (also confirmed in \mbox{[D4]}).
While considering the needs of transparent rendering \mbox{[D5]}, we decided on using a ray-casting approach that treats the 3D medical structure as a transparent object that could transmit, emit and absorb light~(like a cloud).
Although \textbf{D}irect \textbf{V}olume \textbf{R}endering~(DVR) and \textbf{M}aximum \textbf{I}ntensity \textbf{P}rojection~(MIP) are commonly used approach to render transparent volumes, MIP has a well known limitation regarding depth perception~(\eg~ambiguous front and back relations)~\cite{Ney1990, Mady2020}.
We thus decided on using DVR.
As we focus on understanding the merits that the 3D UI could offer to contouring, and the tuning of rendering parameters is important \mbox{[D6]} yet beyond our scope, we therefore pre-adjusted the rendering parameters for each evaluation task before final evaluation (see \mbox{Step 1} below and Figure~\ref{fig::vrcontour::rendering}). 
This process was done during \mbox{[I4]} with P2.
The implementations of rendering pipelines include three steps:

\vspace{+0.1in}
\noindent{\bf Step 1: Data Prepossessing and Sampling.}
In general, DVR considers the 3D structure as the semitransparent and light-emitting medium, which could emit, transmit and absorb the light (like a cloud), and therefore allowing participants to see through the data.
We represent the 3D medical structure using a 3D array, by stacking all \mbox{DICOM} slices.
To emulate practical contouring, we retain the original resolution of each slices without either upsampling or downsampling.
The physical spacing between voxels along each dimension is determined by the scanning machine and is typically on the order of $mm$.
We used min--max normalization to scale the density at each voxels to $[0, 1]$ and the resultant view of 3D visualizations are called {\it ``full density range view''}.
During \mbox{[I4]}, we adjusted such range (\ie~vary the lower and upper density bound) to make sure that the target structures are visible (\ie~{\it ``fine density range view''}, see also Figure~\ref{fig::vrcontour::rendering}).
The ray, cast from camera, would then sample the volume into $256$~equal-distance steps that were chosen based on visual effects and system performance. 

\vspace{+0.1in}
\noindent{\bf Step 2: Classification.}
Classification refers to the step of using transfer functions to assign each discrete step with color and opacity.
For simplicity, we only consider the simplest 1D transfer function to map the normalized (or adjusted) density to a RGBA vector ($f:\mathbb{R} \rightarrow \mathbb{R}^4$).
While some prior works advocate the ideas of offering participants flexibility on tuning the transfer functions (\eg~\cite{Zhang2021, Zhang2024AcuVR}) or increase their orders (\eg~\cite{unityvolumetricrendering}), in order to see more details of underlying soft tissues, our preliminary study demonstrated that the lack of understanding of transfer function and overwhelm tuning options might degrade usability and increase learning cost.

\vspace{+0.1in}
\noindent{\bf Step 3: Composition.}
Finally, we aim to generate a composited value by blending the RGBA at each step along the ray, which would be eventually rendered on the 2D projection.
We used the \textit{over} operator~\cite{Porter1984} to compute the final RGBA vector.
The algorithm for front-to-back traversal could be referred to equations~(\ref{eq::c}--\ref{eq::alpha}), where $\mathbf{c}_i$ and $\mathbf{\alpha}_i$ refer to the color and opacity at $i^{th}$ step ($i \in \{1, 2, 3 ... 256\}$).

\begin{equation}
    \mathbf{c}_i = \mathbf{c}_{i - 1} + \mathbf{c}_i\mathbf{\alpha}_i(1 - \mathbf{\alpha}_i)
    \label{eq::c}
\end{equation}

\begin{equation}
    \mathbf{\alpha}_i = \mathbf{\alpha}_{i - 1} + \mathbf{\alpha}_i(1 - \mathbf{\alpha}_i)
    \label{eq::alpha}
\end{equation}

More details in terms of volumetric rendering techniques are outlined by Lichtenbelt \etal~\cite{Lichtenbelt1998}.

\subsection{Interaction Techniques}\label{sec::vrcontour::system::interaction}

\begin{figure}[h!]
    \centering
    \includegraphics[width=\linewidth]{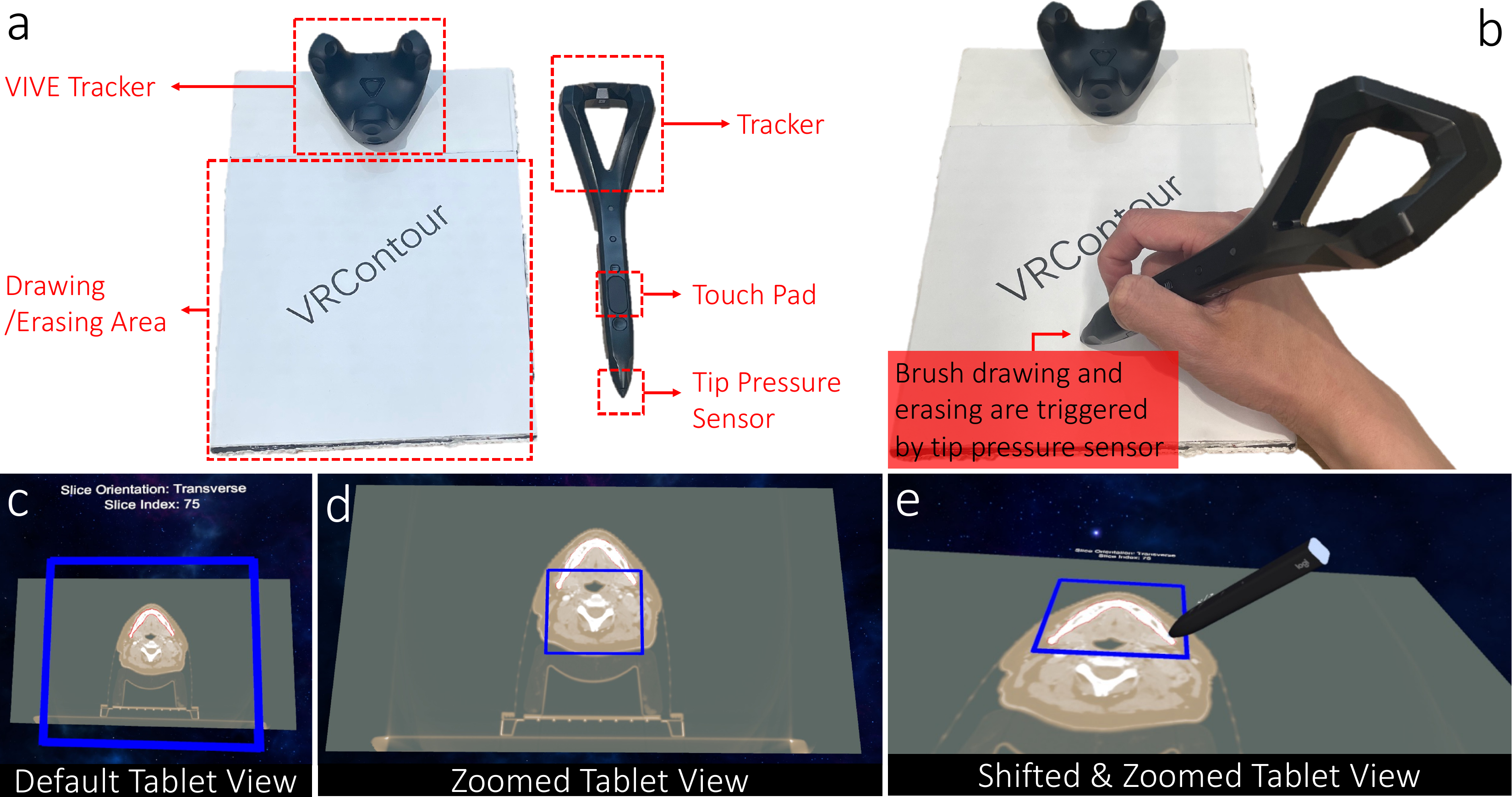}
    \caption[Contour on 2D.]{Contour on 2D. (a) Tracked tablet and VR pen; (b) real-world scene using VR pen and a  tablet;  (c) a blue frame indicating the physical boundary of the tablet; (d) the cutting planes could be zoomed in for gaining a detailed view; (e) oncologists can only draw inside the blue frame due to the physical surface support.}
    \label{fig::vrcontour::tablet_sizing}
\end{figure}

\noindent{\bf Pen + Tablet Interactions and Supporting Precise Contouring.} 
We decided to use the \href{https://www.logitech.com/en-us/promo/vr-ink.html}{Logitech VR stylus}~\cite{LogitechVRInk} as input tool in both 3D and 2D contouring for its high precision, which meets the need for precision drawing in a virtual environment~\cite{Arora2017}~(Figure~\ref{fig::vrcontour::tablet_sizing}a).
We used a tracked physical board, with same size as a $10.2''$~iPad, since having a physical support should increase the drawing precision~\cite{Arora2017}~(Figure~\ref{fig::vrcontour::tablet_sizing}a - b). 
Participants could scale the size of the virtual tablet to inspect the image at higher resolution~(Figure~\ref{fig::vrcontour::tablet_sizing}c - d), in which case the boundary of the physical tablet are highlighted to help visually identify the drawing area (Figure~\ref{fig::vrcontour::tablet_sizing}d - e).

\vspace{+0.1in}
\noindent{\bf Cutting-Planes and Cross-Dimension Contouring.}
We found that participants enjoyed the capabilities of viewing and contouring in 3D, yet they still wanted to preserve the capability of contouring on the 2D surface, due to its merits of precision and ease of observing detailed structures. 
Additionally, 2D contouring leverages existing skills of 2D medical images interpretation~\mbox{(\mbox{[N3]}, \mbox{[D1]})}.
Therefore, while designing C4, we proposed a cross-dimension contour workflow, by combining fast and direct contouring in 3D and precise 2D contouring on a tracked 2D tablet.
This process includes three aspects:
{\bf(a)}~Upon finding the target, oncologists first perform contouring \textit{directly} in the 3D volume using a 3D brush~(as the design in \mbox{[I2]}, see Section~\ref{sec::vrcontour::prelim::procedure} and Figure~\ref{fig::vrcontour::drawing}e - f).
The brush-based contour would result in a 3D colored volume.
{\bf(b)} Oncologists would then select a cutting-plane with one of three orientations (Figure~\ref{fig::vrcontour::example_interface}e).
We did not include an arbitrary orientation, as understanding cross-sectional slices with arbitrary orientation is challenging for domain experts~\mbox{[D3]}.
The contours on each slices were generated based on the selected slicer orientation~(Figure~\ref{fig::vrcontour::drawing}g).
{\bf(c)} Oncologists would finally inspect and revise contours mirrored on the tablet (Figure~\ref{fig::vrcontour::drawing}a - d). 
The inter-slice interpolation will be applied as needed to reduce the lead time (see Section~\ref{sec::vrcontour::method::implementations}).

\begin{figure}[!ht]
    \centering
    \includegraphics[width=0.9\linewidth]{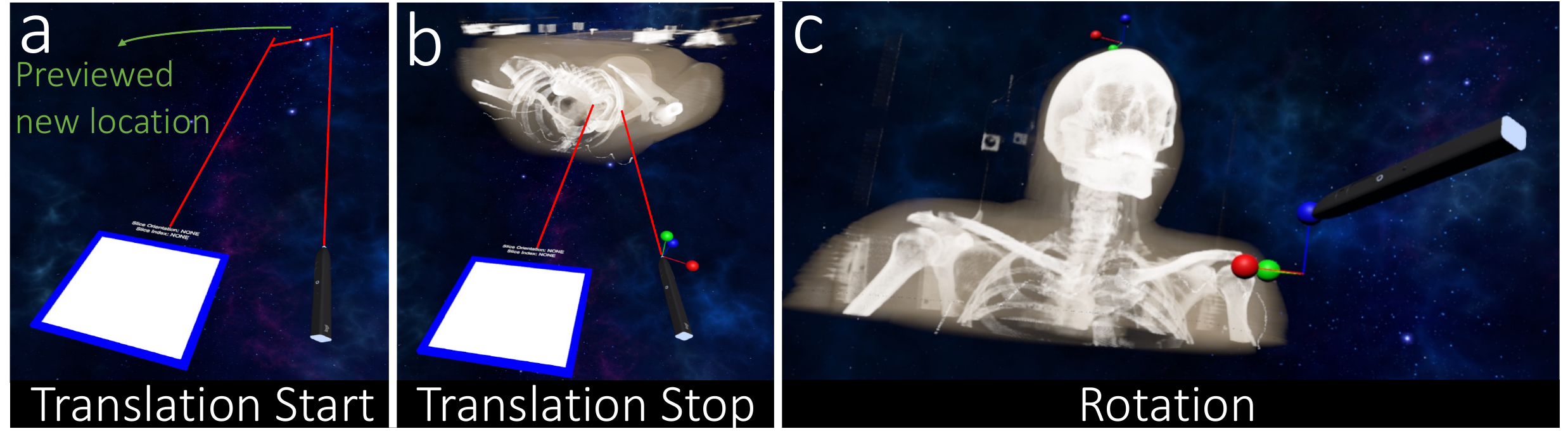}
    \caption[Transforming 3D medical structure.]{Transforming 3D medical structure. (a) Two rays from the VR pen and tablet are initiated after translation start; (b) after the translation stop, the 3D medical structures are anchored to the new place; (c) during rotation mode, the quarternion of the near--hand anchor is projected onto the 3D medical structure.}
    \label{fig::vrcontour::transformation}
\end{figure}

\vspace{+0.1in}
\noindent{\bf Transforming 3D Medical Structures.}
During C3 and C4, participants were required to see the 3D medical structures while contouring, as well as to draw directly inside the 3D volume. 
It is therefore critical to design efficient ways for manipulating 3D medical structures inside VR.
To address this, we revisited and revised the design techniques of workspace translation and scaling in \mbox{SymbiosisSketch~\cite{Arora2018}}, which was evaluated among six design professionals.
After initiating the transforming mode, two rays are used to determine the new position where the 3D volume will be placed.
The new position is decided by the midpoint of the line segment whose length is the shortest between two rays~(Figure~\ref{fig::vrcontour::transformation}a).
To avoid visual clutter, a 3D cursor is used to indicate the center of the newly placed 3D volume.
A second touchpad click allows the 3D volume to be moved into the target place~(Figure~\ref{fig::vrcontour::transformation}b).
Oncologists can rotate the volume by clicking-and-dragging the near-hand anchor (Figure~\ref{fig::vrcontour::transformation}c). 
The quaternion of the near-hand anchor is then mapped to the 3D medical structure.

\vspace{+0.1in}
\noindent{\bf Support of Coarse- and Fine-grained Tuning of Parameters.}
Finding optimal parameters has been identified as a critical factor that affects contour delineations~(\eg~locating the slices, changing the tablet and brush size, see [D7], [D8]). 
For coarsely tuning parameters, oncologists can use the scrollbar shown in front of them.
They can also click on the pen's touchpad for fine-grained parameter adjustments.
Although we enabled both tuning methods for all parameters during \mbox{[I2]}, oncologists suggested only adding fine-grained tuning to the {\it slice selections} due to its frequent uses during contouring.

\vspace{+0.1in}
\noindent{\bf Spatially Distributed Control Panels.}
While only retaining essential features of general computer-supported contouring applications, we found that it was challenging for participants to promptly locate target UI elements if aggregated onto only one panel~\mbox{[D2]}. 
Therefore, we decided to dispatch UI elements onto three sub-control panels for C2 - C4. 
Each panel has elements with similar features: {\it tablet and volume control}, {\it stylus control}, and {\it interpolation control}.
Through this design, oncologists are able to re-organize the UI layout and/or hide specific panels as needed.
Participants were satisfied with these design enhancement during \mbox{[I4]}.

\section{Study Methods}\label{sec::vrcontour::method}
\subsection{Participants}\label{sec::vrcontour::method::participants}
\begin{figure}[h!]
    \centering
    \includegraphics[width=\linewidth]{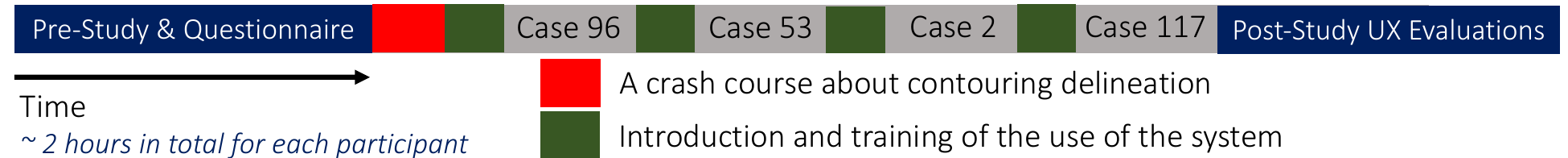}
    \caption{Horizontal view of the study timeline.}
    \label{fig::vrcontour::timeline}
\end{figure}

We recruited eight participants~(two females and six males, age: $M$ = $25.6$, $SD$ = $1.2$) who have taken human anatomy related curricula, but who {\it do not} yet had formal training on 2D medical image interpretation.
Our rationale behind this decision is based on two factors; 
First, since we are evaluating the affordances of a specific modality, extensive past learning experience of 2D medical image interpretation could affect UX evaluations and system preferences;
Second, participants taking human anatomy classes are able to identify and contour basic structures, even though clinical contouring is performed by senior residents and attending physicians.
To accommodate participants' anatomy knowledge, we intentionally selected fundamental structures with domain experts and made sure that the task was feasible for any students who have taken anatomy classes.
During our pre-study phase (see Figure~\ref{fig::vrcontour::timeline}), participants overall {\it agreed} on a $5$-point Likert scale that they were able to identify key structures from 3D anatomy~($MED$ = $4$), yet {\it disagreed} that they could interpret 2D medical images~($MED$ = $2.5$).
Any participant that was involved with our design work (Section~\ref{sec::vrcontour::prelim}) was excluded from this evaluation.

\subsection{Contouring Tasks}\label{sec::method::tasks}
We decided to use the dataset from \href{https://econtour.org}{eContour}~\cite{Sherer2019}, which contains $119$~anonymized cases with expert contouring guidance.
Since investigating 3D rendering is out of our scope, we decided on four structures from four different patients with oncology educators that could be well visualized with our rendering pipeline~(see Section~\ref{sec::vrcontour::sys::rendering}).
Figure~\ref{fig::vrcontour::rendering} demonstrates the 3D views of the selected tasks (with fine and full density ranges), along with one example transverse cutting-plane, as well as an expert contour (noted by the red outlines).

\subsection{Procedures}\label{sec::vrcontour::method::procedure}
We structured our within-subject study into three phases~(see Figure~\ref{fig::vrcontour::timeline}).

\vspace{+0.1in}
\noindent{\bf Phase 1: Pre-Study Orientations \& Questionnaire -}
Participants were first asked to completed a pre-study questionnaire, aiming to collect demographic information, VR experience, and the skills for identify key structures in 3D anatomy and from 2D scans. 
Participants were then introduced to the concept of contouring and inter--slice interpolation.

\vspace{+0.1in}
\noindent{\bf Phase 2: VR Data Collection.}
We aim to evaluate participants' contouring performance of four conditions (see~Figure~\ref{fig::vrcontour::design-space-explorations}).
To avoid the confounding factors of learning experience, a Latin square design~\cite{Dekking2005} was used to counterbalance the order of conditions.
This means that the contouring task (see Figure~\ref{fig::vrcontour::rendering}) performed by each participants using a specific condition would be varied and balanced across participants.
Participants were given opportunities before each session to familiarize themselves with the system.

\vspace{+0.1in}
\noindent{\bf Phase 3: Post-Study UX Evaluations.}
Participants were finally asked to fill out a questionnaire, and think-aloud while recording their responses.
\textbf{S}ystem \textbf{U}sability \textbf{S}cale~(SUS)~\cite{Brooke1996sus} and NASA TLX~\cite{nasatlx} were used to evaluate their perceived usability and task load.
Participants were also asked to rate the usability of using all four conditions as tools to {\it learn} contouring, on a $5$-point Likert scale.
Brief semi-structured interviews were also conducted focusing on the merits and setbacks of seeing and directly drawing inside 3D medical structures.

\begin{figure*}[h]
    \centering
    \includegraphics[width=\linewidth]{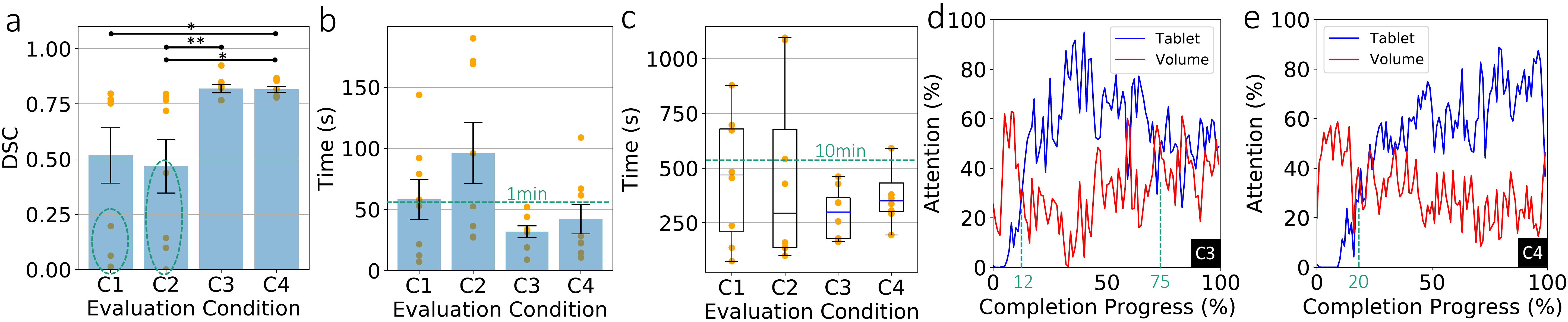}
    \caption[Quantitative results.]{Quantitative results~(\mbox{$* = p < .05$}, \mbox{$** = p < .01$}, \mbox{$*** = p < .001$}) for DSC (a), time for initial explorations (b), and overall TCT (c) measured on four conditions; standard error is used to visualize the error bar and the overlaid yellow scatters indicate individual measurements. Panels (d) and (e) show the average of attention ($\%$) that participants allocated to tablet and 3D volume while contouring using C3 and C4 respectively.}
    \label{fig::vrcontour::quant_results}
\end{figure*}

\subsection{Quantitative Measures}\label{sec::vrcontour::measures}

Participants' contouring performances were evaluated from temporal and spatial domain.
For C3 and C4, we analyzed participants' gaze to understand potential usefulness of inspecting and drawing directly inside 3D medical structure.

\noindent{\bf Consensus Measurement.}
To evaluate the quality of participants' contour, we compute the \textbf{D}ice \textbf{S}imilarity \textbf{C}oefficient~(DSC)~\mbox{\cite{Zijdenbos1994}} between participants' and expert's contours. 
As contouring on transverse plane is used for all testing cases~\cite{eContour}, DSC was computed by converting all contouring onto transverse slices.
Equation~(\ref{eqn::dsc}) shows the computations of DSC, where $X_i$ and $Y_i$ refer to the contour masks of expert's and user's contour at $i^{th}$ slice.

\begin{equation}
    DSC = \frac{2 \sum_{i = 0}^{n} |X_i \cap  Y_i|}{\sum_{i = 0}^{n} |X_i| + \sum_{i = 0}^{n} |Y_i|}
    \label{eqn::dsc}
\end{equation}

\vspace{+0.1in}
\noindent{\bf Temporal Domain Evaluations.}
We measure the time for initial explorations and overall \textbf{T}ask \textbf{C}ompletion \textbf{T}ime~(TCT).
The {\it time for initial exploration} refers to the duration \after the spatial UI being anchored and \before the first stroke being drawn.
During this step, participants need to identify and localize the majority of the target.
On the other hand, the {\it overall TCT} refers to the duration that participants take for completing the entire contouring procedures.

\vspace{+0.1in}
\noindent{\bf Gaze Analysis.}
We analyzed the gaze direction to analyze {\it where did participant focus on during a particular fixation instant?}).
We first filter the frames with the gaze angular speed $\ge 150$$^\circ$$/$s, which are usually considered as saccade~\cite{Britannica1987}.
We then compute the \% of attention that participants allocated during each sliding window on {\it tablet}~(2D) and {\it volume}~(3D), quantified by the \% of frames when gaze collides with tablets and volume, respectively.
Since the overall TCT varies among individuals, we used the \% of the completion progress as the independent variable.
For simplicity, $1$\% of overall TCT is used as the width of sliding window, and the $\%$ of attention was computed upon every unit increment of task completion progress. 
Gaze analysis are only used to analyze C3 and C4, as only 2D interfaces are included in C1 and C2.

\subsection{Implementation}\label{sec::vrcontour::method::implementations}
\begin{figure}[t]
    \centering
    \includegraphics[width=\linewidth]{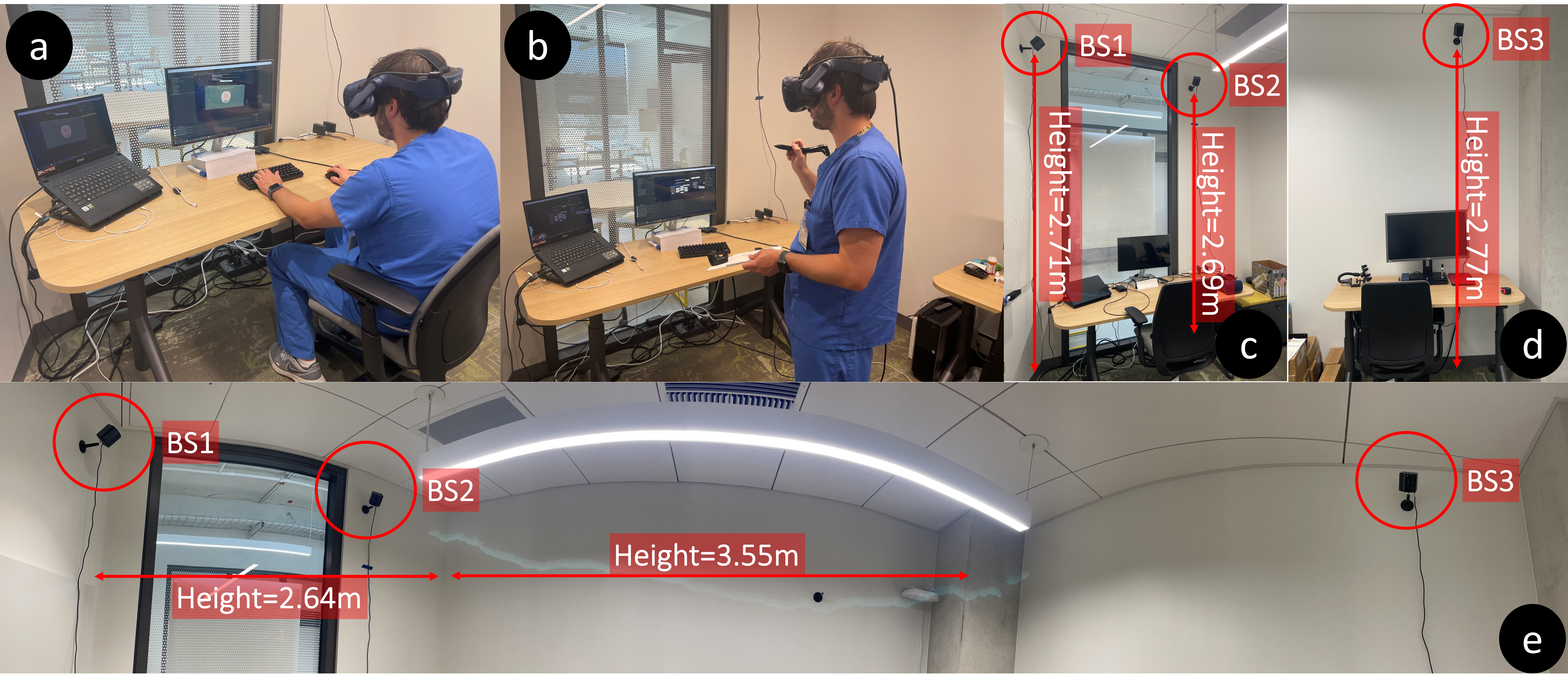}
    \caption[Experimental setups.]{Experimental setups. (a -- b) Participant could sit or stand while performing designated contouring tasks. (c) The base station mounted at the front of the room. (d) The base stations are mounted at the back of the room. (e) A panoramic view of the experimental office.}
    \label{fig::vrcontour::experimental_room}
\end{figure}

We implemented all testbeds on HTC VIVE Pro Eye~\cite{HTCVivePro} using Unity~(v2020.3.6f1) and SteamVR~(v1.21.2).
Tobii XR SDK (v3.0.1.179)~\cite{tobiixr} was used to track gaze direction.
The rendering ran on a gaming laptop ($32$GB RAM, $12$-core CPU, and a RTX~$3080$ GPU).
Inter-slice interpolation was also integrated~\cite{Albu2008, Zukic2016nd}.
%
For C1, we used a keyboard~\cite{hi_keyboard} and mouse~\cite{logitech_mouse} as the input device, to emulate contouring on today's 2D display.
For C2 - C4, we used a Logitech VR Pen~\cite{LogitechVRInk} and a tracked tablet as the input device (see Section~\ref{sec::vrcontour::system::interaction}).
we used three SteamVR Base Station (v2.0) for tracking purposes.
As investigations of tracking performance is out of our scope, we have adjusted the placements of base station and workspace and ensure that the pen and tablet are tracked during at least $99\%$ of the frames (see Figure~\ref{fig::vrcontour::experimental_room}).
We implemented a vertex and fragment shader based on \cite{unityvolumetricrendering}, for rendering cutting--planes and 3D volumes.
A compute shader was implemented for coloring each pixel/voxel.
For C3 and C4, participants' drawn contours would be mirrored on 3D medical structure while drawing on tablet, and vice versa.

\section{Results}\label{sec::vrcontour::results}
\subsection{Quantitative Measurements}\label{sec::vrcontour::results::quant}
Figure~\ref{fig::vrcontour::quant_results} shows quantitative results.
\textbf{R}epeated \textbf{M}easures \textbf{A}nalysis of \textbf{V}ariance (\mbox{RM-ANOVA}) were used to analyze various measures ($\alpha$ = $.05$).
We used \textbf{A}ligned \textbf{R}ank \textbf{T}ransformation~(ART)~\cite{Wobbrock2011} for those that failed to pass the normality check, and ART procedures for multi-factor contrast test for post-hoc test, with Bonferroni corrections~\cite{artc2021}.

\vspace{+0.1in}
\noindent{\bf Overall DSC.}
We first show that changing the evaluation condition (C1, C2, C3, or C4) might change the overall DSC~(\mbox{$F_{3, 28}=5.79$}, \mbox{$p = .005$}, \mbox{$\eta^2=0.45$}).
The subsequent post-hoc test demonstrates a statistical significance between C1 and C4~($p = .04$), C2 and C4~($p = .04$), as well as C2 and C3~($p = .02$).
By looking at the overlaid scatter plot in Figure~\ref{fig::vrcontour::quant_results}a, we observed that three and four participants failed to locate the target structure in C1 and C2 (\textit{without} 3D visualizations), leading to $<50$\% of DSC~(\mbox{$M_1 = 51.80$\%}, \mbox{$M_2 = 46.76$\%}), and higher standard deviations compared to their counterpart (\mbox{$SD_1 = 13.65$\%}, \mbox{$SD_2 = 12.10$\%}~\vs~\mbox{$SD_3 = 1.93$\%}, \mbox{$SD_4 = 1.37$\%}). 
All participants could successfully identify the majority of the targets with C3 and C4 (\textit{with} 3D visualization), leading to $>75$\% of DSC~(\mbox{$M_3 = 81.98$\%}, \mbox{$M_4 = 81.61$\%}).
It is worth mentioning that the DSC measured from our study should not be considered as a reference for clinical purposes~(\eg~usually $>80$\% of DSCs were measured in contouring of cervical cancer~\cite{Viswanathan2014}), as all participants are pre-med or junior MD students without practical contouring experience.
Additionally, tracking inaccuracies could also hinder the consensus measurements.

\vspace{+0.1in}
\noindent{\bf Time for Initial Exploration.}
RM-ANOVA (\mbox{$F_{3, 28}=3.03$}, \mbox{$p = .04$}, \mbox{$\eta^2=0.25$}) shows a marginal $p$--value, but large effect size, implying that our results show an effect, but are likely limited by the small sample size.
No statistical significance was detected during post-hoc test.
All participants located the target using C3 within $1$~min.

\vspace{+0.1in}
\noindent{\bf Overall TCT.}
No statistical significance was detected when it comes to overall TCT. To illustrate our results we therefore visualize the box plot shown in Figure~\ref{fig::vrcontour::quant_results}c.
Results indicate all participants were able to finish the instructed contour tasks within $10$~min with C3.
While it seems that the capability of direct contouring in 3D could accelerate the overall task by minimizing the procedures of delineating outlines, more time was spent for refining the outlines, compensating for the merits brought by the novel affordance given by direct 3D drawing.

\vspace{+0.1in}
\noindent
{\bf Gaze Analysis.}
Figure~\ref{fig::vrcontour::quant_results}d - e approximate the average of \% of attention that participants allocated on tablet~(2D) and volume~(3D) while progressing towards task completion.
With C3, participants heavily relied on the 3D volume for initial explorations~($\sim$ first $12$\% of progress).
When task completion progress reached more than $75$\%, both tablet and 3D volume were used for finalizing the contouring.
With C4, participants mainly used the 3D volume during initial explorations and 3D contours~($\sim$ first $20$\%).

\begin{figure}[t]
    \centering
    \includegraphics[width=\linewidth]{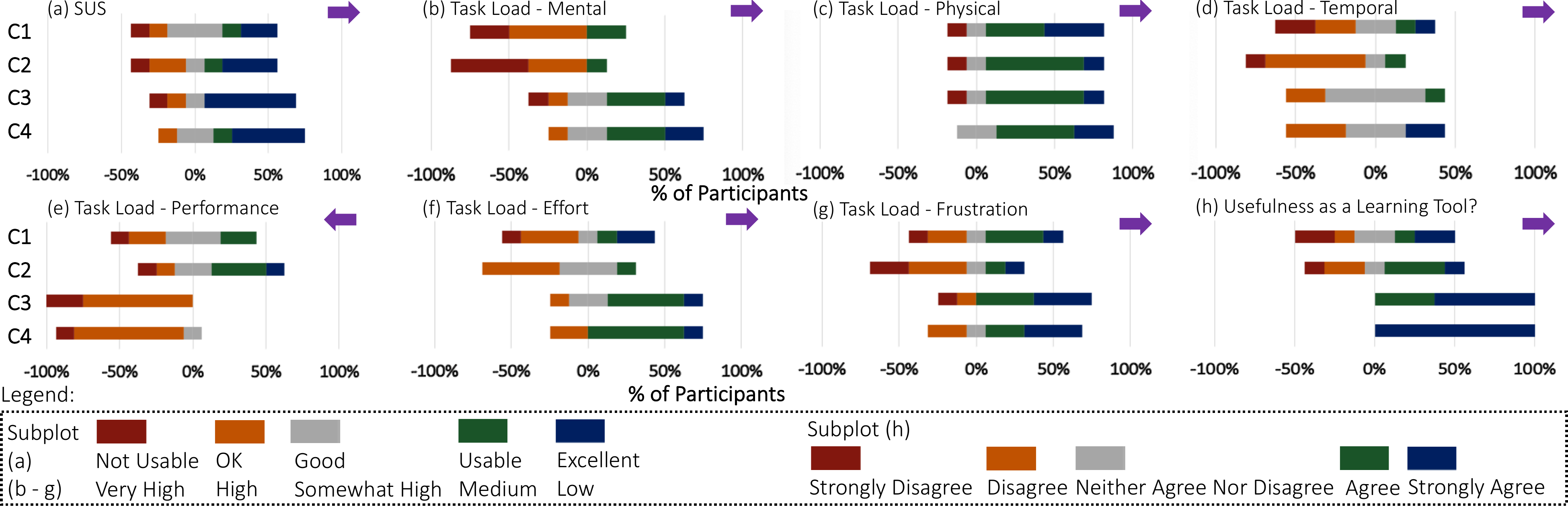}
    \caption[Responses to post-study surveys.]{Responses to post-study surveys. Purple arrow (top right) indicates the direction where a {\it good} system should be designed toward.}
    \label{fig::vrcontour::ux}
\end{figure}

\subsection{Post-Study UX Evaluations}\label{sec::vrcontour::results::ux}
We report here on participants' responses to \textbf{S}ystem \textbf{U}sability \textbf{S}cale~(SUS) (Figure~\ref{fig::vrcontour::ux}a), task load (Figure~\ref{fig::vrcontour::ux}b - g), and level of agreement of \textit{usefulness as a learning tool} (Figure~\ref{fig::vrcontour::ux}h).
We used Sauro's~\cite{Sauro2018} and Hart~\etal's~\cite{nasatlx} recommendations to interpret SUS and task load scores respectively.
Overall, participants rated C3 ({\it with} 3D visualization, {\it without} 3D contouring) as the most usable prototype (\mbox{$MED_3$ = $87.5$} \vs~\mbox{$MED_4$ = $78.8$}, \mbox{$MED_2$ = $72.5$}, \mbox{$MED_1$ = $71.3$}).
``Excellent'' is considered for \mbox{SUS $\geq 80.8$} ~\cite{Brooke1996sus, Sauro2018}.
Four takeaways are generated through thematic analysis:

\vspace{+0.1in}
\noindent{\bf Seeing 3D Medical Structures.}
All participants appreciated the merits of seeing the 3D structure inside VR, and think {\it ``the 3D was very useful''} [P5].
With the 3D visualization (C3 and C4), more participants reported less mental load (Figure~\ref{fig::vrcontour::ux}b) and higher performance (Figure~\ref{fig::vrcontour::ux}e).
First, participants believed that 3D visualizations are useful for understanding the structure ({\it ``when I just look at the slides, I'm not sure if that's the target or not, but if I see it in 3D, I'm more confident.''} [P3]), reduce mental load ({\it``It helps to have a bigger picture, and it takes off that cognitive load to actually imagine the 3D representation, which is a very high load.''} [P2]), and locating targets [P5].
P7 pointed out that the 3D visualization addresses one of the hardest challenges in anatomy class: {\it ``I think that the hardest part of anatomy is trying to look at pictures on a 2D surface, and then make them 3D, which is like exactly what this is accomplishing''} [P7].
P3 and P7 emphasized that the 3D medical structure helped them tremendously on deciding the boundary of the target structures (\eg~heart and lungs).
%
Some participants appreciated the transparent volumetric rendering: {\it``we use an app called Complete Anatomy, but I like [the transparent rendering] a lot better. You can see more detail and get depth. It's really hard to get depth for non-transparent rendering.''} [P7].
Second, with 3D structures, participants could better leverage the inter-slice interpolation by deciding the slice position where the rapid structure changes occurred 
(\eg~{\it ``it will help in locating the cutting-plane.''} [P5], {\it ``you can see the plane that you put in your virtual cadaver, and it helps feel where the rapid changes occur.''} [P7]).

\vspace{+0.1in}
\noindent{\bf Drawing Contours in 3D.}
Participants appreciated the capabilities of direct drawing in 3D while attempting to locate the target (\eg~{\it``A huge part of the work gets done way faster''} [P2], {\it ``I like the mapping that map my annotation in 3D to 2D''} [P6]).
Some participants also appreciated the precision offered by the VR pen ({\it``[the pen] is intuitive for precise hand control''} [P3]).
However, we also observed participants expressing slightly higher frustration by comparing C3 and C4~($MED$: $12.5$\% \vs~$20$\%);
potential reasons include difficulties of drawing precisely in 3D 
(\eg~{\it ``It was very hard for me to be precise when drawing in 3D, I didn't see where I was starting''} [P2]), 
lack of capabilities for inspecting inside 3D volumes 
(\eg~{\it ``You cannot look inside the body. When you look inside the body, you cannot see very clearly where you are. [...] I think a better thing is as your pen is approaching, in the 3D model, you're looking at a slice of the 3D model, so that you know exactly where you are''} [P3]),
and small size (\eg~{\it ``if you could blow it up more, to be able to get finer precision, that would make it a lot easier''} [P7]).

\vspace{+0.1in}
\noindent{\bf Learning Cost.}
Echoing \mbox{[N2]} and \mbox{[N3]}, some participants also mentioned that {\it``medical school students mostly never used VR''} [P3].
While trained to use the system before each session, all participants found it not hard to learn and get familiar with system operations (\eg~{\it``a week for learning probably would be sufficient''} [P8]).
This echos the fact that seven participants {\it agree} or {\it strongly agree} with the statement that {\it ``I would imagine that most people would learn to use this system very quickly''} in SUS responses.

\vspace{+0.1in}
\noindent{\bf Contouring Education.}
%
All participants {\it agreed} or {\it strongly agreed} that seeing and drawing in 3D is extremely helpful for learning contouring (\eg~{\it``if you could put this into like, even our anatomy labs, that'd be amazing''} [P7]).
While comparing the experience of using C3 and C4, P1 outlined additional setbacks of learning anatomy with cadavers, including the unpleasant smell of formaldehyde, the mental stress triggered by a real body, and the reduced accessibility and resulting practicing opportunities.
%
P7 appreciated the transparent rendering while comparing it with \emph{Complete Anatomy}~\cite{complete_anatomy_2020}, a mobile application designed for anatomy learning purpose, because of the capability of {\it ``seeing more details and getting depth information''}. 

\section{Discussion and Future Work}\label{sec::vrcontour::discussion}

\subsection{Seeing and Contouring in 3D}\label{sec::vrcontour::discuss:::viz_contour_3D}
Our results show that there is great potential for radiation oncology and contouring when integrating 3D visualizations into existing contouring workflows through VR.
While our focus is on RT treatment planning, many findings are more general and could be transferred to other similar applications that require annotations on 3D volumes (\eg~annotations during surgical planning~\cite{ARTEMIS2021}).

Since some of our participants felt that it was less intuitive and overall harder to directly draw in 3D - mostly because of lack of depth perception and fine grained control - one natural follow up would be to {\it investigate the design of novel interaction techniques where participants can draw precisely in 3D in these contexts}.

Another direction that future work might consider is the inherent challenges raised by interacting with 3D medical structures.
While the metaphor of pen $+$ tablet could offer precise drawing, some participants believe that they are less intuitive when it comes to 3D object manipulations.
Future investigations of interactions with 3D volume might be needed to uncover more insights.

\subsection{Rethinking the Contouring Training Curriculum}\label{sec::vrcontour::dicuss::curriculum}
Contouring is a challenging skill to master, and it is typically trained during residency programs.
While interpreting 2D scans is usually difficult, it is a necessary prerequisite for learning contouring.
However, Dmytriw~\etal~\cite{Dmytriw2015} surprisingly showed that $< 10$\% of students are \textit{very} or \textit{completely} confident in interpreting basic radiology images.
Although additional training during residency program might address such gaps, many residents might not have that opportunity, especially in less developed areas~\cite{Holt2001, Sura2017, Eansor2021}.
With \mbox{VRContour} we unveil an alternative solution - bringing the contouring workflow into a VR-based 3D UI - that could possibly help radiation oncologists (and others) to massively scale contouring education. 

Although participants recognized that getting familiar with VRContour is not challenging, we also found that half participants did not have extensive experience on using VR (rated as {``only tried it out once or twice before''} in our pre-study questionnaire), and exhibited less experience with exploring 3D user interfaces.
One future direction is to rethink and modernize today's oncology educations by fully integrating VR as part of future curricula.

\subsection{Limitations}\label{sec::vrcontour::discussion::limitation}
The limitation of our work can be summarized into three areas.
First, our results are based on $8$ participants, and the small sample size together with the lack of background/expertise/preference diversity might affect some results.
While we only considered participants without extensive experience in interpreting 2D scans to minimize the confounding factors of learning experience, such participants will also not have prior contouring experience, limiting the application of our results to different levels of contouring proficiency.
Second, since examining participants' anatomy knowledge was beyond our scope, our study was based on only four cases extracted from \href{https://econtour.org}{eContour}~\cite{Sherer2019} during \mbox{[I4]}, with the goal to have baseline scenarios that would be familiar for any students with basic anatomy training.
However, practical contouring usually involves tumors and finer-grained anatomy, which might require prior experience on oncology and pathology, which would have been challenging for current participants.
Future work might introduce more realistic tasks, and include residents and attending physicians as part of the evaluation.
Finally, we intentionally excluded procedures for adjusting volumetric rendering parameters from our study.
Instead, we pre-tuned parameters and offered {\it full} and {\it fine} density range view for participants to inspect the target and overall view of the 3D volume. 
However, adjusting such parameters is an inevitable step in practical contouring, and it might be considered in future work.
\section{Conclusion}\label{sec::vrcontour::conclusion}
We presented {\it VRContour}, and investigated how to bring radiation oncology contouring into VR.
We explored and defined three design spaces by considering the dimensionality for consuming and inputting information. 
In a within-subject study \mbox{(N = $8$)}, we found that the inclusion of a 3D visualization could increase contour precision and reduce mental load, frustration, and overall effort, as well as introduce important benefits for learning and educational purposes.
We believe that our work will benefit future research in designing VR-based medical software tools for radiation oncology and beyond.
\section{Acknowledgment}

The research presented in this chapter was partially funded by the \textbf{A}gency for \textbf{H}ealthcare \textbf{R}esearch and \textbf{Q}uality (AHRQ).
We acknowledge the helps and resources from Larry Hernandez and Scott Lundy.

Chapter~\ref{sec::vrcontour}, in full, is a reprint of the material as it appears in the Proceedings of the 2022 IEEE \textbf{I}nternational \textbf{S}ymposium on \textbf{M}ixed and \textbf{A}ugmented \textbf{R}eality (ISMAR 2022). The dissertation author was the primary investigator and author of this conference paper. Co-authors includes Matin Yarmand, Varun Singh, Mechael V. Sherer, James D. Murphy, Yang Zhang and Nadir Weibel. 
This work was collaborated with the School of Medicine at University of California San Diego and the Department of Electrical and Computer Engineering at University of California Los Angeles.
The previously published manuscript can be referred to \cite{Chen2022VRContour}.
Readers are encouraged to refer to \cite{Chen2022VRContourWIP, Chen2022PrecisionDrawing} for the related publications.

\chapter{Discussion and Future Work}\label{chapter::discussion}

Grounded in our experience designing, implementing, and evaluating the interactive system presented in Chapter~\ref{sec::memovis}, Chapter~\ref{sec::papertoplace} and Chapter~\ref{sec::vrcontour}, this section discusses key lessons learned and practical implications for designing seamless and efficient interactions within mixed-dimensional information spaces.

\section{Reflections on Interaction Experiences within the Mixed-Dimensional Information Space}\label{discuss::system_reflection}

The representations of both 2D and 3D have its own merits~\cite{Thoravi2022Interactive}.
2D entities, such as images and text, are easy to create, understand, and share, whereas can be less intuitive to represent complex information.
On the other hand, 3D entities, whether presented through Web3D on a 2D display or rendered within XR environments, offer intuitive and realistic experiences but require users to \emph{actively} explore different perspective to fully understand the information:
users may need to use a keyboard and mouse to navigate views on a 2D display or rely on dedicated designed 3D UI to manipulate volumetrically rendered objects.
The \emph{seams} within the mixed-dimensional information space often arises from challenges such as prospective mismatching, spatial misalignment, and dimensional incongruent.
The interactive systems presented in this dissertation highlight the user experiences and associated challenges that emerge when 2D and 3D representations coexist within the information space.

In MemoVis~\cite{Chen2024MemoVis}, we focused on the experience when feedback providers creating reference images alongside textual comments for 3D designs.
In this workflow, users must explore and identify relevant views for anchoring feedback comments by interacting with both the textual comments and the 3D objects rendered on the display.
While using online reference images is common, finding images with views that match the anchored view of the 3D model for conveying feedback can be tedious.
MemoVis demonstrates how VLFMs and GenAI can augment this workflow by allowing users to focus on the primary task - creating and typing textual comments.

In PaperToPlace~\cite{Chen2023PaperToPlace}, we explored how users interact within 3D space while consuming 2D media.
We focused on instructional procedural tasks, where novice users must complete real-world activities guided by instruction documents.
The seams between instructional documents and the physical 3D space can lead to unnecessary context switching, confusion, and inefficiencies during task completion.
In a MR environment, PaperToPlace introduced a computational approach that optimally places virtual instruction steps adjacent to key objects - positioned close enough to read easily, but not too close to occlude the view.

In VRContour~\cite{Chen2022VRContour, Chen2022VRContourWIP}, we investigated the challenge of cognitive disconnection that arises when users perform contouring tasks on medical images.
While exploring and interpreting medical images, users often need to mentally reconstruct anatomical structures in 3D. 
On the other side, although volumetrically rendered 3D anatomy provides an intuitive overview, contouring workflows still require users to focus on 2D cross-sectional slices when annotating target structures.
The study on VRContour demonstrates how 2D (cross-sectional slices) and 3D (volumetrically rendered anatomical structure) can assist users in exploring and contouring medical images, improving precision, task completion time, and cognitive load, as evaluated through both quantitative and qualitative measures.

\section{Context Awareness}\label{sec::discuss::context}

\begin{boxH}

{\it ``the challenge for future human-centered computer systems is not to deliver more information `to anyone, at anytime, and from anywhere,' but to provide `the \textbf{right} information, at the \textbf{right}
time, in the \textbf{right} place, in the \textbf{right} way, to the \textbf{right} person''}~\cite{Fischer2012} --  Gerhard Fischer, 2012

\end{boxH}

Although interacting with both 2D and 3D is often required, the experience should allow users to stay focused on their primary tasks without being deviated by the limitations of dimensional representations or transitions between entities.
We demonstrated the importance of adapting interface design to the \emph{contexts}.
The ultimate goal is to realize the vision of context-aware systems, as envisioned by Fischer~\cite{Fischer2012}, which deliver {\it ``the \underline{right} information, at the \underline{right} time, in the \underline{right} place, in the \underline{right} way, to the \underline{right} person.''}~

For example, PaperToPlace introduced a novel pipeline that computationally determines the placement of virtually rendered procedural instruction steps in MR by leveraging contextual clues from 3D spaces (\eg~anchoring surfaces of key objects) and user behaviors (\eg~gaze and hand gestures).
In the case of MemoVis, we contributed to a novel real-time viewpoint suggestion feature, based on VLFMs, to help feedback providers anchor a textual comments with a relevant camera viewpoint.
However, our design solely relied on the typed textual comments provided by feedback providers.
While this \emph{implicit} cue can be seen as a form of context, future work may explore richer contextual signals - such as mouse and keyboard activity or behavioral gestures like gaze - to better understand the key information users attend to when composing feedback comments.
Similarly, although some of our experimental scenarios in VRContour allowed users to freely choose between 3D volumetric renderings and 2D cross-sectional slices to view, explore and delineate medical images, future work could explore a context-aware interface that suggests the suitable dimensional representation based on the task or user behavior.

\section{Balancing Information Richness and Cognitive Load}\label{sec::discuss::info_and_load}
Designing a seamless and efficient interaction experience within the mixed-dimensional information space needs balancing information richness and cognitive load across different stages of the task.
This implication can be seen as a concrete realization of the ``right information'' principle in the design of context-aware systems, as envisioned by Fischer~\cite{Fischer2012} (see Section~\ref{sec::discuss::context}).
A similar point was emphasized by Darken~\etal~\cite{Darken2005}, who advocated the principal of ``dimensional congruence'', emphasizing aligning the dimensionality of interaction techniques with the dimensional requirements of the task, although their work primarily focused on basic 3D interaction tasks such as selection and positioning.
Thoravi~\etal~\cite{Thoravi2022Interactive} also advocates {\it ``carefully balance the tradeoff between the density of information presented to the users, and the cognitive load required to perceive and interact with them.''}

Our research around VRContour showed that while 3D-rendered anatomical structures are preferred during initial and final stage of contouring, participants held a strong preference for using the VR stylus and 2D-tracked tablet for creating and refining contours.
During the workflow of creating and refining contours - after identifying the anatomical structure and target location - additional 3D representations alongside depth information of non-target anatomical structure may distract users from primary workflow and result in unnecessary mental load.
From a different perspective, the interaction experience with 2D instructional documents and the physical 3D space in PaperToPlace demonstrated how to establish spatial connections between documents and key objects in the environment.
By designing adaptive placement of instructional guides, novice users can stay focused on the task, as irrelevant 3D entities in the 3D space may otherwise distract them from the primary workflow.

\section{Limitations and Future Work}\label{sec::discuss::future}

We demonstrated how seamless and efficient interaction within mixed-dimensional information systems can look like by presenting the design, implementation, and evaluation of three systems, including MemoVis~\cite{Chen2024MemoVis} introduced in Chapter~\ref{sec::memovis}, PaperToPlace~\cite{Chen2023PaperToPlace} introduced in Chapter~\ref{sec::papertoplace}, and VRContour~\cite{Chen2022VRContour, Chen2022VRContourWIP}.
We highlight key limitations and implications for future research.

\vspace{+0.1in}
\noindent{\bf Integrating interaction workflows into real-world applications and deploying them in ecologically valid settings.}
All introduced systems in this dissertation have been evaluated by real-world stakeholders in controlled lab settings.
However, the affordances of the proposed interaction workflows within mixed-dimensional information spaces need be more thoroughly understood using real-world application and deployment.
Future researchers might consider integrating the proposed interaction workflows into real-world applications and deploying them in ecologically valid settings.
A real-world deployment enables the collection of deeper insights and valuable lessons learned, beyond user studies in controlled lab settings.
For example, our design of MemoVis~\cite{Chen2024MemoVis} could be extended into a lightweight web-based application or integrated as a plugin within existing instant messaging tools like Slack~\cite{Slack}, to assist users in efficiently generating 3D design feedback that includes reference images and textual comments.
In the design of PaperToPlace~\cite{Chen2023PaperToPlace}, we focused solely on procedural instruction tasks within the context of cooking in an office kitchen, due to the resolution limitations of the Quest Pro~\cite{QuestPro} at the time of the user study.
However, with today’s high-quality video see-through MR headsets, such as the Vision Pro~\cite{visionProVision}, PaperToPlace~\cite{Chen2023PaperToPlace} can be extended into a service that guides museum visitors through spatially distributed exhibits or assists travelers as they settle into a temporary rental apartment (\eg~by providing essential instructions related to physical objects such as dishwashers and laundry machines).
Similarly, VRContour~\cite{Chen2022VRContour} can be extended and deployed in modern medical offices and educational settings, such as classrooms for medical training.
Deploying these applications in real-world settings can enable future researchers and practitioners to better understand the affordances of the our proposed interaction workflows.

\begin{figure}[t]
    \centering
    \includegraphics[width=\linewidth]{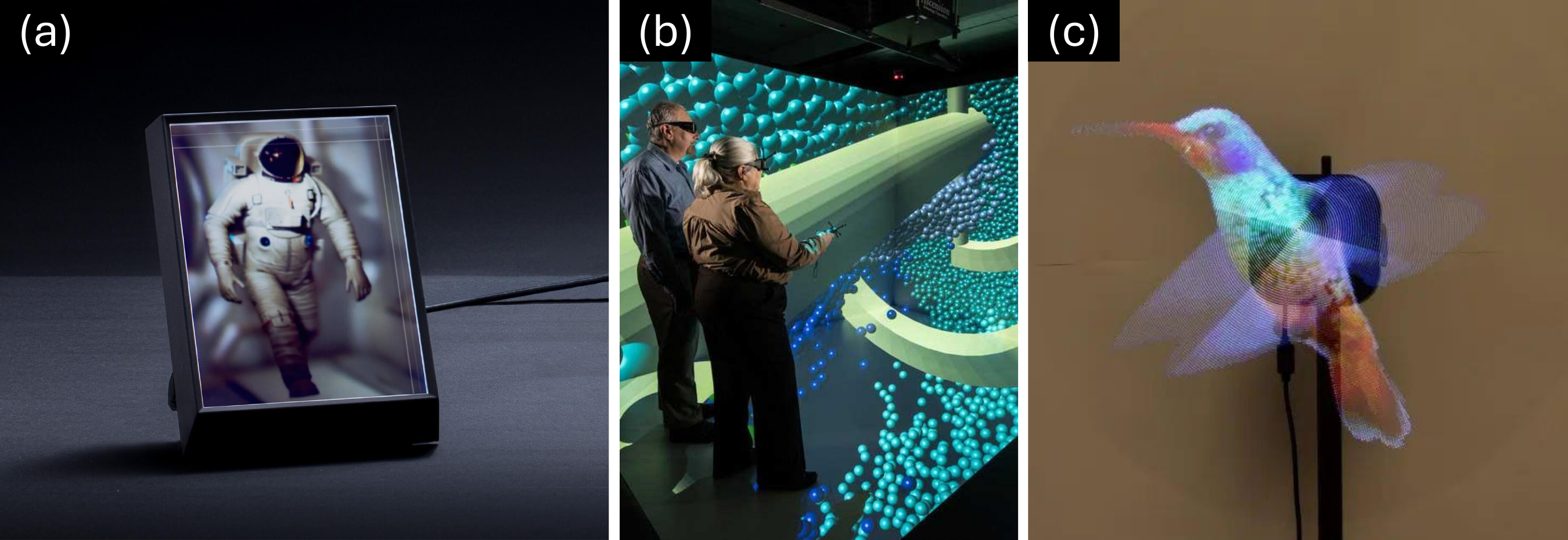}
    \caption[Examples of delivering 3D content using emergent display technologies.]{Examples of delivering 3D content using emergent display technologies: (a) spatial display by Looking Glass Factory~\cite{LookingGlass}; (b) projector-based CAVE environment~\cite{Cruz1992CAVE}; (c) 3D holographic fan.}
    \label{fig::discuss::displays}
\end{figure}

\vspace{+0.1in}
\noindent{\bf Exploring more diverse interactions within mixed-dimensional information spaces using a broader range of display technologies.}
Mixed-dimensional information spaces suggest the inclusion of both 2D and 3D visual representations.
While it seems straightforward to bring 2D representations into the information space (such as through printed documents and/or 2D display), deliver 3D representations can be involved.
Although the 3D representations in MemoVis are presented using Web3D~\cite{Web3D} on everyday 2D displays and those in PaperToPlace and VRContour are delivered through head-mounted XR devices, future researchers can explore the design and affordances of a broader range of display technologies. 
For example, future researchers may explore how interaction experiences within mixed-dimensional information spaces may look like when 3D content is delivered through emerging spatial displays (Figure~\ref{fig::discuss::displays}a), projector-based \textbf{C}ave \textbf{A}utomatic \textbf{V}irtual \textbf{E}nvironment~(CAVE) environments (Figure~\ref{fig::discuss::displays}b)~\cite{Cruz1992CAVE} and novel holographic displays such as 3D holographic fan~ (Figure~\ref{fig::discuss::displays}c).

\vspace{+0.1in}
\noindent{\bf Designing for a broader range of tasks and user populations.}
We focus on designing seamless and efficient interaction experiences within the mixed-dimensional information space, grounded on three specific applications: creating reference images for 3D design feedback, performing procedural tasks with instructions, as well as exploring and contouring medical images.
Future work can explore a broader range of tasks, focusing on the challenges of interaction within mixed-dimensional information spaces and how to design techniques or revise existing workflows to address them.
While the systems presented in this dissertation target both general users and specialized users such as medical professionals, future work may explore a broader range of user populations.
Ultimately, this future direction aims to contribute to a generalizable framework for designing interaction workflows involving multiple information entities represented across different dimensions. 
This framework may serve as a `blueprint' for future researchers and practitioners when developing interactive applications tailored to specific user populations.

\chapter{Conclusion}\label{sec::conclusion}

Grounded in the user-centered design approach and demonstrated through three applications - creating reference images for 3D design feedback, performing procedural tasks with instructions, and exploring and contouring medical images - this dissertation contributes to the design of \emph{seamless} and \emph{efficient} interaction experiences within the \emph{mixed-dimensional information space}, where users engage with entities represented in both 2D and 3D.
Our contribution was demonstrated through three interactive systems: MemoVis introduced in Chapter~\ref{sec::memovis}, PaperToPlace introduced in Chapter~\ref{sec::papertoplace}, and VRContour introduced in Chapter~\ref{sec::vrcontour}.
The interactive experiences introduced in this dissertation includes those rendered on traditional 2D display and those experienced through emergent extended reality headset.
In the long run, we believe that these approaches can inform future work in designing interactive workflows that enable users more efficiently engage with mixed-dimensional information spaces comprising content represented across different dimensions.

\appendix

\chapter{Ethical Disclaimer}
This dissertation contains multiple research that was involved with human subjects.
All research described in this dissertation has been approved by the \textbf{I}nstitutional \textbf{R}eview \textbf{B}oard (IRB) at University of California San Diego.
All \textbf{P}ersonal \textbf{I}dentifiable \textbf{I}nformation~(PII), such as the face has been intentionally removed (\eg~being pixelized or blurred).
Before each user study, we have obtained participants' consent on video and audio recordings, as well as heterogeneous behavior data collections. 
All participants have consented and acknowledged the data and results presented in this dissertation, to be published and presented publicly for research purposes.
Monetary or non-monetary incentives were provided as per described in the related chapters.
All participants have been informed with our research progress, related follow-up works and publications.

\chapter{Supplementary Materials for Chapter ~\ref{sec::memovis}}

\section{Copyright Disclaimer}\label{sec::memovis::app::ethical_disclaimer}
For research and demonstration purposes, we used multiple internet searched images in Figure~\ref{fig::memovis::polycount-example} and Figure~\ref{fig::memovis::study-2-baseline-example}, copyrights of these images belong to the original content creators. 
While copyrights of the synthesized images created by GenAI are still a challenging research topics~\cite{Samuelson2023, Murray2023}, we do not claim the copyrights of all synthesized images by GenAI. 
These images are only used for research and demonstration purposes.

\section{Supplementary Algorithm Design}\label{sec::memovis::app::algo}

Algorithm~\ref{alg::remove_hidden_obj} shows the algorithm to remove the residual pixels when a new objects are added by the feedback providers, with text + scribble and grab'n go modifiers.
Full system design details could be referred to Section~\ref{sec::memovis::system::modifiers}.

\begin{algorithm}[h!]
   \small 
   \caption{Approximate initial design image without hidden objects.} \label{alg::remove_hidden_obj}
   \begin{algorithmic}[1]

        \PROCEDURE{SelectMeshPrimitives}{$\bm{I}_{seg}$, $\bm{v}$, $r_{th} = 0.7$}
            \STATE $cam =$ \textbf{new} $OrbitCamera()$
            \STATE $cam.setViewAngle(\bm{v})$
            \STATE $scene.render()$
            \STATE $hits \gets$ \textbf{new} $Set()$
            \FOR{{sampled pixels \textbf{in}} $\bm{I}_{seg}$}
                \IF{$\bm{I}_{seg}(r, c) == 0$}
                   \STATE \textbf{continue}
                \ENDIF
                \STATE $mesh \gets scene.raycast(r, c, cam)$
                \IF{$mesh != null$}
                   \STATE $hits.add(mesh)$
                \ENDIF
            \ENDFOR
            \FOR{$mesh$ \textbf{in} $hits$}
                \STATE $DepthTextureRenderer.renderList \gets [mesh]$
                \STATE $DepthTextureRenderer.render()$
                \STATE $depth \gets DepthTextureRenderer.getImage()$
                \STATE $\bm{I}_{mesh} \gets (depth < depth.max())$
                \STATE $r \gets sum({\bm{I}_{mesh} \cap \bm{I}_{seg}}) /  sum(\bm{I}_{mesh})$
                \IF{$r \leq r_{th}$}
                   \STATE $hits.remove(mesh)$
                \ENDIF
            \ENDFOR
            \STATE \textbf{return} $hits$
        \ENDPROCEDURE

        \PROCEDURE{GetInitialImage}{$\bm{I}_{seg}$, $\bm{v}$, $r_{th} = 0.5$}
            \STATE $meshes \gets SelectMeshPrimitives(\bm{I}_{seg}, \bm{v}, r_{th} = 0.5)$
            \STATE $RGBTextureRenderer.renderList \gets meshes.toList()$
            \STATE $RGBTextureRenderer.render()$
            \STATE \textbf{return} $RGBTextureRenderer.getImage()$
        \ENDPROCEDURE
    \end{algorithmic}
\end{algorithm}
\clearpage
\section{Examples of ControlNet with Depth and Scribble Conditions}\label{sec::memovis::app::compare_controlnet_depth_scribble}
\begin{figure*}[h]
    \centering
    \includegraphics[width=\textwidth]{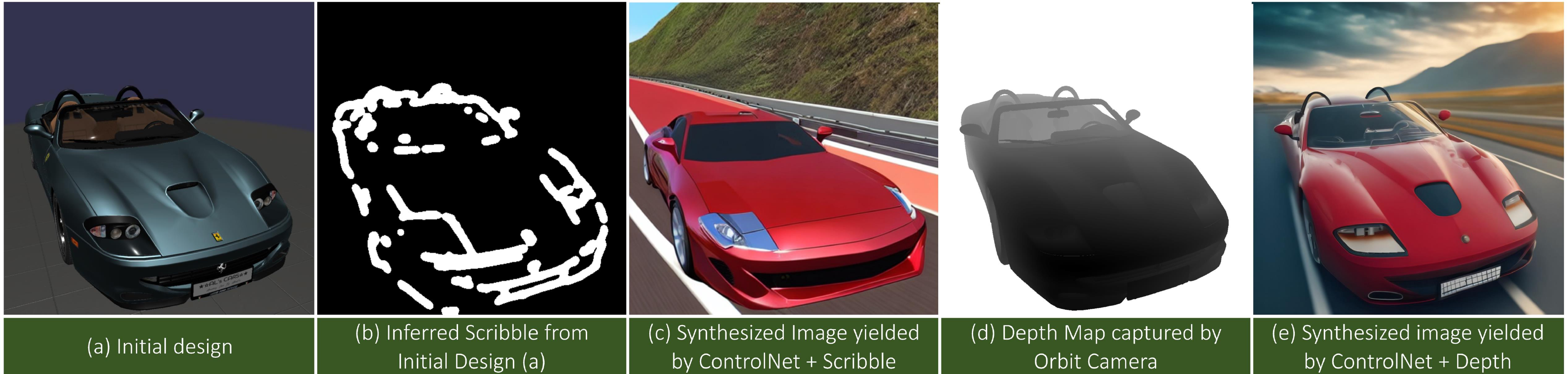}
    \caption[Examples of synthesized images created by ControlNet conditioned on scribble and depth.]{Examples of synthesized images created by ControlNet conditioned on scribble and depth. (a) The viewpoint of the initial 3D design; (b) Users' scribble inferred from (a); (c) Synthesized image yielded by ControlNet conditioned on scribble; (d) Depth map captured by orbit camera with (a); (e) Synthesized image yielded by ControlNet conditioned on depth. For both conditions, we used the prompt {``a red car driving on the freeway''}.}
    \label{fig::memovis::controlnet-scribble-depth-compare}
\end{figure*}

Designing of MemoVis leveraged the year-2023's pretrained ControlNet conditioned on depth and scribble~\cite{Zhang2023ControlNet}.
This section provides additional supplementary example demonstrating how depth-conditioned ControlNet could better anchor the generated image based on the original design.
Figure~\ref{fig::memovis::controlnet-scribble-depth-compare}c and Figure~\ref{fig::memovis::controlnet-scribble-depth-compare}e shows an example of how the synthesized images look like guided by the textual prompt {\it ``a red car driving on the freeway''}.
Notably, Figure~\ref{fig::memovis::controlnet-scribble-depth-compare}b and Figure~\ref{fig::memovis::controlnet-scribble-depth-compare}d shows the inferred scribbles with HED annotator~\cite{Xie2015HED} and the depth map of the 3D model.

\clearpage
\section{Pre-Trained Models}\label{sec::memovis::app::models}
MemoVis was prototyped based on a set pre-trained model, and deployed on a cloud server with four A10G GPUs.
This section provides supplementary details of the VLFMs we used.
Full design and implementation details could be referred to Sec.~\ref{sec::memovis::system}.

\begin{itemize}[noitemsep, leftmargin=*]

\item {\bf CLIP}. We used the pre-trained \texttt{clip-ViT-B-32} due to its inference performance and our available computing resources. Other variant CLIP pre-trained model would lead to similar results.

\item {\bf SAM}. We used the checkpoint of \texttt{sam\_vit\_h\_4b8939.pth}. All hyper-parameters that were used during inference are consistent with the default suggestions~\cite{Kirillov2023SAM}.

\item {\bf ControlNet}. We used two different pre-trained ControlNet models. For depth-conditioned synthesis, we used an internal pre-trained model developed by our organization. 
Similar to the original ControlNet model, our internal model generates realistic images from the depth maps. For our use cases, we opted to use this internal model since it was optimized to generate high quality realistic images. 
While the open-source equivalent of depth-conditioned ControlNet (\texttt{lllyasviel/sd-controlnet-depth}) might work, the quality of the synthesized images might be degraded.
For depth- and scribble-conditioned synthesis, we used the open source pre-trained ControlNet under the depth (\texttt{lllyasviel/sd-controlnet-depth}) and scribble (inferred by HED annotator~\cite{Xie2015HED}) conditions, based on original Stable Diffusion (runwayml/stable-diffusion-v1-5). 
For all inference tasks, we used {\it ``realistic, high quality, high resolution, 8k, detailed''} and {\it ``monochrome, worst quality, low quality, blur''} as the positive and negative prompts, respectively. 
The inference step was set to $30$ to balance the quality of synthesized image and inference latency.

\item {\bf Inpainting}. The model \texttt{kandinsky-community/kandinsky-2-2-decoder-inpaint} was used to support inpainting feature. The prompt {\it ``background''} was used to realize the feature of removing object(s) from the reference image~\cite{inpaintRemove}.

\end{itemize}

\clearpage
\section{Participants Recruitment for User Studies}\label{sec::memovis::app::evaluation_participant}
\begin{figure}[h!]
    \centering
    \includegraphics[width=\textwidth]{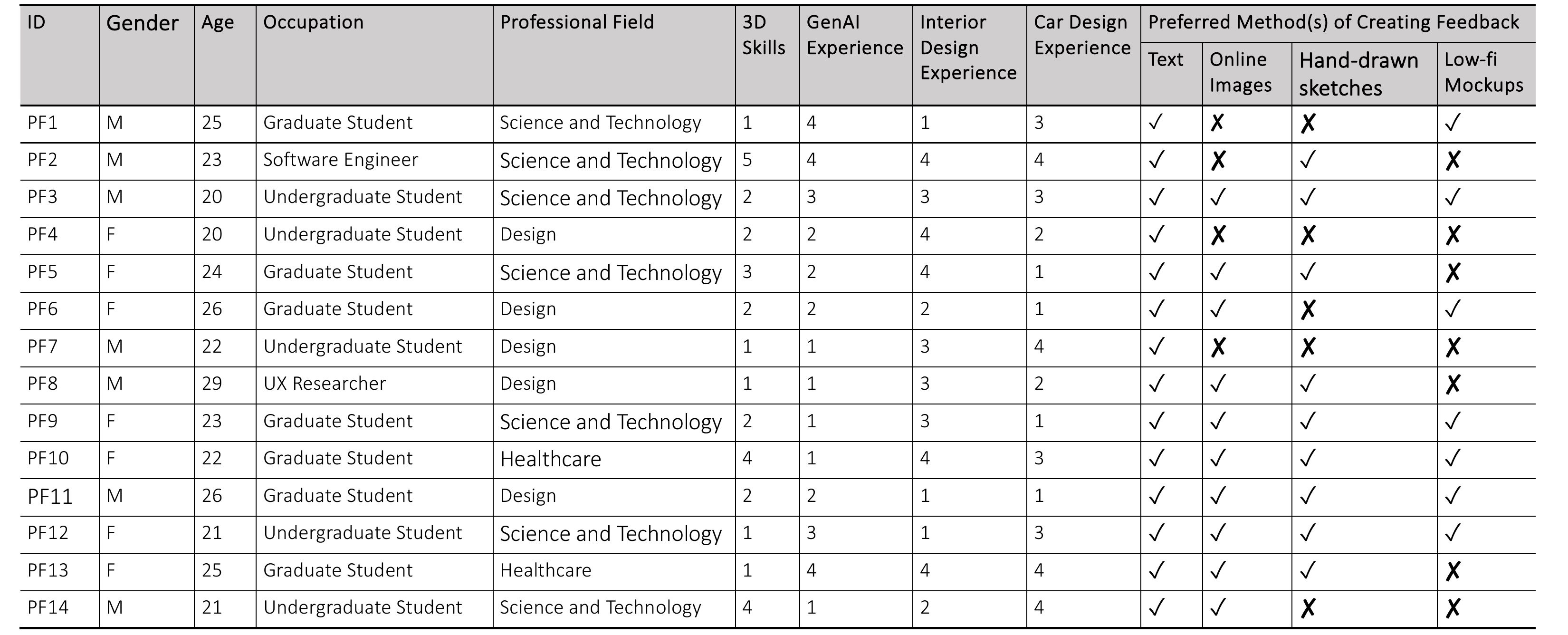}
    \caption[Participants' demographic background of the Study 1 (Section~\ref{sec::memovis::eval::create_feedback}).]{Participants' demographic background of the Study 1 (Section~\ref{sec::memovis::eval::create_feedback}). In a 5-point Likert scale, participants were instructed to self-evaluate their skills of using 3D software and prior experience of GenAI, interior design as well as car design. The methods of ``text'', ``online images'', ``hand-drawn sketches'', and ``low-fi mockups'' refer to the feedback communication method of using typed texts, online-searched reference images, hand-drawn sketches, and the reference images created using image editing tools.}
    \label{fig::memovis::study1-participant}
\end{figure}

Figure~\ref{fig::memovis::study1-participant} shows the participants' demographic background of the Study 1 (Section~\ref{sec::memovis::eval::create_feedback}).
Participants were recruited from various social media platforms and a university instant messaging application group, \incl~ faculty, students and alumni, through convenience sampling.
Even though all participants viewed themselves as experienced in providing design feedback, we reported their preferred methods for creating feedback in the past.
Figure~\ref{fig::memovis::study1-participant} shows that all participants used texts to communicate the feedback.
Participants may also use reference images searched from online, drawn by hand sketches, and/or created using image editing tools, which are referred as ``online images'', ``hand drawn sketches'', and ``low-fi mockups'' respectively in Figure~\ref{fig::memovis::study1-participant}.
The power analysis of the sample size of $14$ demonstrates a power of $80.06\%$, with the significance level ($\alpha$), number of groups, and effect size being set to $0.05$, $2$, and $0.4$, respectively. 
We used the $\eta_p^2$ to compute the effect size, and chose a large effect size due to the lengthy duration of the Study 1.

\begin{figure}[h!]
    \centering
    \includegraphics[width=\textwidth]{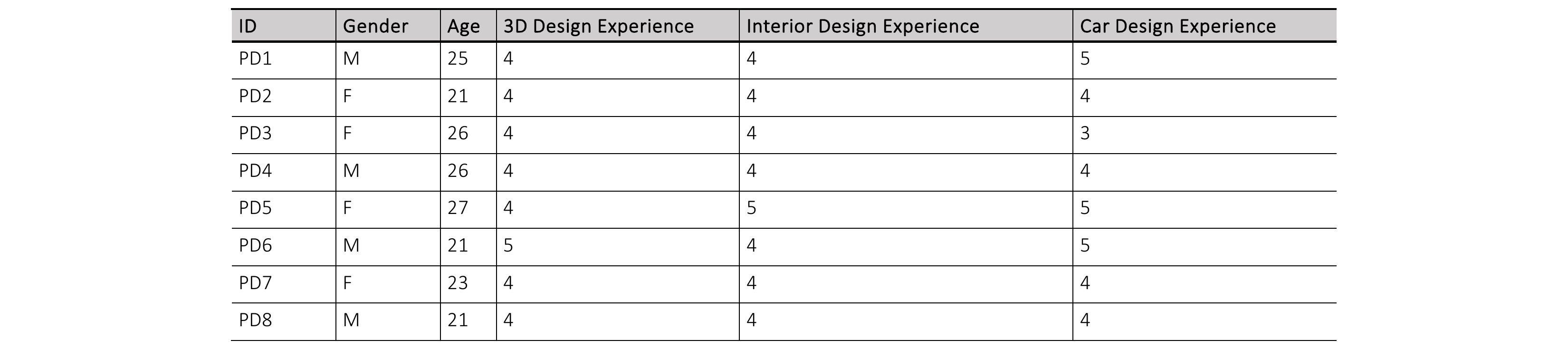}
    \caption[Participants' demographic background of the Study 2~(Section~\ref{sec::memovis::eval::eval_feedback}).]{Participants' demographic background of the Study 2~(Section~\ref{sec::memovis::eval::eval_feedback}). In a scale of 1 to 5, participants were instructed to self-evaluate their prior experience of 3D design in general as well as experience of interior and car design.}
    \label{fig::memovis::study2-participant}
\end{figure}

Figure~\ref{fig::memovis::study2-participant} shows the participants' demographic background of the Study 2 (Section~\ref{sec::memovis::eval::eval_feedback}).
In a scale of $1$ to $5$, participants were instructed to self-rate their prior experience of 3D design in general, interior as well as car design.

\clearpage
\section{Study Tasks}\label{sec::memovis::app::study_tasks}
\begin{figure}[h]
    \centering
    \includegraphics[width=\textwidth]{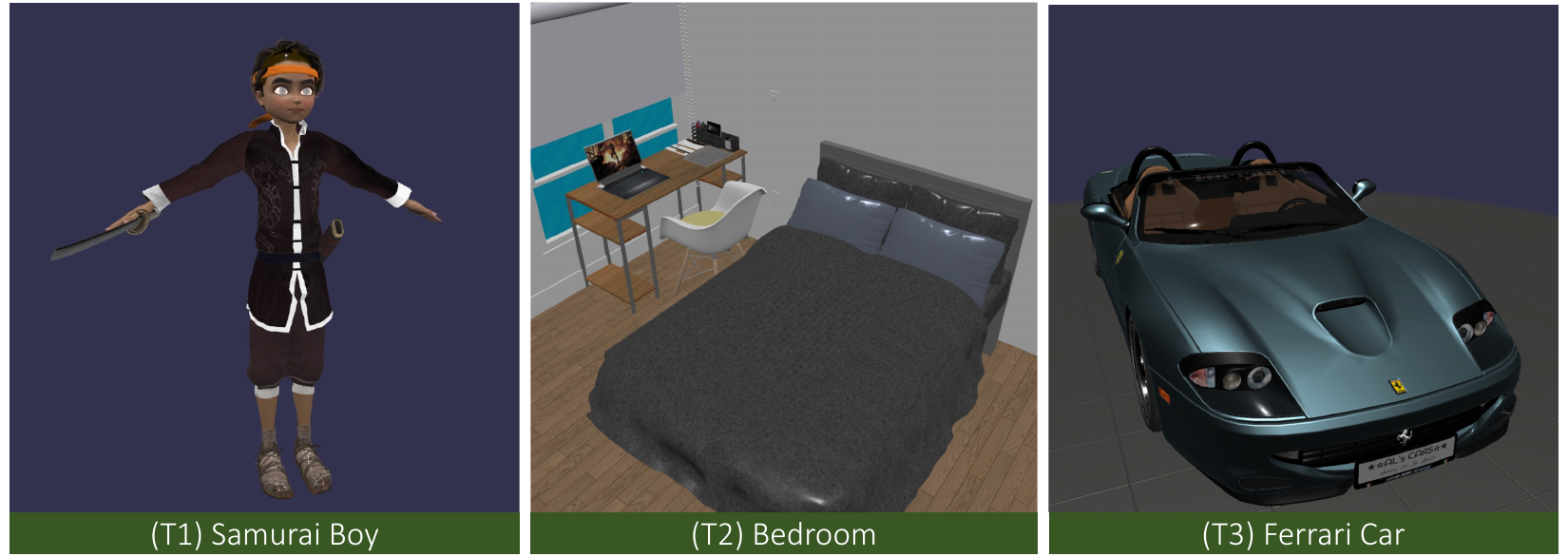}
    \caption{3D models used in the user study, including a samurai boy (for task T1), a bedroom (for task T2), and a car (for task T3). Notably, T1 was only used for training.}
    \label{fig::memovis::study-models}
\end{figure}

This section provides supplementary materials of the study tasks that were used in final evaluations (Section~\ref{sec::memovis::evaluation}).
Figure~\ref{fig::memovis::study-models} shows three models we used for feedback providers to review and create textual feedback comments with companion visual reference image(s).
With T1, participants were required to provide feedback for a samurai boy design (Figure~\ref{fig::memovis::study-models}a).
This tasks were used to help feedback provider participants to get familiar with both interface conditions.
All data collected from T1 was \emph{excluded} from the final analysis.
With T2, feedback provider participants were instructed to enhance the bedroom design in Figure~\ref{fig::memovis::study-models}b, such that the bedroom is comfortable to live in.
With T3, feedback provider participants were instructed to improve the design of a car in Figure~\ref{fig::memovis::study-models}c, such that the car is comfortable to drive with (in terms of aesthetics, ergonomics, and functionalities).
Full study details could be referred to Section~\ref{sec::memovis::evaluation}.

\section{Usages of the Image Modifiers}\label{sec::memovis::app::modifier_usages}
\begin{figure}[ht]
    \centering
    \includegraphics[width=\textwidth]{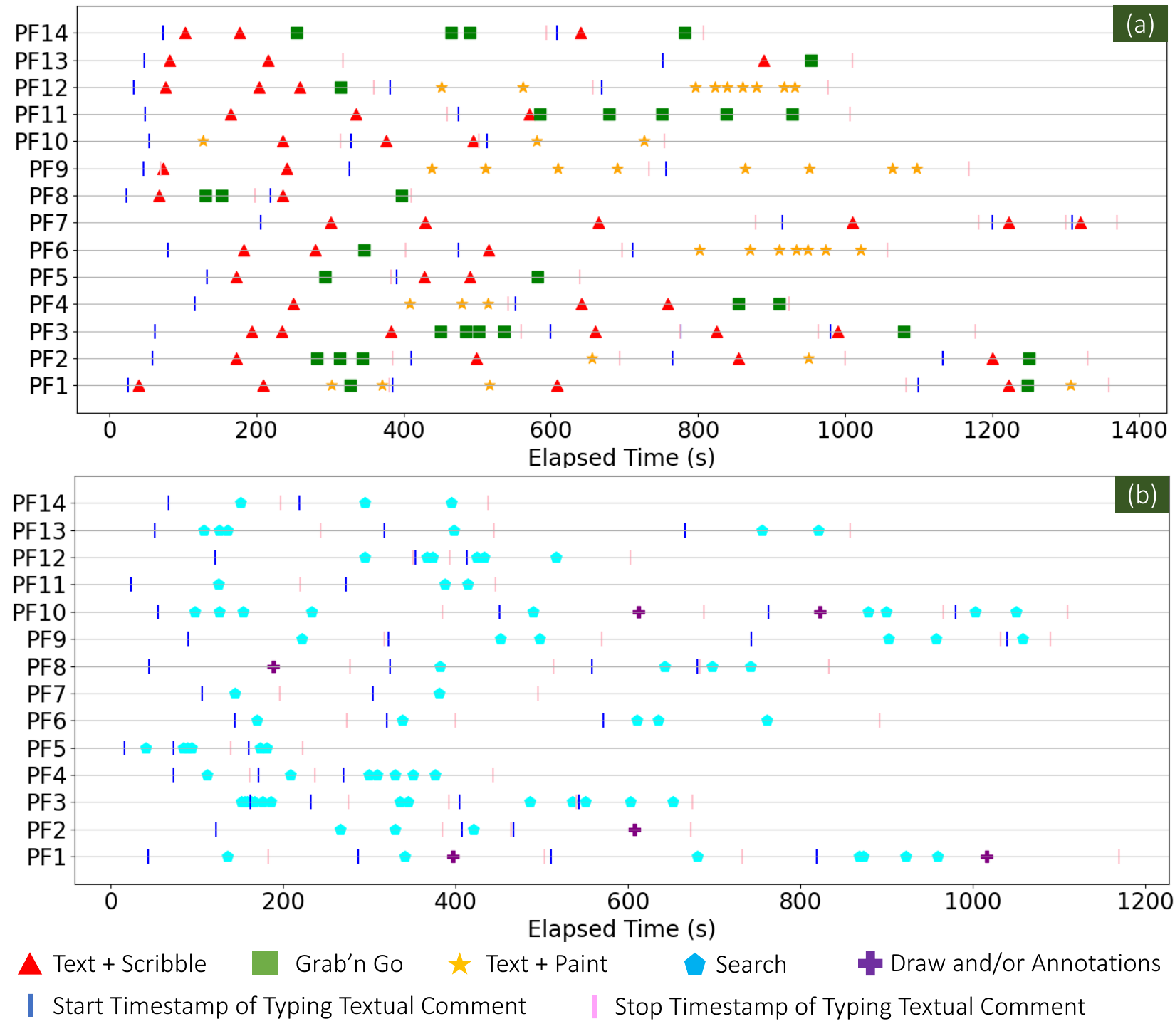}
    \caption{Visualizations of how each participants interact with one (or multiple) image modifiers, with MemoVis (a) and baseline~ (b) interface conditions.}
    \label{fig::memovis::study-1-modifier-usages}
\end{figure}

As a supplementary material for Section~\ref{sec::memovis::eval::create_feedback}, Figure~\ref{fig::memovis::study-1-modifier-usages} shows the visualizations of how each participants used the one (or multiple) of the image modifiers, provided by MemoVis.
For the baseline condition, we visualize the instants when the feedback providers use search, as well as draw and/or annotations, to create reference images.

\section{Codebook and Themes from Qualitative Data Analysis}\label{sec::memovis::app::codebook}
\begin{figure}[ht]
    \centering
    \includegraphics[width=0.65\textwidth]{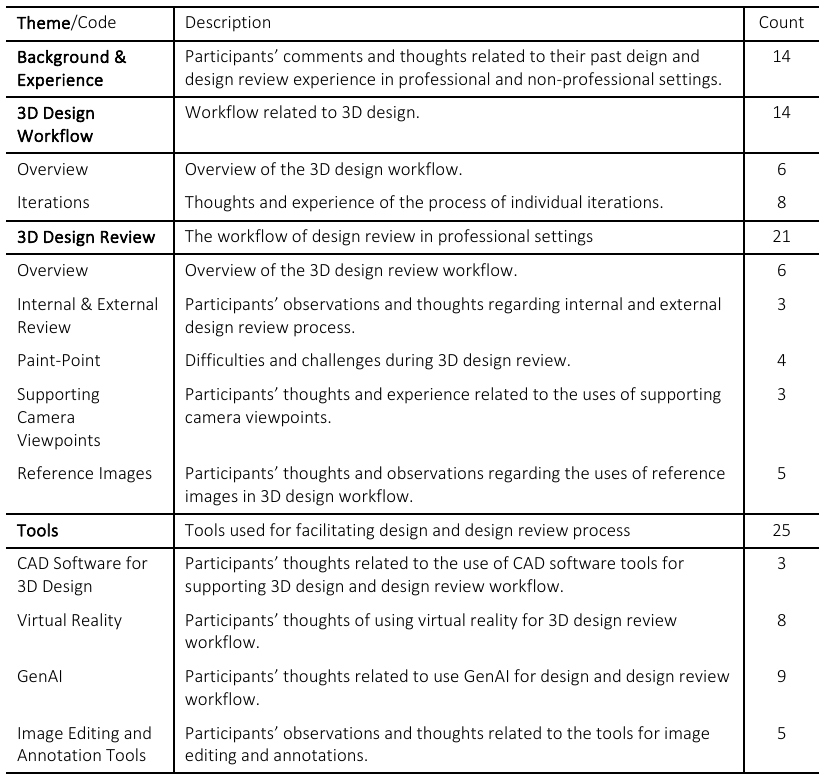}
    \caption{Themes and codes for Formative Study 1. Notably, it is possible that multiple codes are assigned to one quote.}
    \label{fig::memovis::codebook-online-dataa-nalysis-formative-study-1}
\end{figure}

\begin{figure}[t]
    \centering
    \includegraphics[width=0.65\textwidth]{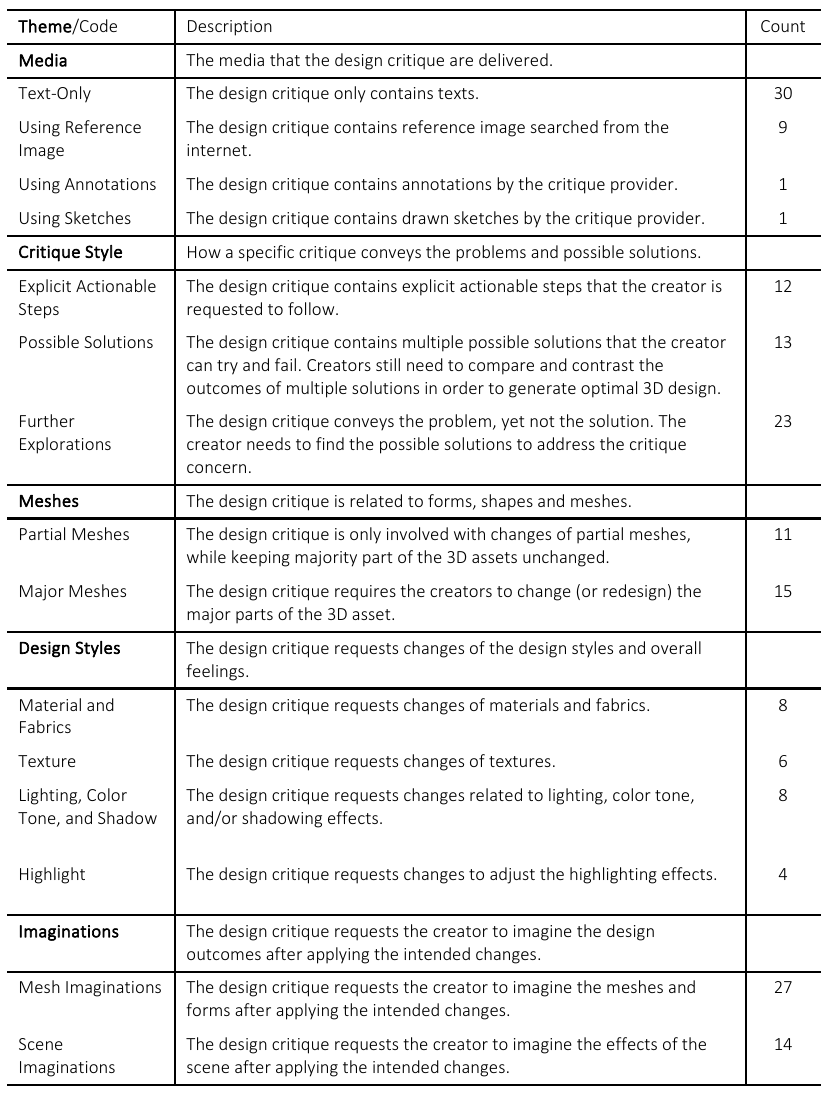}
    \caption{Themes and codes for analyzing real-world 3D design feedback data from Formative Study 2 . Notably, same feedback might be labeled by multiple codes.}
    \label{fig::memovis::codebook-online-data-analysis}
\end{figure}

\begin{figure}[t]
    \centering
    \includegraphics[width=\textwidth]{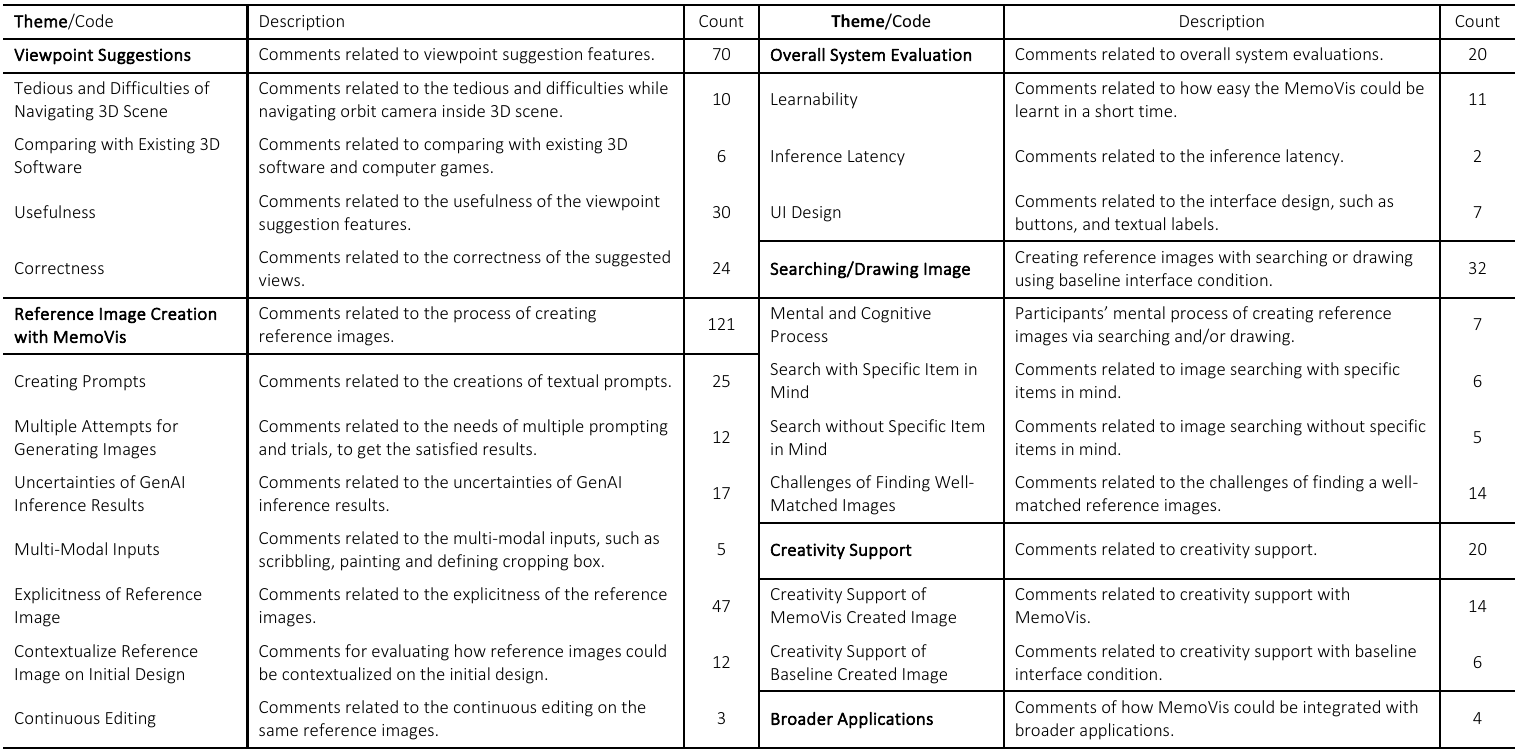}
    \caption{Themes and codes for Study 1. Notably, it is possible that multiple codes are assigned to one quote.}
    \label{fig::memovis::codebook-study-1}
\end{figure}

\begin{figure}[t]
    \centering
    \includegraphics[width=0.65\textwidth]{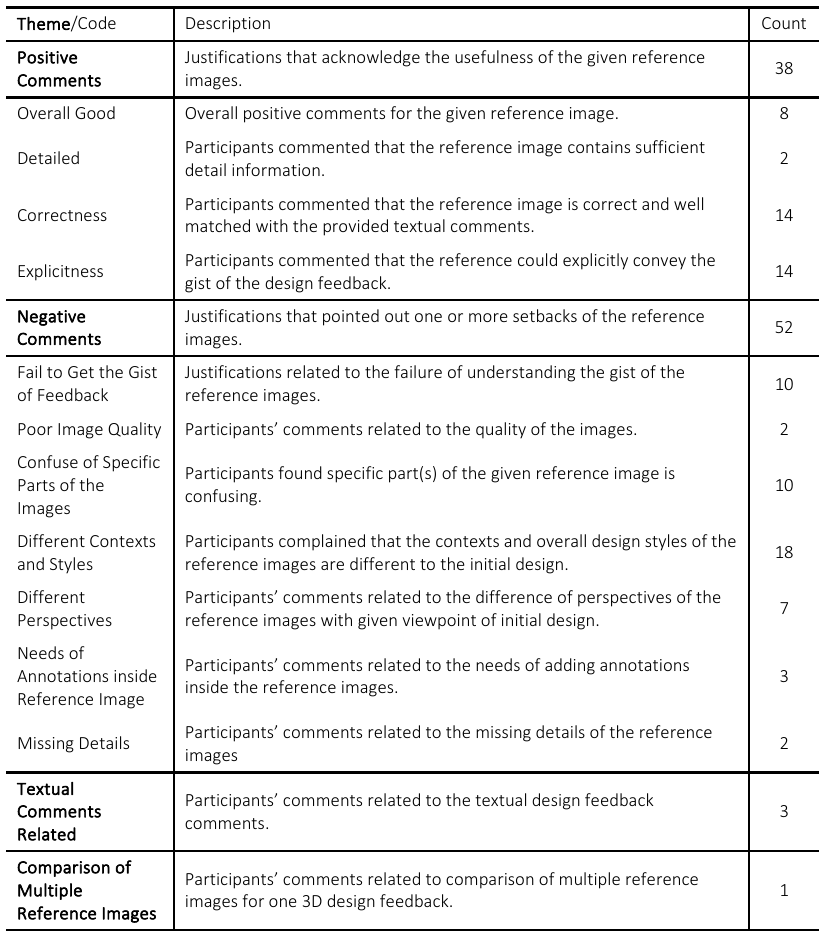}
    \caption{Themes and codes for participants' survey responses from Study 2. Notably, it is possible that multiple codes are assigned to one response.}
    \label{fig::memovis::codebook-study-2}
\end{figure}

As part of supplementary material, we attached the resultant codebook for qualitative analysis for Formative Study 1 (Figure~\ref{fig::memovis::codebook-online-dataa-nalysis-formative-study-1}), Formative Study 2 (Figure~\ref{fig::memovis::codebook-online-data-analysis}), and final user studies with feedback provider participants (Figure~\ref{fig::memovis::codebook-study-1}) and designer participants (Figure~\ref{fig::memovis::codebook-study-2}).
Notably, multiple codes might be assigned to each observation (\eg~participants' quote, 3D design feedback, and survey responses from Study 2).

\chapter{Supplementary Materials for Chapter ~\ref{sec::papertoplace}}

\section{Experimental Tasks}\label{sec::papertoplace::app::tasks}
The selected tasks for final user study (Section~\ref{sec::papertoplace::study}) include:

\begin{itemize}[leftmargin=*]

\item \textbf{(T1)} Microwave Scrambled Eggs~\cite{ScrambleEggs};

\item \textbf{(T2)} Quick Microwave-Poached Eggs on Avocado Toast~\cite{AvocadoToast};

\item \textbf{(T3)} Instant Mac `n' Cheese~\cite{MacNCheese}; 

\end{itemize}

Table~\ref{table::task1}, Table~\ref{table::task2} and Table~\ref{table::task3} offers the supplementary material regarding the specific instruction steps of the experimental tasks T1, T2, and T3 respectively.
Notably, T1 was used as the training tasks for participants to learn and get familiar with the interfaces (Table~\ref{table::task1}).
Assuming the results from OCR is fully correct (\ie~all texts of all instruction steps could be successfully extracted), the overall accuracies of the associated key objects predicted by our pre-trained language model for each experimental tasks are $50\%$~(T1), $84.62\%$~(T2), and $80.00\%$~(T3).
Notably, the overall accuracies for T2 and T3 are closed to our benchmark results while fine-tuning the BERT model in Section \ref{sec::papertoplace::authoring::edit}, which is $82.13\%$.
This ensures the results yielded by the evaluation of authoring pipeline (Section \ref{sec::papertoplace::study::authoring}) is generalizable to some extend.
Although the overall accuracy of T1 is far lower than our benchmark results due to the relative short of instruction document, the instruction of T1 was only used for participants to familiarize themselves with the given interfaces (either on iPad or inside MR), and the data yielded by T1 was excluded from our evaluation results in Section \ref{sec::papertoplace::study}.

\begin{table}
    \small
    \centering
    \caption{Experimental cooking recipe for making basic microwave scrambled eggs (T1).}
    \begin{tabular}{l} 
     \hline
     \textbf{Step Descriptions} \\
     \hline\hline 

     \makecell[l]{$\bullet$~Spray microwave-safe container (\eg~mug, ramekin, or egg cooker) \\ with cooking spray or wipe lightly with vegetable oil.} \\ 
    
     \makecell[l]{$\bullet$~Whisk eggs, milk, salt and pepper in container (or whisk ingredients \\ in another bowl and pour into microwave container). If using a mug \\ or ramekin, cover with plastic wrap, pulling back small area for \\ venting. If using an egg cooker, place lid on cooker base, lining up \\ notches. Twist to secure.}  \\
     
     \makecell[l]{$\bullet$~Microwave on Medium-High ($70$\% power) for 90 seconds, stirring \\ several times during cooking.} \\
     
     \makecell[l]{$\bullet$~Cover and let stand for 30 seconds to 1 minute before serving. Eggs \\ will look slightly moist, but will finish cooking upon standing.} \\
     
     \hline
    \end{tabular}

    \label{table::task1}
\end{table}

\begin{table}
    \small
    \centering
    \caption{Experimental cooking recipe for making an avocado toast with a microwave-poached egg (T2).}
    \begin{tabular}{l} 
     \hline
     \textbf{Step Descriptions} \\
     \hline\hline 

     \makecell[l]{$\bullet$~In a small bowl, combine the basil leaves and sea salt, and set aside.} \\ 
    
     \makecell[l]{$\bullet$~Take a tomato from the fridge.}  \\
     
     \makecell[l]{$\bullet$~And gently clean a tomato with your hand to help remove dirt and \\ bacteria. Do not use detergent, soap, or bleach.} \\
     
     \makecell[l]{$\bullet$~Then cut the tomato into parallel thin slices working from the top \\ of the tomato towards to the bottom and set aside.} \\
     
     \makecell[l]{$\bullet$~Take an avocado from the fridge.} \\
     
     \makecell[l]{$\bullet$~Cut and mash the avocado in a small bowl. Squeeze the lemon into \\ the avocado and spread liberally on the toast. And put tomato on top \\ of the toast.} \\
     
     \makecell[l]{$\bullet$~Pour $1/2$ cup ($118$~ml) water into a microwave-safe coffee mug. } \\
     
     \makecell[l]{$\bullet$~Crack 1 egg into the mug, cover with a small plate, and microwave on \\ high for 30 seconds. } \\
     
     \makecell[l]{$\bullet$~Take the mug out of the microwave, lift the plate carefully (to let \\ steam escape) and check the egg. } \\
     
     \makecell[l]{$\bullet$~If the white is not completely set, cover and continue to microwave in \\ $10$-second intervals until the egg white is opaque. (The time varies with \\ the power of the microwave and may take up to 60 seconds).} \\
     
     \makecell[l]{$\bullet$~Carefully pour off the water in the mug, using a slotted spoon to keep \\ the egg from falling out. } \\
     
     \makecell[l]{$\bullet$~Transfer the egg to one of the slices of avocado toast. } \\
     
     \makecell[l]{$\bullet$~Sprinkle the toasts with the seed mixture and serve immediately.} \\
     
     \hline
    \end{tabular}

    \label{table::task2}
\end{table}

\begin{table}
    \small
    \centering
    \caption{Experimental cooking recipe for making microwaved mac `n' cheese (T3).}
    \begin{tabular}{l} 
     \hline
     \textbf{Step Descriptions} \\
     \hline\hline 

     \makecell[l]{$\bullet$~Find a mug that holds twice the volume of your dry pasta – the \\ bigger, the better.} \\ 
    
     \makecell[l]{$\bullet$~Add the macaroni.}  \\
     
     \makecell[l]{$\bullet$~Add some water.} \\
     
     \makecell[l]{$\bullet$~Cover with cling film and pierce $3$ times.} \\
     
     \makecell[l]{$\bullet$~Stand the mug in a microwave-proof bowl to catch any spillages, and \\ cook in the microwave on high for $2$ minutes. The liquid will bubble \\ up and over the sides, so tip any liquid from the bowl back into \\ the mug (be careful as it will be very hot) and give it a good stir.  } \\
     
     \makecell[l]{$\bullet$~Leave to stand for $1$ minute.} \\
     
     \makecell[l]{$\bullet$~Repeat twice more or until the pasta is cooked (it may take longer \\ depending on the pasta).} \\
     
     \makecell[l]{$\bullet$~Then remove from the microwave. } \\
     
     \makecell[l]{$\bullet$~Repeat twice more or until the pasta is cooked (it may take longer \\ depending on the pasta).} \\
     
     \makecell[l]{$\bullet$~Stir through the butter, cheese and spinach or Marmite, if using. } \\
     
     \makecell[l]{$\bullet$~The heat from the pasta should melt the cheese and wilt the spinach, \\ but if not, pop back in the microwave for $30$ seconds.} \\
     
     \hline
    \end{tabular}

    \label{table::task3}
\end{table}

\section{User Study Results}\label{sec::papertoplace::app::study_results}
\begin{figure}
    \centering
    \includegraphics[width=0.90\textwidth]{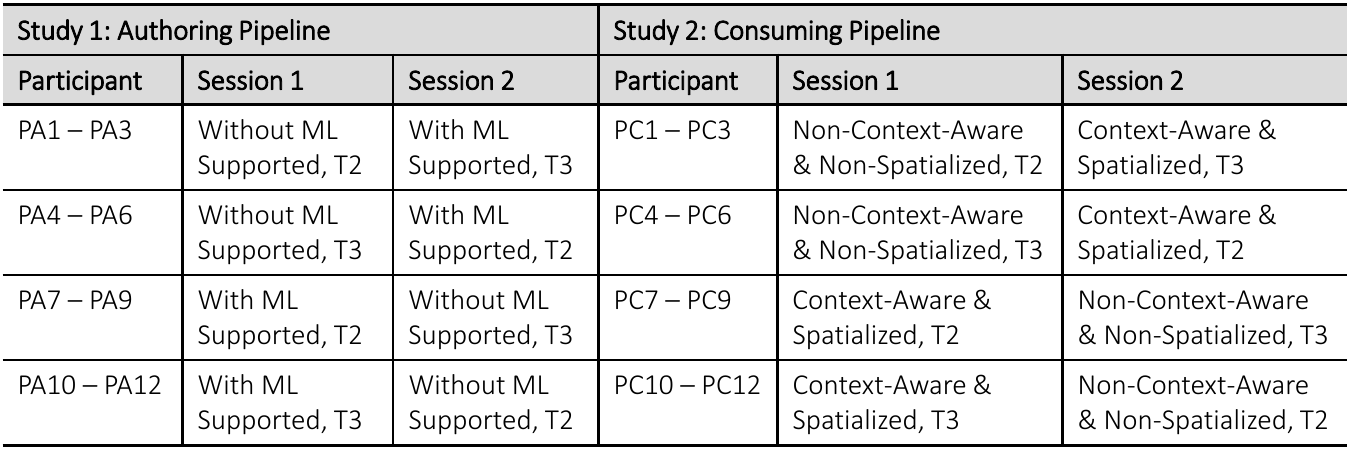}
    \caption[Study design for evaluating PaperToPlace.]{Study design for evaluating PaperToPlace. Each participant needs to conduct session 1 and session 2 in order. T2 and T3 were used for formal evaluation while T1 was used for training purposes.}
    \label{fig::papertoplace::study_design}
\end{figure}

This section presents supplementary material for Section \ref{sec::papertoplace::study}.
Figure~\ref{fig::papertoplace::study_design} shows the specific tasks and interface conditions that were assigned while evaluating authoring and consumption pipeline.
Notably, T1 was used for training purposes.

We also provide visualizations of survey responses for authoring pipeline evaluations (Section \ref{sec::papertoplace::app::study_results::authoring}) as well as consumption evaluations (Section \ref{sec::papertoplace::app::study_results::consuming}).

\begin{figure}
    \centering
    \includegraphics[width=0.52\textwidth]{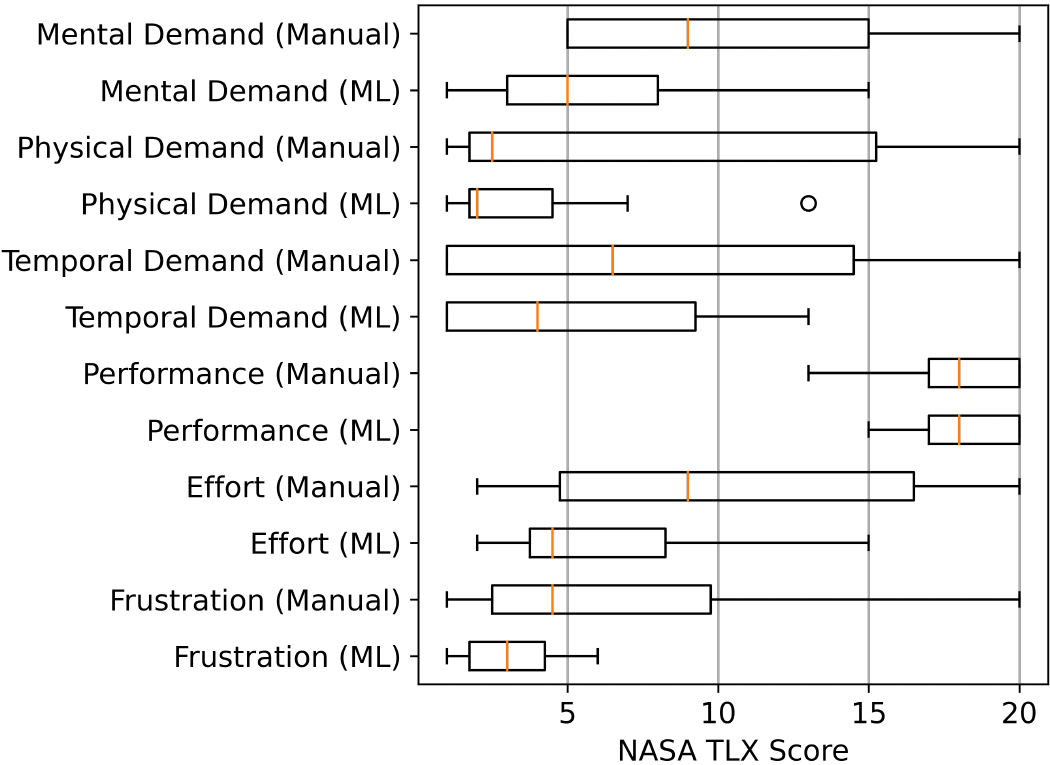}
    \caption[Survey results from NASA TLX questionnaires collected during the instruction authoring workflow..]{Survey results from NASA TLX questionnaires collected during the instruction authoring workflow. We use ``Manual'' and "ML" to indicate the interface condition while manual and ML supported approaches are used for extracting associated key objects from each instruction steps, respectively}
    \label{fig::papertoplace::appenx::authoring_results::tlx}
\end{figure}

\begin{figure}
    \centering
    \includegraphics[width=0.52\textwidth]{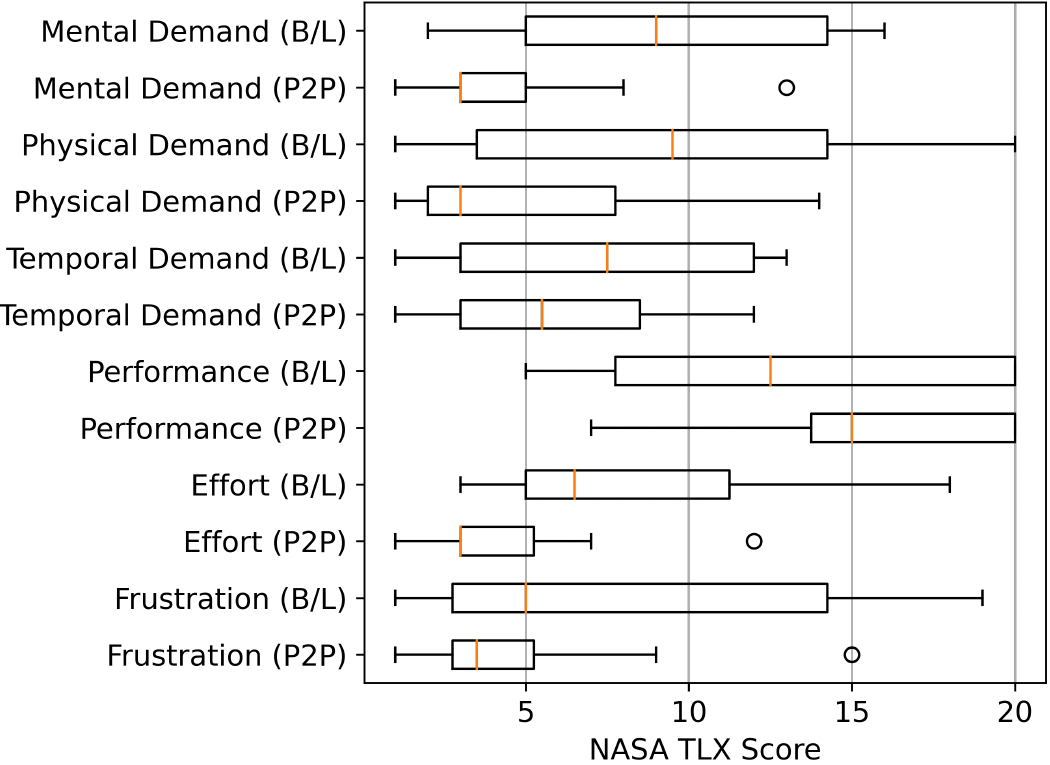}
    \caption[Survey results from NASA TLX questionnaires collected during the instruction consuming workflow. ]{Survey results from NASA TLX questionnaires collected during the instruction consuming workflow. We use ``B/L'' to refer to the baseline interface, and ``P2P'' to indicate PaperToPlace, which delivers spatialized and context-aware instruction step.}
    \label{fig::papertoplace::appenx::consumption_results::tlx}
\end{figure}

\begin{figure}
    \centering
    \includegraphics[width=\textwidth]{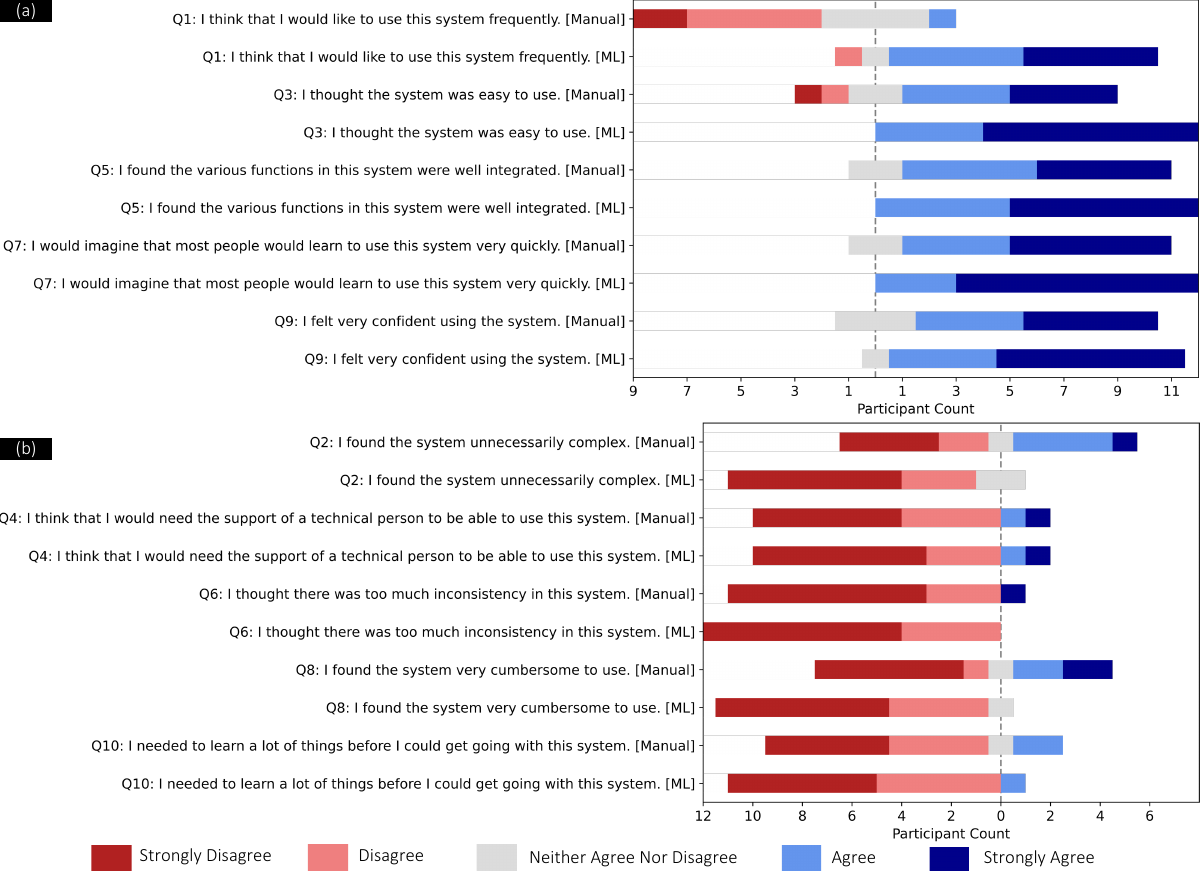}
        \caption[Survey results of SUS questionnaires of authoring pipeline evaluations.]{Survey results of SUS questionnaires of authoring pipeline evaluations. We use ``Manual'' and ``ML'' to indicate the interface condition while manual and ML support approaches are used for extracting associated key objects from each instruction steps, respectively. To increase readability, we cluster the survey results of positive statements~(Q1, Q3, Q5, Q7, Q9) into subplot~(a), where a higher level of agreement indicates a better user experience. The survey results of negative statements~(Q2, Q4, Q6, Q8, Q10) are clustered into subplot~(b), where a lower level of agreement indicates a better user experience.}
    \label{fig::papertoplace::appenx::authoring_results::sus}
\end{figure}

\begin{figure}
    \centering
    \includegraphics[width=\textwidth]{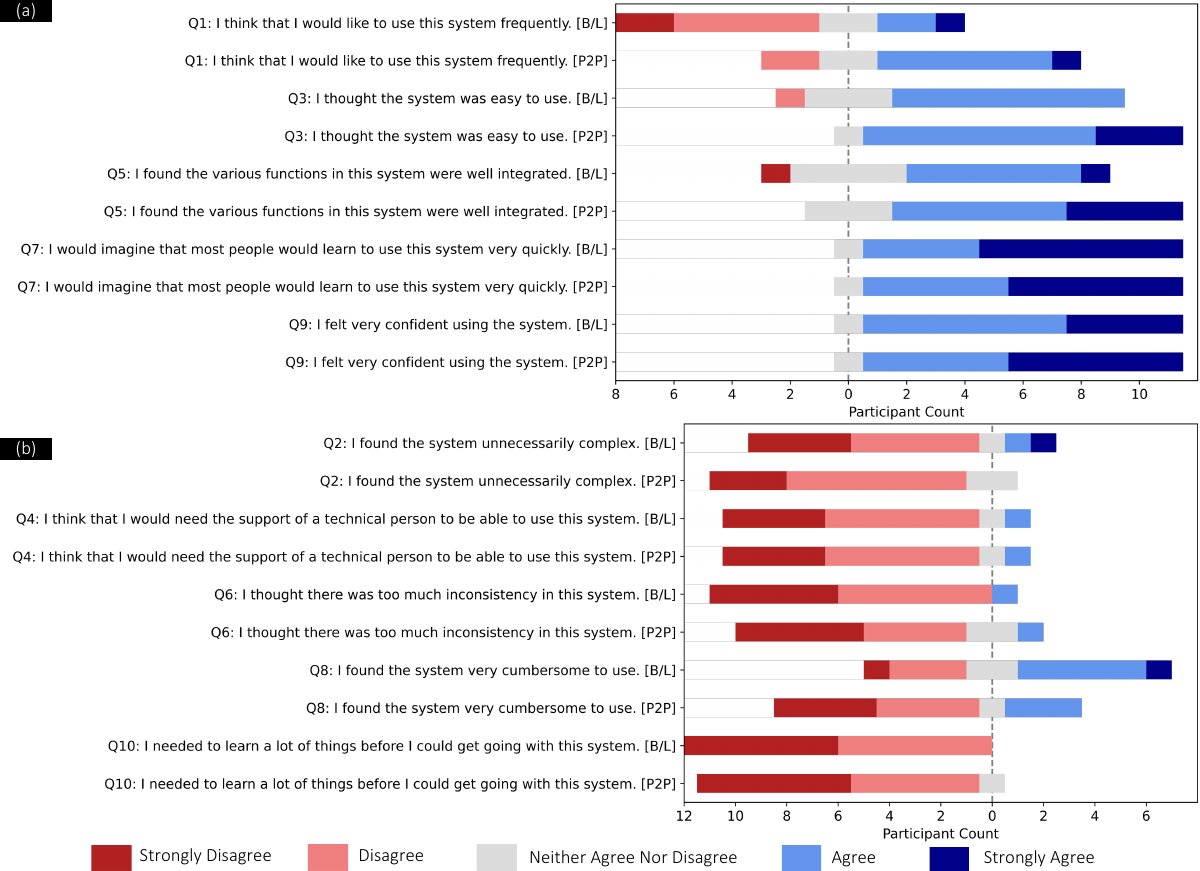}
    \caption[Survey results of SUS questionnaires of consumption pipeline evaluations.]{Survey results of SUS questionnaires of consumption pipeline evaluations. We use ``B/L'' to refer to the baseline interface, and ``P2P'' to indicate PaperToPlace, which delivers spatialized and context-aware instruction step. To increase readability, we cluster the survey results of positive statements~(Q1, Q3, Q5, Q7, Q9) into subplot~(a), where a higher level of agreement indicates a better user experience. The survey results of negative statements~(Q2, Q4, Q6, Q8, Q10) are clustered into subplot~(b), where a lower level of agreement indicates a better user experience.}
    \label{fig::papertoplace::appenx::consumption_results::sus}
\end{figure}

\begin{figure}
    \centering
    \includegraphics[width=\textwidth]{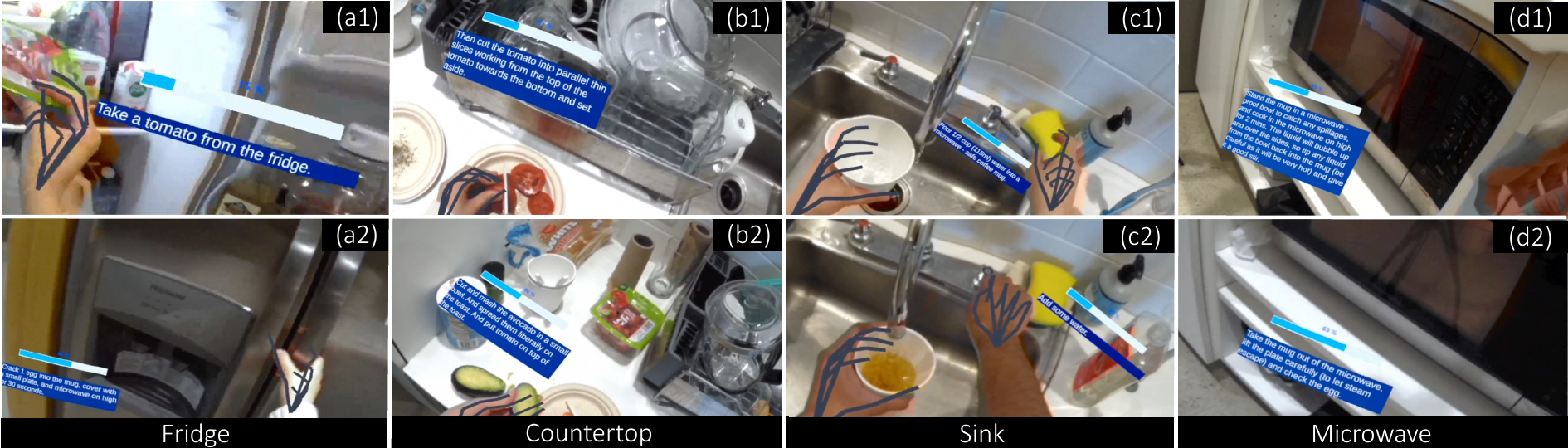}
    \caption[First-person view through MR of the examples of placing instruction step next to the key objects with our consumption pipeline.]{First-person view through MR of the examples of placing instruction step next to the key objects with our consumption pipeline. Example key objects include fridge (a), countertop (b), sink (c), and microwave (d).}
    \label{fig::papertoplace::consumption::examples}
\end{figure}

\subsection{Evaluations of Authoring Pipeline}\label{sec::papertoplace::app::study_results::authoring}
Figure~\ref{fig::papertoplace::appenx::authoring_results::tlx} demonstrates the NASA TLX responses of each perceived workloads from participants PA1 - PA12, while using manual and ML-assisted interfaces to extract document profile from designated paper instruction.
Figure~\ref{fig::papertoplace::appenx::authoring_results::sus} provides supplementary material of survey results of SUS questionnaires.
For Figure~\ref{fig::papertoplace::appenx::authoring_results::tlx} and Figure~\ref{fig::papertoplace::appenx::authoring_results::sus}, we use ``Manual'' and ``ML'' to indicate the interface condition while manual and ML supported approaches are used while extracting the associated key objects from each instruction steps, respectively.

\subsection{Evaluations of Consumption Pipeline}\label{sec::papertoplace::app::study_results::consuming}
Figure~\ref{fig::papertoplace::appenx::consumption_results::tlx} demonstrates the NASA TLX responses of each perceived workloads from participants PC1 - PC12, while using baseline and PaperToPlace interfaces to perform the designated tasks.
Figure~\ref{fig::papertoplace::appenx::consumption_results::sus} provides supplementary material of survey results of SUS questionnaires.
To be consistent with the remaining of this paper, we use ``B/L'' to refer to the baseline interface, and ``P2P'' to indicate PaperToPlace, which delivers spatialized and context-aware instruction step.
Finally, Figure~\ref{fig::papertoplace::consumption::examples}e - h showed examples of how the virtual instructions steps would be anchored on the key objects with the consumption pipeline of PaperToPlace, which are easy to read and would not occlude the consumer's sight while completing the tasks.


\section{Codebook and Themes from Qualitative Data Analysis}\label{sec::papertoplace::app::codebook}
\begin{figure*}[t]
    \centering
    \includegraphics[width=\textwidth]{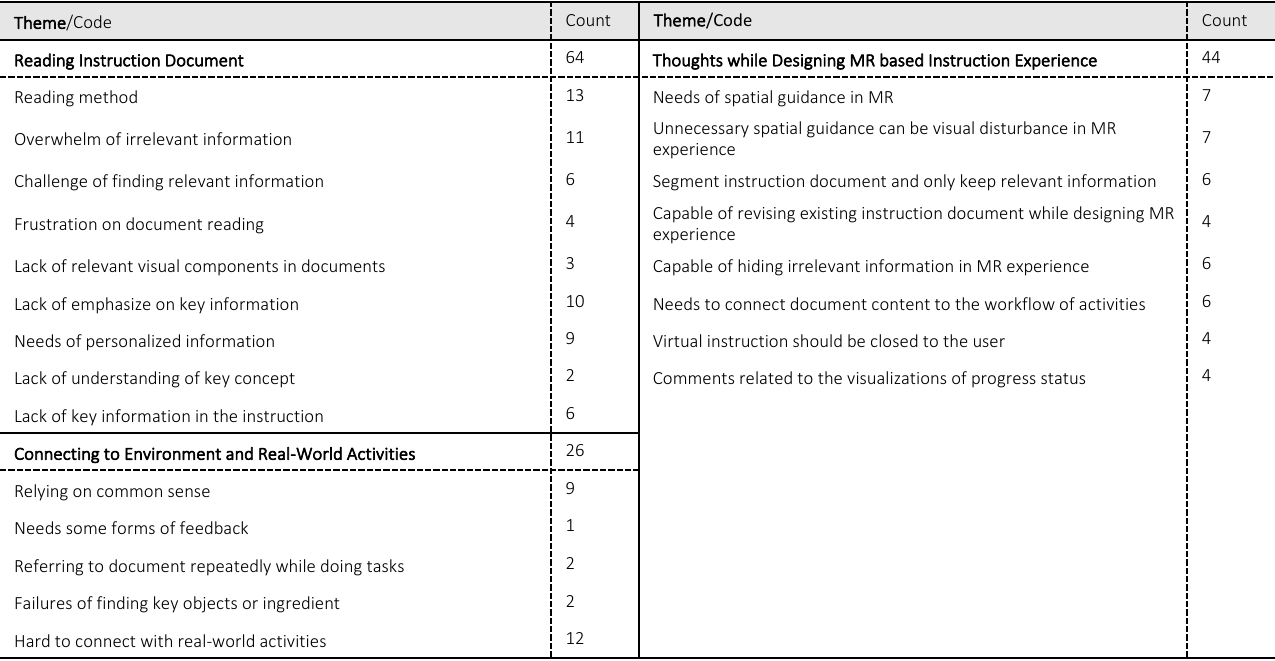}
    \caption[The codebook that resulted from our qualitative analysis of interview data for preliminary needs-finding study.]{The codebook that resulted from our qualitative analysis of interview data for preliminary needs-finding study. ``Count'' refers to the number of quote for each theme or code. It is possible that multiple codes are assigned to one quote.}
    \label{fig::papertoplace::appenx::codebook::prelim}
\end{figure*}

\begin{figure*}[t]
    \centering
    \includegraphics[width=\textwidth]{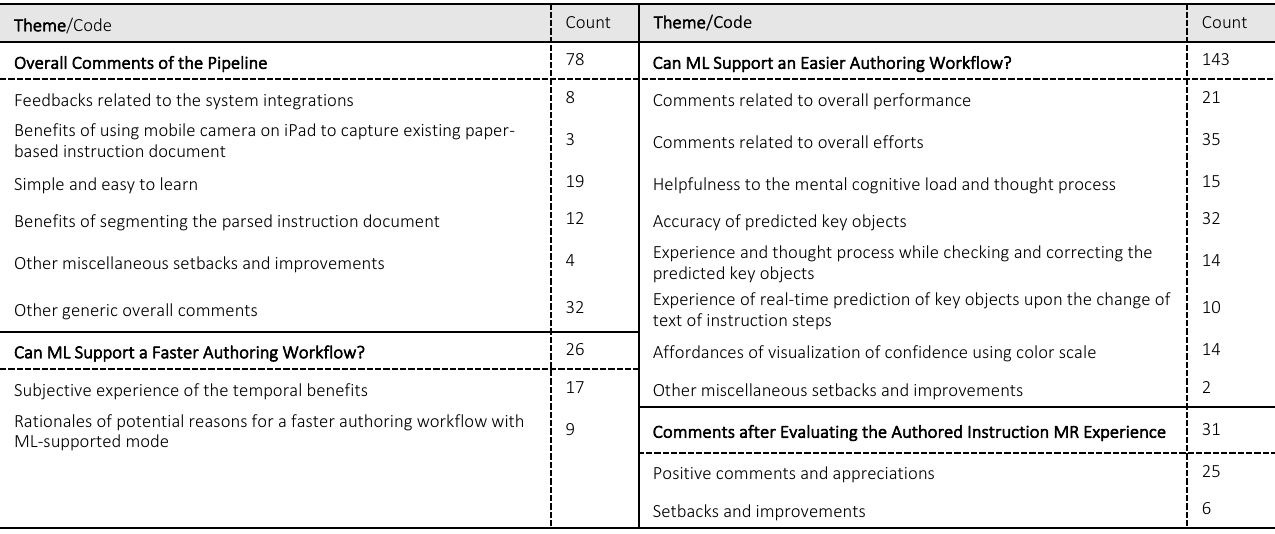}
    \caption[The codebook that resulted from our qualitative analysis of interview data for authoring pipeline evaluations.]{The codebook that resulted from our qualitative analysis of interview data for authoring pipeline evaluations. ``Count'' refers to the number of quote for each theme or code. It is possible that multiple codes are assigned to one quote.}
    \label{fig::papertoplace::appenx::codebook::authoring}
\end{figure*}

\begin{figure*}[t]
    \centering
    \includegraphics[width=\textwidth]{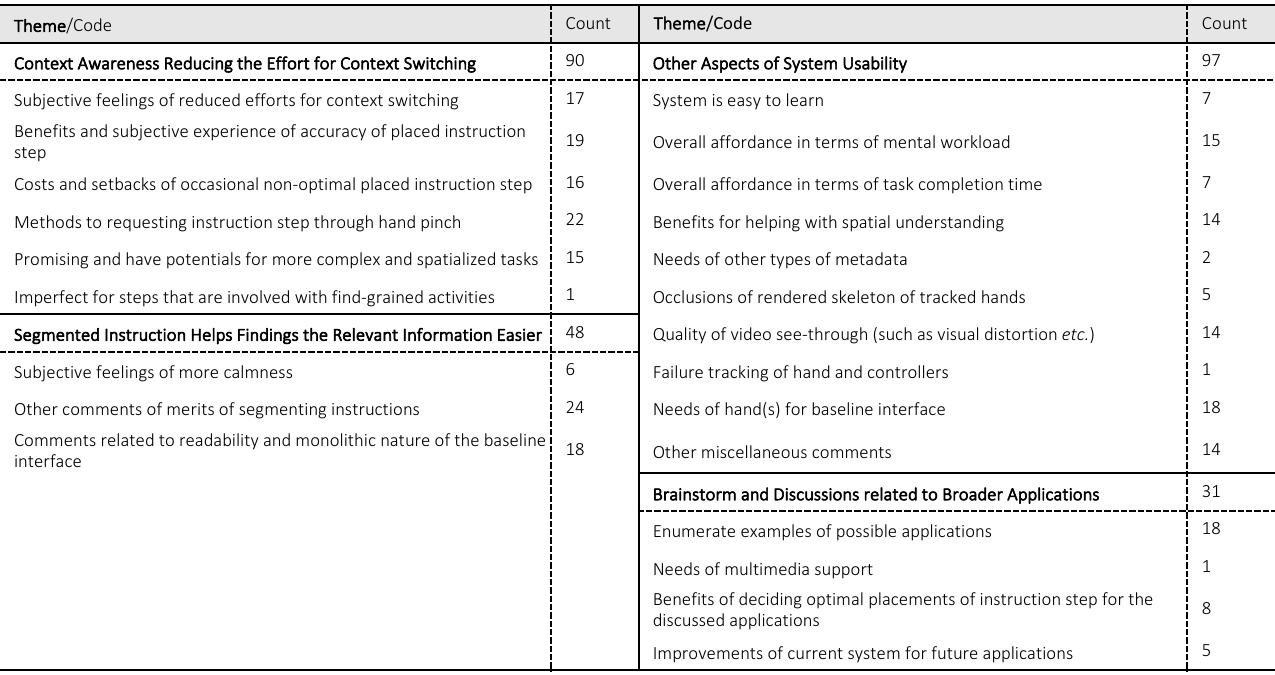}
    \caption[The codebook that resulted from our qualitative analysis of interview data for consumption pipeline evaluations.]{The codebook that resulted from our qualitative analysis of interview data for consumption pipeline evaluations. ``Count'' refers to the number of quote for each theme or code. It is possible that multiple codes are assigned to one quote.}
    \label{fig::papertoplace::appenx::codebook::consumption}
\end{figure*}

We used thematic analysis~\cite{Braun2012}, and deductive and inductive coding~\cite{Elo2008} to analyze qualitative data, collected from preliminary needs-finding study (Section \ref{sec::papertoplace::prelim}) and final user study (Section \ref{sec::papertoplace::study}).
As part of supplementary material, we attached the resultant codebook in Figure~\ref{fig::papertoplace::appenx::codebook::prelim},  Figure~\ref{fig::papertoplace::appenx::codebook::authoring} and Figure~\ref{fig::papertoplace::appenx::codebook::consumption}, respectively.
Notably, ``Count'' refers to the number of quote for each theme or code.
It is also possible that multiple codes are assigned to one quote.

\chapter{Glossory of Terms}\label{sec::glossory}

This section describes the list of key abbreviations and acronyms used throughout this dissertation.
This list of glossary serves as a reference and checklist to assist readers in understanding the terminology discussed within the text.
The following list of key glossary of terms are ordered alphabetically.

\begin{itemize}[leftmargin=*,noitemsep,nolistsep]

    \item {\bf AR}: \textbf{A}ugmented \textbf{R}eality

    \item {\bf ART}: \textbf{A}ligned \textbf{R}ank \textbf{T}ransform

    \item {\bf BERT}: \textbf{B}idirectional \textbf{E}ncoder \textbf{R}epresentations from \textbf{T}ransformers
    
    \item {\bf BLIP}: \textbf{B}ootstrapping \textbf{L}anguage-\textbf{I}mage \textbf{P}re-training Model

    \item {\bf CAVE}: \textbf{C}ave \textbf{A}utomatic \textbf{V}irtual \textbf{E}nvironment

    \item {\bf CLIP}: \textbf{C}ontrastive \textbf{L}anguage-\textbf{I}mage \textbf{P}re-training Model

    \item {\bf CT}: \textbf{C}omputer \textbf{T}omography Imaging

    \item {\bf DC}: \textbf{D}esign \textbf{C}onsiderations

    \item {\bf DoF}: \textbf{D}egree \textbf{o}f \textbf{F}reedom

    \item {\bf DFM Tools}: \textbf{D}esign \textbf{F}or \textbf{M}anufacturability Tools

    \item {\bf DICOM}: \textbf{D}igital \textbf{I}maging and \textbf{CO}mmunications in \textbf{M}edicine

    \item {\bf DSC}: \textbf{D}ice \textbf{S}imilarity \textbf{C}oefficient

    \item {\bf DVR}: \textbf{D}irect \textbf{V}olume \textbf{R}endering
    
    \item {\bf GAN}: \textbf{G}enerative \textbf{A}dversarial \textbf{N}etwork
    
    \item {\bf GenAI}: \textbf{G}enerative \textbf{AI}

    \item {\bf HMD}: \textbf{H}ead-\textbf{M}ounted \textbf{D}isplay

    \item {\bf IM}: \textbf{I}nstant \textbf{M}essaging

    \item {\bf IRB}: \textbf{I}nstitutional \textbf{R}eview \textbf{B}oards

    \item {\bf MIP}: \textbf{M}aximum \textbf{I}ntensity \textbf{P}rojection~(MIP)

    \item {\bf ML}: \textbf{M}achine \textbf{L}earning
    
    \item {\bf MR}: \textbf{M}ixed \textbf{R}eality

    \item {\bf MRI}: \textbf{M}agnetic \textbf{R}easonance \textbf{I}maging

    \item {\bf PET}: \textbf{P}ositron \textbf{E}mission \textbf{T}omography Imaging

    \item {\bf PII}: \textbf{P}ersonal \textbf{I}dentifiable \textbf{I}nformation

    \item {\bf RM-ANOVA}: \textbf{R}epeated \textbf{M}easures \textbf{A}nalysis of \textbf{V}ariance

    \item {\bf RV Continuum}: \textbf{R}eality-\textbf{V}irtuality Continuum

    \item {\bf RQ}: \textbf{R}esearch \textbf{Q}uestion
    
    \item {\bf RT Treatment Planning}: \textbf{R}adiation \textbf{T}herapy (\aka~Radiotherapy) Treatment Planning

    \item {\bf SAM}: \textbf{S}egment \textbf{A}nything \textbf{M}odel

    \item {\bf SOTA}: \textbf{S}tate-\textbf{O}f-\textbf{T}he-\textbf{A}rt

    \item {\bf SUS}: \textbf{S}ystem \textbf{U}sability \textbf{S}cale

    \item {\bf TCT}: \textbf{T}ask \textbf{C}ompletion \textbf{T}ime

    \item {\bf UI}: \textbf{U}ser \textbf{I}nterface
    
    \item {\bf UX}: \textbf{U}ser \textbf{E}xperience

    \item {\bf VE}: \textbf{V}irtual \textbf{E}nvironment

    \item {\bf VLFM}: \textbf{V}ision-\textbf{L}anguage \textbf{F}oundation \textbf{M}odel

    \item {\bf VR}: \textbf{V}irtual \textbf{R}eality
    
    \item {\bf XR}: e\textbf{X}tended \textbf{R}eality

\end{itemize}

\backmatter
\bibliographystyle{ACM-Reference-Format}
\bibliography{references}

\end{document}